\documentclass[phd, titlesmallcaps, examinerscopy, copyrightpage, foronline]{mqthesis}

\usepackage{rotating}  
\usepackage{theorem}   
\usepackage{bm}  
\usepackage{amsmath,amsfonts,amssymb}
\usepackage{booktabs}  
\usepackage{pdfsync}   
\usepackage{epstopdf}
\usepackage{color}
\ifpdf
    \pdfcatalog { /PageMode (/UseOutlines) /OpenAction (fitbh) }
\fi

\begin{document}

\frontmatter

\title{Control and characterization of nano-structures with the symmetries of light}

\ifthenelse{\boolean{foronline}}{
  \author{\href{mailto:xavier.zambranapuyalto@mq.edu.au}{Xavier Zambrana-Puyalto}}
  \department{Physics \& Astronomy}
}{
  \author{Xavier Zambrana-Puyalto}
  \department{Physics \& Astronomy}
}

 \submitdate{September 2014}


\titlepage

\chapter{Acknowledgements}
Our lives are such a confluence of random decisions. All that surrounds us influences us in a way or another. Nonetheless, it looks like there is something inherent in ourselves that lets us kind of control the relevance that we assign to any external input. That is, we might be somehow programmed to let some of those inputs influence us, and dismiss the others. A stupid example: I will never understand why I find fruit so repulsive. I am not allergic to it. It is healthy, cheap, and filling. My family loves it and I was grown up in a region where good fruit is produced. I have been surrounded by fruit since I was a baby. Yet here I am, a 28 years old guy who does not eat fruit. Nonsense. This fact leads me to believe that indeed there might be something there in my genes which has made me avoid my beloved fruits for 28 year. However, I would not doubt for even one second that the fact that I am writing these words now is a mere coincidence. Not to mention what hopefully you are going to read or skim through in the next 200 pages. That, I believe, is one the most serious convolution of random decisions ever made. I was definitely not programmed to find out about Mie resonances, multipolar fields, optical vortices, Spatial Light Modulators and circular dichroism. I still remember that in one of those moments of enlightenment, I once told my good friend Pere Pl\`{a} Junc\`{a} that I was certain about something: I wasn't ever going to do a full-time PhD. The reason: every time I went to uni on a Sunday, I would see those PhD students from the Construction Department. Those guys seemed to be there forever. Moreover, their supervisors were this bunch of cranky, full-of-themselves civil engineers who used to grace us with their divine presence in our undergrad courses from time to time. To me, it looked like their job was not really teaching us, but rather recalling how little we knew and how much students used to know in the old good times. Young as a I was, I used to think that PhD students did their PhDs at the same university as they carried out their undergrad studies. And I was not going to do that in my Saint Civil Engineering Faculty, for the aforementioned reasons. A crucial event that made me consider the possibility of doing a PhD happened while I did an exchange year in the south of Spain, specifically in Granada. I was lucky to share a course on Statistical Mechanics with a very cheerful and friendly crowd. Lu\'{i}s, Javi, Juan, Paco, Bilel, Virginia and Salva. They let me know about a great webpage called TIPTOP Physics\footnote{\textit{Requiescat in pacem}.}, where lots of PhD offers from all over the world used to be posted. I had always fancied the idea of living in a different country, and that seemed a good excuse/opportunity to go on an adventure. Then, I also found out that PhD students travel quite often to conferences and stuff. So my evil plan was starting to get some shape: I was going to go to a different country, live there, and travel around the globe for free thanks to this PhD business. I just needed to find a friendly supervisor who would not constantly whip my back. When I met Gabriel via TIPTOP Physics, I knew I had found one. I also still remember one of the first things that he told me when we first met: `I'll have to pay you something so that you can work here for some weeks. It will be nothing, it'll just buy you a couple of drinks on the weekends'. As you can see, I am not that hard to convince. Jokes apart, Gabriel, I am really grateful that you offered me the possibility to come to Australia and work with you. You have truly been a great mentor and I will always remember my time spent in your group with a fond big smile. \textit{Muchas gracias, de verdad!} The yellow bricks road to my PhD Thesis would have been completely different if I had not shared it with Ivan. Ivan started his PhD with Gabriel one month before me and it has been a pleasure to share this journey with him. You have been a great friend; I have learnt tons from you and I have enjoyed all our conversations about politics, economy, and physics. Not sure if we will ever see the Nature dynasty fall, but I will certainly campaign for some open public arXiv-type publication system, where unfair referee comments such as `I do not believe this, it must be an artefact' or `This is in Jackson' are not given by anonymous peers. \textit{Moltes gr\`{a}cies tio!}. Then, due to Gabriel's ability of getting all the grants that he applies for, new people have kept joining in and contributing to my development as a researcher. In particular, I warmly thank Xavi for all the hours spent with me in the lab. Not only that, but also for his virtue of letting me do things at my own pace, even though that was detrimental for his interests at some points. If it was not for you, my lab table would probably still be as empty as it was when you arrived. \textit{Moltes gr\`{a}cies Xavi, ha estat un plaer treballar amb tu!} I would like to thank Mathieu as well. Even though we were assigned different projects, you have always had the time to help me out in the lab or explain me things in the office. \textit{Muchas gracias Mathieu!}. The rest of members of Gabriel's group have been easy-going and friendly, so I thank you all guys: Nora, Alex, Alexander, Eugene, Rich and Andrea. And finally, I cannot leave off of this paragraph my office mates from the quantum group, with whom I have shared hours of fun, culture and physics. Together we have all learnt from each other either talking at lunch, attending PhD seminars, or going to houseparties: Aharon, Johann, Thorn, Mike, Lauri, Mauro (you are The Dude), Tommaso, Sukhi, Hossein, Keith, Ayenni.  \\ \\
Now, going backwards a little bit, it is impossible for me to track all those people whose actions, comments, ideas or thoughts influenced me to study a Bachelors in Physics, which 8 year later resulted in the writing of this thesis. However, I think that it is my duty to list some of them. First of all, I want to sincerely thank my parents. They triggered my hunger to learn and fed it with knowledge in the early days. As I grew up, they gave me everything, and invested a lot of time in me so that I could receive a good education in all sort of aspects. They never refused to pay more bills so that I could study more. Certainly, I would not be here in Sydney if they had not allowed me to continue my studies in Physics once I was an engineer and was ready to work. For all this and more, \textit{moltes gr\`{a}cies mama i papa, us estimo.} Secondly, I must thanks Prof. Francisco Marqu\'{e}s for being a truly inspirational lecturer and a perfect supervisor. Although I had always liked Physics, I do not think that I would have ever decided to study a degree in Physics if it was not for his lectures in my first year of Civil Engineering. Next, I want to thank my family and friends to make my life easier and fulfil it with joy, good moments and laughs. Even though I have been living far away from you guys, I always have you in mind. Also, I want to thank my favourite Canadian, Kayla. Thanks for loving me and making me a more responsible and thoughtful person. Thanks for teaching me new recipes and correct me when I make up words. You are great, and I love you. Last but not least, I want to thank all those anonymous people in my country who fought and worked so that a child born in an average family like me could have a quality free education. And at the same time, I want to condemn those who are attacking this system and whose decisions are already resulting in an enlargement of the current existing outrageous gap between the rich and the poor. All newborns should have the same opportunities that I had.

\chapter{List of Publications}

\begin{itemize}
\item[$\bullet$] \textbf{[A]} R. Bowman, N. Muller, \textbf{X. Zambrana-Puyalto}, O. Jedrkiewicz, P. Di Trapani, and M.J. Padgett. \emph{Efficient generation of Bessel beam arrays by means of an SLM}. Eur. Phys. J. Special Topics \textbf{199}, 159-166 (2011).
\item[$\bullet$] \textbf{[B]} I. Fernandez-Corbaton, \textbf{X. Zambrana-Puyalto}, and G. Molina-Terriza. \emph{Helicity and angular momentum: A symmetry-based framework for the study of light-matter interactions}. Phys. Rev. A \textbf{86}, 042103-14 (2012).
\item[$\bullet$] \textbf{[C]} \textbf{X. Zambrana-Puyalto}, X. Vidal, and G. Molina-Terriza. \emph{Excitation of single multipolar modes with engineered cylindrically symmetric fields}. Opt. Express \textbf{20}, 24536--24544 (2012).
\item[$\bullet$] \textbf{[D]} N. Tischler, \textbf{X. Zambrana-Puyalto}, and G. Molina-Terriza. \emph{The role of angular momentum in the construction of electromagnetic multipolar fields}. Eur. J. Phys \textbf{33}, 1099 (2012).
\item[$\bullet$] \textbf{[E]} \textbf{X. Zambrana-Puyalto}, and G. Molina-Terriza. \emph{The role of the angular momentum of light in Mie scattering. Excitation of dielectric spheres with Laguerre-Gaussian modes}. J. Quant. Spectrosc. Radiat. Transfer \textbf{126}, 50-55 (2013).
\item[$\bullet$] \textbf{[F]} \textbf{X. Zambrana-Puyalto}, I. Fernandez-Corbaton, M.L. Juan, X. Vidal, and G. Molina-Terriza. \emph{Duality symmetry and Kerker conditions}. Opt. Lett. \textbf{38}, 1857-1859 (2013).
\item[$\bullet$] \textbf{[G]} \textbf{X. Zambrana-Puyalto}, X. Vidal, M.L. Juan, and G. Molina-Terriza. \emph{Dual and anti-dual modes in dielectric spheres}. Opt. Express \textbf{38}, 1857-1859 (2013).
\item[$\bullet$] \textbf{[H]} I. Fernandez-Corbaton, \textbf{X. Zambrana-Puyalto}, N. Tischler, X. Vidal, M.L. Juan, and G. Molina-Terriza. \emph{Electromagnetic Duality Symmetry and Helicity Conservation for the Macroscopic Maxwell's Equations}. Phys. Rev. Lett. \textbf{111}, 060401 (2013).
\item[$\bullet$] \textbf{[N]} R. Neo, S.J. Tan, \textbf{X. Zambrana-Puyalto}, S. Leon-Saval, J. Bland-Hawthorn, and G. Molina-Terriza. \emph{Correcting vortex splitting in higher order vortex beams}. Opt. Express \textbf{22}, 9920-9931 (2014).
\item[$\bullet$] \textbf{[I]} N. Tischler, I. Fernandez-Corbaton, \textbf{X. Zambrana-Puyalto}, A. Minovich, X. Vidal, M.L. Juan, and G. Molina-Terriza. \emph{Experimental control of optical helicity in nanophotonics}. Light: Sci. Appl. \textbf{3}, e183 (2014).
\item[$\bullet$] \textbf{[J]} \textbf{X. Zambrana-Puyalto}. \emph{Probing the nano-scale with the symmetries of light}. J. Proc. R. Soc. New South Wales \textbf{147}, 55-63 (2014)
\item[$\bullet$] \textbf{[K]} I. Fernandez-Corbaton, \textbf{X. Zambrana-Puyalto}, and G. Molina-Terriza. \emph{On the transformations generated by the electromagnetic spin and orbital angular momentum operators}. J. Opt. Soc. Am. B \textbf{31}, 2136-2141 (2014).
\item[$\bullet$] \textbf{[L]} \textbf{X. Zambrana-Puyalto}, X. Vidal, and G. Molina-Terriza. \emph{Angular momentum-induced circular dichroism in non-chiral nanostructures}. Nat. Commun. \textbf{5}, 4922 (2014).
\item[$\bullet$] \textbf{[M]} N. Tischler, M.L. Juan, I. Fernandez-Corbaton, \textbf{X. Zambrana-Puyalto}, X. Vidal,  and G. Molina-Terriza. \emph{Topologically robust optical position sensing}. (submitted)

\end{itemize}

\chapter{Statement of contribution}
This thesis contains material that has been published, accepted, submitted or prepared for publication, as follows:\\\\
\large \textbf{Introduction}\\
\normalsize Some of the contents of Introduction have been published in \textbf{[J]}. My contributions to \textbf{[J]} have been the writing, calculations and simulations. Some of the ideas in \textbf{[J]} were discussed with Gabriel Molina-Terriza.\\\\
\large \textbf{Chapter 2}\\
\normalsize The contents of Chapter \ref{Ch1} have not been published as a whole. Nevertheless, some parts of it have been extracted from an initial version of the Appendices of \textbf{[B]}, which I had entirely written. Some others have been used in \textbf{[J]}. \\\\
\large \textbf{Chapter 3}\\
\normalsize The contents of Chapter \ref{Ch2} have been partially extracted from \textbf{[E]}. My contributions to \textbf{[E]} are: 
All the simulations and most of the analytical calculations. Some of the analytical calculations were done by Gabriel Molina-Terriza. The initial idea came jointly from Gabriel Molina-Terriza and I. I did all the writing, with feedback from Gabriel Molina-Terriza. \\\\
\linebreak
\large \textbf{Chapter 4}\\
\normalsize Most of the results contained in Chapter \ref{Ch3} have been published in \textbf{[C]} and \textbf{[E]}. My contributions to \textbf{[C]} are: All the analytical calculations and simulations. The initial idea came jointly from Gabriel Molina-Terriza and I. Gabriel Molina-Terriza and I did the writing, and Xavier Vidal gave us feedback.\\\\
\large \textbf{Chapter 5}\\
\normalsize Most of the results presented in Chapter \ref{Ch4} have been published in \textbf{[F]} and \textbf{[G]}. My contributions to these two articles are the following ones. In \textbf{[F]}, Ivan Fernandez-Corbaton gave the symmetry arguments regarding K1 and Xavier Vidal made the figure. Then, I had the idea of the paper and carried out all the analytical discussion. I also wrote the paper, with feedback from all the co-authors. In \textbf{[G]}, the very original concept came from a discussion with Mathieu L. Juan, Xavier Vidal, and myself. Then, I did all the calculations, simulations, and analysis. I also wrote the paper, and all the co-authors gave me feedback.\\\\
\large \textbf{Chapter 6}\\
\normalsize The material presented in Chapter \ref{Ch5} has not been published anywhere, except for the characterisation of the circular nano-apertures, which has been used in \textbf{[L]}. The SEM images of the circular nano-apertures and the dark-field images of the spherical particles were done by Xavier Vidal. Xavier Vidal also wrote the code to analyse the images and retrieve their size.\\\\
\large \textbf{Chapter 7}\\
\normalsize The material presented in Chapter \ref{Ch6} has been mostly published in \textbf{[L]}. In \textbf{[L]}, Xavier Vidal and I set up the experiment and collected the data. Gabriel Molina-Terriza had the main idea of the paper. I analysed the data and wrote the paper, and both co-authors gave me feedback.\\\\
\linebreak
\large \textbf{Chapter 8}\\
\normalsize The material presented in Chapter \ref{Ch7} has not been published anywhere yet. Xavier Vidal and I set up the experiment, and then I collected all the data and analysed it. 

\chapter{Abstract}

Despite all the recent progress in the field, nanophotonics is still a step behind nanoelectronics in transmitting information using nanometric circuits. A lot of effort is being put into making very elaborate structures that can guide light and control light-matter interactions at the nano-scale. The field of plasmonics has been especially successful in this. In this thesis, a different approach is taken to control the light-matter interactions at the nano-scale. The approach is based on considering light and sample as a whole system and exploiting its symmetries. Thanks to this new perspective, new phenomena have been unveiled. These new phenomena have been developed theoretically and/or experimentally and are scattered across this thesis. \\
In chapter \ref{Ch1}, the theoretical grounds of this thesis are settled. Even though every physicist is familiar with the concept of symmetry, a formalism to systematically describe the symmetries of electromagnetic fields is explained. With this formalism, some well-known symmetry considerations can be as easily retrieved as some much less intuitive. For example, it can be demonstrated that a linearly polarised Bessel beam is not cylindrically symmetric; whilst a circularly polarised Bessel beam is both cylindrically and dual symmetric. Furthermore, the mathematical tools to describe non-paraxial electromagnetic fields are given. Due to the fact that this work deals with sub-wavelength scatterers, the light-matter interaction cannot usually be described within the paraxial approximation. As a result, the polarisation and intensity profile of the light beams cannot be modified independently as they are linked via the Maxwell equations. \\
Chapters \ref{Ch2}, \ref{Ch3} and \ref{Ch4} deepen in the study of Generalized Lorenz-Mie Theory. Using the formalism developed in chapter \ref{Ch1}, various new effects are discovered. In chapter \ref{Ch3}, the excitation of WGM modes on micron-sized spheres is described. Indeed, using cylindrically symmetric beams, light can be coupled into spherical resonators without the use of evanescent coupling. Furthermore, it is shown that the use of cylindrically symmetric modes also allows for the enhancement of the ripple structure in scattering. Finally, chapter \ref{Ch4} generalizes the Kerker conditions and uses cylindrically and dual symmetric beams to control the helicity content in scattering. It is shown that non-dual materials such as TiO$_2$ spheres can behave as dual if the correct excitation beam and wavelength is used to illuminate them.\\
Chapters \ref{Ch5}, \ref{Ch6} and \ref{Ch7} are devoted to experiments. In chapter \ref{Ch5}, a description of the experimental techniques used in chapters \ref{Ch6} and \ref{Ch7} is carried out. In particular, the basics of Spatial Light Modulators and Computer Generated Holograms are given. Spatial Light modulators are used in chapters \ref{Ch6} and \ref{Ch7} to create vortex beams. In chapter \ref{Ch6}, the symmetries of these vortex beams turn out to be crucial to induce a giant circular dichroism in a non-chiral sample. Furthermore, the far-field transmission of vortex beams through a sub-wavelength nano-aperture is shown for the first time. Finally, chapter \ref{Ch7} presents the dependence of scattering measurements on the wavelength and the topological charge of the incident vortex beam. As predicted in chapter \ref{Ch3}, it is seen that some scattering resonances are hidden under a Gaussian beam excitation. These resonances can be unveiled when the illumination is a vortex beam. \\
Overall, this work shows a number of new effects (theoretical and/or experimental) produced by the excitation of symmetric structures with symmetric light. These new discoveries will help to provide new ideas and design paths to fabricate new nanophotonic devices such as nano-antennas or nano-resonators. A study of the symmetries of the system should always be kept in mind for any photonic device where the spatial degrees of freedom and the polarisation cannot be decoupled.

\tableofcontents
\chapter{List of Abbreviations}
\begin{list}{}{%
\setlength{\labelwidth}{30mm}
\setlength{\leftmargin}{35mm}}

\item[AM\dotfill] angular momentum
\item[BC\dotfill] boundary conditions
\item[BNS\dotfill] Boulder Non-Linear (name of a company)
\item[BS\dotfill] beam splitter
\item[CCD\dotfill] charge-coupled device
\item[CD\dotfill] circular dichroism
\item[CGH\dotfill] computed generated hologram
\item[EM\dotfill] electromagnetic
\item[FIB\dotfill] focused ion beam
\item[GLMT\dotfill] generalized Lorenz-Mie theory
\item[HWP\dotfill] half-wave plate
\item[K1\dotfill] first Kerker condition
\item[K2\dotfill] second Kerker condition
\item[LCP\dotfill] left circularly polarised
\item[LG\dotfill] Laguerre-Gaussian
\item[LP\dotfill] linear polariser
\item[LUT\dotfill] lookup table
\item[MDR\dotfill] morphological-dependent resonance
\item[MO\dotfill] microscope objective
\item[NA\dotfill] numerical aperture
\item[QWP\dotfill] quarter-wave plate
\item[RCP\dotfill] right circularly polarised
\item[RGB\dotfill] red green blue
\item[SEM\dotfill] scanning electron microscope
\item[SLM\dotfill] spatial light modulator
\item[STED\dotfill] stimulated emission depletion 
\item[TE\dotfill] transverse electric
\item[TM\dotfill] transverse magnetic
\item[WD\dotfill] working distance
\item[WGM\dotfill] whispering gallery mode
\end{list}
\mainmatter

\begin{savequote}[10cm] 
\sffamily
``Growing up in the place I did I never was aware of any other option but to question everything.'' 
\qauthor{Noam Chomsky}
\end{savequote}

\chapter{Introduction}
\graphicspath{{ch_intro/}} 
\label{Intro}

\newcommand\krho{k_{\rho}}
\newcommand\sphat{\mathbf{\hat{\sigma}_+}}
\newcommand\smhat{\mathbf{\hat{\sigma}_-}}

\begin{figure}[tbp]
\centering
\includegraphics[width=\columnwidth]{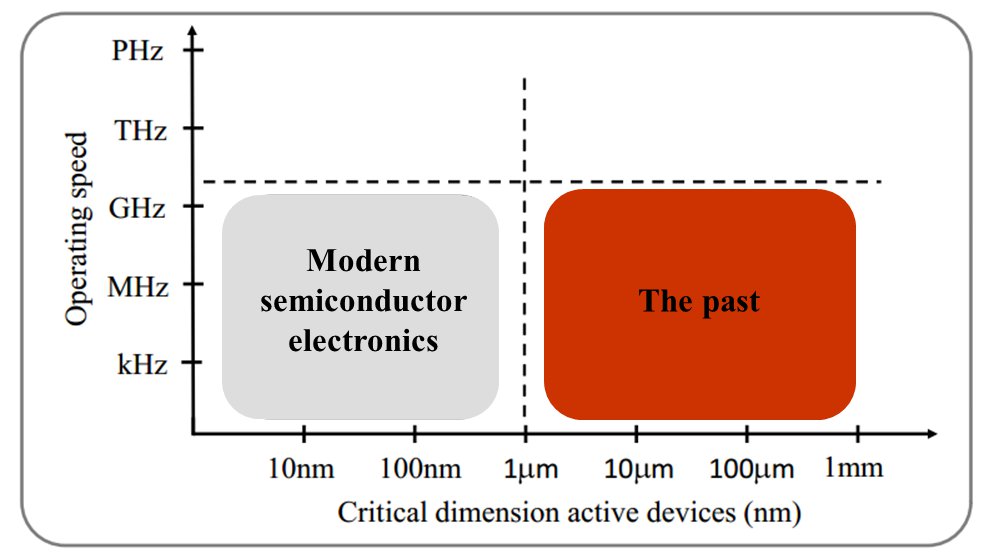} 
\caption{Transmission data rates of different technologies and critical dimension of their active devices. The past refers to the semiconductor electronics and vacuum tubes used before the 1980's, whilst modern semiconductor electronics is the technology used in nano-electronics.  \label{Fig11}}
\end{figure}

\begin{figure}[tbp]
\centering
\includegraphics[width=\columnwidth]{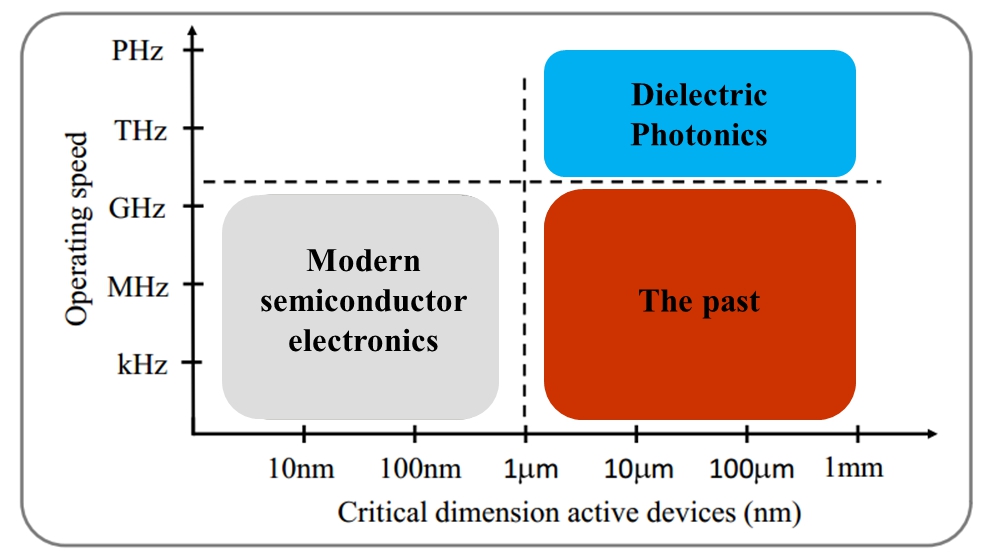} 
\caption{Transmission data rates of different technologies and critical dimension of their active devices. Dielectric photonics refers to fibre optics technologies. It can be seen that photonics is much faster than electronics.  \label{Fig12}}
\end{figure}

\begin{figure}[tbp]
\centering
\includegraphics[width=\columnwidth]{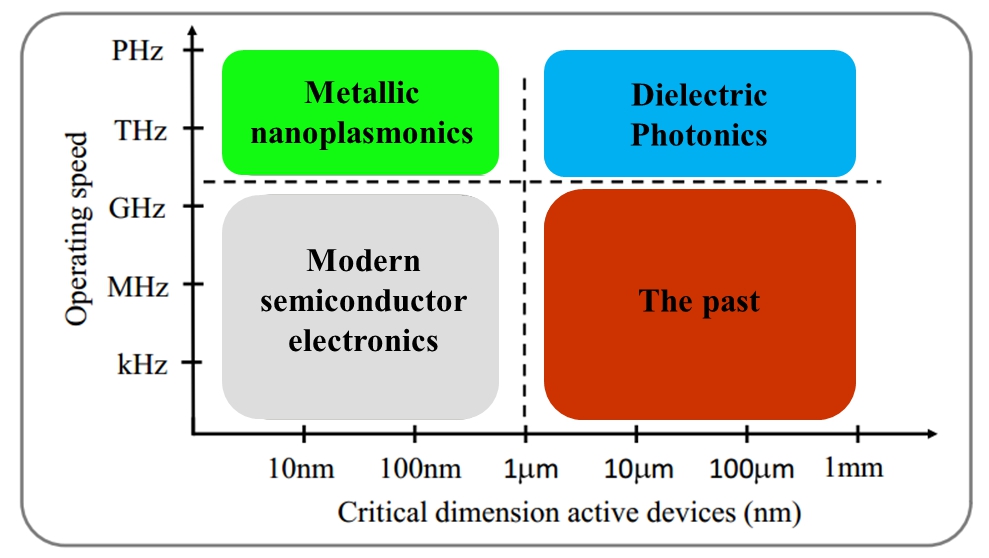} 
\caption{Transmission data rates of different technologies and critical dimension of their active devices. Metallic nano-plasmonics is expected to be the technology that will bring photonics to the nano-scale.  \label{Fig13}}
\end{figure}

\begin{figure}[tbp]
\centering
\includegraphics[width=14.5cm]{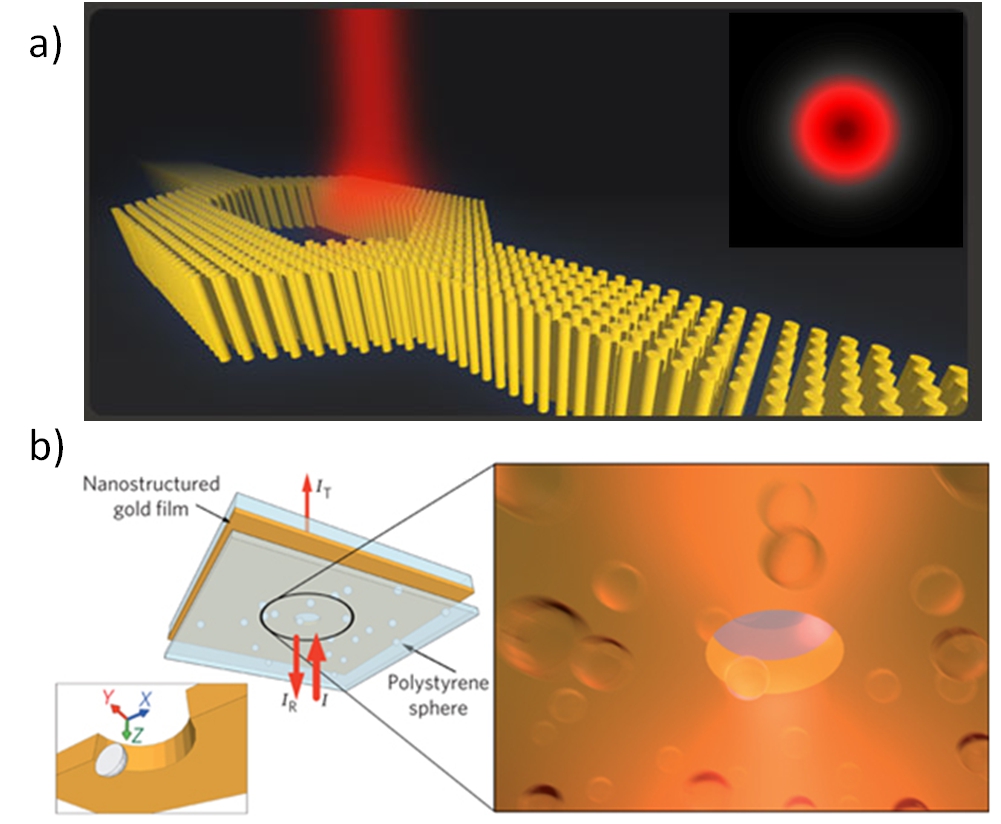} 
\caption{a) Plasmonic wave-guide made of metallic nano-rods. An incident Gaussian beam shone upon the structure is carried along the plasmonic nano-rods. The inset is an intensity plot of a transverse section of the excitation beam. The image has been taken from activeplasmonics.org with permission of Anatoly Zayats. Credits to Dr. Ryan McCarron, King's College, London. b) A metallic film with a nano-hole is excited with a Gaussian beam to excite a surface plasmon. The surface plasmon is used to trap a polystyrene particle. The image has been taken from \cite{Mathieu2009}, with permission of Mathieu L. Juan. \label{Fig14}}
\end{figure}

\begin{figure}[tbp]
\centering
\includegraphics[width=\columnwidth]{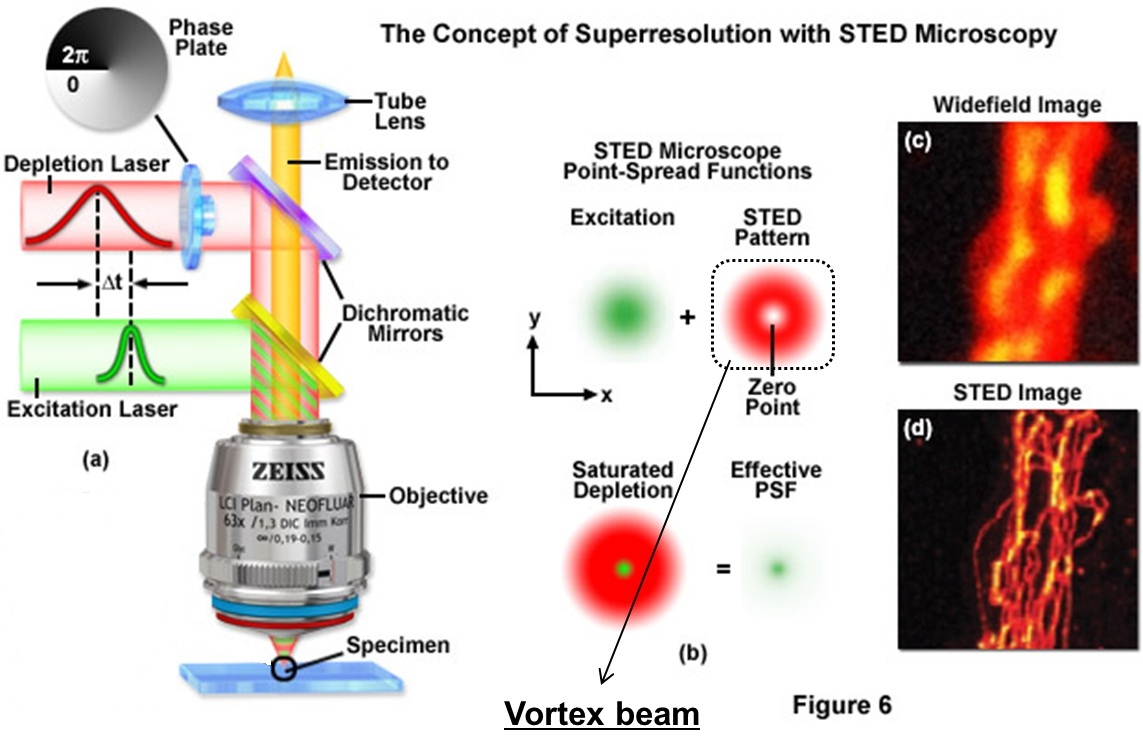} 
\caption{Schematics of STED microscopy. a) A depletion laser beam (red) is added to a typical microscopy set-up. b) The use of a vortex beam considerably reduces the Point Spread Function of the system. The Point Spread Function describes the response of an imaging system to a point source. The largest the point spread function, the blurrier the image will be. c) Standard optical microscopy image of a sub-wavelength specimen. d) STED image of the same sub-wavelength specimen. Image from www.zeiss.com/campus. Copyright: Mike Davidson, FSU, Tallahassee. \label{Fig15}}
\end{figure}

\begin{figure}[tbp]
\centering
\includegraphics[width=\columnwidth]{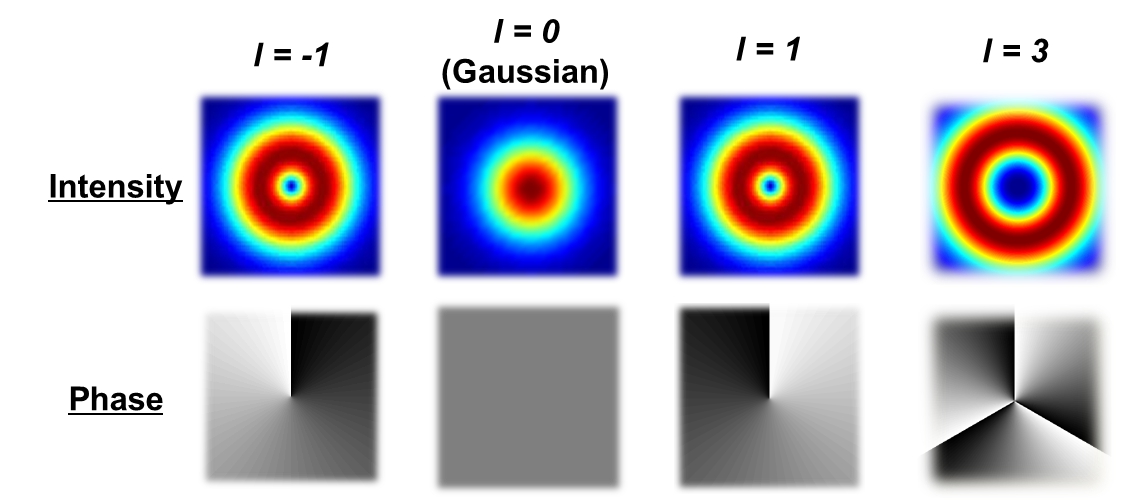} 
\caption{Intensity and phase profiles of vortex beams with topological charges $l=-1,0,1,3$. For the intensity plots, red indicates maximum and dark blue minimum. For the phase, white means $0$ phase, and black means $2\pi$. \label{Fig16}}
\end{figure}

Invented by the Sumerians in ancient Mesopotamia around 3000 BC, \cite{Schmandt1978,Glassner2003,Butler1998}, writing  may be considered one of the greatest human inventions of all-times. Since then, it has been used to store, transmit and manipulate information. Concurrently, the technology employed in writing has changed over the years. The development of computers, perhaps one of mankind's most significant technological revolutions in the last century, was intrinsically linked to writing. Despite the principles of modern computers having been described by Alan Turing in 1936 \cite{Turing1936}, the first programmable \textbf{electronic} computer was not built until 1946, at Pennsylvania University \cite{Braun1982}. In the following 6 years, John von Neumann and co-workers laid the ground work upon which modern computers rest to this day \cite{Burks1946,Goldstine1948,Bowden1953}. The first generation of commercial electronic computers was made available in the US in 1951, but the giant leap towards our current technologies occurred in 1955, when the so-called `second generation computers' reached the market \cite{Lavington1975}. Since that time, the evolution of computers has been linked to the advent of semiconductor electronics. Simply put, this is the technology associated with the control of how the electrons move in semiconductor materials\footnote{The development of semiconductor electronics was made possible thanks to the advances in quantum mechanics, and in particular in solid-state physics.}. The majority of our current information devices function because of our ability to control electron currents from one spot to another one, and re-interpret this current as chains of 0's and 1's, which are used to encode information. Since the 1970's, our means of re-interpreting electron currents as chains of 0's and 1's have remained fundamentally unchanged. However, the techniques and devices to move electrons around in a controlled manner have changed drastically. Of particular note, the size of the electronic devices has decreased by many orders of magnitude. This can be observed in Figure \ref{Fig11}.
`The past' includes the electronic circuits (using semiconductor electronics, or vacuum tubes) used up until the 1980's. From then onwards, the semiconductor manufacturing processes achieved sizes below 1$\mu$m – currently known as nano-electronics. In fact, nano-electronics has advanced at a very fast pace, consistently following Moore's Law \cite{Schaller1997}. At present, \hyperlink{http://www.intel.com/content/www/us/en/silicon-innovations/intel-22nm-technology.html}{Intel}\footnote{http://www.intel.com/content/www/us/en/silicon-innovations/intel-22nm-technology.html} is fabricating transistors at 22nm, and current projections are that the 11nm technology will be obtained by 2015. \\\\
The fact that electronics has been so successful over the past 60 years does not imply that it is the ultimate technology to carry out digital writing or information processing. \textbf{Photonics} has been following a delayed but similar
development. The term photonics began to be used in the late 1960's, once it became
evident that photons could be used as bits of information in the same way as electrons
were being used. However, it did not reach the common use until the 1980's, when the IEEE Lasers and Electro-Optics Society established an archival journal named Photonics Technology Letters. Some of the early crucial developments in the field were the invention of the laser in 1958 \cite{Schawlow1958,Siegman1986}; the diode laser in the 1962 \cite{Holonyak1962}; the first demonstration of optical communication using optical fibres in 1966, done at the University of Ulm, Germany \cite{Borner1966,Borner1974}; and the invention of the optical fibre amplifier in the early 1980's \cite{Mears1986}. Photonics presents some significant advantages over electronics. The most important one is speed. Due to its much larger carrier frequencies, photonics has a much larger bandwidth (or data transmission rate) than electronics. This is depicted in Figure \ref{Fig12}, which shows that the operating speed of photonic devices can be up to five orders of magnitude higher than electronics.
Another important feature of photons is their weak interaction with the environment. Photons can travel long distances almost without changing their state. This implies much lower losses in transmission. Actually, this feature is also exploited in quantum information, in particular in the field of quantum cryptography \cite{Bennett1984} to create secure protocols to data exchange. As a consequence, photonic integrated circuits have been gaining a lot of attention in recent times \cite{Coldren2012}. The most ambitious goal is to eventually usurp the predominant role of electronic integrated circuits in information devices. However, the goal seems to be unrealistic at the moment. Photonic circuits suffer from a lack of storage capacity, \textit{i.e.} it is difficult to stop and store light. Hence, the storage needs to be done in an electronic medium. Moreover, integrated electronics is non-linear, whereas non-linear integrated photonics still needs to evolve to be competitive in the market. Last but not least, the typical size of photonic components is limited by the wavelength of light. Due to the diffraction limit of light, dielectric photonic circuit cannot resolve or retrieve information smaller than a fraction of the wavelength (200nm approximately) \cite{Maier2001,Gramotnev2010}, as opposed to nano-electronics where 22nm transistors are already a reality. \\\\
Most of the efforts to overcome the diffraction limitations in photonics have been happening in the field of plasmonics. In fact, lots of authors consider that nanophotonics should be based on metallic nano-plasmonics (see Figure \ref{Fig13}) \cite{Brongersma2010}.
Plasmonics is the science that studies the interaction between light and free electrons on a metal. The first theoretical studies in plasmonics were done in the 1950's \cite{Bohm1951,Pines1952,Bohm1953,Ritchie1957}, and the first experiments with metallic surfaces in the 1970's \cite{Otto1968,Kretschmann1971}. In general, the strategy followed in plasmonics to bring photonics down to the nano-scale has a sample-based perspective. That is, given a fixed external excitation beam (see inset in Figure \ref{Fig14}(a) for a typical example), samples are engineered (shape and materials) so that they produce certain effects.  
For example, in Figure \ref{Fig14}(a), an array of metallic nano-rods is fabricated to guide light from one location to another. Similarly, in Figure \ref{Fig14}(b) a circular nano-hole drilled in a gold film deposited on top of a glass layer is inserted into a liquid chamber. The excitation of the structure enabled the authors of \cite{Mathieu2009} to trap 50nm particles that had been dissolved in the water solution. There is no doubt that fabrication and characterization of samples play a huge role in the current plasmonic technologies. Interestingly, the spatial profile of the incident beam also plays a very important role. However, most of the work in the field is done with Gaussian beams (the beam on the inset of Figure \ref{Fig14}). In this thesis, it will be shown that more elaborate beams of light can be used to retrieve plenty of additional information from nano-structures. An illustrative example is the Stimulated Emission Depletion (STED) microscopy. STED microscopy was invented by Stefan Hell and co-workers in 1994 \cite{Hell1994}. It is one of the so-called super-resolution microscopy methods, as it can resolve defects as tiny as 30nm. A very complete schematic of its working principle is shown in Figure \ref{Fig15}, and its working principle can be found in \cite{Hell1994,Hell2007}.
Even though Figure \ref{Fig15} is complicated, the message should be clear. Probing a sub-wavelength specimen with a Gaussian beam (green beam in Figure \ref{Fig15}) results in a blurry image (Figure \ref{Fig15}(c)). Nevertheless, the combined use of a Gaussian and a doughnut-shaped beam (red) results in a much neater image (see Figure \ref{Fig15}(d)). There are different kinds of doughnut-shaped beams, but the ones used in STED are called vortex beams \cite{Gabi2007,Yao2011,Richard2011b}. \\\\
Vortex beams are defined by their topological charge $l$, which is an integer number. The topological charge $l$ accounts for the number of times that the phase of the beam wraps around its centre in a $2\pi$ circle. Examples of four different vortex beams are shown in Figure \ref{Fig16}. 
It can be seen that when $l=-1$, the phase goes from $0$ to $2\pi$ one time in counter-clockwise direction. In contrast, when $l=3$, the phase goes three times from $0$ to $2\pi$ in a clockwise direction. In chapter \ref{Ch5}, sections \ref{Ch5_CGH} and \ref{Ch5_SLM} have been dedicated to explain how to characterize and generate vortex beams in the laboratory. Note that the definition of vortex beam and topological charge is independent of the polarisation. However, for many applications in nano-photonics (STED microscopy is one of them), vortex beams need to be tightly focused. When that is the case, vortex beams become much more complex. In fact, in that regime, the definition of the phase singularity is polarisation-dependent. This is what many authors have described as one of the manifestations of the spin-to-orbit conversion \cite{Zhao2007,Vuong2010,Rodriguez-Herrera2010,Bliokh2010,Bliokh2011}. Then, characterising vortex beams in terms of their symmetries becomes enormously useful. It allows for a systematic study of the beam without the need of considering different focusing regimes\footnote{For the advanced reader: note the parallelism between the two focusing regimes (paraxial/non-paraxial) in optics with the two regimes in quantum mechanics (non-relativistic/relativistic). In the paraxial approximation, light can be described by a scalar function and polarization is added as a different degree of freedom. Similarly, the scalar wave-function of a particle is obtained with Sch\"{o}dinger equation, and the vectorial behaviour given by the spin is added by hand. In contrast, when light is focused a lot, Maxwell equations have to be used, and the polarization and spatial degrees of freedom are intermingled. Analogously, when a quantum particle is boosted and its speed is relativistic, Dirac equations must be used. In Dirac equations, the spin degrees of freedom cannot be modified without modifying the spatial properties of the field.}. \\\\
Sections \ref{Ch1_basis} and \ref{Ch1_multipoles} are dedicated to a systematic characterisation of beams in terms of their symmetries. Thanks to the understanding and analysis of the different transformations under which light beams are symmetric, different results have been obtained, both theoretically and experimentally. In the succeeding chapters, three symmetries have been specially studied: mirror symmetry, rotational symmetry, and duality \cite{Tung1985,Messiah1999,Ivan2013}. Both samples and beams of light which are (or are not) symmetric under these transformations have been used and as a consequence new fundamental phenomena have been found\footnote{Saying that a system is symmetric under a transformation means that when that transformation is applied to the system, its physical properties remain unchanged.}. In chapter \ref{Ch3}, the interaction between single spheres and cylindrically symmetric beams has been theoretically studied. Among others, a technique to excite Whispering Gallery Modes (WGMs) has been developed. In chapter \ref{Ch4}, the cylindrical symmetry of a single sphere and the duality symmetry of a light beam have been used to cancel the forward scattering of a sample. Furthermore, it has been proven that a sample can effectively be dual in a controlled manner if its dimensions are changed in a certain way with respect to an incident cylindrically symmetric beam. In chapter \ref{Ch6}, a mirror and cylindrically symmetric sample has been used to induce a giant Circular Dichroism (CD). Again, the use of dual, cylindrically and mirror symmetric modes has been a key point to be able to achieve it. Finally, chapter \ref{Ch7} merges all the knowledge accumulated in the previous chapters, both theoretical and experimental. Using various dual cylindrically symmetric beams, the resonant scattering behaviour of a dielectric sphere is unveiled. This resonant behaviour is hidden under the Gaussian beam (or plane wave) excitation, and the use of vortex beams is crucial to find it. Due to time constraints, this last study is not as thorough as the others presented in the previous chapters. Nevertheless, it allowed for an experimental observation of some of the theoretical predictions of the thesis. \\\\
Now, even though the details of what has been done in each chapter may be too technical for a non-specialised reader, these should only be considered to emphasise the point made before: nanophotonics can also be done from another perspective different from the prevailing sample-based view. In particular, the consideration of the sample and the exciting light as a whole system, along with symmetry considerations, allows for the discovery of new phenomena at the nano-scale. The future in this unexplored direction looks as exciting as promising. In the next chapters, the reader will find a taste of it.

\begin{savequote}[10cm] 
\sffamily
``If I could explain it to the average person, I wouldn't have been worth the Nobel Prize.'' 
\qauthor{Richard P. Feynman}
\end{savequote}

\chapter{Theoretical methods in nano-optics}
\graphicspath{{ch1/}} 
\label{Ch1}

\newcommand\kz{k_z}
\newcommand\xhat{\mathbf{\hat{x}}}
\newcommand\yhat{\mathbf{\hat{y}}}
\newcommand\rhat{\mathbf{\hat{r}}}
\newcommand\lhat{\mathbf{\hat{l}}}
\newcommand\zhat{\mathbf{\hat{z}}}
\newcommand\ahat{\mathbf{\hat{a}}}
\newcommand\nhat{\mathbf{\hat{n}}}
\newcommand\spphat{\mathbf{\hat{\sigma}_p}}
\newcommand\sppmhat{\mathbf{\hat{\sigma}_{-p}}}

\newcommand\shat{\widehat{\mathbf{s} }}
\newcommand\phat{\widehat{\mathbf{p} }}
\newcommand\phik{\phi_k}
\newcommand\thetak{\theta_k}
\newcommand\plane{\exp(i \textbf{k} \cdot \textbf{r})}
\newcommand\planep{ \dfrac{\shat + \phat}{\sqrt{2}} \plane }
\newcommand\planem{ \dfrac{\shat - \phat}{\sqrt{2}}  \plane }
\newcommand\kx{k_x}
\newcommand\ky{k_y}
\newcommand\hel{ \Lambda }
\newcommand\helop{ \frac{\nabla \times }{k} }
\newcommand\expp{\exp (i \phik)}
\newcommand\expm{\exp (-i \phik)}
\newcommand\dkrho{\delta(\krho-\krho')}
\newcommand\emphik{\exp(-im\phik)}
\newcommand\epphik{\exp(im\phik)}
\newcommand\ekz{\exp(i \kz z)}
\newcommand\bbp{J_{m+1}(\krho \rho)}
\newcommand\bbm{J_{m-1}(\krho \rho)}
\newcommand\bb{J_{m}(\krho \rho)}
\newcommand\emphi{\exp(-i\phi)}
\newcommand\epphi{\exp(i\phi)}
\newcommand\ephi{\exp(im\phi)}
\newcommand\kzk{\frac{\kz}{k}}
\newcommand\rotation{\mathbf{R} (\phik, \thetak)}
\newcommand\ep{\mathbf{e_{+}}}
\newcommand\emm{\mathbf{e_{-}}}
\newcommand\alphak{\alpha(\phik, \thetak)}
\newcommand\betak{\beta(\thetak, \phik)}
\newcommand\cp{\cos \phik}
\newcommand\ct{\cos \thetak}
\newcommand\spp{\sin \phik}
\newcommand\st{\sin \thetak}
\newcommand\Einf{\mathbf{E}_{\infty}(\phik,\thetak)}
\newcommand\nt{\mathbf{n}_{\theta}}
\newcommand\npp{\mathbf{n}_{\phi}}
\newcommand{\E}{\mathbf{E}}
\newcommand{\Hh}{\mathbf{H}}
\newcommand{\TDt}{T_{\Delta t}}
\newcommand{\psixt}{ \psi(\mathbf{r},t) }
\newcommand{\MTE}{\mathbf{M}^{\text{TE}}}
\newcommand{\MTM}{\mathbf{M}^{\text{TM}}}
\newcommand\ephat{\mathbf{\hat{e}}_p}
\newcommand\epphat{\mathbf{\hat{e}}_+}
\newcommand\epmhat{\mathbf{\hat{e}}_-}
\newcommand{\eppphat}{\hat{\mathbf{e}}_{-p}}
\newcommand{\BTE}{\mathbf{B}_{m\krho}^{\text{TE}}}
\newcommand{\BTM}{\mathbf{B}_{m\krho}^{\text{TM}}}
\newcommand{\Bx}{\mathbf{B}_{m\krho}^{(y)}}
\newcommand{\Bmp}{\mathbf{B}_{m\krho}^{p}}
\newcommand{\Bmpp}{\mathbf{B}_{m\krho}^{+}}
\newcommand{\Bmpm}{\mathbf{B}_{m\krho}^{-}}
\newcommand{\Am}{\mathbf{A}_{jm_z}^{(m)}}
\newcommand{\Ae}{\mathbf{A}_{jm_z}^{(e)}}
\newcommand{\Ay}{\mathbf{A}_{jm_z}^{(y)}}
\newcommand{\dd}{\text{d}}
\newcommand{\Djmp}{D^j_{m_zp}(\phik,\thetak)}
\newcommand{\Ajmp}{\mathbf{A}_{jm_z}^{+}}
\newcommand{\Ajmm}{\mathbf{A}_{jm_z}^{-}}
\newcommand{\Ajmpp}{\mathbf{A}_{jm_z}^{p}}
\newcommand{\Sconv}{\mathbf{\overline{S}}}
\newcommand\Ein{\mathbf{E}^\mathrm{in}}
\newcommand\Eout{\mathbf{E}^\mathrm{out}}
\newcommand{\thetahat}{\mathbf{\hat{\theta}}}
\newcommand{\rhohat}{\mathbf{\hat{\rho}}}
\newcommand{\phihat}{\mathbf{\hat{\phi}}}
\newcommand{\Uu}{U(\mathbf{r},t)}
\newcommand{\U}{U(\mathbf{r})}

\section[Maxwell Equations]{Maxwell Equations for a linear, non-dispersive, homogeneous, isotropic and source-free medium} \label{Ch1_HilMax}
The material presented in this thesis can be described using the tools of EM field theory \cite{Jackson1998,Stratton1941,Bo2004}. In particular, the subdivisions of EM field theory that this thesis especially deals with are electromagnetic optics and vectorial scattering theory \cite{Saleh2007,Bohren1983,Morse1953,GLMT_book,Novotny2006,Rose1955,Rose1957,Born1999}. In this chapter, I will introduce all the necessary concepts and tools to understand the theoretical developments scattered across the next chapters. For completeness, the starting point will be the macroscopic Maxwell Equations for a source-free medium \cite{Jackson1998,Saleh2007}:
\renewcommand{\arraystretch}{1.5}
\begin{equation} \begin{array}{cccccr}
\nabla \cdot \mathcal{D} & = & 0 &\qquad \nabla \times \mathcal{E} & = &-\dfrac{\partial \mathcal{B}}{\partial t} \\
\nabla \cdot \mathcal{B} & = & 0 & \qquad \nabla \times \mathcal{H} & = & \dfrac{\partial \mathcal{D}}{\partial t}  
\end{array} \end{equation} 
where $\mathcal{E}$ and $\mathcal{H}$ are the electric and magnetic fields, $\mathcal{D}$ is the electric displacement, and $\mathcal{B}$ is the magnetic induction. The relations between $\mathcal{D}$ and $\mathcal{E}$, as well as between $\mathcal{B}$ and $\mathcal{H}$ are given by the properties of the medium:
\renewcommand{\arraystretch}{1}
\begin{eqnarray}
\mathcal{D} &=& \epsilon_0 \mathcal{E} + \mathcal{P} \\
\mathcal{B} &=& \mu_0  \mathcal{H} + \mu_0 \mathcal{M} 
\end{eqnarray}
with $\epsilon_0, \mu_0$ being the electric permittivity and magnetic permeability of vacuum. $\mathcal{P}$ is defined as the polarization density, which is the density of molecular dipole moments induced by the electric field \cite{Saleh2007,Jackson1998}. Similarly, $\mathcal{M}$, defined as the magnetization density, is the density of molecular magnetic dipole moments induced by the magnetic field. I will be particularly interested in linear, non-dispersive, homogeneous\footnote{I may consider also  linear, non-dispersive, inhomogeneous media formed by different linear, non-dispersive, homogeneous, and isotropic submedia.} and isotropic media. In these cases, a linear relation can be established between $\mathcal{P}$ and $\mathcal{E}$, and therefore the following relation holds \cite{Saleh2007}: 
\begin{equation}
\mathcal{D} = \epsilon \mathcal{E}
\end{equation}
where $\epsilon$ is a scalar quantity defining the electrical properties of the medium called electric permittivity. In the same way, a linear relation can be cast between $\mathcal{M}$ and $\mathcal{H}$, yielding the following linear relation between $\mathcal{B}$ and $\mathcal{H}$
\begin{equation}
\mathcal{B} = \mu \mathcal{H}
\end{equation}
with $\mu$ being the so-called magnetic permeability constant of the medium. Hence, the macroscopic Maxwell Equations for a linear, non-dispersive, homogeneous, isotropic and source-free medium can be written only as functions of $\mathcal{E}$ and $\mathcal{H}$ \cite{Saleh2007}:
\renewcommand{\arraystretch}{1.5}
\begin{equation}\begin{array}{cccccr}
\nabla \cdot \mathcal{E} & = & 0 &\qquad \nabla \times \mathcal{E} & = &-\mu  \dfrac{\partial \mathcal{H}}{\partial t} \\
\nabla \cdot \mathcal{H} & = & 0 & \qquad \nabla \times \mathcal{H} & = &\epsilon \ \dfrac{\partial \mathcal{E}}{\partial t}  \label{E_Max_EH}
\end{array} \end{equation}
Here, I will be especially interested in the monochromatic solutions of these equations. That is, I will suppose that all the fields have a $e^{-i\omega t}$ time dependence, with $\omega$ being the frequency of light. Then, any field can be expressed as the real part of a product of a purely temporal part and a purely spatial part: $\mathcal{A}(\mathbf{r},t) = \text{Re} \left[ \mathbf{A}(\mathbf{r})e^{-i\omega t} \right]$, where $\mathbf{A}(\mathbf{r})$ represents a complex-amplitude vector. Then, equations (\ref{E_Max_EH}) can be re-written just in terms of their spatially dependent parts:
\renewcommand{\arraystretch}{1}
\begin{equation} \begin{array}{cccccr}
\nabla \cdot  \mathbf{E} & = & 0 & \quad \nabla \times \mathbf{E} & = & i \omega \mu \mathbf{H} \\
\nabla \cdot \mathbf{H} & = & 0 & \quad \nabla \times \mathbf{H} & = & -i \omega \epsilon \mathbf{E}
\label{E_Max_EH_mono}
\end{array} \end{equation}
Now, applying the curl operation to equations (\ref{E_Max_EH_mono}) and making use of the vector identity $\nabla \times (\nabla \times \mathbf{A}) = \nabla (\nabla \cdot \mathbf{A}) - \nabla^2 \mathbf{A}$, one obtains that both $\E$ and $\Hh$ fulfil the vectorial Helmholtz Equation \cite{Jackson1998}:
\begin{equation}\begin{array}{lcc}
\nabla^2 \E + k^2 \E & = & 0 \\
\nabla^2 \Hh + k^2 \Hh & = & 0 
\label{E_Max_Helm}
\end{array}\end{equation}
where $k = \omega \sqrt{\epsilon \mu}$ is known as the wave-number. Nonetheless, equations (\ref{E_Max_EH_mono}, \ref{E_Max_Helm}) cannot be uniquely solved unless some boundary conditions (BC) are applied. Firstly, the EM fields must have a proper decay at infinity and be finite everywhere else. Secondly, when equations (\ref{E_Max_EH_mono}) are applied between two different media, the following BC need to hold: 
\begin{equation}\begin{array}{llllll}
\left( \epsilon_2 \E_2 - \epsilon_1 \E_1 \right) \cdot \nhat & = & \sigma \qquad &  \left( \mu_2 \Hh_2 - \mu_1 \Hh_1 \right) \cdot \nhat & = & 0 \\
\nhat \times \left( \E_2 -  \E_1 \right) & = & 0 \qquad & \nhat \times \left(\Hh_2 - \Hh_1 \right) & = & \mathbf{j}_s
\label{E_BC}
\end{array} \end{equation}
where $\sigma$ is the surface charge density at the surface; $\mathbf{j}_s$ is the surface current density; $\nhat$ is the unitary normal vector to the interface between 1 and 2; and $\left\lbrace \E_1,\Hh_1  \right\rbrace$ and $\left\lbrace \E_2,\Hh_2  \right\rbrace$ account for the electric and magnetic fields at two different media. It is observed that the tangential components of $\E$ and $\Hh$ must be continuous, while the normal components of $\mathbf{D}$ and $\mathbf{B}$ can be discontinuous depending on the properties of the surface. Equations (\ref{E_Max_EH_mono}) plus the BC listed above will be the framework in which this thesis will be grounded on. 


\section{Fields, operators and symmetry groups}\label{Ch1_symm}
In the previous section, I have introduced the monochromatic Maxwell Equations for linear, non-dispersive, homogeneous, isotropic and source-free media (see equation (\ref{E_Max_EH_mono})). I will impose that all the theoretical developments done in this thesis fulfil equations (\ref{E_Max_EH_mono}). That is, I will require that any meaningful \textbf{EM field} fulfils equations (\ref{E_Max_EH_mono}), BCs (\ref{E_BC}), it decays properly at infinity and it is finite everywhere else.\\\\
Another framework to do optical physics is the paraxial equation \cite{Saleh2007,Lax1975}, which is an approximation of equations (\ref{E_Max_EH_mono}) but it is much simpler to use. In paraxial optics, a complex-amplitude electric field $\mathbf{E}$ can be expressed as $\mathbf{E(\mathbf{r})}=E(\mathbf{r})\mathbf{u}$, with $E(\mathbf{r})$ being a complex-wave amplitude scalar function that must fulfil the paraxial equation
\begin{equation}
\left(\partial_x^2 + \partial_y^2  +2ik\partial_z \right) E(\mathbf{r}) =0
\end{equation}
and $\mathbf{u}$ being an arbitrary polarization vector. Then, it is clear that in paraxial optics the vectorial character of the field (given by $\mathbf{u}$) is decoupled from its spatial properties (given by $E(\mathbf{r})$). The importance of the paraxial equation is that collimated beams, which are very frequently used in the laboratory, can be accurately described within this framework. In those beams, if the propagation direction of the beam is the $z$ axis, a plane wave decomposition of the wave yields partial waves whose $\krho \ll k_z $, and $k_z \approx k$ (see \ref{Ch1_planewaves} to check the definitions of $\krho$ and $k_z$). This implies that the complex amplitude function $E(\mathbf{r})$ can be expressed as $ E(\mathbf{r}) = A(\mathbf{r}) \exp (i k z)$, where $A(\mathbf{r})$ is a magnitude called the envelope of the wave which varies slowly with $z$:
\begin{equation}
\dfrac{\partial A}{\partial z} \ll k A
\end{equation}
However, the paraxial approximation does not properly work in nano-optics. In nano-optics, light beams are either tightly focused or interact with sub-wavelength objects. In those cases, the polarization degrees of freedom of the beams cannot be decoupled from their spatial properties \cite{Novotny2006,Ivan2012PRA}. Then, imposing that all EM fields must be solutions of the full Maxwell equations rules out all those beams that fulfil the paraxial equation, but are not solutions of the full Maxwell equations given by equation (\ref{E_Max_EH_mono}) \cite{Lax1975,Agrawal1979,Hall1996}.\\\\
Now, I will define an \textbf{operator} as an entity that picks up any EM field fulfilling equations \ref{E_Max_EH_mono} and transforms it into a different one that still fulfils equations \ref{E_Max_EH_mono}. Finally, I will define a \textbf{symmetry group} as a set of operators with group structure. A group $G$ is defined as a set of elements $\{ a,b,c,...\}$ along with an inner product. The inner product ($\cdot$) is such that the product of two arbitrary elements of the group yields an element of the group \cite{Tung1985}. Furthermore, the product must satisfy the following conditions \cite{Tung1985}:
\begin{itemize}
\item It must be associative, \textit{i.e.} $a \cdot (b \cdot c) = (a \cdot b) \cdot c$, $\forall a,b,c \in G$. 
\item There must be an element $e \in G$, called the identity, which is such that $a \cdot e = a$, $\forall a \in G$.
\item For each $a \in G$, there must exist an element $a^{-1} \in G$, called the inverse of $a$, such that $a \cdot a^{-1} = e$.
\end{itemize}
Groups can be subdivided into two categories - discrete, and continuous. In physics, most of continuous groups of interest are part of a subdivision called `linear Lie groups' \cite{Tung1985}. Linear Lie groups can be labelled by one or more continuous variables. In fact, each label is related to a generator. The generators of the group define the local behaviour of the group near the identity, and any element of the group can be obtained as a smooth function of the generator. For example, for the $SO(2)$ group of 2-dimensional rotations, any rotation $R(\phi)$ can be obtained as $R(\phi)=\exp(-i\phi J)$, with $J$ the generator of the group. Note that this implies that if the element of the group is an operator, so must be the generator. In fact, group generators are very good candidates to compose a complete set of commuting operators. A complete set of commuting operators of a vector space is a set of operators that commute with each other and whose eigenvalues completely specify the state of a system \cite{Sakurai1995,Gasiorowicz2007}. The idea is that if an EM field is an eigenvector of a generator of a symmetry, then it is also invariant under the symmetry transformation generated by that generator, and vice versa\footnote{This statement is a consequence of Noether's theorem, which states that every continuous symmetry of a physical system corresponds to a conservation law of the system.} \cite{Noether1918,Tung1985}. Then, using generators to make a set of commuting operators will be useful as the fields will be characterized in terms of their symmetries. Now, I will differentiate between EM fields and \textbf{EM modes}. EM fields are solutions of Maxwell equations (\ref{E_Max_EH_mono}), whereas EM Modes are elements of a basis of solutions of Maxwell equations (\ref{E_Max_EH_mono}). Thus, in general, any EM field can be decomposed into EM modes. In addition, EM modes will be chosen so that they are eigenstates of some generators, \textit{i.e.} they will be invariant under certain symmetries. Next, I will list the most common operators that I will use to describe EM modes and their differential representation will be given. Some of them are generators of symmetries, so in those cases I will also comment on the symmetries that they give rise to. 
\begin{itemize}
\item[i)] \underline{Hamiltonian}, $H=-i \dfrac{\partial}{\partial t}$. The hamiltonian is the generator of temporal translations (or time evolution) \cite{Messiah1999}. That is, the transformation $T_{\Delta t}=\exp \left( -i H \Delta t\right)$ applied to an EM field $ \psi(\mathbf{r},t)  $  the state evolve from $t=t_0$ to $t'=t_0+\Delta t$:
\begin{equation}
\TDt  \left[ \psixt \right] = \psi(\mathbf{r},t+\Delta t) 
\end{equation}

\item[ii)] \underline{Linear momentum operator,} $\mathbf{P}=-i\nabla$. This is a vectorial operator, \textit{i.e.} it is composed of three operators in three different directions: $\mathbf{P}=P_x \xhat + P_y \yhat + P_z \zhat$. Its three components commute with each other $\left[ P_i,P_j \right] =0$. Each of the components generates a translation in its corresponding axis. Hence, the vectorial addition of all of them is the generator of linear translations in any direction of space: $T_{\Delta \mathbf{r}} = \exp \left( -i \mathbf{P} \cdot \Delta \mathbf{r} \right)$ \cite{Tung1985}. That is, 
\begin{equation}
T_{\Delta \mathbf{r}} \left[ \psixt \right] =  \psi(\mathbf{r} +\Delta \mathbf{r} ,t) 
\end{equation}

\item[iii)] \underline{Angular momentum operator}, $\mathbf{J}=\mathbf{L}+\mathbf{S}$, with $\mathbf{L}=-i (\mathbf{r} \times \nabla)$ and $\mathbf{S} = \\ -i \left( \epsilon_{1nm}\xhat  + \epsilon_{2nm}\yhat  + \epsilon_{3nm}\zhat \right)$ respectively, where $\epsilon_{lnm}$ is the total anti-symmetric tensor and $l,m,n=1,2,3$. It is also composed of three operators in three components, $\mathbf{J}=J_x \xhat + J_y \yhat + J_z \zhat$, but they do not commute with each other: $\left[ J_i,J_j \right] =i \epsilon_{ijk} J_k$. The reason why the three components of the angular momentum (AM) $\mathbf{J}$ do not commute is that they are the generators of rotations on three different axis, and rotations do not commute (whereas spatial translations do, see above) \cite{Rose1957}. A general rotation of an angle $\varphi$ along an axis on the direction $\nhat$ is given by: $R_{\nhat}(\varphi)=\exp \left( -i \mathbf{J} \cdot \nhat \varphi \right)$. Then, the application of a rotation onto a general EM field yields:
\begin{equation}
R_{\nhat}(\varphi) \left[ \psixt \right] = \mathbf{M}_{\nhat}(\varphi)  \cdot \psi(\mathbf{M}_{\nhat}^{-1}(\varphi) \cdot \mathbf{r},t) 
\label{E_rotpsi}
\end{equation}
where $\mathbf{M}_{\mathbf{n}}(\varphi)$ is a rotation matrix \cite{Tung1985} (see Appendix \ref{Appendix}). Finally, note that even though the $\mathbf{L}$ and $\mathbf{S}$ operators used to define $\mathbf{J}$ are usually called orbital and spin AM respectively, they are not proper AM momentum operator on their own \cite{Lifshitz1982,Cohen1997,Ivan2012PRA,Ivan2014}. None of them generates proper rotations, only their addition does. Furthermore, they are not proper operators, as their application to an EM field fulfilling Maxwell equations yields a field that does not fulfil Maxwell equations \cite{Lifshitz1982,Cohen1997,Ivan2012PRA,Ivan2014}.

\item[iv)] \underline{Angular momentum squared}, $J^2=J_x^2+J_y^2+J_z^2$, where $J_i$ have been defined in iii). $J^2$ is called a Casimir operator, and commutes with all the operators. Even though it does not generate any meaningful transformation, it is very useful for problems with spherical symmetry.

\item[v)] \underline{Helicity operator}, $\Lambda = \dfrac{ \mathbf{J }\cdot \mathbf{P} }{\vert \mathbf{P} \vert}$. The helicity operator is defined as the projection of the AM onto the direction of the linear momentum. It was proven in \cite{Calkin1965,Zwanziger1968} that $\Lambda$ generates generalized duality transformations \cite{Jackson1998,Ivan2013}: $D_{\varphi} = \exp \left( -i \Lambda \varphi \right)$. Generalized duality transformations mix the electric and magnetic parts of an EM field. That is, if we express $\psixt$ as $\psixt = \left( \E(\mathbf{r},t), \Hh(\mathbf{r},t) \right)^{T}$, then
\begin{equation}
D_{\varphi} \left[ \psixt \right] = D_{\varphi} \left[ \left( \begin{array}{l} \E(\mathbf{r},t) \\  \Hh(\mathbf{r},t) \end{array} \right) \right] =  \left( \begin{array}{l} \E(\mathbf{r},t) \cos \varphi - \Hh(\mathbf{r},t) \sin \varphi \\   \E(\mathbf{r},t) \sin \varphi + \Hh(\mathbf{r},t) \cos \varphi \end{array} \right)
\end{equation}
Its differential representation for monochromatic fields is $\Lambda = \displaystyle\helop$. Unlike all the rest of symmetries listed hereby, duality symmetry is a non-geometrical symmetry. It can be proven that its preservation only depends on the material properties of the medium in consideration \cite{Ivan2013}. A very thorough study of duality symmetry and its generator (helicity) can be found in \cite{IvanThesis}.

\item[vi)] \underline{Parity operator}, $\Pi$. Parity transformations $(\mathbf{r} \rightarrow - \mathbf{r})$ are discrete, therefore they are not generated by any other operator. The application of $\Pi$ onto a Maxwell field yields \cite{Messiah1999}:
\begin{equation}
\Pi \left[  \psixt \right] =  \psi(-\mathbf{r},t) 
\end{equation}

\item[vii)] \underline{Mirror symmetry operator}, $M_{\nhat}$. A mirror transformation of the kind $M_{\nhat}$ indicates that the spatial coordinates are reflected off a plane given by its normal vector $\nhat$\footnote{Throughout this thesis, another notation will also be used for mirror transformations: $M_{\left\lbrace \nhat \right\rbrace }$. The difference is that in $M_{\nhat}$ the reflection is off a plane whose normal vector is $\nhat$, whereas $M_{\left\lbrace \nhat \right\rbrace }$ indicates that the reflection is done upon a plane that contains the axis $\nhat$.}. Actually, it can be re-written as $M_{\nhat} = \Pi R_{\nhat}(\pi)$, \textit{i.e.} a mirror symmetry operator can be obtained by multiplying a parity and rotation operators. Like parity, it is a discrete transformation, so it is not generated by any other operator. Its application onto $\psixt$ gives \cite{Messiah1999}:
\begin{equation}
M_{\nhat} \left[ \psixt \right] =  \psi(\mathbf{r} - 2 \nhat (\nhat \cdot \mathbf{r}),t) 
\end{equation}
\end{itemize}
Last but not least, I will give the most useful commutation rules between the operators listed above \cite{Messiah1999}. These commutation rules are crucial in order to define a set of commuting operators. In fact, it is known that only four commuting operators are needed to describe EM fields, as they are elements of the Poincar\'{e} group \cite{Tung1985,Schwinger1998}: 
\begin{itemize}
\item $T_{\Delta t}$ commutes with all the other operators of the list above.
\item $\mathbf{P}$ commutes with $T_{\Delta t}$, $J^2$, $\Lambda$, and 
$M_{\nhat}$. It does not commute with $\mathbf{J}$, their commutator is $\left[J_i, P_j \right] = i \epsilon_{ijk}P_k$. It anti-commutes with parity $\left\lbrace P_i , \Pi   \right\rbrace =0$. 
\item $\mathbf{J}$ commutes with $T_{\Delta t}$, $J^2$, $\Lambda$, $\Pi$ and $M_{\nhat}$ (when $\nhat$ is parallel to $\mathbf{J}$). It does not commute with $M_{\left\lbrace \nhat \right\rbrace}$, in fact it anti-commutes: $J_i M_{\left\lbrace \nhat \right\rbrace} = - M_{\left\lbrace \nhat \right\rbrace} J_i$.
\item $J^2$ commutes with all the rest of operators, as it is a Casimir operator.
\item $\Lambda$ commutes with $T_{\Delta t}$, $\mathbf{P}$, $\mathbf{J}$ and $J^2$. It anti-commutes both with $\Pi$ and $M_{\nhat}$.
\item $\Pi$ commutes with $T_{\Delta t}$, $\mathbf{J}$, $J^2$, and $M_{\nhat}$. It anti-commutes with $\Lambda$ and $\mathbf{P}$. 
\end{itemize}

\section{Basis of solutions of free-source Maxwell equations} \label{Ch1_basis}
The aim of this chapter is to establish the basis on which the works presented on this thesis are grounded. In section \ref{Ch1_HilMax}, I have introduced Maxwell equations. Then, in section \ref{Ch1_symm}, I have discussed that I will require all the EM fields to fulfil Maxwell equations in its form given by equations (\ref{E_Max_EH_mono}). Furthermore, some operators and generators have been introduced. In this section and the next one, six different sets of EM modes that stem from six different sets of complete commuting operators will be described. Each of these sets of EM modes will be especially suited to describe some particular EM interactions. In fact, since generators of symmetries will be used as the operators in a set, a set of modes will be chosen over another depending on the symmetries of the interaction. There are different ways of constructing these sets of EM modes. From a symmetries' point of view, probably the clearest technique is `the induced representation method'. Wu-Ki Tung describes it in chapter 9 of his book \cite{Tung1985}. As in most of the cases, the use of symmetries simplifies the amount of operations tremendously. Unfortunately, this method is not very common in optics. Hence, another equivalent method much better-known in the optics community will be used here. This method is based on some mathematical theorems about the solutions of the vectorial Helmholtz equation (\ref{E_Max_Helm}). The method is described in \cite{Stratton1941,Morse1953,Bohren1983} among others. It is based on the following property. Given a general solution of the scalar Helmholtz equation in a system of coordinates where the solution is in separated variables $\Psi(x_1,x_2,x_3)=X_1(x_1)X_2(x_2)X_3(x_3)$, then it can be proven that if $\ahat$ is a unitary vector in a privileged direction\footnote{Not any vector $\ahat$ is valid. Some conditions apply. These conditions can be found in \cite{Morse1953}.},
\begin{equation} \begin{array}{lll}
\MTE & = & \nabla \times (\mathbf{\ahat} \Psi(x_1,x_2,x_3)) \\
\MTM & = & \left\lbrace (\nabla \times) / k \right\rbrace \left[ \nabla \times (\mathbf{\ahat} \psi(x_1,x_2,x_3))\right]
\label{E_MN}
\end{array} \end{equation}
are general solutions of the vectorial Helmholtz equation and fulfil the divergence equations $\nabla \cdot \E = \nabla \cdot \Hh = 0$ \cite{Stratton1941,Morse1953}. That is, any $\E$ field can be obtained as a superposition of the modes $\MTE$ and $\MTM$ defined in equation (\ref{E_MN}). Then, the magnetic field $\Hh$ can be derived from Maxwell equations (\ref{E_Max_EH_mono})\footnote{As $\Hh$ can always be derived from $\E$, I will only focus on the properties of $\E$, and derive $\Hh$ using equations (\ref{E_Max_EH_mono}). Besides, any symmetry considerations regarding $\E$ will also apply to $\Hh$, except for parity and mirror inversion. It can be shown that when $\E$ is odd, $\Hh$ is even and viceversa \cite{Jackson1998}}. Note that I have differentiated the behaviour of the two general solutions by its TE or TM character. The notation of TE/TM behaviour is a bit arbitrary, as both the electric and magnetic fields are `transverse' in the Maxwell sense, \textit{i.e.} $\nabla \cdot \E = \nabla \cdot \Hh = 0$. In fact, this notation gets even more cumbersome when spherical coordinates are used, as TE modes are usually referred to as `magnetic' and TM as 'electric'. Here, I will stick to the most common notation and denote the first solution in equation (\ref{E_MN}) as TE and the second one as TM for cartesian and cylindrical coordinates \cite{Stratton1941,Morse1953,GLMT_book}. Then, in spherical coordinates I will refer to the first solution as magnetic and the second one as electric \cite{Rose1955,Rose1957,Jackson1998}, as that is the dominant notation.

\subsection{Plane waves} \label{Ch1_planewaves}
It can be shown that $\Psi(x,y,z)=\plane$ is a general solution of the scalar Helmholtz equation for cartesian coordinates \cite{Stratton1941}, with $\mathbf{r}=(x,y,z)$ being the position vector and $\mathbf{k}=(k_x,k_y,k_z)$ the wave-vector, whose modulus $\vert \mathbf{k} \vert = k = \omega / c$ is the wave-number. Then, using $\ahat = \zhat$ as the symmetry direction, the two following vectorial solutions arise:
\begin{eqnarray}
\label{s}
\MTE  = & \widehat{\textbf{s}}  \plane  = &\frac{i}{k_{\rho}} \left( k_y \widehat{\textbf{x} } - k_x \widehat{\textbf{y} } \right) \exp(i \textbf{k} \cdot \textbf{r})\\
\label{p}
\MTM  = &  \phat  \plane  = &  \left[ \frac{ - \kz \left( k_x \xhat + ky \yhat \right) + \krho^2 \zhat} { k \krho} \right] \plane
\end{eqnarray}
where $\left\lbrace \xhat, \yhat, \zhat  \right\rbrace$ are the polarization vectors in cartesian directions. These TE and TM modes are very well-known in optics and are referred to as $\shat$ and $\phat$ plane waves. In fact, they define a trihedron such that $\phat \times \shat = \mathbf{k} / \vert \mathbf{k} \vert $. As previously mentioned, all the fields considered here are monochromatic, therefore both $\shat$ and $\phat$ waves will be eigenvectors of the $H$ operator. In addition, it can be checked that they are eigenvectors of the following operators:
\begin{equation} \begin{array}{llllll}
P_x \left[ \shat \plane \right]& = & k_x \ \shat \plane & \qquad P_x \left[ \phat \plane \right] & = & k_x \ \phat \plane \\
P_y \left[ \shat \plane \right] & = & k_y \  \shat \plane & \qquad P_y \left[ \phat \plane \right] & = & k_y \  \phat \plane \\
M_{\shat} \left[ \shat \plane \right] & = & - \  \shat \plane  & \qquad M_{\shat} \left[ \phat \plane \right] & = & + \ \phat \plane
\label{E_sp_eigen}
\end{array} \end{equation} 
Then, $\shat$ and $\phat$ are EM modes invariant under temporal translations, spatial translations along $x$ and $y$ axis, and mirror symmetry transformations. Hence, they are eigenvectors of $H$, $P_x$, $P_y$, and $M_{\shat}$. Now, if we look at equations (\ref{E_MN}), we can see that the $\MTM$ mode ($\phat$ wave) is obtained by applying $(\nabla \times) / k$ to the $\MTE$ mode ($\shat$ wave). In section \ref{Ch1_symm}, I have shown that the operation $(\nabla \times) / k$ is actually the differential form of the helicity operator for monochromatic fields:
\begin{equation}
\hel = \dfrac{\mathbf{J} \cdot \mathbf{P} }{ \mathbf{P}  } = \dfrac{\nabla \times}{k}
\end{equation}
Then, it can immediately be seen that:
\renewcommand{\arraystretch}{1.5}
\begin{eqnarray}
\hel \left[  \shat \plane  \right]  & = & \phat \plane \\
\hel \left[ \phat \plane \right] & = & \shat \plane 
\end{eqnarray}
Thus, additions and subtractions of $\shat$ and $\phat$ waves yield states with a well-defined helicity\footnote{Saying that a state has a well-defined property (helicity, for example) means that the state is an eigenstate of the operator representing that property.}:
\begin{eqnarray}
\label{hel:1}
\hel \left[ \planep  \right]  & = & + \left[ \planep \right] \\
\hel \left[ \planem  \right] & = &   - \left[ \planem \right] 
\end{eqnarray}
I will denote these combinations as $\ephat = (\shat + p \phat)/\sqrt{2}$, with $p=\pm 1$ depending on the value of helicity:
\renewcommand{\arraystretch}{1.5}
\begin{eqnarray}
\epphat = \dfrac{\shat + \phat}{\sqrt{2}} = \dfrac{1}{2} \left[ \left(1 - \dfrac{k_z}{k}  \right) e^{i \phi_k} \ \smhat +  \left(-1 - \dfrac{k_z}{k}  \right) e^{-i \phi_k} \ \sphat + \dfrac{\sqrt{2}\krho}{k} \ \zhat \right] \\
\epmhat = \dfrac{\shat - \phat}{\sqrt{2}} = \dfrac{1}{2} \left[ \left(1 + \dfrac{k_z}{k}  \right) e^{i \phi_k} \ \smhat +  \left(\dfrac{k_z}{k} -1 \right) e^{-i \phi_k} \ \sphat - \dfrac{\sqrt{2}\krho}{k} \ \zhat \right]
\end{eqnarray}
where $\krho$ and $\phi_k$ are the wave-vector components in cylindrical coordinates, \textit{i.e.} $k_x = \krho \cos \phi_k$, $k_y = \krho \sin \phi_k$ and $\left\lbrace \sphat,\smhat \right\rbrace$ are the circular polarizations vectors: $\sphat = (\xhat + i \yhat)/\sqrt{2}$ is the left circular polarization and $\smhat = (\xhat - i \yhat)/\sqrt{2}$ is the right one. Then, a new basis of EM modes can be created with $\epphat$ and $\epmhat$. This will be symmetric under temporal translations, spatial translations along $x$ and $y$ axis, and generalized duality transformations:
\renewcommand{\arraystretch}{1}
\begin{equation} \begin{array}{llllll}
P_x \left[ \epphat \plane \right]& = & k_x \ \epphat \plane & \qquad P_x \left[ \epmhat \plane \right] & = & k_x \ \epmhat \plane \\
P_y \left[ \epphat \plane \right] & = & k_y \  \epphat \plane & \qquad P_y \left[ \epmhat \plane \right] & = & k_y \  \epmhat \plane \\
\Lambda \left[ \epphat \plane \right] & = & + \  \epphat \plane  & \qquad \Lambda \left[ \epmhat \plane \right] & = & - \ \epmhat \plane 
\label{E_ep_eigen}
\end{array} \end{equation} 
Finally, it can be checked that the two sets of EM modes given here are normalized in the following way:
\renewcommand{\arraystretch}{1.5}
\begin{equation}\begin{array}{lll}
 \displaystyle\int_{\mathbb{R}^3} (\shat \plane)^* \cdot \shat \exp(i \mathbf{k'} \cdot \textbf{r}) \dd^3 \mathbf{r}  & = &  \delta(\mathbf{k}-\mathbf{k'}) \\
\displaystyle\int_{\mathbb{R}^3} (\phat \plane)^* \cdot \phat \exp(i \mathbf{k'} \cdot \textbf{r}) \dd^3 \mathbf{r}  & = &  \delta(\mathbf{k}-\mathbf{k'}) \\
 \displaystyle\int_{\mathbb{R}^3} (\ephat \plane)^* \cdot \mathbf{\hat{e}}_{p'} \exp(i \mathbf{k'} \cdot \mathbf{r}) \dd^3 \mathbf{r}  & = & \delta(\mathbf{k}-\mathbf{k'}) \ \delta_{pp'}
\end{array} 
\end{equation}
where $\delta(\mathbf{k}-\mathbf{k'}) $ is a Dirac delta, and $\delta_{pp'}$ is a Kronecker delta \cite{Morse1953}.

\subsection{Bessel beams}
Cylindrically symmetric interactions will play a very significant role in this thesis. Therefore, a basis that exploits this symmetry needs to be found. A way of attacking the problem is to use equations (\ref{E_MN}) in cylindrical coordinates with $\ahat = \zhat$ as well. This method, used by \cite{Stratton1941,Morse1953,Hall1996,Jauregui2005} and several others, naturally gives rise to the so-called TE/TM Bessel beams. Nevertheless, the symmetries of the problem are a bit blurred by the methodology. Then, in order to highlight the symmetries of the Bessel beams, I will use an alternative method to derive them. This method is based on the fact that any field can be decomposed into plane waves. Then, the superpositions are chosen so that the symmetry requirements are fulfilled. A Bessel beam can be defined as an EM mode symmetric under temporal translations, spatial translations and rotations along a symmetry axis (usually, the $z$ is chosen). This implies that Bessel beams need to be eigenstates of $H$, $P_z$, and $J_z$. As mentioned earlier, since I am only considering monochromatic fields, they are immediately eigenstates of $H$. Then, having a set $\omega$ fixes a relation between $k_z$ and $\krho$: $\omega^2/c^2 = k^2 = k_z^2 + \krho^2 $. Therefore, the $z$ component and the transverse component of the linear momentum are linked via the frequency. Thus, having a well-defined $P_z$ is analogous of having a well-defined $P_{\rho}$. Consequently, adding plane waves on a circular cone symmetric under the $z$ axis will yield a field with a well-defined $P_{\rho}$. Then, I will be able to find the conditions on the superposition coefficients by imposing that the beam is an eigenvector of $J_z$. To start, let me re-write the four plane waves derived in the previous sub-section in terms of rotations (see Appendix \ref{Appendix}):
\begin{eqnarray}
\shat \plane &=& - \mathbf{R} (\phik, \thetak) i \yhat \exp(ikz)\\
\phat \plane &=& - \mathbf{R} (\phik, \thetak)  \xhat \exp(ikz) \label{E_p-x}\\
\epphat \plane & = & - \mathbf{R} (\phik, \thetak) \sphat \exp(ikz)\label{E_ep-l} \\
\epmhat \plane & = & - \mathbf{R} (\phik, \thetak) \smhat \exp(ikz) \label{E_ep-r}
\end{eqnarray}
where $\rotation$ is the operator of rotations defined in Appendix \ref{Appendix}. Then, due to the symmetry considerations given by equations (\ref{E_sp_eigen}, \ref{E_ep_eigen}), a superposition of $\shat$ waves will yield a mode with a well-defined frequency and a defined TE/TM character. In general, it will no longer be an eigenstate of $P_x$ and $P_y$, but if we only add $\shat$ waves on a cone, as mentioned in the paragraph above, we can retrieve a mode with a well defined $P_z$ (or $P_{\rho}$). This can be given by the general superposition:
\renewcommand{\arraystretch}{1.5}
\begin{equation} \begin{array}{lll}
\mathbf{B}_{\krho}^{\text{TE}} & = & \displaystyle\int_0^{\pi} \sin \thetak \dd \thetak \int_0^{2\pi} \dd \phik  \alphak \shat \plane \\
& = & - \displaystyle\int_0^{\pi} \sin \thetak \dd \thetak \int_0^{2\pi} \dd \phik  \alphak  \mathbf{R} (\phik, \thetak) i \yhat \exp(ikz)
\end{array} \label{E_Bxrot}
\end{equation}
where $\alphak = f(\phik) \delta (\thetak - \thetak')$. By construction, the field $\mathbf{B}_{\krho}^{\text{TE}}$ is an eigenstate of $H$ and  $P_{\rho}$. In order to make it a Bessel beam, a right function $f(\phik)$ needs to be chosen so that $\mathbf{B}_{\krho}^{\text{TE}}$ is an eigenstate of $J_z$. It can be proven that $f(\phik)= \frac{1}{2\pi} \epphik $ does exactly this, and it also assures the unitarity of the transformation:
\begin{equation} 
\begin{split} 
J_z [\mathbf{B}_{\krho}^{\text{TE}}]  = &  \dfrac{\text{Id}-\mathbf{R}(\dd \phi,0)}{i\dd \phi} [\mathbf{B}_{\krho}^{\text{TE}}] \\
 =  & - \int_0^{\pi} \sin \thetak \dd \thetak \int_0^{2\pi} \dd \phik f(\phik) \delta (\thetak - \thetak') \dfrac{\rotation - \mathbf{R}(\phik + \dd \phi, \thetak) }{i\dd \phi} i \yhat \exp(ikz) \\ 
 = & \int_0^{\pi} \sin \thetak \dd \thetak \int_0^{2\pi} \dd \phik \dfrac{\partial f(\phik)}{\partial \phik} \delta (\thetak - \thetak') \rotation  \yhat \exp(ikz)  =  m \mathbf{B}_{\krho}^{\text{TE}}
 \label{E_JzB}
\end{split} \end{equation}
where an integration by parts has been carried out on the last step. It is straightforward to see that similar Bessel beams could have been obtained if instead of rotating $i\yhat \exp(ikz)$, the other three plane waves listed on equations (\ref{E_p-x}-\ref{E_ep-r}) had been used. This fact allows for the definition of two sets of EM modes. I will denote the first set as $\Bx$, with $(y)= \text{TE/TM}$, or Bessel beams\footnote{Due to the fact that these are the Bessel beams that the community generally uses, I denote them as Bessel beams, instead of Bessel beams with a TE/TM character.}. Besides $H$, they are eigenstates of the following operators:
\renewcommand{\arraystretch}{1}
\begin{equation} \begin{array}{lll}
P_{\rho} \left[ \Bx \right]& = & \krho \ \Bx \\ 
J_z \left[ \Bx  \right] & = & m \  \Bx 
\label{E_Bx_eigen}
\end{array} \end{equation}
Their expressions as functions of $\shat$ and $\phat$ waves are:
\renewcommand{\arraystretch}{1.5}
\begin{equation} \begin{array}{lll}
\BTE = \dfrac{1}{2\pi}\displaystyle\int_0^{\pi} \sin \thetak \dd \thetak \int_0^{2\pi} \dd \phik  \delta (\thetak - \thetak') \epphik  \shat \plane \\
\BTM = \dfrac{1}{2\pi}\displaystyle\int_0^{\pi} \sin \thetak \dd \thetak \int_0^{2\pi} \dd \phik  \delta (\thetak - \thetak') \epphik  \phat \plane
\label{E_BBplane}
\end{array} \end{equation}
Whilst their expressions in the real space are:
\begin{equation} \begin{array}{lll}
\BTE & = & i^m \sqrt{\dfrac{\krho}{4\pi}} \left[ i J_{m+1}(\krho \rho) e^{i (m+1) \phi} \ \smhat + i J_{m-1}(\krho \rho) e^{i (m-1) \phi} \ \sphat  \right] e^{ik_z z}\\
\BTM & = & i^m \sqrt{\dfrac{\krho}{4\pi}} \left[ - \dfrac{k_z}{k} \left( i J_{m+1}(\krho \rho) e^{i (m+1) \phi} \ \smhat - i J_{m-1}(\krho \rho) e^{i (m-1) \phi} \ \sphat \right) \right. \\
&& \left. + \sqrt{2} \dfrac{\krho}{k} J_m(\krho \rho) e^{im\phi} \ \zhat \right] e^{ik_z z} \\
\label{E_BBTEM}
\end{array} \end{equation}
where $\krho = k \sin \thetak'$ and the following equality has been used \cite{Stratton1941}:
\renewcommand{\arraystretch}{1}
\begin{equation}
\plane = \sum_n i^n J_n(\krho \rho ) \exp(i n (\phi - \phik)) \exp(i\kz z)
\end{equation}
with $J_n(\krho \rho)$ being a Bessel function of the first kind \cite{Abramowitz1970,Watson1995}. An analogous procedure can be followed to derive the Bessel beams with a well-defined helicity, $\Bmp$. Their expressions as functions of the $\ephat$ waves can be compacted into:
\begin{equation} 
\Bmp = \dfrac{1}{2\pi}\displaystyle\int_0^{\pi} \sin \thetak \dd \thetak \int_0^{2\pi} \dd \phik  \delta (\thetak - \thetak') \epphik  \ephat \plane 
\end{equation}
Which yields the following expression for the real space fields:
\begin{equation} \begin{array}{lll}
\Bmp & = & \sqrt{\dfrac{\krho}{2\pi}}i^m e^{i\kz z} \left[ \dfrac{i}{\sqrt{2}} \left(  (1-p\dfrac{\kz}{k})J_{m+1}(\krho\rho)e^{i(m+1)\phi} \ \smhat \right. \right. \\
&& \left. \left. +(1+p\dfrac{\kz}{k})J_{m-1}(\krho\rho)e^{i(m-1)\phi} \ \sphat \right) +p\dfrac{\krho}{k}J_m(\krho\rho)e^{im\phi} \ \zhat \right] \\
\end{array} \label{E_Bmp} \end{equation}
As for the symmetries of $\Bmp$, they are eigenstates of $H$, and the following operators:
\begin{equation} \begin{array}{lll}
P_{\rho} \left[ \Bmp \right]& = & \krho \ \Bmp \\
J_z \left[ \Bmp  \right] & = & m \  \Bmp  \\ 
\Lambda \left[ \Bmp \right] & = & p \  \Bmp   
\label{E_Bp_eigen}
\end{array} \end{equation}
It is clear, then, that the relation between the parity and helicity modes is the following one:
\begin{equation}
\Bmp = \dfrac{\BTE + p \BTM}{\sqrt{2}}
\end{equation}
As with the plane waves, the two sets of Bessel beams are also normalized:
\renewcommand{\arraystretch}{1.5}
\begin{equation}\begin{array}{lll}
\displaystyle\int_{\mathbb{R}^3} (\Bx)^* \cdot \mathbf{B}_{m'\krho'}^{(y')} \dd^3 \mathbf{r} & = & \delta(\krho-\krho')\delta(k_z-k_z')\delta_{mm'}\delta_{yy'} \\
\displaystyle\int_{\mathbb{R}^3} (\Bmp)^* \cdot \mathbf{B}_{m'\krho'}^{(p')} \dd^3 \mathbf{r}  & = & \delta(\krho-\krho')\delta(k_z-k_z')\delta_{mm'}\delta_{pp'} \\
\end{array} 
\end{equation}
Finally, for completeness, the decomposition of plane waves into Bessel beams is also given:
\renewcommand{\arraystretch}{1.5}
\begin{eqnarray}
\shat \plane & = & \int_0^{\pi} \sin \thetak' \dd \thetak' \sum_{m=-\infty}^{\infty} \delta(\thetak' - \thetak) \emphik \BTE \\
\phat \plane & = & \int_0^{\pi} \sin \thetak' \dd \thetak' \sum_{m=-\infty}^{\infty} \delta(\thetak' - \thetak) \emphik \BTM \\
\ephat \plane & = & \int_0^{\pi} \sin \thetak' \dd \thetak' \sum_{m=-\infty}^{\infty} \delta(\thetak' - \thetak) \emphik \Bmp
\end{eqnarray}

\section{Multipolar fields}\label{Ch1_multipoles}
The multipolar fields\footnote{Here, I will call multipolar fields to those multipolar fields with a well-defined parity. I will show that multipolar fields can also have a well-defined helicity, and those will be denoted as multipolar fields with a well-defined helicity.} are EM modes specially useful to describe EM interactions with spherical symmetry. They are widely used in antenna theory \cite{Jorio2002}, Mie Theory (see sections (\ref{Ch2}-\ref{Ch4})), nuclear \cite{Alladio1986}, atomic and molecular physics \cite{Porsev2006,Afanasev2013,Brasselet1998}, astrophysics \cite{Antiochos1999}, and many more fields. The fact that they are used in such vast variety of fields makes a unique notation almost impossible. Sometimes, even within the same field, different notations are used, making it hard to reconcile calculations from different sources \cite{Nora2012}. My notation is very similar to Rose's \cite{Rose1955,Rose1957}, and it is based on the symmetries of the modes. I will denote a multipolar field as $\Ay$, where $(y)=(m),(e)$ accounts for the parity of the field\footnote{Remember that a TE parity (or mode) is also called `magnetic'; while a TM parity can also be called 'electric'.}. The notation becomes clear when we look at the symmetries of the modes. Besides the hamiltonian $H$, these modes are also symmetric under:
\renewcommand{\arraystretch}{1}
\begin{equation} \begin{array}{llllll}
J^2 \left[ \Am \right]& = & j(j+1) \ \Am & \qquad J^2 \left[ \Ae \right]& = & j(j+1) \ \Ae \\
J_z \left[ \Am  \right] & = & m_z \  \Am  & \qquad J_z \left[ \Ae  \right] & = & m_z \ \Ae \\
\Pi \left[ \Am \right] & = & (-1)^j \  \Am  & \qquad \Pi \left[ \Ae \right] & = & (-1)^{j+1} \  \Ae 
\label{E_Ax_eigen}
\end{array} \end{equation}
Like Bessel beams, they can also be expressed as general superpositions of plane waves. Their expressions can be found in \cite{Cohen1997,Tung1985,Ohtsu1999}: 
\begin{equation} \begin{array}{lll}
\Am & = & \dfrac{2j+1}{4\pi }\displaystyle\int_0^{\pi} \sin \thetak \dd \thetak \int_0^{2\pi} \dd \phik  \Djmp \dfrac{\left( \shat + \phat \right) }{\sqrt{2}} \plane \\
\Ae & = & \dfrac{2j+1}{4\pi}\displaystyle\int_0^{\pi} \sin \thetak \dd \thetak \int_0^{2\pi} \dd \phik  \Djmp \dfrac{\left( \shat - \phat \right) }{\sqrt{2}} \plane
\label{E_Ame_sp}
\end{array} \end{equation} 
where $\Djmp$ is the Wigner rotation matrix \cite{Rose1957,Tung1985}. Note that the symmetry requirements defined by equations (\ref{E_Ax_eigen}) do not imply anything about the radial dependence of the multipoles. Nonetheless, in the plane-wave decomposition given by equations (\ref{E_Ame_sp}), the radial function has been chosen so that these multipoles are regular at the origin\footnote{Notice that the radial dependence determines the behaviour of the multipolar field at infinity. In the next couple of pages, another radial function will be introduced. This radial function, known as Hankel function, is frequently used to model emission or scattering problems.}. Thus, they have the following expressions in the real space \cite{Rose1955,Bohren1983,Mishchenko2002,Nora2012}:
\renewcommand{\arraystretch}{1.5}
\begin{equation} \begin{array}{lll}
\Am & = & C_{jm_z} \left[ i \dfrac{ m_z}{\sin \theta} P_j^{m_z}(\eta) j_j(kr) e^{im_z\phi} \ \mathbf{\hat{\theta}}  -  \dfrac{\dd P_j^{m_z}(\eta) }{\dd \theta} j_j(kr) e^{im_z\phi} \ \mathbf{\hat{\phi}}   \right]\\
\Ae & = & C_{jm_z} \left[ j(j+1)P_j^{m_z}(\eta) \dfrac{j_j(kr)}{kr} e^{im_z\phi} \ \rhat + \dfrac{1}{kr} \dfrac{\dd \left[ kr j_j(kr)  \right]}{\dd (kr)}   \dfrac{\dd P_j^{m_z}(\eta) }{\dd \theta} e^{im_z\phi} \ \mathbf{\hat{\theta}} \right. \\
&& \left.  + ie^{im_z\phi} \dfrac{m_z }{\sin \theta} P_j^{m_z}(\eta) \dfrac{1}{kr} \dfrac{\dd \left[ kr j_j(kr)  \right]}{\dd (kr)}  \ \mathbf{\hat{\phi}} \right]
\label{E_Ame_real}
\end{array} \end{equation}
where $j_j(kr)$ is a spherical Bessel function \cite{Abramowitz1970,Watson1995}, $P_j^{m_z}(\eta)$ is a generalized Legendre polynomial \cite{Morse1953,Abramowitz1970}, $\eta=\cos \theta$, $\left\lbrace \rhat, \mathbf{\hat{\phi}}, \mathbf{\hat{\theta}}    \right\rbrace$ are the polarization vectors in spherical coordinates, and $C_{jm_z}$ is a constant so that the modes are normalized. Its expression is given by: 
\renewcommand{\arraystretch}{1}
\begin{equation}
C_{jm_z}^2 = \dfrac{4 \pi (2j+1)(j+m_z)!}{2k^2 j (j+1)(j-m_z)!}
\label{E_Mul_norm}
\end{equation}
Indeed, with $C_{jm_z}$ the multipolar fields are an orthonormal basis:
\begin{equation} \begin{array}{lll}
\displaystyle\int_{\mathbb{R}^3} (\Ay)^* \cdot \mathbf{A}_{j'm_z'}^{(y')} \dd^3 \mathbf{r}  & = &  \delta(k-k')\delta_{jj'}\delta_{m_zm_z'}\delta_{yy'} \\
\end{array} \end{equation}
Similarly, to the plane waves and Bessel beams, a new set of multipolar fields can be defined by adding and subtracting the electric and magnetic modes:
\begin{equation}
\Ajmp = \dfrac{\Am + i \Ae} {\sqrt{2}} \qquad \Ajmm = \dfrac{\Am - i \Ae} {\sqrt{2}}
\end{equation}
This implies that their plane wave decomposition is \cite{Tung1985}:
\renewcommand{\arraystretch}{1.5}
\begin{equation} \begin{array}{lll}
\Ajmp & = & \dfrac{2j+1}{4\pi}\displaystyle\int_0^{\pi} \sin \thetak \dd \thetak \int_0^{2\pi} \dd \phik  \Djmp \epphat \plane \\
\Ajmm & = & \dfrac{2j+1}{4\pi}\displaystyle\int_0^{\pi} \sin \thetak \dd \thetak \int_0^{2\pi} \dd \phik  \Djmp \epmhat \plane
\label{E_Ajmp_ep}
\end{array} \end{equation}
which yields the following real space expression:
\begin{equation} \begin{array}{lll}
\Ajmpp & = & \dfrac{C_{jm_z}}{\sqrt{2}} \left[ ip j(j+1)P_j^{m_z}(\eta) \dfrac{j_j(kr)}{kr} e^{im_z\phi} \ \rhat \right. \\
&& + i e^{im_z\phi} \left( \dfrac{ m_z}{\sin \theta} P_j^{m_z}(\eta) j_j(kr) + p  \dfrac{1}{kr} \dfrac{\dd \left[ kr j_j(kr)  \right]}{\dd (kr)}   \dfrac{\dd P_j^{m_z}(\eta) }{\dd \theta} \right)  \mathbf{\hat{\theta}}  \\
&& \left. - e^{im_z\phi} \left( \dfrac{\dd P_j^{m_z}(\eta) }{\dd \theta} j_j(kr) e^{im_z\phi} + p \dfrac{m_z }{\sin \theta} P_j^{m_z}(\eta) \dfrac{1}{kr} \dfrac{\dd  \left[ kr j_j(kr)  \right]}{\dd (kr)} \right)  \mathbf{\hat{\phi}}   \right]
\label{E_Ajmp_real}
\end{array} \end{equation}
Besides of the $H$ operator, the multipolar fields with well-defined helicity $\Ajmpp$ are eigenstate of the following three operators:
\renewcommand{\arraystretch}{1}
\begin{equation} \begin{array}{llllll}
J^2 \left[ \Ajmpp \right]& = & j(j+1) \ \Ajmpp \\
J_z \left[ \Ajmpp  \right] & = & m_z \ \Ajmpp \\
\Lambda \left[ \Ajmpp \right] & = & p \  \Ajmpp
\end{array} \end{equation}
Finally, by construction, $\Ajmpp$ are also normalized:
\begin{equation} \begin{array}{lll}
\displaystyle\int_{\mathbb{R}^3} (\Ajmpp)^* \cdot \mathbf{A}_{j'm_z'}^{p'} \dd^3 \mathbf{r}  & =&   \delta(k-k')\delta_{jj'}\delta_{m_zm_z'}\delta_{pp'} \\
\end{array} \end{equation}
Now, as it will be clear in chapter \ref{Ch2}, there is a need to define other multipolar fields. Indeed, the two sets of multipolar fields defined by equations (\ref{E_Ame_real}, \ref{E_Ajmp_real}) have a far-field ($kr \rightarrow \infty$) behaviour which is not desired for scattering problems \cite{Bohren1983,GLMT_book,Jackson1998}. This behaviour stems from the spherical Bessel function:
\begin{equation}
\lim_{kr \rightarrow \infty}  j_j(kr) \rightarrow \dfrac{\sin \left( kr - \dfrac{j\pi}{2}  \right)}{kr} \label{E_jj}
\end{equation}
That is, the multipolar fields given by equations (\ref{E_Ame_real}, \ref{E_Ajmp_real}) behave as stationary spherical waves in the far-field. Nevertheless, lots of EM problems with spherical symmetry have an emitting nature, \textit{e.g.} radiation emitted from antennas \cite{Gibson1979}, quantum dots \cite{Pelton1999,Boz2010}, nitrogen-vacancies \cite{Gubanov1996}, electrons \cite{Frank1991}, among others. One of these problems, which lays in the core of this thesis, is the Mie scattering problem, where a sphere embedded in a homogeneous medium scatters light that is illuminated on it \cite{Bohren1983,GLMT_book,Mishchenko2002,vandeHulst1957}. In all these problems, the spherical Bessel function of first kind must be replaced by a Hankel function of first kind $h^{(I)}_j(kr)$\footnote{If the harmonic dependence of all the fields in the problem had been chosen to be $e^{i\omega t}$ instead of $e^{-i\omega t}$, the Hankel function of second kind $h^{(II)}_j(kr)$ should have been chosen \cite{GLMT_book}.} \cite{Jackson1998,Watson1995}. Because $h^{(II)}_j(kr)$ will not be used throughout this thesis, I denote the Hankel function of the first kind simply as $h_j(kr)$. In the far-field, the Hankel function produces a desired response, \textit{i.e.} it behaves like an outgoing spherical wave:
\begin{equation}
\lim_{kr \rightarrow \infty}  h_j(kr) \rightarrow (-i)^{j+1}\dfrac{e^{ikr}}{kr} \label{E_hj}
\end{equation}
The real expressions of these new multipolar fields is analogous to the ones presented by equations (\ref{E_Ame_real}, \ref{E_Ajmp_real}), where $h_j(kr)$ replaces $j_j(kr)$:
\renewcommand{\arraystretch}{1.5}
\begin{equation} \begin{array}{lll}
 \mathbf{A}_{jm_z}^{(m),h}& = & C_{jm_z} \left[ i \dfrac{ m_z}{\sin \theta} P_j^{m_z}(\eta) h_j(kr) e^{im_z\phi} \ \mathbf{\hat{\theta}}  -  \dfrac{\dd P_j^{m_z}(\eta) }{\dd \theta} h_j(kr) e^{im_z\phi} \ \mathbf{\hat{\phi}}   \right]\\
 \mathbf{A}_{jm_z}^{(e),h} & = & C_{jm_z} \left[ j(j+1)P_j^{m_z}(\eta) \dfrac{h_j(kr)}{kr} e^{im_z\phi} \ \rhat + \dfrac{1}{kr} \dfrac{\dd \left[ kr h_j(kr)  \right]}{\dd (kr)}   \dfrac{\dd P_j^{m_z}(\eta) }{\dd \theta} e^{im_z\phi} \ \mathbf{\hat{\theta}} \right. \\
&& \left.  + ie^{im_z\phi} \dfrac{m_z }{\sin \theta} P_j^{m_z}(\eta) \dfrac{1}{kr} \dfrac{\dd \left[ kr h_j(kr)  \right]}{\dd (kr)}  \ \mathbf{\hat{\phi}} \right] \\
 \mathbf{A}_{jm_z}^{p,h} & = & \dfrac{C_{jm_z}}{\sqrt{2}} \left[ ip j(j+1)P_j^{m_z}(\eta) \dfrac{h_j(kr)}{kr} e^{im_z\phi} \ \rhat \right. \\
&& + i e^{im_z\phi} \left( \dfrac{ m_z}{\sin \theta} P_j^{m_z}(\eta) h_j(kr) + p  \dfrac{1}{kr} \dfrac{\dd \left[ kr h_j(kr)  \right]}{\dd (kr)}   \dfrac{\dd P_j^{m_z}(\eta) }{\dd \theta} \right)  \mathbf{\hat{\theta}}  \\
&& \left. - e^{im_z\phi} \left( \dfrac{\dd P_j^{m_z}(\eta) }{\dd \theta} h_j(kr) e^{im_z\phi} + p \dfrac{m_z }{\sin \theta} P_j^{m_z}(\eta) \dfrac{1}{kr} \dfrac{\dd  \left[ kr h_j(kr)  \right]}{\dd (kr)} \right)  \mathbf{\hat{\phi}}   \right]
\end{array} \end{equation}
However, the plane wave decomposition of $\mathbf{A}_{jm_z}^{(m),h}, \mathbf{A}_{jm_z}^{(e),h}, \mathbf{A}_{jm_z}^{p,h}$ changes dramatically. The Hankel function is singular in the origin, $\displaystyle\lim_{kr \rightarrow \infty} \vert h_j(kr) \vert \rightarrow \infty$, and consequently it cannot be expanded only as a superposition of propagating plane waves. Evanescent waves must appear in the superposition. Moreover, it is necessary to split the domain into two subspaces, $z \gtrless 0$. The decomposition is given by the following equation \cite{Wittmann1988}: 
\begin{equation} \begin{array}{lll}
\mathbf{A}_{jm_z}^{(m),h} & = & \dfrac{2j+1}{2\pi}\displaystyle\int_{-\infty}^{\infty} \dd k_x \int_{-\infty}^{\infty} \dd k_y \  \Djmp \ \dfrac{\left( \shat + \phat \right) }{\sqrt{2}} \dfrac{\plane}{\kappa k} \\
\mathbf{A}_{jm_z}^{(e),h} & = & \dfrac{2j+1}{2\pi}\displaystyle\int_{-\infty}^{\infty} \dd k_x \int_{-\infty}^{\infty} \dd k_y \ \Djmp \ \dfrac{\left( \shat - \phat \right) }{\sqrt{2}} \dfrac{\plane}{\kappa k} \qquad z > 0\\
\mathbf{A}_{jm_z}^{p,h} & = & \dfrac{2j+1}{2\pi}\displaystyle\int_{-\infty}^{\infty} \dd k_x \int_{-\infty}^{\infty} \dd k_y \ \Djmp \ \ephat \dfrac{\plane}{\kappa k} 
\label{E_Ajm_Hank}
\end{array} \end{equation}
where $\kappa$ is given by:
\renewcommand{\arraystretch}{1}
\begin{equation}
\kappa = k_z = \left\lbrace \begin{array}{ll} 
\sqrt{k^2 - \krho^2}, & \quad k \geq \krho \\
i \sqrt{k^2 - \krho^2}, & \quad k < \krho 
\end{array} \right. 
\end{equation}
and the spherical momentum variables $\phik$ and $\thetak$ must be computed as functions of $k_x,k_y$:
\renewcommand{\arraystretch}{1.5}
\begin{equation} \begin{array}{lll}
\phik & = & \arctan \left( \dfrac{k_y}{k_x}\right) \\
\thetak & = &  \arccos \left( \dfrac{\sqrt{k^2-k_x^2-k_y^2}}{k} \right) 
\end{array} \end{equation}
The expression for the semi-plane $z<0$ can be obtained substituting the factor $\plane / (\kappa k)$ by $\exp(-i \textbf{k} \cdot \textbf{r})/(\kappa k)$ \cite{Wittmann1988}. Equations (\ref{E_Ajm_Hank}) have large implications in nano-optics. Evanescent waves play a key role in obtaining information beyond the diffraction limit of light \cite{Novotny2006}. Here, it is shown that evanescent waves are an integral part of Hankel multipolar fields. Furthermore, it is seen that, like evanescent waves, Hankel multipolar fields are intrinsically linked to interfaces, as they are not well-defined on the whole space. This fact is largely exploited in the excitation of WGMs (see section \ref{Ch3_WGM}). As described by Oraevsky in \cite{Oraevsky2002}, given a sphere of radius $R$, Hankel multipolar fields have an evanescent dependence for a short range of $r$'s outside the spherical surface:
\renewcommand{\arraystretch}{1}
\begin{equation}
\vert \mathbf{A}_{jm_z}^{(m),h} \vert \propto \exp \left[  - \beta r \right], \qquad r>R
\end{equation}  
Last but not least, some authors have managed to link the evanescent behaviour of plane waves to complex rotations \cite{Zvyagin1998,Bekshaev2013}. In fact, this is what is seen in the expressions (\ref{E_Ajm_Hank}), where complex angles would be needed if the double integral $\int \int \dd k_x \dd k_y$ was to be expressed in spherical coordinates.

\section{Scattering control}\label{Ch1_scatt}
The work presented in this thesis belongs to the field of nano-optics. That is, most of the work deals with light-matter interactions where the light is at optical wavelengths ($\lambda \in [390 - 780 ]$ nm \cite{Saleh2007}) and the matter is a nano-structure. I will be specially interested in single nano-structures, therefore it will be almost mandatory to tightly focus the beam onto the structure, so that the light-matter interaction is stronger. However, this is not the typical approach in nanophotonics or metamaterials, where structures are mostly designed to interact with paraxial Gaussian beams or plane waves \cite{Ferry2008,Fano2010,Sol2011,Schaferling2012}. In general, the approach followed in the design of nano-circuits or nano-materials is the following one. The nano-structure is characterized by its scattering matrix $\Sconv$. This scattering matrix is a function of many geometrical and material properties of the system, $\Sconv(\mathbf{g},\mathbf{m})$, where $\mathbf{g}$ and $\mathbf{m}$ are general sets of variables describing the geometrical and material properties of the structure. However, the properties of the structure do not depend on the incident field. Then, the response of the structure to an incoming field $\Ein$ can be cast as a convolution of $\Sconv(\mathbf{g},\mathbf{m})$ with $\Ein$:
\begin{equation}
\Eout(\mathbf{r}) = \left( \Sconv(\mathbf{g},\mathbf{m}) *  \Ein \right) (\mathbf{r})
\end{equation}
Since the properties of the structure do not depend on $\Ein$ and the incoming field is well-known (a plane wave), the light-matter interaction is reduced to a complete characterization of the scattering matrix $\Sconv(\mathbf{g},\mathbf{m}) $. Then, adjusting the geometry or material of the structure, a controlled interaction can be carried out. In this thesis, I will look at these light-matter interactions from a different perspective \cite{Zambrana2014RNSW}. I will study single highly symmetric nano-structures whose scattering matrix are analytical or easy to compute. Then, I will use different plane waves decompositions to control the scattering. That is, instead of controlling $\Eout$ with the geometry and materials of the structure, the interaction will be controlled with the incoming field $\Ein$:
\begin{equation}
\Eout(\mathbf{r}) = \int \dd^3 \mathbf{k_i'} \ \plane  \left( \Sconv(\mathbf{g},\mathbf{m}) *  \Ein (\mathbf{k_i'}) \right) (\mathbf{r})
\end{equation}
where $\mathbf{k_i'}$ is each of the different plane waves that take part in the superposition, and $\Ein (\mathbf{k_i'})$ is the Fourier transform of the incident field, which modulates the plane wave decomposition. Thus, it will be crucial to know the analytical expression and the symmetries of the incoming beam. In addition, as previously mentioned, due to the fact that single structures will be used, tightly focused light beams will be used to strengthen the light-matter interaction. Tightly focused beams can be analytically described with the aplanatic model of a lens. This is explained in the following section.

\section{Aplanatic lens model}\label{Ch1_apla}
\begin{figure}[tbp]
\centering
\includegraphics[width=13cm]{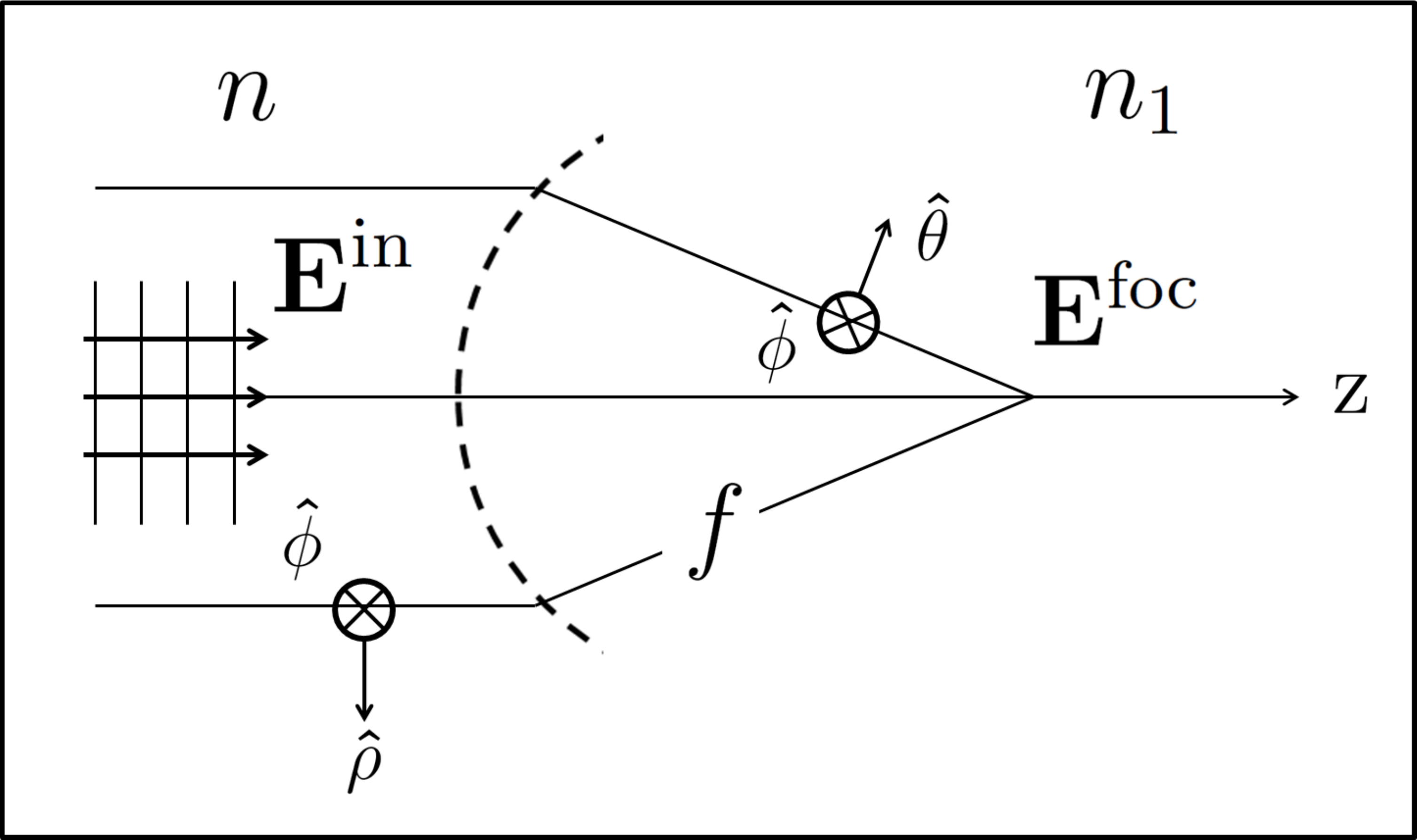} 
\caption{Schematics of the aplanatic lens model. $\Ein$ is the incident collimated electric field; $\mathbf{E}^{\mathrm{foc}}$ is the focused field; $n$, $n_1$ are the refractive indexes of the media embedding the lens before and after the field is focused;  $\left\lbrace \mathbf{\hat{\rho}} , \mathbf{\hat{\phi}} \right\rbrace$ and $\left\lbrace \mathbf{\hat{\phi}}, \mathbf{\hat{\theta}} \right\rbrace$ are polarization vectors in cylindrical and spherical coordinates. \label{Fig21}}
\end{figure}

In this section, I study one of the most frequently used models to describe the focusing of light beams, \textit{i.e.} the aplanatic lens model \cite{Wolf1959I,WolfII1959,Born1999,Youngworth2000,Novotny2006}. In particular, I will be especially interested in the symmetries of this model. I will show that the aplanatic lens model does not change the helicity nor the AM content of the beam as long as some conditions are met. When those are met, the focusing of a Bessel beam with a well-defined helicity such as $\Bmp$ can be easily modelled as $\mathbf{B}_{m\krho}^{p,\mathrm{foc}}=\int \dd \krho' \mathbf{B}_{m\krho'}^{p}$. Thus, $\mathbf{B}_{m\krho}^{p,\mathrm{foc}}$ will still be an eigenvector of $\Lambda$ and $J_z$. But before getting there, the aplanatic model of a lens is explained. The variables playing a role in the model are the collimated incident beam $\Ein$, the focal distance of the lens $f$, and the refractive indexes of the media on the two sides of the lens $n$, $n_1$. The geometrical representation of the aplanatic model is given by Figure \ref{Fig21}.
As it can be seen there, the collimated beam $\Ein$ travels in a medium whose refractive index is $n$ and hits the back-aperture of a lens. The analytical description of the collimated beam at the a $z$ plane of the back-aperture of the lens is done in the paraxial approximation, \textit{i.e.} $\Ein = E^{in}  \mathbf{u}$ with $\mathbf{u}$ being the polarisation vector. The incident beam gets refracted and its expression $\Einf$ is obtained as: 
\begin{equation}
\Einf = \left[ t^s(\phik,\thetak) \left( \Ein \cdot \mathbf{\hat{\phi}}   \right) \mathbf{\hat{\phi}}   + t^p(\phik,\thetak) \left( \Ein \cdot \mathbf{\hat{\rho}}  \right) \mathbf{\hat{\theta}}    \right] \sqrt{\frac{n}{n_1}}(\cos \thetak)^{1/2} 
\label{E_Einf}
\end{equation}
where $t^s(\phik,\thetak),t^p(\phik,\thetak)$ are the effective Fresnel coefficients of the lens for $\shat$ and $\phat$ waves respectively \cite{Novotny2006,Born1999}; $\left\lbrace \mathbf{\hat{\rho}} , \mathbf{\hat{\phi}} , \zhat \right\rbrace$ and $\left\lbrace \rhat, \mathbf{\hat{\phi}}, \mathbf{\hat{\theta}} \right\rbrace$ are the polarization vectors in cylindrical and spherical coordinates (see Figure \ref{Fig21}). Note that three assumptions have been made here. First, the sine condition of geometrical optics has been considered. The sine condition states that any ray emerging or converging to the focus of an aplanatic system meets its conjugate ray at the surface of a sphere of radius $f$, at a distance of the optical axis $h=f\sin \theta$, where $\theta$ is the divergence angle \cite{Novotny2006}. Second, the intensity law has also been assumed. The intensity law implies that the energy incident on the aplanatic lens equals the energy leaving it. It can be expressed as $\vert \mathbf{E}^{\mathrm{foc}} \vert = \vert \Ein \vert \sqrt{n \mu/(n_1 \mu_1)} \cos^{1/2}\theta$ \cite{Novotny2006}. Finally, it is imposed that the real space-expression of $\Ein$ on that $z$ plane gives rise to the angular spectrum of the focused beam. That is, when introducing the expression of $\Ein$ in equation (\ref{E_Einf}), the following change of variables needs to be done:
\begin{equation}
\Ein(\rho,\phi)=\Ein(f\sin \thetak, \phik)
\label{E_apl_spectr}
\end{equation}
where no $z$ dependence is supposed as the expression is taken on a plane with a constant $z$. With all this, the electric field at the focus of the lens is obtained as:
\renewcommand{\arraystretch}{1.5}
\begin{equation}
\begin{array}{ll}
\mathbf{E}^{\mathrm{foc}} & = \dfrac{i k f e^{(-ikf)}}{2\pi} \displaystyle\int_0^{\thetak^{M}} \int_0^{2\pi}  \sin \thetak  \dd \thetak \dd \phik \ \Einf e^{(i k z \cos \theta)} e^{(i k \rho \sin \theta \cos (\phi - \phik))} \\ 
& =  \dfrac{i k f e^{(-ikf)}}{2\pi} \displaystyle\int_0^{\thetak^{M}} \int_0^{2\pi}  \sin \thetak  \dd \thetak \dd \phik \ \Einf \plane
\label{E_apl_model}
\end{array} \end{equation}
with $f$ being the focal distance of the lens, and $\thetak^{M} = \arcsin(\text{NA})$ with NA the numerical aperture of the lens.

\subsection{$\Lambda$ preservation}\label{Ch1_apla_hel}
The fact that the aplanatic model preserves the helicity of light is not trivial nor a casualty. In fact, simple lenses made of glass do not preserve helicity. However, it is interesting for different applications that the polarisation of light is preserved under a microscope set-up. In order to achieve that (which implies helicity preservation), microscope objective manufacturers need to apply a special coating to microscope objectives \cite{Novotny2006,Bliok2011}. This fact is translated into a condition relating the effective Fresnel coefficients $t^s,t^p$. In order to show this condition, a general plane wave with a well-defined helicity $p$ will be focused down using the aplanatic lens model. It will be seen that the helicity of the beam is preserved provided a certain condition is met. Consider a general incident paraxial beam with a well-defined helicity $p$ within the paraxial approximation and a $\plane$ dependence:
\begin{equation}
\Ein = E^{in}(\phik,f\sin\thetak)\spphat 
\label{E_Ein_ep}
\end{equation}
where $E^{in}(\rho,\phi)$ is the amplitude of the electric field on the plane of the back-aperture of the lens. Then, the computation of $\Einf$ yields:
\begin{equation}
\Einf = E^{in}(\phik,f\sin\thetak) e^{ip\phik} \left[  t^s(\phik,\thetak) ip \dfrac{\mathbf{\hat{\phi}}}{\sqrt{2}}
 +  t^p(\phik,\thetak) \dfrac{\mathbf{\hat{\theta}}}{\sqrt{2}} \right]  \sqrt{\frac{n}{n_1}}(\cos \thetak)^{1/2}
\label{E_Einf_hel}
\end{equation}
where the following equalities have been used:
\begin{eqnarray}
\spphat \cdot \mathbf{\hat{\rho}} & = &  \dfrac{e^{ip\phik}}{\sqrt{2}}\\
\spphat \cdot \mathbf{\hat{\phi}} & = & ip \dfrac{e^{ip\phik}}{\sqrt{2}}
\end{eqnarray} 
Equation (\ref{E_Einf_hel}) can be expressed in terms of $\shat$ and $\phat$:
\begin{equation}
\Einf = E^{in}(\phik,f\sin\thetak) e^{ip\phik} \left[  t^s(\phik,\thetak)  \dfrac{\shat}{\sqrt{2}}
 +  t^p(\phik,\thetak) p \dfrac{\phat}{\sqrt{2}} \right]  \sqrt{\frac{n}{n_1}}(\cos \thetak)^{1/2}
\label{E_Einf_sp}
\end{equation}
where the two following relations have been used:
\begin{eqnarray}
\shat & = & - i \mathbf{\hat{\phi}} \label{E_sphi}\\
\phat & = & - \mathbf{\hat{\theta}} \label{E_ptheta}
\end{eqnarray}
Hence, examining equation (\ref{E_Einf_sp}), it is straightforward to conclude that the aplanatic lens model preserves helicity if and only if 
\begin{equation}
t^s(\phik,\thetak) = t^p(\phik,\thetak) = t(\phik,\thetak)  \quad \forall \phik,\thetak
\label{E_tstp}
\end{equation} 
I will consider that equation (\ref{E_tstp}) is valid for the rest of my theoretical calculations.

\subsection{$J_z$ preservation}\label{Ch1_apla_jz}
Looking at Figure \ref{Fig21}, one notices that the geometrical representation of the aplanatic model is cylindrically symmetric around the $z$ axis. Thanks to Noether's theorem, in principle it follows that the model preserves $J_z$. Nevertheless, in the same way as it happened in the previous subsection, the Fresnel coefficients play an important role. Indeed, next it will be proven that either the Fresnel coefficients  fulfil a certain condition, or the model is not cylindrically symmetric and therefore does not preserves $J_z$. Using equation (\ref{E_apl_model}), the expression for the electric field at the focus of the aplanatic lens when an incoming beam given by equation (\ref{E_Ein_ep}) can be computed:
\begin{equation} \begin{array}{lll}
\mathbf{E}^{\mathrm{foc}} & = & \dfrac{i k f e^{(-ikf)}}{2\pi} \displaystyle\int_0^{\thetak^{M}} \int_0^{2\pi}  \sin \thetak  \dd \thetak \dd \phik \ \Einf \plane \\
& = & \dfrac{i k f e^{(-ikf)}}{2\pi} \displaystyle\int_0^{\thetak^{M}} \int_0^{2\pi}  \sin \thetak  \dd \thetak \dd \phik \ E^{in} t e^{ip\phik}  \sqrt{\frac{n}{n_1}}(\cos \thetak)^{1/2} \ephat  \plane
\label{E_Efoc_ep}
\end{array} \end{equation}
It is interesting to see that equation (\ref{E_Efoc_ep}) can be expressed as a superposition of rotated plane waves with a well-defined helicity. The way to construct such a superposition is intuitive and reflects very well the definition of the helicity of a beam. We take a circularly polarized plane wave propagating along the z axis with helicity $p$. We rotate it in all directions and we sum all the different contributions with a certain weight $g(\phik, \thetak)$. This rotation does not change the polarization in the system of reference of the plane wave (it maintains transversality), therefore the helicity has still a well-defined value $p$. 
\begin{equation}
\label{E_gen_hel}
\mathbf{A}^p= \int_0^{\pi} \sin (\thetak) \dd \thetak \int_0^{2\pi} \dd \phik \ g ( \phik,\thetak ) \rotation \spphat \exp\left( ikz \right)
\end{equation}
where $\mathbf{A}^p$ is a general field with a well-defined helicity. Then, using equations (\ref{E_ep-l}, \ref{E_ep-r}), $\mathbf{A}^p$ can be re-written as:
\begin{equation}
\mathbf{A}^p= - \int_0^{\pi} \sin (\thetak) \dd \thetak \int_0^{2\pi} \dd \phik \ g ( \phik,\thetak ) \ephat \plane
\label{E_Ap}
\end{equation}
At this point, a direct comparison between equations (\ref{E_Ap}) and (\ref{E_Efoc_ep}) yields a function $g(\phik,\thetak)$ that makes both equations equal:
\begin{equation}
g(\phik, \thetak) = - \dfrac{ikf e^{(-ikf)}}{2\pi} E^{in}(f\sin\thetak,\phik) t(\phik,\thetak) e^{ip\phik} \left( 1-H(\thetak - \thetak^{M}) \right) \sqrt{\dfrac{n}{n_1}}(\cos \thetak)^{1/2}
\label{E_gthphi}
\end{equation}
where $H(\thetak - \thetak^{M})$ is the Heaviside step function. With $g(\phik,\thetak)$ given by equation (\ref{E_gthphi}), the field at the focus of a lens can be given by:
\begin{equation}
\mathbf{E}^{\mathrm{foc}} = \int_0^{\pi} \sin (\thetak) \dd \thetak \int_0^{2\pi} \dd \phik \ g ( \phik,\thetak ) \rotation \spphat \exp\left( ikz \right)
\end{equation}
Now, I will show that $\mathbf{E}^{\mathrm{foc}}$ has a well-defined $J_z$ if the incoming field does, too. That is, I will show that the aplanatic model of a lens preserves $J_z$ by verifying the following condition:
\begin{equation} \begin{array}{llll}
\text{If } & J_z [\Ein] = m \Ein & \Longrightarrow &  J_z [\mathbf{E}^{\mathrm{foc}}] = m \mathbf{E}^{\mathrm{foc}}
\end{array} \end{equation}
A paraxial beam with a well-defined helicity $p$ and an value of $J_z=m$\footnote{This is an abuse of language that will be used throughout the whole thesis. It means that the beam is an eigenstate of $J_z$ with eigenvalue $m$.} is given by any expression of the kind:
\begin{equation}
\mathbf{E}^{\mathrm{par}} = E^{par}(\rho)e^{im\phi}\exp(ikz) \spphat
\end{equation}
If this expression is substituted in equation (\ref{E_gthphi}), the following expression for $g(\phik,\thetak)$ follows:
\begin{equation}
g(\thetak, \phik) = \dfrac{-ikf e^{-ikf}}{2\pi} E^{par}(f\sin\thetak)e^{i(m+p)\phik} t(\phik,\thetak) \left( 1-H(\thetak - \thetak^{M}) \right) \sqrt{\dfrac{n \cos \thetak}{n_1}}
\label{E_g_par}
\end{equation}
Now, in order to prove that the aplanatic lens model does not change the $J_z$ content of $\mathbf{E}^{\mathrm{par}} $, I apply $J_z$ to $\mathbf{E}^{\mathrm{foc}}$ when $g(\phik,\thetak)$ is given by equation (\ref{E_g_par}). The fact that $\mathbf{E}^{\mathrm{foc}}$ can be expressed as a rotation helps to compute the result. As seen by equation (\ref{E_JzB}), $J_z$ can be applied as a partial derivative with respect to $\phik$ under the integral operation. Then, it can be seen that:
\begin{equation}
J_z \left[ \mathbf{E}^{\mathrm{foc}} \right]  =  \int_0^{\pi} \sin (\thetak) \dd \thetak \int_0^{2\pi} \dd \phik \ \dfrac{\partial g(\phik,\thetak)}{\partial \phik} \rotation i \spphat \exp\left( ikz \right)
\end{equation}
with $g(\phik,\thetak)$ only depending on $\phik$ via two terms, $e^{i(m+p)\phik}$ and $t(\phik,\thetak)$. As one could expect, it is easy to see that the aplanatic model preserves $J_z$ of $\Ein$ if and only if $t(\phik,\thetak) = t(\thetak)$. Throughout this thesis, I will consider that all lenses (as long as they are centred with the beam axis) preserve $J_z$ and therefore $t(\thetak) = t^s(\thetak)=t^p(\thetak)$.

\section{Overview}\label{Ch1_over}
In this chapter, different theoretical methods to describe light-matter interactions have been explained. They will be used in the remaining of this thesis, both to understand the experimental results, as well as to carry out different theoretical developments. Next, I present a summary of the methods that have been described above as well as their applications in the remaining of the thesis.\\\\
In section \ref{Ch1_HilMax}, Maxwell Equations for a linear, non-dispersive, homogeneous, isotropic and source-free medium have been introduced. These equations will be used in chapters \ref{Ch2}, \ref{Ch3} and \ref{Ch4} to deepen in understanding of the interaction of vortex beams with spherical particles.\\\\
In section \ref{Ch1_symm}, the paraxial approximation, which is a very relevant limit of Maxwell equations, has been explained. This approximation is going to be used in all the experimental chapters (\ref{Ch5}, \ref{Ch6}, and \ref{Ch7}). Then, the concepts of EM field, operator, and symmetry have been given. Of particular interest is the concept of symmetry, which is used throughout this thesis to predict theoretical results or interpret experimental results. A complete list of symmetries with their respective generators can also be found in section \ref{Ch1_symm}. Many of these generators will be used to characterize the light beams as well as the samples that will be used in this thesis. In particular, chapters \ref{Ch3}, \ref{Ch4}, \ref{Ch6}, \ref{Ch7} describe different phenomena that arise when beams with duality and cylindrical symmetry interact with samples with mirror and cylindrical symmetry. \\\\
In section \ref{Ch1_basis}, a general method to construct solutions of the vectorial Helmholtz equation is sketched. Then, plane waves are constructed using this method. Plane waves are characterised in terms of their symmetries, and they are used to construct Bessel beams, which are symmetric under some different symmetries. Both plane waves and Bessel beams are later used in section \ref{Ch5_CGH}. \\\\
Section \ref{Ch1_multipoles} introduces two different basis of multipolar fields with two different radial functions each. The understanding of the symmetries of these beams will be crucial in chapters \ref{Ch2}, \ref{Ch3} and \ref{Ch4}, where they are used to solve Maxwell equations in a spherical domain. \\\\
Section \ref{Ch1_scatt} presents in a formal way the key idea behind the work done in this thesis, which was introduced in \ref{Intro}. That is, controlling and/or characterising the scattering of nano-structures using symmetric light. This method is exploited both theoretically and experimentally in chapters \ref{Ch3}, \ref{Ch4}, \ref{Ch6} and \ref{Ch7}. It is seen that symmetric beams can be used to effectively turn a non-dual particle into dual, or inducing circular dichroism in a non-chiral smaple.\\\\
Finally, in section \ref{Ch1_apla}, the description of the aplanatic model of a lens is given. The aplanatic model is used in chapter \ref{Ch3} to theoretically control the multipolar content of a beam. Furthermore, the symmetry conditions proven in \ref{Ch1_apla_hel} and \ref{Ch1_apla_jz} are then assumed in chapters \ref{Ch6} and \ref{Ch7} to understand the experimental results.

\begin{savequote}[10cm] 
\sffamily
``I learned very early the difference between knowing the name of something and knowing something.'' 
\qauthor{Richard P. Feynman}
\end{savequote}

\chapter{Generalized Lorenz-Mie Theory}
\graphicspath{{ch2/}} 
\label{Ch2}

\newcommand\Ei{\mathbf{E}^\mathrm{i}}
\newcommand\Hi{\mathbf{H}^\mathrm{i}}
\newcommand\Eint{\mathbf{E}^\mathrm{int}}
\newcommand\Hint{\mathbf{H}^\mathrm{int}}
\newcommand\Es{\mathbf{E}^\mathrm{s}}
\newcommand\Hs{\mathbf{H}^\mathrm{s}}
\newcommand\Et{\mathbf{E}^\mathrm{t}}
\newcommand\Ht{\mathbf{H}^\mathrm{t}}
\newcommand\Ss{\mathbf{S}^\mathrm{s}}
\newcommand\St{\mathbf{S}^\mathrm{t}}
\newcommand\Si{\mathbf{S}^\mathrm{i}}
\newcommand\Sext{\mathbf{S}^\mathrm{ext}}

\newcommand{\Sop}{\overline{S}}

\newcommand{\gjm}{g_{jm_z}^{(m)}}
\newcommand{\gje}{g_{jm_z}^{(e)}}
\newcommand{\ajm}{a_{j,m_z}}
\newcommand{\bjm}{b_{j,m_z}}
\newcommand{\cjm}{c_{j,m_z}}
\newcommand{\djm}{d_{j,m_z}}

\section{Introduction}\label{Ch2_Intro}
\begin{figure}[tbp]  
\begin{center}
\setlength{\unitlength}{1cm}
\includegraphics[width=14cm]{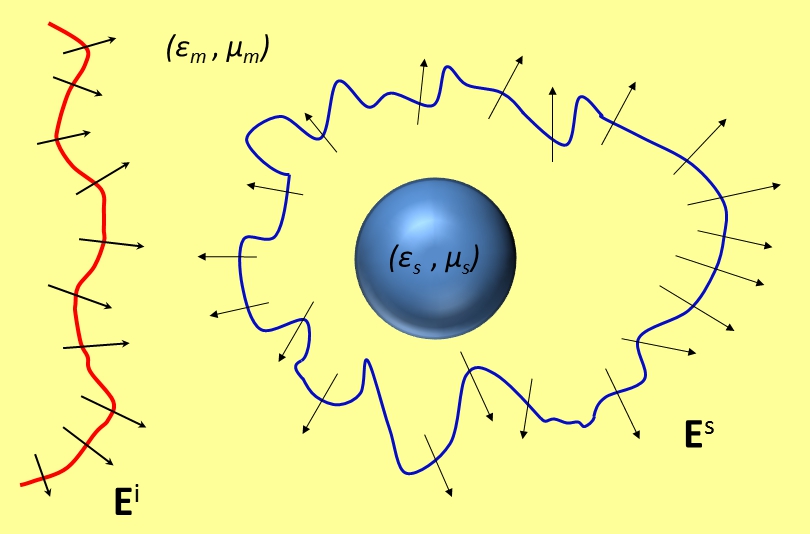}
\end{center}
\caption{Schematics of the Generalized Lorenz-Mie problem. An arbitrary EM field $\Ei$ (red) propagates through a lossless, isotropic and homogeneous medium with optical constants $(\epsilon_m,\mu_m)$. The field interacts with an isotropic, homogeneous sphere with optical constants $(\epsilon_s,\mu_s)$. As a result, the incident field is scattered in all directions giving rise to $\Es$ (blue).}
\label{F_GLMT_schematics}
\end{figure}

In this chapter, I will use the tools described in the preceding chapter to solve the electromagnetic problem described in Figure \ref{F_GLMT_schematics}. An arbitrary electromagnetic field $\Ei$ excites a single homogeneous, isotropic sphere embedded in a homogeneous, isotropic and lossless medium\footnote{Note that denoting the incident field as $\Ei$ is an abuse of language. The incident electromagnetic field has also a magnetic component $\Hi$, and it is indeed used to solve the Maxwell's equations.}. 
As mentioned in the introduction, the scattering of a sphere illuminated by a plane wave was independently solved by Ludvig Lorenz in 1890 and Gustav Mie in 1908 \cite{Mie1908, Lorenz1898}. Unfortunately for Lorenz, Mie's name has stuck in the literature and nowadays the problem is mostly known as the Mie scattering problem \cite{Logan1965}. A very complete review about the history of this problem can be found in \cite{GLMT_book}. Before the invention of lasers in 1958 \cite{Schawlow1958,Siegman1986}, Mie theory had already been successfully applied to a very wide range of fields, \textit{e.g.} chemistry, material science or atmospheric physics \cite{vandeHulst1957, Stratton1941}. However, once Mie theory started being tested with lasers, it became clear that a more general formulation of the problem had to be established. Although in many cases the light beam produced by a laser could be modelled as a plane wave with a good approximation, a more general theory that took into account the spatial non-uniformity of light beams was needed. In this context, the Generalized Lorenz-Mie Theory (GLMT) was created in the 1980's \cite{Gouesbet2009}. The GLMT development was especially led by G\'{e}rard Gouesbet and G\'{e}rard Gr\'{e}han, both Professors at the University of Rouen, France. In its first ten years, the progress of GLMT was mainly focused on the theoretical part. In 1985, Gouesbet \textit{et al.} managed to solve the scattering problem of a free propagating Gaussian beam interacting with an on-axis arbitrary sphere using Bromwich potentials \cite{Gouesbet1985}. And then, in 1988, a seminal paper was also published by Gouesbet and co-workers where the scattering of a sphere located in an arbitrary position with respect to the axis of propagation of a Gaussian beam was solved \cite{Gouesbet1988}. This paper laid the foundations of the GLMT, as it solved the scattering problem of a single sphere excited with an arbitrary beam of light \cite{Maheu1988,Gouesbet2011}. Since then, as the theoretical framework had already been laid, the development of the theory happened in the applications and computation side. Due to computation limitations at that time, a lot of effort was put into finding ways of efficiently computing the so-called beam shape coefficients of an arbitrary beam \cite{Gouesbet1990}. In addition to that, lots of applications were found, including phase-Doppler instruments, imaging, optical characterization and optical manipulation, among others \cite{Gouesbet2009}. In this chapter, I will summarize the main theoretical results of GLMT and interpret them from a symmetry-based point of view. 

\section{Symmetry considerations}\label{Ch2_Symmetries}
The wide variety of applications of GLMT is a proof of the success of the theory. However, due to the typical method of solving the problem via vector potentials (Bromwich, Debye or Hertz) \cite{Stratton1941,GLMT_book}, the symmetries of the system are usually blurred. In order to gain a bit of insight about the symmetries in the problem, let's take a look at Figure \ref{F_Mie_symmetries}, where the different symmetries of the Mie problem are presented. I have ruled out the incident beam of any consideration, as I consider it as an excitation to the system, but not part of the system itself. The system is formed by the sphere and the surrounding medium. The origin of the frame of reference in which the problem is described is placed at the centre of the sphere. This is necessary for the three last symmetry considerations to hold:
\begin{figure}[tbp]  
\begin{center}
\setlength{\unitlength}{1cm}
\includegraphics[width=14cm]{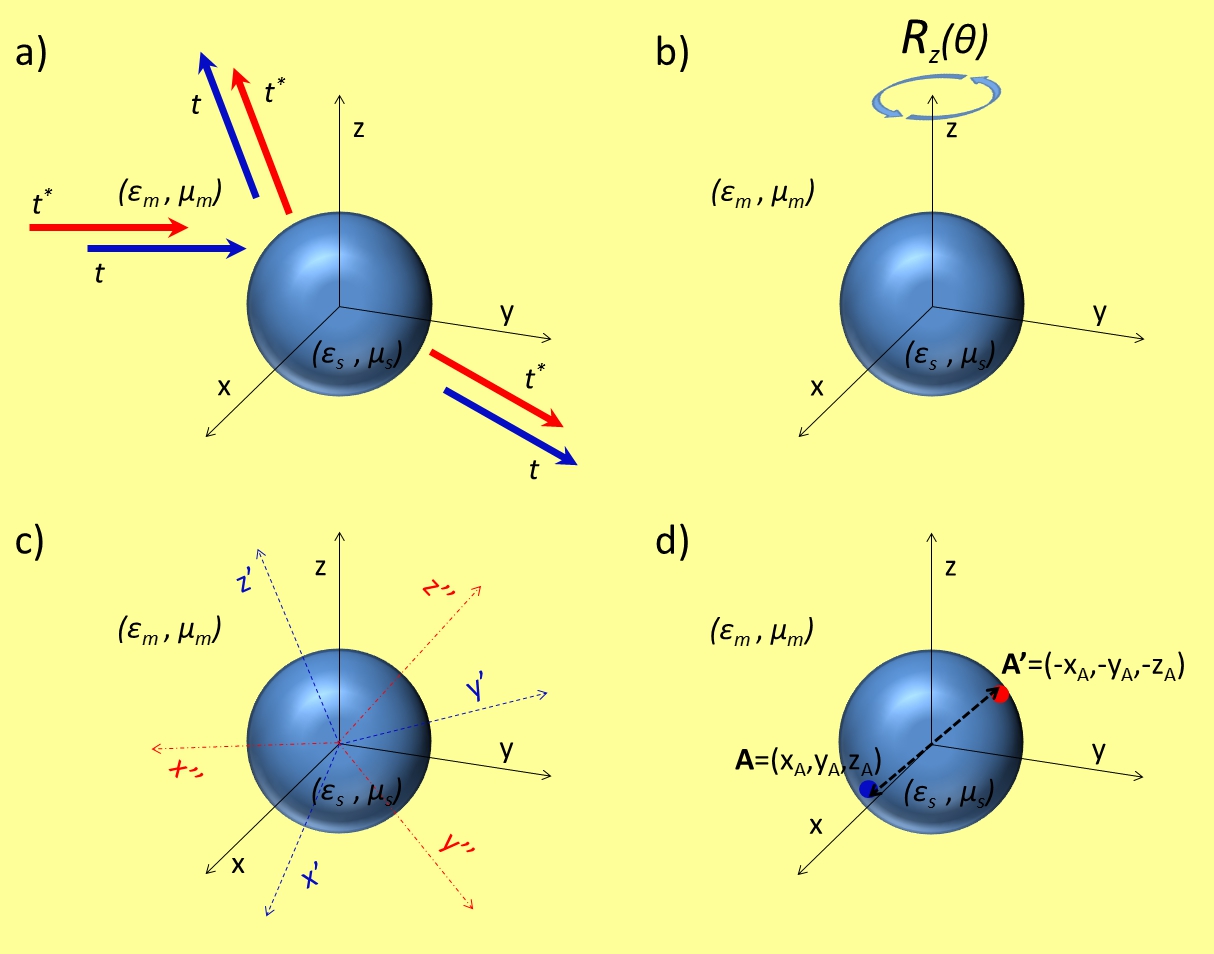}
\end{center}
\caption{Symmetries in Mie Theory. a) Temporal translational symmetry. Two exact plane waves interact with the sphere at different instants ($t$ for the blue wave, $t^*$ for the red one). The interaction is the same one for both. b) Rotations around the $z$ axis. The system remains invariant when it is rotated around the $z$ axis. c) Independence of orientation of reference frame. The problem is cylindrically symmetric around the $z$ axis independently of the orientation of the reference frame. d) Parity. A parity $\mathbf{r} \rightarrow - \mathbf{r}$ change of coordinates leaves the problem invariant.}
\label{F_Mie_symmetries}
\end{figure}
\begin{itemize}
\item \emph{Temporal translations, $T_{\Delta t}$}. Because both the sphere and the medium surrounding it are assumed to be linear media, the excitation of the sphere is independent of the instant when the sphere is excited. 
\item \emph{Rotations around the $z$ axis, $R_z$}. A rotation around the $z$ axis leaves the system invariant. This will also be referred to as cylindrical symmetry.  
\item \emph{Independence of orientation of the reference frame}. As long as the origin is the center of the sphere, the orientation of the axis will be irrelevant. That is, any axis is an axis of cylindrical symmetry.
\item \emph{Parity, $\Pi$}. The system is invariant under parity transformations, \textit{i.e.} changes of $\mathbf{r} \rightarrow - \mathbf{r} $.
\end{itemize}
These four transformations commute with each other. Thus, the natural modes of the system will be eigenmodes of these four transformations. Now, due to the fact that continuous symmetries can be expressed as a function of their generators, I will use the generators of the three first symmetries to classify the natural modes of the system (see section \ref{Ch1_symm}):
\begin{itemize}
\item Hamiltonian, $H$. Temporal translations are generated by the Hamiltonian $T_{\Delta t} = \exp(-i H \Delta t)$.
\item Projection of the AM in the $z$ axis, $J_z$. The $z$ component of the AM generates $R_z(\theta) = \exp(-i J_z \theta )$.
\item The AM squared, $J^2$. The transformation $T_J(\theta)= \exp (-i J^2 \theta)$ is meaningless. Nevertheless, the independence of orientation of axis implies that the eigenmodes of the system need to be eigenstates of $J^2$.
\end{itemize}
Parity ($\Pi$) is not a continuous transformation, therefore I will use the transformation itself to classify the eigenmodes of the system. Now, in \ref{Ch1_multipoles} we have seen that one of the complete basis of solutions of source-free Maxwell equations are the multipolar modes $\lbrace \mathbf{A}_{jm_z}^{(m)},  \mathbf{A}_{jm_z}^{(e)} \rbrace$. As shown in there, these modes are eigenmodes of $H$, $J_z$, $J^2$, and $\Pi$ respectively. Thus, the natural modes of the system depicted in Figures \ref{F_GLMT_schematics}, \ref{F_Mie_symmetries} will be the multipolar modes, as they fulfil all the symmetry requirements. A harmonic time dependence $e^{-i\omega t}$ will be assumed throughout the rest of the chapter, as non-linear phenomena are not going to be studied. 

\section{The scattering problem}\label{Ch2_Scatt}
In this section I will solve the problem sketched in Figure \ref{F_GLMT_schematics}. That is, I will find the EM field ($\E$, $\Hh$) in all space for a given arbitrary incident field $\Ei$. In order to achieve that, I will describe all the EM fields as a superposition of multipolar modes. This will simplify all the calculations, as they are the natural modes of the system. The procedure will be the following one. First, I will break down the domain into two sub-domains which will be separated by the surface of the sphere. The EM field inside the sphere will be given by $\Eint$ and $\Hint$, the interior electric and magnetic fields. The expression of the EM field outside the sphere will be given by the total field, which can be computed as the addition of the scattered field ($\Es$, $\Hs$) and the incident field ($\Ei$, $\Hi$), \textit{i.e.} $(\Et, \Ht) = (\Es + \Ei, \Hs + \Hi)$. Then, I will impose that the tangential components of the fields inside and outside the sphere are continuous \cite{Bohren1983}. To make the problem as general as possible, I will suppose that the incident electric field is the most general superposition of multipolar modes:
\begin{equation}
\Ei = \sum_{j=1}^{\infty} \  \sum_{m_z=-j}^{j} \gjm \Am + \gje \Ae  
\label{Ei}
\end{equation}
where $\gjm$ and $\gje$ are some general coefficients modulating the multipolar superposition. Then, the magnetic field can be obtained using the Maxwell equations (\ref{E_Max_EH_mono}) and the relations between multipolar modes given in \cite{Rose1957}:
\begin{equation}
\Hi = - \dfrac{ \nabla \times \Ei}{i\omega \mu k} =  \dfrac{-1}{\omega \mu} \left[ \sum_{j=1}^{\infty} \  \sum_{m_z=-j}^{j} \gjm \Ae - \gje \Am \right] 
\end{equation}
The scattered and interior fields are also supposed to be a general superposition of multipolar fields, whose expression are given by:
\begin{eqnarray}
\Es & = & \sum_{j=1}^{\infty} \  \sum_{m_z=-j}^{j} \bjm \Am + \ajm \Ae \label{E_all_fields1}\\
\Hs & = & \dfrac{-1}{\omega \mu} \left[ \sum_{j=1}^{\infty} \  \sum_{m_z=-j}^{j} \bjm \Ae - \ajm \Am \right] \\
\Eint & = & \sum_{j=1}^{\infty} \  \sum_{m_z=-j}^{j} \cjm \Am + \djm \Ae \\
\Hint & = & \dfrac{-1}{\omega \mu_1} \left[ \sum_{j=1}^{\infty} \  \sum_{m_z=-j}^{j} \cjm \Ae - \djm \Am \right]  \label{E_all_fields4}
\end{eqnarray}
where $\mu_1$ and $\mu$ are the magnetic permeabilities of the sphere and the surrounding medium respectively. However, as mentioned in \ref{Ch1_multipoles}, even though we use the same notation for the multipolar modes in scattering and free propagation, their radial functions are different. For scattering, the radial function is a Hankel function of first kind, $h_j^{(1)}$. Whereas for the interior and incident fields, the radial function is a Bessel function, $j_j$ \cite{GLMT_book,Abramowitz1970}. Once the general expression of the fields has been found, we apply the tangential Maxwell boundary conditions, which yield the following four equalities:
\begin{equation}\begin{array}{cccccl}
E_{\theta}^{\mathrm{int}}  & = &   E_{\theta}^{\mathrm{s}} + E_{\theta}^{\mathrm{i}} & \qquad E_{\phi}^{\mathrm{int}}  & = &   E_{\phi}^{\mathrm{s}} + E_{\phi}^{\mathrm{i}} \\
H_{\theta}^{\mathrm{int}}    & = &   H_{\theta}^{\mathrm{s}} + H_{\theta}^{\mathrm{i}}  & \qquad H_{\phi}^{\mathrm{int}}  & = &   H_{\phi}^{\mathrm{s}} + H_{\phi}^{\mathrm{i}}
\label{Tangential}
\end{array} \end{equation}
Equations (\ref{Tangential}) can be solved with Cramer's rule, yielding the following results:
\begin{equation} \begin{array}{ccccl}
\ajm & = & a_j \gje \qquad  \bjm & = & b_j \gjm \\
\cjm & = & c_j \gjm \qquad  \djm & = & d_j \gje
\label{General}
\end{array}
\end{equation}
where $\left\lbrace a_j,b_j,c_j,d_j\right\rbrace$ are the so-called Mie coefficients, whose expressions are given by: 
\begin{equation}
\renewcommand{\arraystretch}{2}
\begin{array}{ccccl}
\displaystyle a_j  =  \frac{\mu n_r^2 j_j(n_rx) [xj_j(x)]'-\mu_1 j_j(x)[n_r x j_j(n_rx)]'}{\mu n_r^2 j_j(n_r x)[x h_j^{(1)}(x)]' - \mu_1 h_j^{(1)}(x) [n_r x j_j(n_rx)]'} \\
\displaystyle b_j  =  \frac{\mu_1 j_j(n_r x) [xj_j(x)]'- \mu j_j(x) [n_r x j_j(n_r x)]'}{\mu_1 j_j(n_r x)[x h_j^{(1)}(x)]' - \mu h_j^{(1)}(x) [n_r x j_j(n_rx)]'} \\
\displaystyle c_j=\frac{\mu_1 j_j(x) [xh_j^{(1)}(x)]'-\mu_1 h_j^{(1)}(x) [xj_j(x)]'}{\mu_1 j_j(n_r x)[x h_j^{(1)}(x)]' - \mu h_j^{(1)}(x) [n_r x j_j(n_rx)]'} \\
\displaystyle d_j=\frac{\mu_1 n_r j_j(x) [xh_j^{(1)}(x)]'-\mu_1 n_r h_j^{(1)}(x) [xj_j(x)]'}{\mu n_r^2 j_j(n_r x)[x h_j^{(1)}(x)]' - \mu_1 h_j^{(1)}(x) [n_r x j_j(n_rx)]'}
\label{mie_coeffs}
\end{array}
\end{equation}
where $x=2\pi R / \lambda$ is the so-called size parameter, with $R$ being the radius of the sphere in consideration and $\lambda$ the excitation wavelength, inversely proportional to the frequency $\omega$; $n_r$ is the relative refractive index of the sphere with respect to the surrounding medium, $n_r=\sqrt{\mu_r \epsilon_r}$. It can be observed from equations (\ref{General}) that once the coefficients $\{ \gjm, \gje \}$ have been found for the incident field, the solution of the scattering problem is straightforward: the expression for the scattered and interior fields are a mere formal copy of the incident field, where each multipole is modulated by a Mie coefficient. Hence, the key point to solve the Generalized Lorenz-Mie problem is finding the coefficients $\{ \gjm, \gje \}$. Both $\gjm$ and $\gje$ are usually known as beam shape coefficients, except for a subtlety. In GLMT, the beam shape coefficients are defined as $A^m_j$ and $B^m_j$, and they are equal to $\gje$ and $\gjm$ divided by a factor \cite{GLMT_book}:
\begin{eqnarray}
B^m_j =\dfrac{\gjm}{c^{pw}_j} \qquad A^m_j =\dfrac{\gje}{c^{pw}_j}
\end{eqnarray}
where $c^{pw}_j$ are the plane-wave coefficients:
\begin{equation}
c^{pw}_j = \dfrac{1}{ik} (-i)^j \dfrac{2j +1}{j (j+1)}
\end{equation}
Now, given $\Ei$, we can find them applying the techniques shown in (\ref{Ch1_basis}). That is, due to the orthonormality relations between $\Am$ and $\Ae$, we can find $\gjm$ and $\gje$ projecting $\Ei$ onto the multipolar modes, yielding a unique result. Thus, from equation (\ref{Ei}), we obtain:
\begin{eqnarray}
\gjm = \int_{-1}^{1}\int_0^{2\pi} (\mathbf{A}_{jm_z}^{(m)})^* \cdot \Ei \ {\dd (\cos \theta)} {\dd \phi} \label{int_gjm}\\
\gje =  \int_{-1}^{1}\int_0^{2\pi} (\mathbf{A}_{jm_z}^{(e)})^* \cdot \Ei \ {\dd (\cos \theta)} {\dd \phi} \label{int_gje}
\end{eqnarray}
Equations (\ref{int_gjm}, \ref{int_gje}) give us the formal solution of the problem. Hence, solving the GLMT problem basically consists in computing two sets of two-dimensional angular integrals. With our current computational power, $\gjm$ and $\gje$ can be computed using standard numerical integration techniques, without too much of a struggle. Nevertheless, as mentioned in \ref{Ch2_Intro}, that was not the case when GLMT was firstly formulated, and so a lot of effort was put into computing the beam shape coefficients. A very complete review of all the different available techniques to compute the beam shape coefficients can be found in \cite{Gouesbet2011}. I will not explicitly comment further on those techniques, as in the proceeding sections I will simplify (\ref{int_gjm}, \ref{int_gje}) thanks to some symmetry considerations. 

\section{Cross sections}
In the previous section I have given the expressions of the EM field in all points of space. Nonetheless, it is difficult to measure the EM field in experiments at optical frequencies. Therefore, we need to derive some quantities that we can experimentally determine, even though they might provide a less complete description of the scattering process. One of these quantities is the EM power carried by the radiation, since many light detectors are sensitive to it. Thus, we need to mathematically characterize how much power is carried by the three fields in the problem. In order to do that, we first define the time-averaged Poynting vector\footnote{Remember that a harmonic $\exp(-i \omega t)$ dependence is supposed} \cite{Mishchenko2002}:
\begin{equation}
\langle \St \rangle = \dfrac{1}{2} \text{Re} \left[ \Et \times (\Ht)^* \right]
\label{Poyting} \end{equation} 
Now, the total fields are obtained adding the scattered and incident fields, \textit{i.e.} $(\Et, \Ht) = (\Es + \Ei, \Hs + \Hi)$. Hence, we can split the time-averaged Poynting vector in three parts:
\begin{eqnarray}
\langle \St \rangle & = & \langle \Si \rangle + \langle \Ss \rangle + \langle \Sext \rangle \\
\langle \Si \rangle & = & \dfrac{1}{2} \text{Re} \left[ \Ei \times (\Hi)^*  \right] \\
\langle \Ss \rangle & = & \dfrac{1}{2} \text{Re} \left[ \Es \times (\Hs)^*  \right] \\
\langle \Sext \rangle & = & \dfrac{1}{2} \text{Re} \left[ \Ei \times (\Hs)^* + \Es \times (\Hi)^*  \right]
\end{eqnarray}
The flux of the Poynting vector across a surface gives us the energy flux per unit of area and unit of time, which accounts for the loss of energy inside the surface. We will denote this flux as minus the absorption cross-section:
\begin{equation}
- C_{abs} = \int_S \langle \St \rangle \cdot \mathbf{n} \ \dd \Omega \label{C1}
\end{equation}
where $\dd \Omega $ is a differential surface element, with $\nhat$ being the normal vector to this element at every point. Due to the distributive property of integrals, this can be also split in three terms, to which we will refer as incident, scattering and extinction\footnote{The extinction cross section is defined as energy loss as well, that is why define it with a minus sign.} cross-sections:
\begin{eqnarray}
-C_{abs} & = & C_{i} + C_{s} + C_{ext}\\
C_i & = & \int_S \langle \Si \rangle \cdot \nhat \ \dd \Omega  \\
C_s & = & \int_S \langle \Ss \rangle \cdot \nhat \ \dd \Omega  \\
- C_{ext} & = & \int_S \langle \Sext \rangle \cdot \nhat \ \dd \Omega 
\label{C5}
\end{eqnarray}
Clearly, if we consider the flux of the Poynting vector across a spherical surface, $C_i = 0$, as the surrounding medium is non-absorbing. The choice of a spherical surface simplifies the calculations without any loss of generality, therefore I will use it for the rest of the calculations. This will directly imply that the surface element $\dd \Omega = \dd \left( \cos \theta \right) \dd \phi$, and the normal vector $\nhat$ will point in the radial direction $\rhat$.
I will start by computing $C_s$. In order to do this calculation, the following orthonormality relations of the multipolar fields will be considered \cite{Rose1955}:
\begin{eqnarray}
\int \left(\textbf{A}_{jm_z}^{(y)} \times (\textbf{A}_{jm_z}^{(y)})^* \right) \cdot \rhat \, \,  \dd \Omega = 0 \\ 
\int  \, \left( \textbf{A}_{jm_z}^{(y)} \right)^* \cdot \textbf{A}_{j'm_z'}^{( y' )} \, \dd \Omega=F\left( r \right) \delta_{jj'} \delta_{m_zm_z'} \delta_{yy'}
\end{eqnarray}
where $F(r)$ is 
\begin{equation}
F(r)= \left\lbrace \begin{array}{cccl} \vert \xi_j \vert^2 & \text{for} & y=m \\ \dfrac{j \vert \xi_{j+1} \vert^2 + (j+1) \vert \xi_{j-1} \vert^2}{2j+1} & \text{for} & y=e  \end{array} \right. \label{Mult_Bessel}
\end{equation}
and $\xi_j$ is a Bessel function for the incident and interior field and a Hankel function of the first kind for the scattered field. Then, it can be proven that
\begin{equation}
C_s =  \sum_{j=1}^{\infty} \sum_{m_z=-j}^j  \left( \vert
a_j \gje \vert^2 + \vert b_j \gjm \vert^2 \right)
\label{E_Cs}
\end{equation}
where the two following relations have been used \cite{Rose1955,Jackson1998}:
\begin{eqnarray}
\nabla \times \Am = i\Ae, \quad \nabla \times \Ae = -i\Am \\
\nabla \cdot \left( \mathbf{A} \times \mathbf{B} \right) = \mathbf{B} \cdot \left( \nabla \times \mathbf{A} \right) - \mathbf{A} \cdot \left( \nabla \times \mathbf{B} \right)
\end{eqnarray}
Note that equation (\ref{E_Cs}) differs from the expression given in \cite{GLMT_book} in some factors due to the fact that the normalisation of the multipolar modes given by equation \ref{E_Mul_norm} is not considered in \cite{GLMT_book} using the Bromwich potentials. Very similarly, the following expression can be obtained for $C_{ext}$ \cite{GLMT_book}:
\begin{equation}
C_{ext} =  2 \text{Re} \left[ \sum_{j=1}^{\infty} \sum_{m_z=-j}^j  \left( a_j  \vert
\gje \vert^2 + b_j \vert  \gjm \vert^2 \right) \right]
\label{E_Cext}
\end{equation}
Consequently, $C_{abs}$ can be obtained subtracting expressions (\ref{E_Cext}) and (\ref{E_Cs}): $C_{abs}=C_{ext} - C_s$.\\

\section{Efficiency factors}\label{Ch2_eff}
As I have shown above, the expressions for the interior and scattered field depend on the incident field. In particular, their expressions depend on the intensity carried by $\Ei$. This fact naturally translates into the cross sections, \textit{i.e.} the cross sections depend on the intensity of the incident beam. In order to remove the intensity dependence from the cross-section expression, we normalize the cross sections over the intensity of the beam across a transverse section \cite{GLMT_book,Stout2013}:
\begin{equation}
Q_{v}= \frac{1}{\int \vert \mathbf{E}^i \vert^2 d\Omega}\int \langle \textbf{S}^{v} \rangle \cdot \rhat \dd \Omega 
\label{E_efficiency}
\end{equation}
where $v$ can take three values $v=s,i,ext$. I will not give general expressions for the efficiencies. Instead, I will either explicitly compute them (see chapters \ref{Ch3} and \ref{Ch4}), or measure them (see chapter \ref{Ch7}). A subtlety about the measurements in chapter \ref{Ch7} is in order at this point. Despite normalising the scattering intensity over the power of the incident beam ($I^{\text{norm}}$), the result is modified by the limited NA of the detector. That is, the definitions of cross section and efficiency factors given by equations (\ref{C1}-\ref{C5}) and (\ref{E_efficiency}) suppose that the scattering in all angles $\left( \theta, \phi \right)$ is collected. However, this is not the case in the experiments done in chapter \ref{Ch7}. There, only scattered light in the backward semi-space is collected. Now, even though backward cross sections do not generally have the same $\lambda$ dependence as proper cross sections \cite{Bohren1983}, the features that I will be especially interested in will still remain. Namely, as it will be shown in section \ref{Ch3_WGM}, the excitation of spheres with plane waves hides some resonant behaviour that can be unveiled using vortex beams.

\begin{savequote}[10cm] 
\sffamily
``I think it only makes sense to seek out and identify structures of authority, hierarchy, and domination in every aspect of life, and to challenge them; unless a justification for them can be given, they are illegitimate, and should be dismantled, to increase the scope of human freedom.'' 
\qauthor{Noam Chomsky}
\end{savequote}

\chapter{Excitation of single multipolar resonances}
\graphicspath{{ch3/}} 
\label{Ch3}
\newcommand{\ms}{m_z^*}
\newcommand{\Bm}{\mathbf{B}_{mk_{\rho}}}
\newcommand{\A}{\mathbf{A}_{jm_z}}
\newcommand{\gjms}{g_{jm_z^*}^{(m)}}
\newcommand{\gjes}{g_{jm_z^*}^{(e)}}
\newcommand{\Ams}{\mathbf{A}_{jm_z^*}^{(m)}}
\newcommand{\Aes}{\mathbf{A}_{jm_z^*}^{(e)}}

\newcommand{\Cjmp}{C_{jm_zp}}
\newcommand{\Cjmsp}{C_{jm_z^*p}}
\newcommand{\R}{\mathbf{R}}
\newcommand{\Rk}{\mathbf{R}(\thetak,\phik)}

\newcommand{\djmp}{d^j_{m_zp}(\thetak)}
\newcommand{\djmsp}{d^j_{m_z^*p}(\thetak)}
\newcommand{\Einc}{E_{inc}}

\section{GLMT with cylindrically symmetric beams} \label{Ch3_cylsymm}
In the previous chapter, I have solved the Generalized Lorenz-Mie problem. That is, I have found the EM fields in the whole space when a homogeneous and isotropic sphere is excited by an arbitrary monochromatic light beam $\Ei$. In the present chapter, I will particularize the solution given by equations (\ref{E_all_fields1}-\ref{int_gje}) for an incident monochromatic cylindrically symmetric beam. The definition of a general cylindrically symmetric beam stems from the definition of cylindrical symmetry. As it has been described in chapters \ref{Ch1}-\ref{Ch2}, a light beam is symmetric under a transformation $\mathbf{T}$ when its EM field is an eigenstate of the transformation generated by $\mathbf{T}$: $\mathbf{T} \mathbf{u} = c_u \mathbf{u}$. I define a cylindrically symmetric beam as an eigenstate of $R_z(\theta)$, the operator of rotations around the $z$ axis. Now, because $R_z(\theta)$ can be expressed as $R_z(\theta) = \exp(-i J_z \theta )$, a cylindrically symmetric beam can be alternatively defined as an eigenstate of $J_z$. As we have seen in chapter \ref{Ch1}, there are different families of monochromatic modes that have a well-defined $J_z$. Both Bessel beams and multipolar fields are basis of modes that are eigenstates of $J_z$ with eigenvalues $m$ and $m_z$ respectively (see equations (\ref{E_Bx_eigen}, \ref{E_Ax_eigen})):
\begin{eqnarray}
J_z \Bm = m \Bm \\
J_z \A = m_z \A 
\end{eqnarray}
An arbitrary superposition of them (keeping $m$ or $m_z$ constant) will also be an eigenstate of $J_z$. Now, both basis are valid to describe the problem, but I will use the multipolar fields as they are the normal modes of a sphere (see section \ref{Ch2_Symmetries}). In this basis, the expression for the electric field of a general cylindrically symmetric beam is:
\begin{equation}
\Ei=\sum_{j=\vert m_z^*\vert}^{\infty} \gjms\Ams +  \gjes \Aes
\label{E_Ei_c}
\end{equation}
where $\gjes$ and $\gjms$ are proportional to the beam shape coefficients, (see section \ref{Ch2_Scatt}). Note that a general superposition of multipoles is given by equation (\ref{Ei}). That decomposition contains two summations, on $j=1,..,\infty$ and $m_z=-j,..,j$ respectively. In contrast, expression (\ref{E_Ei_c}) only involves a summation in $j=m_z^*,..,\infty$. First, $m_z$ needs to be fixed to $m_z^*$ so that $\Ei$ is still an eigenstate of $J_z$. And secondly, due to the fact that $J^2 \geq J_z^2 $, $m_z^*$ bounds $j$ to values $j \geq m_z^*$. \\
Last but not least, I will also impose that the cylindrically symmetric beams are eigenstates of the helicity operator, $\Lambda$. In this way, they will be symmetric under duality transformations and it will be easier to study the duality properties of the system (see chapter \ref{Ch4}). Imposing that the incident field $\Ei$ has a well-defined helicity has further implications on expression (\ref{E_Ei_c}). As it has been demonstrated in chapter \ref{Ch1}, a mode of light with a well-defined helicity can be expressed as a superposition of electric and magnetic modes. Consequently, the two families of beam shape coefficients can be reduced to one:
\begin{equation}
\Cjmp = \gjm = - i p \gje
\label{cjmgjm}
\end{equation}
where $\Cjmp$ is the only family of beam shape coefficients remaining and $p= \pm 1$ are the helicity values for the beam. Thus, the multipolar decomposition of a cylindrically symmetric beam with a well-defined helicity is:
\begin{equation}
\Ei=\sum_{j=\vert m_z^*\vert}^{\infty} \Cjmsp \left( \Ams +  ip \Aes \right)
\label{E_Ei_cp}
\end{equation}
Equation (\ref{E_Ei_cp}) will be the starting point for this chapter. I will use it to solve the GLMT problem with paraxial cylindrically symmetric beams in \ref{Ch3_paraxial} and with general non-paraxial cylindrically symmetric beams in \ref{Ch3_non}. Then, in \ref{Ch3_WGM}, I will show how the use of these beams of light along with the proper choice of ratio between the particle size and the excitation wavelength is enough to excite WGMs \cite{Zambrana2012}. 

\section{Paraxial excitation}\label{Ch3_paraxial}
\begin{figure}[htbp]
\centering\includegraphics[width=\columnwidth]{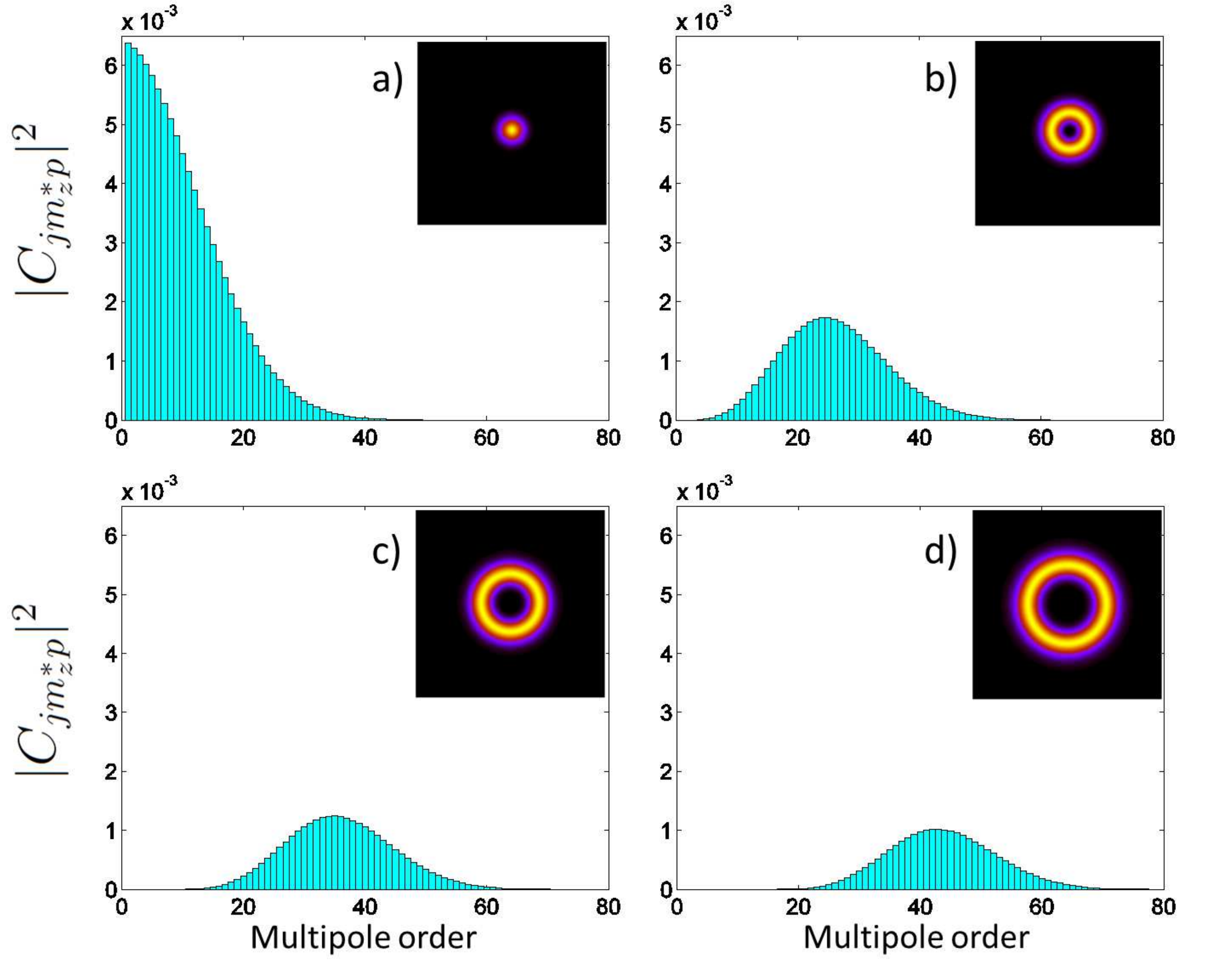}
\caption{$\vert C_{jm_z^*p} \vert^2$ for different LG$_{l,q}$ beams. a) LG$_{0,0}$ b) LG$_{2,0}$ c) LG$_{4,0}$ d) LG$_{6,0}$. The relation between the width ($w_0$) of the beam and the wavelength is $w/\lambda \approx 4$. The helicity is $p=1$ for the four different cases. The insets are intensity plots of the multipolar decomposition for each of the four cases. The intensity of each of the LG modes is then retrieved.}
\label{F_multipolar}
\end{figure}

In this section I solve the GLMT when the incident field $\Ei$ is a Laguerre-Gaussian beam, LG$_{l,q}$\footnote{LG beams are also named as vortex or doughnut beams. However, there are some subtleties. The `vortex beam' term usually refers to beams of light whose phase dependence is of the kind $\exp \left[ il\phi \right]$. This phase dependence generally gives rise to an intensity profile that resembles that of a doughnut. Nevertheless, a beam can have a doughnut-like intensity profile without having a $\exp \left[ il\phi \right]$ phase dependence. The LG modes are a well-defined set of modes (see equation (\ref{E_LGbeam})). Their phase dependence is that of a vortex beam, but their intensity profile is not always doughnut-like. In fact, their intensity profile is only doughnut-like when the radial parameter $q=0$.}. The LG modes are the normal modes of a cylindrical cavity and can also be obtained as superpositions of Hermite-Gaussian modes, which are the normal modes of spherical-mirror resonator\footnote{In fact, LG modes are also normal modes of this kind of resonator.} \cite{Saleh2007}. They are a basis of solutions of the paraxial equation in cylindrical coordinates \cite{Siegman1986,prePampaloni2004}, \textit{i.e.} any paraxial beam can be described as a superposition of LG modes \cite{Gabi2001}. The LG modes had been known by the optics community for a long time, but their use highly escalated after a seminal paper by Allen and co-workers in 1992 \cite{Allen1992}. In that paper, it was established that there was a relation between the phase and the AM content of these modes. This laid out the grounds of the field of the angular momentum of light. Since then, the AM of light has been used in many diverse fields such as quantum optics \cite{Mair2001,Gabi2007}, optical manipulation \cite{He1995prl,Simpson1996,Friese1996,Friese1998}, optical communications \cite{Bo2007,Tamburini2012} or astrophysics \cite{Harwit2003,Tamburini2011}. The general expression for a LG modes is given by: 
\begin{equation}
\text{LG}_{l,q}=N_{l,q} \exp\left[ {\dfrac{-\rho^2}{w(z)^2}}\right] \rho^l L^l_q(\dfrac{2\rho^2}{w^2(z)})\left( \dfrac{\sqrt{2}}{w(z)} \right)^{l+1} e^{  i \left( \frac{-k\rho^2 z}{2(z^2+z_0^2)} + (2q+l+1) \tan^{-1}(z/z_0) + l\phi \right) }
\label{E_LGbeam} 
\end{equation} 
where $q \geq 0$ is the radial index, and $l$ the azimuthal one. The azimuthal number $l$ describes how the phase winds around the center of the beam, whereas $q$ has to do with the number of intensity nodes in the radial direction. $(\rho,\phi,z)$ are the cylindrical coordinates; $L^l_q$ are the generalized-Laguerre functions; $w(z)$ is the waist of the beam at a plane of $z$ constant; $N_{l,q} = \left[ \dfrac{q!}{(\pi (l+q)!} \right]^{1/2}$ is a normalization constant; and $k$ is the wave-vector. Note that due to the fact that a LG mode is a paraxial beam, the polarization is added to the beam independently of its spatial shape (see section \ref{Ch1_symm})\cite{Lax1975}. Next, I will decompose a circularly polarized LG beam into multipoles. In fact, a circularly polarized paraxial beam has a well-defined helicity within the paraxial approximation \cite{Gabi2008,Ivan2012PRA}. That is, if we apply the helicity operator $\Lambda$ to a circularly polarized LG mode, and we only consider paraxial terms, we obtain the same mode times the helicity value of the beam: 
\begin{equation}
\Lambda \left[ \text{LG}_{l,q} \spphat \right] \approx p \left[ \text{LG}_{l,q} \spphat \right]
\end{equation}
with $\spphat$ being a general circularly polarized vector defined as $\spphat=(\xhat +i p \yhat)/ \sqrt{2}$, and $\xhat$ and $\yhat$ the horizontal and vertical polarization vectors. The differential form of $\Lambda$ is given by the expression $\Lambda=(\nabla \times)/k$ \cite{Messiah1999,Ivan2012PRA}. Then, the multipolar decomposition of LG$_{l,q}\spphat$ is given by equation (\ref{E_Ei_cp}) and only one set of beam shape coefficients is needed, $\Cjmp$. Before computing them, one comment needs to be made. A beam such as LG$_{l,q}\spphat$ does not fulfil Maxwell equations. However, its decomposition into multipoles using equation (\ref{E_Ei_cp}) does. This is a known paradox in optics that was solved by Lax in 1975 \cite{Lax1975,GLMT_book}. The idea is that the Maxwell field given by the multipolar decomposition is equal to the paraxial beam only in the paraxial limit. Furthermore, Lax also showed that a paraxial field gives rise to many different fields fulfilling the Maxwell equations. The general formalism to find one of the possible multipolar decompositions of a circularly polarized LG mode was given by my supervisor in \cite{Gabi2008}. Then, in \cite{Zambrana2013JQSRT}, we applied that formalism to give an analytical formula for the beam shaping coefficients $\Cjmp$. Our findings are shown next. If $\Ei = \text{LG}_{l,q}\spphat$, then we can re-express it as:
\renewcommand{\arraystretch}{2.5}
\begin{equation}
\begin{array}{ccl}
\Ei &= &\displaystyle\sum_{j=\vert m_z^* \vert}^{\infty} i^j (2j+1)^{1/2} \Cjmsp \left[ \Ams +ip\Aes \right] \\
\Cjmsp &= & \displaystyle\int_0^{\pi}d_{m_z^* p}^j(\thetak)f_{l,q}(k\sin\thetak) \ \sin\thetak  \dd \thetak 
\end{array}
\renewcommand{\arraystretch}{1} 
\label{E_general}
\end{equation}
where $m_z^*=l+p$ stands for value of $J_z$\footnote{Remember that as described in chapter \ref{Ch1}, the AM content of a paraxial beam can be obtained by adding the topological charge of the beam ($l$ for a LG$_{l,q}$ beam) and the value of the helicity $p$. This is what is usually described as AM $=$ OAM $+$ SAM, \textit{i.e.} the total angular momentum can be obtained adding its orbital and spin parts}. The reduced rotation matrix $d_{m_z^*p}^j(\thetak)$ can be found in \cite{Rose1957}, and the function $f_{l,q}(k_r)$ is related to the Fourier transform of the incident field, with $\krho$ being the transverse momentum, \textit{i.e.} $\krho=\sqrt{k_x^2+k_y^2}=k\sin\thetak$. Note that the function $\Cjmsp$ is still defined as an integral, although it is no longer a double one, unlike (\ref{int_gjm}, \ref{int_gje}). This is a consequence of the fact that $\Ei$ and the multipolar fields are eigenstates of $J_z$ with value $m_z^*$, therefore projections are only needed on $J^2$ values. Also, note that the definition of $\Cjmsp$ in equation (\ref{E_general}) is slightly different from its definition in (\ref{E_Ei_cp}): a factor $i^j (2j+1)^{1/2}$ has been made explicit. The definition of $\Cjmsp$ given by equation (\ref{E_general}) will be the one used for the rest of the thesis. \\\\
Actually, equation (\ref{E_general}) is valid for any beam whose $J_z$ is well-defined. For the specific case of a LG$_{l,q}$ beam, the integral in equation (\ref{E_general}) can be solved analytically thanks to the expression of $f_{l,q}$ and provided the integral \ref{E_general} is extended to infinity. The following expression is obtained:
\begin{equation}
\begin{split}
C_{jm_z^*p} = (-1)^l \left[ (j+p)!(j-p)!(j+l+p)!(j-l-p)! \right]^{1/2} \\
\sum_s \frac{(-1)^s}{(j-l-p-s)!(j+p-s)!(s+l)!s!} \frac{M_{l+2s}(f_{l,q})}{2^{l+2s}}
\end{split}
\label{Cjlp}
\end{equation}
where the number $M_a(f_{l,q})=\int_0^{\infty} \krho \dd \krho \krho^a f_{l,q}(\krho)$ is the momentum of order $a$ of the function $f_{l,q}$. The expression for $f_{l,q}(\krho)$ when the incident beam is a LG$_{l,q}$ is:
\begin{equation}
f_{l,q}(\krho)= \left( \frac{w_0^2q!}{2\pi (\vert l \vert + q )!} \right)^{1/2} \left(w_0 k_r /\sqrt{2} \right)^{\vert l \vert } L_q^{\vert l \vert}(w_0^2k_r^2/2)\exp \left( -w_0^2k_r^2/4 \right)
\label{F(LG)}
\end{equation}
with $w_0=w(z=0)$ being the beam width of the LG mode in real space. Finally, I give the expression of $M_a(f_{l,q})$ when the function $f_{l,q}$ is given by the expression (\ref{F(LG)}). As it is shown, in this particular case, the integral is analytical. Also, I particularize the result for $q=0$: 
\begin{equation}
\begin{split}
M_a(f_{l,q})= \Gamma \left( \dfrac{a+\vert l \vert + 2}{2} \right) \sqrt{\frac{w_0^2 (\vert l \vert +q)!}{2\pi ( \vert l \vert ! ) ^2 q! }} \ 2^{a+\frac{\vert l \vert}{2}+1} \times \\ \times w_0^{-2-a} \ _2F_1\left( -q, \frac{a+\vert l \vert + 2}{2}, \vert l \vert +1; 2 \right) \\
\end{split}
\label{Ma}
\end{equation}
\begin{equation}
M_{l+2s}(f_{l,0})=  (s+ \vert l \vert )! w_0^{-1-\vert l \vert -2s}  \sqrt{\frac{1}{2\pi \vert l \vert!}} \  2^{\vert l\vert +2s+\frac{\vert l \vert}{2}+1}  
\label{Ma_LG}
\end{equation}
where $\Gamma(z)$ is the Gamma function and $_2F_1(a,b,c:z)$ is the hypergeometric function. The problem, then, is solved. Using equations (\ref{E_general}, \ref{Cjlp}) and (\ref{Ma}) the decomposition of the incident beam is found. Then, using the GLMT formulation introduced in the preceding chapter, the scattered and the interior fields are obtained (see section \ref{Ch2_Scatt}): 
\begin{equation}
\begin{array}{ccl}
\Es =&&\displaystyle\sum_{j=\vert m_z^* \vert}^{\infty} i^j (2j+1)^{1/2} \Cjmsp \left[ b_j \Ams +ip a_j \Aes \right]  \\
\Eint =&&\displaystyle\sum_{j=\vert m_z^* \vert}^{\infty} i^j (2j+1)^{1/2} \Cjmsp \left[ c_j \Ams +ip d_j \Aes \right] \\
\end{array}
\label{E_fields}
\end{equation}
As already mentioned in chapter \ref{Ch2}, it is clear that the solution of the scattering problem only depends on the beam shape coefficients $\Cjmsp$. The beam shape coefficients depend on the parameters of the incident LG modes, \textit{i.e.} the azimuthal and radial numbers $(l,q)$, the helicity $p$, and the waist of the beam $w_0$. That means that once the LG mode and its polarization have been chosen, the only way of tailoring the multipolar content of the beam is with the width of the beam, $w_0$. However, $w_0$ cannot get arbitrary values, as the beam still needs to fulfil the paraxial equation \cite{Lax1975}. As a consequence, the multipolar content of the beam is almost fixed for every single mode.\\\\
In Figure \ref{F_multipolar}, the multipolar content for four different LG modes is depicted. It can be observed that over 40 multipolar modes of each parity are needed to describe a LG mode of an arbitrary order\footnote{Multipolar modes of both parities are excited as $\Cjmsp$ gives rise to the two families of beam shape coefficients (see equation (\ref{cjmgjm})).}. This fact will be further explored in the next section. I advance that using high NA lenses will enable us to describe a cylindrically symmetric beam with a handful of multipoles. This fact will be a key point to excite single multipolar resonances in section \ref{Ch3_WGM}.

\section{Non-paraxial excitation}\label{Ch3_non}
In this section, I will extend the results of the previous section to the non-paraxial case. That is, in this case the incident field $\Ei$ will fulfil Maxwell equations. Mathematically, the only difference is the way the beam shape coefficients $\Cjmp$ are computed. However, physically there is a very big difference: the use of an aplanatic lens enables us to control the multipolar decomposition at the focal plane with a much higher precision. This fact is of great interest to control multipolar resonances both in nanostructures and atoms \cite{Curto2010,Zambrana2012,Curto2013,Afanasev2013,vanBoxem2013}. I will further develop this idea on section \ref{Ch3_WGM}. In addition, I will show that by adjusting the focal distance of a lens, the magnitude of the scattering efficiency can be increased by orders of magnitude.\\\\
To start, I will show how the aplanatic model of a lens tailors the multipolar content of a beam. In section \ref{Ch1_apla}, I have demonstrated the conditions under which the aplanatic lens model does not change the helicity and the angular momentum content ($J_z$) of the incident fields. In order to do that, I have expressed the aplanatic lens transformation \cite{WolfII1959,Novotny2006} as a rotation in the momentum space. Next, I will use the same mathematical technique to retrieve the beam shape coefficients $\Cjmp$ as a function of the focal distance of the lens, the NA, and the media surrounding the lens. On one hand, I will consider a general superposition of circularly polarized propagating plane waves with the same helicity: 
\begin{equation}
\E_1 = \int_0^{\pi}  \sin\thetak \dd \thetak \int_0^{2\pi} \dd \phik \ g(\thetak,\phik) \Rk \spphat e^{ikz}
\label{E_Eint}
\end{equation}
On the other hand, the aplanatic model can be modelled as the integral (\ref{E_apl_model}) of a collimated (or paraxial) beam $\Einc$ over the surface of a sphere \cite{Novotny2006}:
\begin{equation}
\label{E_apl_model2}
\begin{split}
\E_2 & = \frac{i k f \exp (-ikf)}{2\pi} \int_0^{\thetak^{M}}  \sin \thetak  \int_0^{2\pi}  d \thetak d\phik \ \mathbf{E}_{\infty}(\thetak,\phik) e^{(ikz \cos \theta)} e^{(ik \rho \sin \theta \cos (\phi - \phik))}\\ & =  \frac{i k f \exp (-ikf)}{2\pi} \int_0^{\thetak^{M}} \sin \thetak \int_0^{2\pi} d \thetak d\phik \ \mathbf{E}_{\infty}(\thetak,\phik) \plane
\end{split}
\end{equation}
In chapter \ref{Ch1}, I proved that the polarization transformation produced by equations (\ref{E_apl_model2}) and (\ref{E_Eint}) is the same one provided the two transmissivities of the lens are equal, $t^s(\thetak)=t^p(\thetak)$ (see section \ref{Ch1_apla}). I also gave the value of the function $g(\phik,\thetak)$ that makes $\E_1 = \E_2$, which is (see equation \ref{E_gthphi}):
\begin{equation}
g(\thetak, \phik) = - \dfrac{ikf e^{(-ikf)}}{2\pi} E^{in}(f\sin\thetak,\phik) t(\thetak) e^{ip\phik} \left( 1-H(\thetak - \thetak^{M}) \right) \sqrt{\dfrac{n}{n_1}}(\cos \thetak)^{1/2}
\label{E_gthetaphi}
\end{equation}
where $f$ is the focal of the lens; $\thetak^M=\arcsin(\mathrm{NA})$, with NA the numerical aperture of the lens; $n_1$ and $n$ are the index of refraction of the lens and the medium surrounding it respectively; $H(\thetak - \thetak^{M})$ is the Heaviside step function; and $E^{in}$ is the value of the paraxial incident field at the back-aperture of the lens. Note $E^{in}(\rho,\phi)=E^{in}(f\sin\thetak,\phik)$. That is, there is a match between the cylindrical coordinates in the real space of the paraxial incident beam, and the spherical coordinates of its angular spectrum in the momentum space. As mentioned in section {\ref{Ch1_apla}}, this fact is a key factor to prove that the aplanatic model does not change the helicity nor the AM of the incident fiend. Now, using the properties of the multipolar fields under rotations \cite{Rose1957}, the expression (\ref{E_Eint}) can be re-written as:
\begin{equation}
\begin{split}
\E = & \int_0^{\pi}  \sin\thetak \dd \thetak \int_0^{2\pi} \dd \phik \ g(\thetak,\phik) \sum_{j,m_z} i^j (2j+1)^{1/2} \Djmp \left[ \Am +ip \Ae \right] \\ = &\int_0^{\pi}  \sin\thetak \dd \thetak \int_0^{2\pi} \dd \phik \ g(\thetak,\phik) \sum_{j,m_z} i^j (2j+1)^{1/2} \djmp e^{-im\phik} \left[ \Am +ip \Ae \right] \\ = & \sum_{j,m_z} i^j (2j+1)^{1/2} \Cjmp \left[ \Am +ip \Ae \right]
\end{split} \label{E_Emulti}
\end{equation}
with
\begin{equation}
\Cjmp = \int_0^{\pi}  \sin\thetak \dd \thetak \int_0^{2\pi} \dd \phik \ g(\thetak,\phik) \djmp e^{-im\phik}
\label{E_cjmp}
\end{equation}
If the expression for $g(\thetak,\phik)$ found in (\ref{E_gthetaphi}) is used in the previous expression (\ref{E_cjmp}), the value for the $\Cjmp$ is the following one:
\begin{equation}
\Cjmp = \int_0^{\thetak^{M}} \sin \thetak \dd \thetak \dfrac{fe^{-ikf}}{2\pi} \sqrt{\dfrac{n_1}{n_2}} \cos^{1/2}\thetak \djmp \int_0^{2\pi} d\phik e^{-im\phik} e^{ip\phik} E^{in}(f\sin\thetak,\phik)
\label{E_cjmp_non}
\end{equation}
As I mentioned in the beginning of the chapter, I will be interested in cylindrically symmetric beams. The helicity has been set to $p$ already, so the only other requirement on the incident paraxial field $E^{par} (\rho,\phi)=E^{par}(f\sin\thetak,\phik)$ is having an azimuthal dependence of the kind $e^{il\phi}$. In general, this can be achieved by a superposition of LG beams with a fixed azimuthal number $l$:
\begin{equation}
E_l^{par}(f\sin\thetak,\phik)= \sum_{q=1}^{\infty} c_q\text{LG}_{l,q}(f\sin\thetak)e^{il\phik}
\end{equation}
However, for simplicity, the calculations will be done for a single LG$_{l,q}$ mode. Now, using the following orthogonality property of the exponentials,
\begin{equation}
\int_0^{2\pi} dx \ e^{i(m-m')x} = \delta_{mm'}
\end{equation}
it can be seen that the second integral in (\ref{E_cjmp_non}) can be analytically solved, setting $m_z=m_z^*=l+p$. Then, $\Cjmsp$ can be expressed as a 1-dimensional integral. Its expression, as well as the one for the incident field $\Ei$, is given by the following equation:
\begin{equation}
\renewcommand{\arraystretch}{2.5}
\begin{array}{lll}
\Ei &=& \displaystyle\sum_{j=\vert m_z^* \vert}^{\infty} i^j (2j+1)^{1/2} \Cjmsp \left[ \Ams +ip\Aes \right] \\
\Cjmsp & = & \displaystyle \int_0^{\thetak^{M}} \sin \thetak \dd \thetak \dfrac{fe^{-ikf}}{2\pi} \sqrt{\dfrac{n_1}{n_2}} \cos^{1/2}(\thetak) \djmsp N_{l,q}   \\ 
& & \exp\left[ {\dfrac{-(f\sin\thetak)^2}{w(z)^2}}\right](f\sin\thetak)^l L^l_q(\dfrac{2(f\sin\thetak)^2}{w^2(z)})\left( \dfrac{\sqrt{2}}{w(z)} \right)^{l+1} \\
& & \exp \left[ i \left( \dfrac{-k(f\sin\thetak)^2 z}{2(z^2+z_0^2)} + (2q+l+1) \tan^{-1}(z/z_0) \right) \right] \end{array}
\label{E_Eifinal}
\renewcommand{\arraystretch}{1} 
\end{equation}
\begin{figure}[htbp]
\centering\includegraphics[width=\columnwidth]{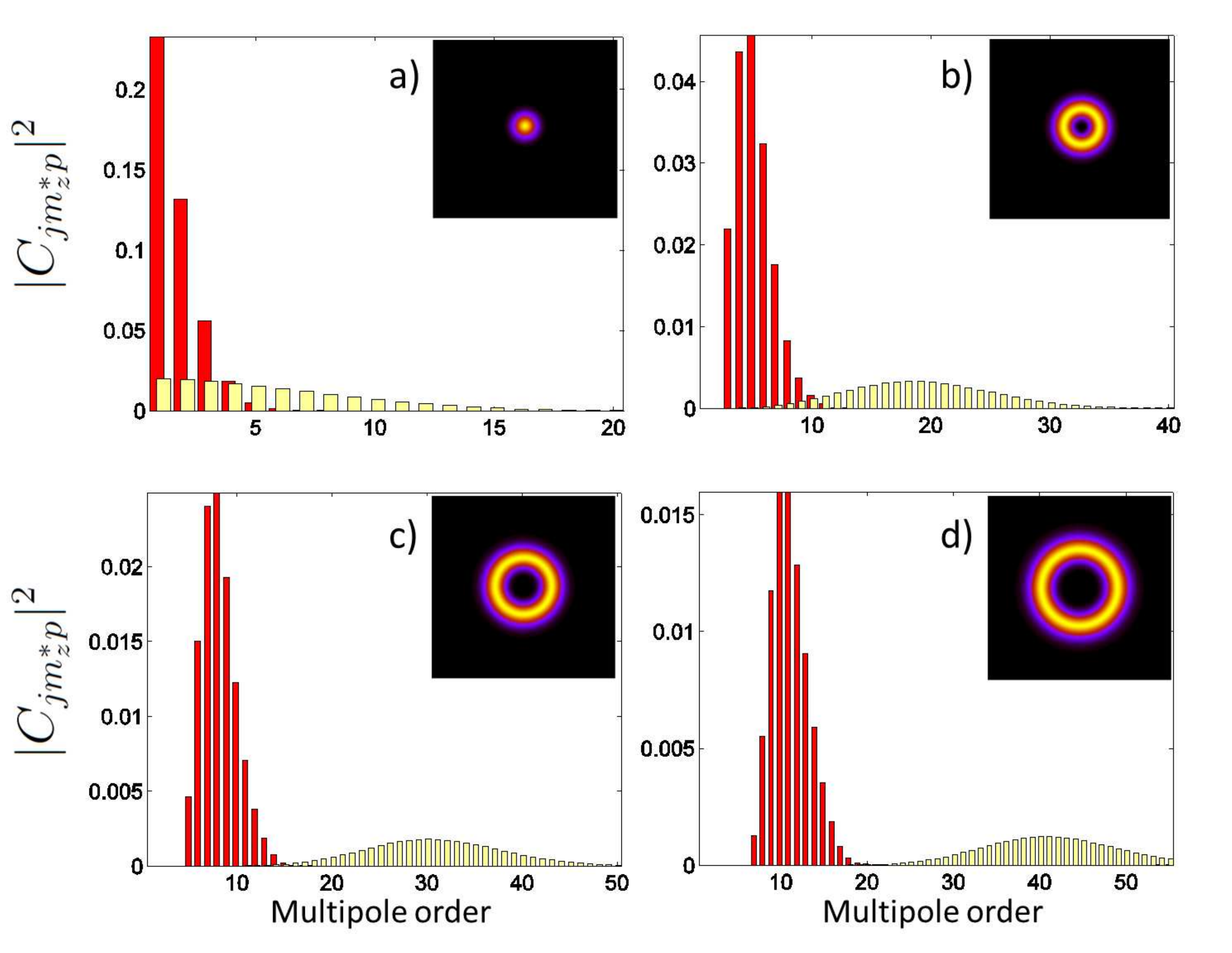}
\caption{Beam shape coefficients ($\vert \Cjmsp \vert^2$) for different focusing scenarios. The insets represent the intensity plots of the paraxial LG modes at entrance pupil of the lens. These paraxial modes are focused with two lenses of NA$=0.9$ and NA$=0.25$, and their multipolar decomposition given by $\Cjmsp$ is found at the focal plane. The red coloured bars indicate NA$=0.9$, and the yellow ones NA$=0.25$. The paraxial beams used are a) LG$_{0,0}$, b) LG$_{2,0}$, c) LG$_{4,0}$, and d) LG$_{6,0}$. Note that $\Cjmsp $ can be described with very few multipoles when a high NA lens is used.}
\label{F_CJMP}
\end{figure}
Next, I show the influence of the NA of a lens in the beam shape coefficients. In Figure \ref{F_CJMP}, I have plotted the multipolar decomposition of the same four LG beams used in Figure \ref{F_multipolar}. The four LG modes are plotted for two different focusing scenarios, one of them being focused with NA$=0.25$ (yellow), and the other one with NA$=0.9$ (red). In all cases I have used LG modes with a well-defined helicity $p=1$. The width of all beams in the paraxial approximation is chosen so that the entrance pupil of the lens is filled. The results presented in Figure \ref{F_CJMP} show that the higher the NA, the narrower the distribution of multipolar modes is. For some cases such as Figure \ref{F_CJMP}(a), the beam shape coefficients $\Cjmsp$ can be described with only a handful of multipoles. In those cases, the value of $\vert \Cjmsp\vert^2$ increases considerably, as the incident field needs to be normalised on a transverse plane \cite{Gabi2008,Zambrana2013JQSRT}. As a consequence, the following condition needs to be fulfilled:
\begin{equation}
\sum_j (2j+1) \vert \Cjmsp \vert^2 = 2
\end{equation}
The other parameter to control the distribution of multipoles is the $J_z$ of the incident beam. As mentioned earlier, the distribution of multipolar amplitudes is zero for values of $j<|l+p|$. This second degree of freedom can be also used in the paraxial case, as displayed in Figure \ref{F_multipolar}, but its influence is less important. Also, comparing Figures \ref{F_multipolar} and \ref{F_CJMP}, one can see that in the paraxial case more multipoles are needed to describe the same LG modes, even if a lens with small NA$=0.25$ has been used. Now, similarly to the previous section, once $\Cjmsp$ has been calculated, the fields in the whole space are obtained applying the GLMT, and expressions (\ref{E_fields}) are re-obtained:
\begin{equation}
\begin{array}{lcl}
\Es &=&\displaystyle\sum_{j=\vert m_z^* \vert}^{\infty} i^j (2j+1)^{1/2} \Cjmsp \left[ b_j \Ams +ip a_j \Aes \right]  \\
\Eint &=&\displaystyle\sum_{j=\vert m_z^* \vert}^{\infty} i^j (2j+1)^{1/2} \Cjmsp \left[ c_j \Ams +ip d_j \Aes \right]
\end{array}
\end{equation}
where $\Cjmsp$ is given by expression (\ref{E_Eifinal}). Finally, I want to acknowledge that before I published these results in \cite{Zambrana2012}, based on the work done by my supervisor in \cite{Gabi2008}, N.M. Mojarad \textit{et al.} had observed the influence of the NA of a lens in focusing a plane wave onto plasmonic particles \cite{Mojarad2008,Mojarad2009}. However, their study only dealt with a single focused $\shat$ plane waves, and therefore it could not be applied to any paraxial field. Also, they never applied their formalism to more elaborated beams such as vortex beams. 

\section{Excitation of Whispering Gallery Modes} \label{Ch3_WGM}
\begin{figure}[htbp]
\centering\includegraphics[width=12cm]{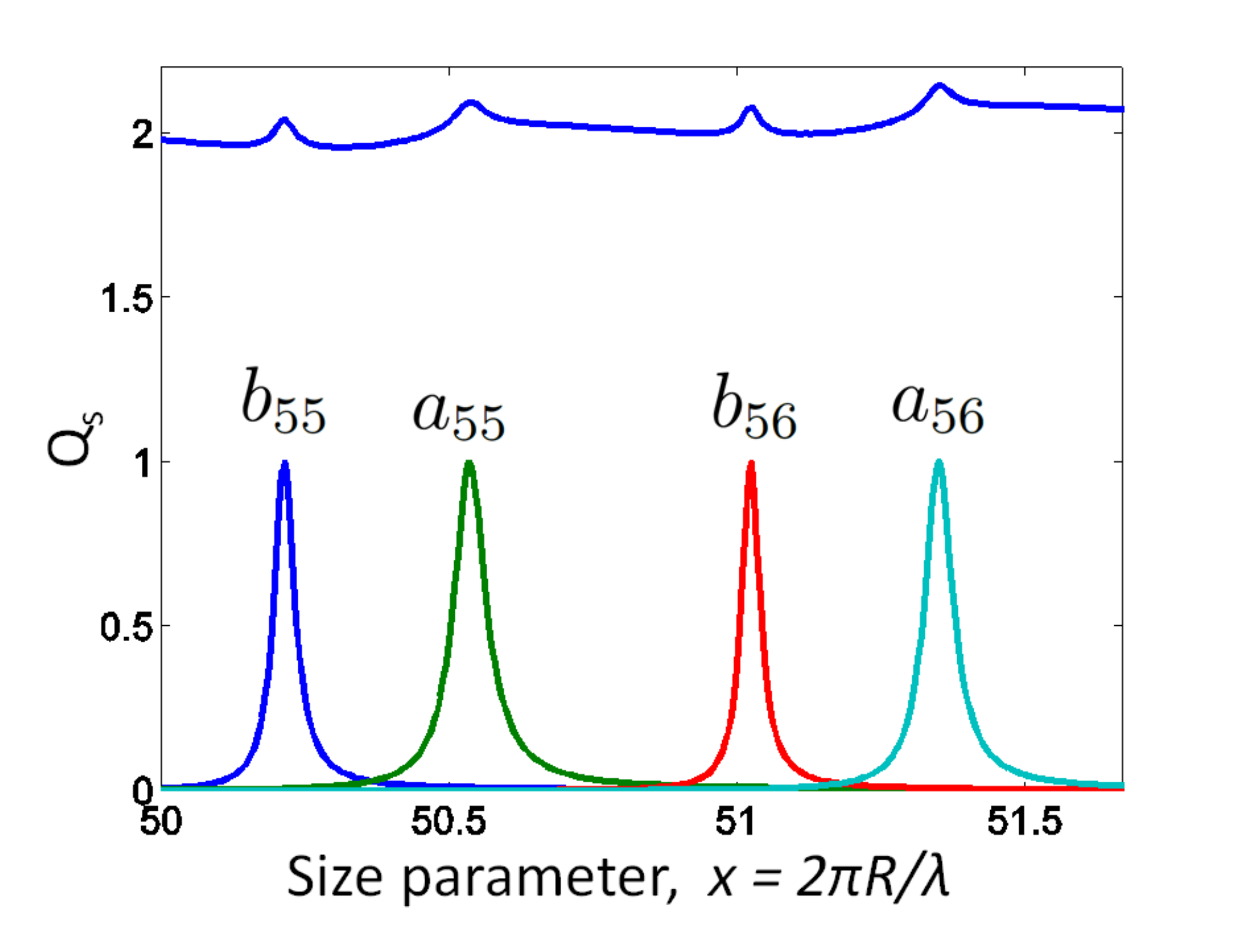}
\caption{$Q_{s}^{Mie}$ (blue curve, on top) for a single sphere with $n_r=1.33$ and $x \in \left[50,52  \right]$. At the bottom, four Mie coefficients ($a_{55}$, $b_{55}$, $a_{56}$ and $b_{56}$) are plotted. It is seen that the ripples in the scattering efficiency are caused by high order Mie resonances.}
\label{F_QsMie}
\end{figure}

\begin{figure}[htbp]
\centering\includegraphics[width=\columnwidth]{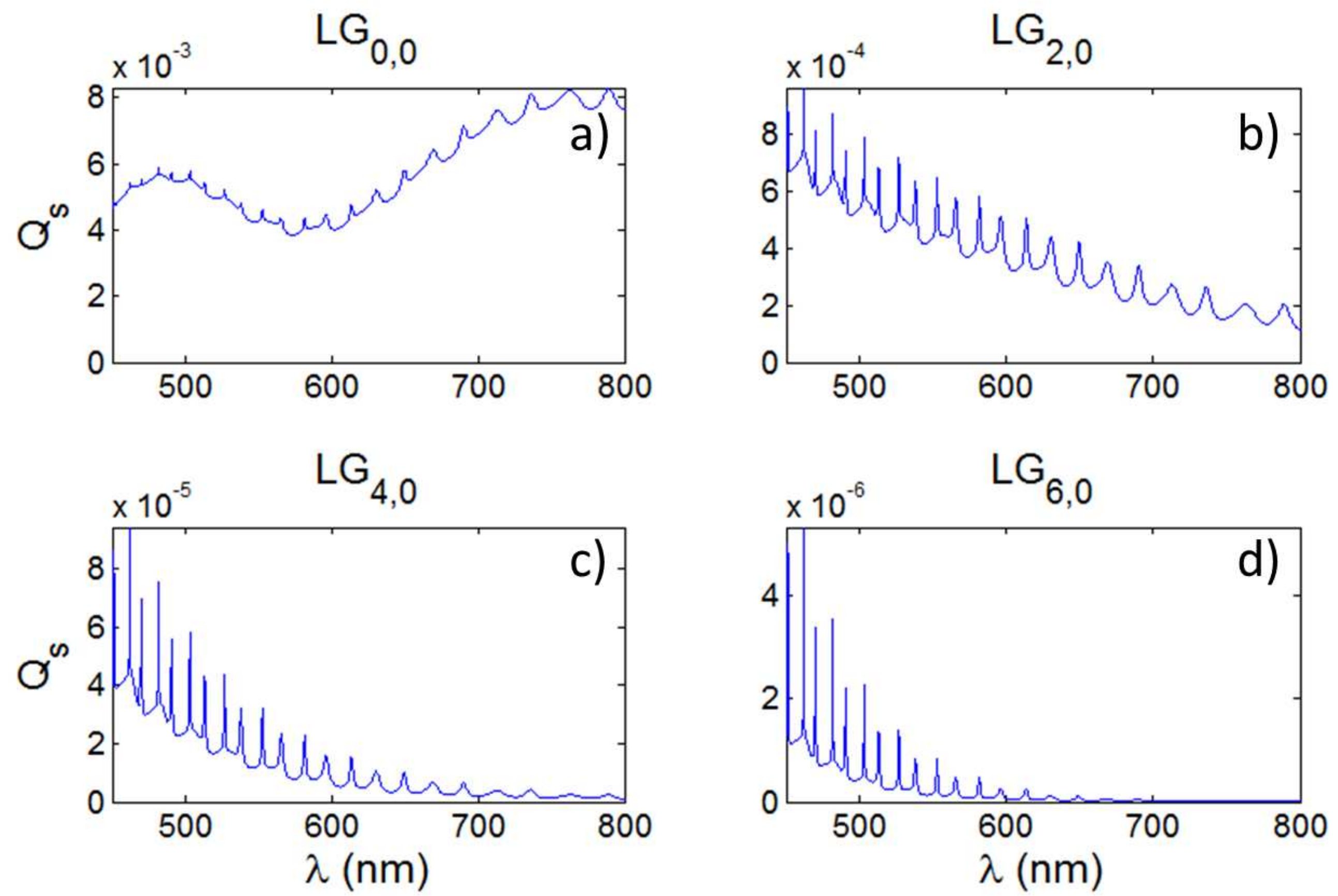}
\caption{$Q_{s}$ as a function of wavelength, $\lambda \in \left[ 450, 800\right]$ nm, for different incident LG beams. a) LG$_{0,0}$ b) LG$_{2,0}$ c) LG$_{4,0}$ d) LG$_{6,0}$. All the rest of parameters do not vary from a) to d): $n_r=1.5$, $p=1$, $R=1.3 \ \mu$m. The multipolar decomposition of each of the LG beams is given in Figure \ref{F_multipolar}.}
\label{F_Qswavelength}
\end{figure}

\begin{figure}[htbp]
\centering\includegraphics[width=\columnwidth]{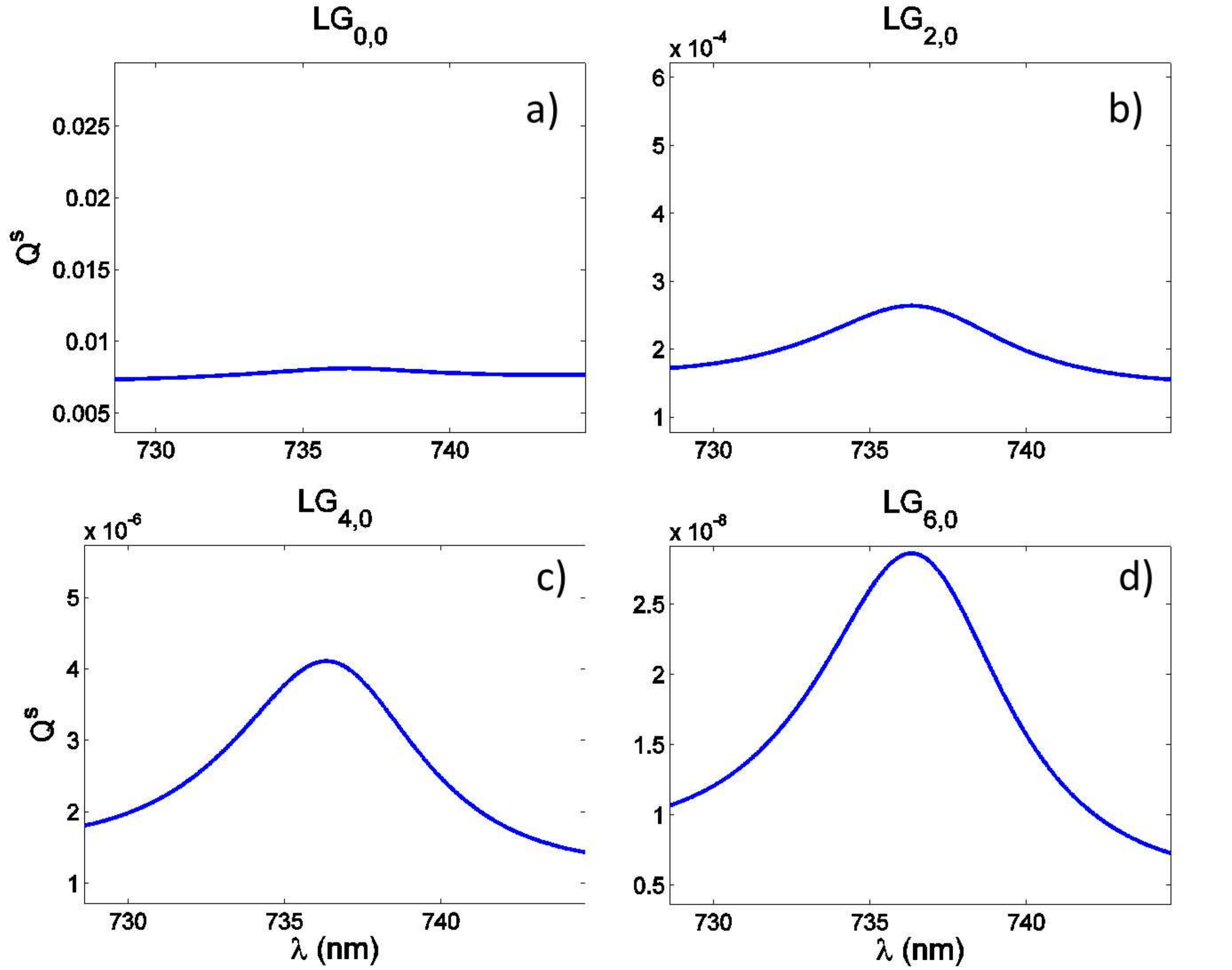}
\caption{Variation of a resonance in $Q_{s}$ as a function of wavelength, $\lambda \in \left[728.6, 744.6 \right]$ nm, and AM of the incident beam. The excitation beam is a) LG$_{0,0}$ b) LG$_{2,0}$ c) LG$_{4,0}$ d) LG$_{6,0}$. All the rest of parameters do not vary: $n_r=1.5$, $p=1$, $R=1.3 \ \mu$m. The multipolar decomposition of each of the LG beams is given in Figure \ref{F_multipolar}. The chosen resonance corresponds to $a_{17}$, that is, an electric multipole with $j = 17$.}
\label{F_Enhancement}
\end{figure}

In this section, I will look into one of the applications of GLMT - the excitation of WGMs, or Morphological-Dependent Resonances (MDRs) \cite{Lock1998,Oraevsky2002}. WGM and MDR are interchangeable expressions, however the term MDR has been typically only used in the context of GLMT, whereas WGM has been used in almost any other branch of physics. Hence, I will use the term WGM throughout this thesis to refer to MDR. In the next pages, I will use paraxial and non-paraxial cylindrically symmetric beams to excite single WGMs in dielectric spheres. I will show that provided an adequate ratio of $R/\lambda$ is obtained the excitation beam, a single WGM can be excited with direct light, without the need of going to evanescent coupling. In order to do that, I will need to compute the scattering efficiency factors defined in section \ref{Ch2_eff} for an incident field given by expression (\ref{E_general}) and a scattered and interior fields given by equations (\ref{E_fields}). As described in chapter \ref{Ch2}, the first step is computing the temporal average of the Poynting vector, $\langle \Ss \rangle$. Then, the flux of it across a spherical surface is computed to obtain the scattering cross section $C_s$. The general result was given in chapter \ref{Ch2} by equation (\ref{E_Cs}): 
\begin{equation}
C_s =  \sum_{j=1}^{\infty} \sum_{m_z=-j}^j  \left( \vert
a_j \gje \vert^2 + \vert b_j \gjm \vert^2 \right)
\label{hola}
\end{equation}
Then, if the result is particularized for $\gjm=i^j (2j+1)^{1/2} \Cjmsp$ and $\gje = i^{j+1} p (2j+1)^{1/2}  \Cjmsp$ (see equations (\ref{cjmgjm}, \ref{E_general})), the scattering efficiency factor $Q_s$ can be computed \cite{Zambrana2012,Zambrana2013JQSRT}:
\begin{equation}
Q_s = \sum_{j=\vert m_z^* \vert}^{\infty} \dfrac{ 2j+1}{x^2} \vert \Cjmsp \vert^2 \left( \vert a_j \vert ^2 + \vert b_j \vert^2 \right)
\label{E_Qs}
\end{equation}
where $m_z^* = l+p$ , $p$ being the helicity of the incident beam, and $l$ the topological charge of the LG mode used to produce the focused beam; and $x = 2\pi R /\lambda$ is the size parameter of the problem. It is insightful to compare equation (\ref{E_Qs}) with the expression for $Q_s$ in Mie theory. This expression is \cite{Bohren1983}:
\begin{equation}
Q_s^{Mie} = \sum_{j=1}^{\infty} \dfrac{ 2j+1}{x^2} \left( \vert a_j \vert ^2 + \vert b_j \vert^2 \right)
\label{E_Qsmie}
\end{equation}
Expressions (\ref{E_Qs}) and (\ref{E_Qsmie}) give rise to very different scattering efficiency as a function of the wavelength. I showed these differences in \cite{Zambrana2013JQSRT} using paraxial LG beams as $\Ei$. That is, I used the equation (\ref{Cjlp}) to compute $\Cjmsp$. The differences between the scattering efficiency using equations (\ref{E_Qs}) and (\ref{E_Qsmie}) are summarised in Figures \ref{F_QsMie}, \ref{F_Qswavelength}, \ref{F_Enhancement}. Firstly, I use equation (\ref{E_Qsmie}) to compute the scattering efficiency for Mie theory. The result is shown in Figure \ref{F_QsMie}. It can be observed that high order multipoles only give rise to ripples in ths scattering structure. That implies that the scattering is dominated by the low order modes.
Secondly, I use equation (\ref{E_Qs}) to plot the scattering efficiency for the same incident four LG modes whose multipolar decomposition has been given in Figure \ref{F_multipolar}, \textit{i.e.} LG$_{0,0}$, LG$_{2,0}$, LG$_{4,0}$, LG$_{6,0}$. As seen in Figure \ref{F_Qswavelength}, an increase in the AM of the incident beam is translated into an enhancement of the ripple structure with respect to the background produced by the low order modes. This phenomenon is clear when we compare Figure \ref{F_Qswavelength}(a) with (\ref{F_Qswavelength}(d)). In Figure \ref{F_Qswavelength}(a), it is observed that the scattering is dominated by the low multipolar orders, and the high ones only contribute to the scattering as ripples, very similarly to Figure \ref{F_QsMie}\footnote{The underlying reason is that both a plane wave and Gaussian beam are very similar when the sphere is much smaller than the beam waist of the Gaussian beam $w_0$. Furthemore, both of them have the same AM content, $J_z=p$.}. Nevertheless, in Figure \ref{F_Qswavelength}(d) the ripple structure is largely enhanced and the scattering is dominated by the ripple structure itself. 
In order to highlight this effect and quantify it, the influence of the same four excitation beams (LG$_{0,0}$, LG$_{2,0}$, LG$_{4,0}$, LG$_{6,0}$) on a single ripple is plotted in Figure \ref{F_Enhancement}. 
The maximum value of the scattering efficiency, $Q_{max}$, which peaks at $736.6$ nm, and the minimum value are computed for each plot. The minimum value is computed as the average of two values $8$ nm apart in wavelength from the position of the maximum. Then, the ratio $\delta$ can be defined as the relative quotient between the subtraction of the maximum and the minimum, and the minimum.
\begin{equation}
\delta= \dfrac{Q_{max}-Q_{min}}{Q_{min}} 
\label{gamma}
\end{equation}
where $Q_{s}^{max}=Q_{s}^{736.6}$ and $Q_s^{min}=(Q_{s}^{744.6}+Q_{s}^{728.6})/2$. The results are presented in table (\ref{T_Gamma}). It can be seen that the background is highly reduced. When a  LG$_{0,0}$ is used to excite the sphere, the peak only represents 8$\%$ of the background. Nevertheless, when the incident beam is a LG$_{6,0}$, the peak contribution to $Q_s$ is two times larger than the addition of all the other modes. This is easily understood looking at equation (\ref{E_Qs}). Due to the rotational symmetry of the system, the AM needs to be preserved. That is, if $\Ei$ is an eigenstate of has $J_z$ with value $m_z^*$, its multipolar decomposition is such that the first $(m_z^* -1) $ multipolar modes are not included in the scattering. 
\renewcommand{\arraystretch}{1}
\begin{table} 
\caption{Enhancement of the ripple structure using different LG beams. }
\begin{center} 
\begin{tabular}{|c|c|c|c|c|}
\hline  & LG$_{0,0}$ & LG$_{2,0}$ & LG$_{4,0}$ & LG$_{6,0}$ \\ 
\hline $\delta (\%)$ & 8.184 & 61.30 & 153.9 & 218.9 \\ 
\hline
\end{tabular}
\end{center}
\label{T_Gamma}
\end{table}
This way of enhancing the ripple structure by cancelling the contribution of the low order modes has a severe drawback: it reduces a lot the scattered power.
\begin{figure}[htbp]
\centering\includegraphics[width=\columnwidth]{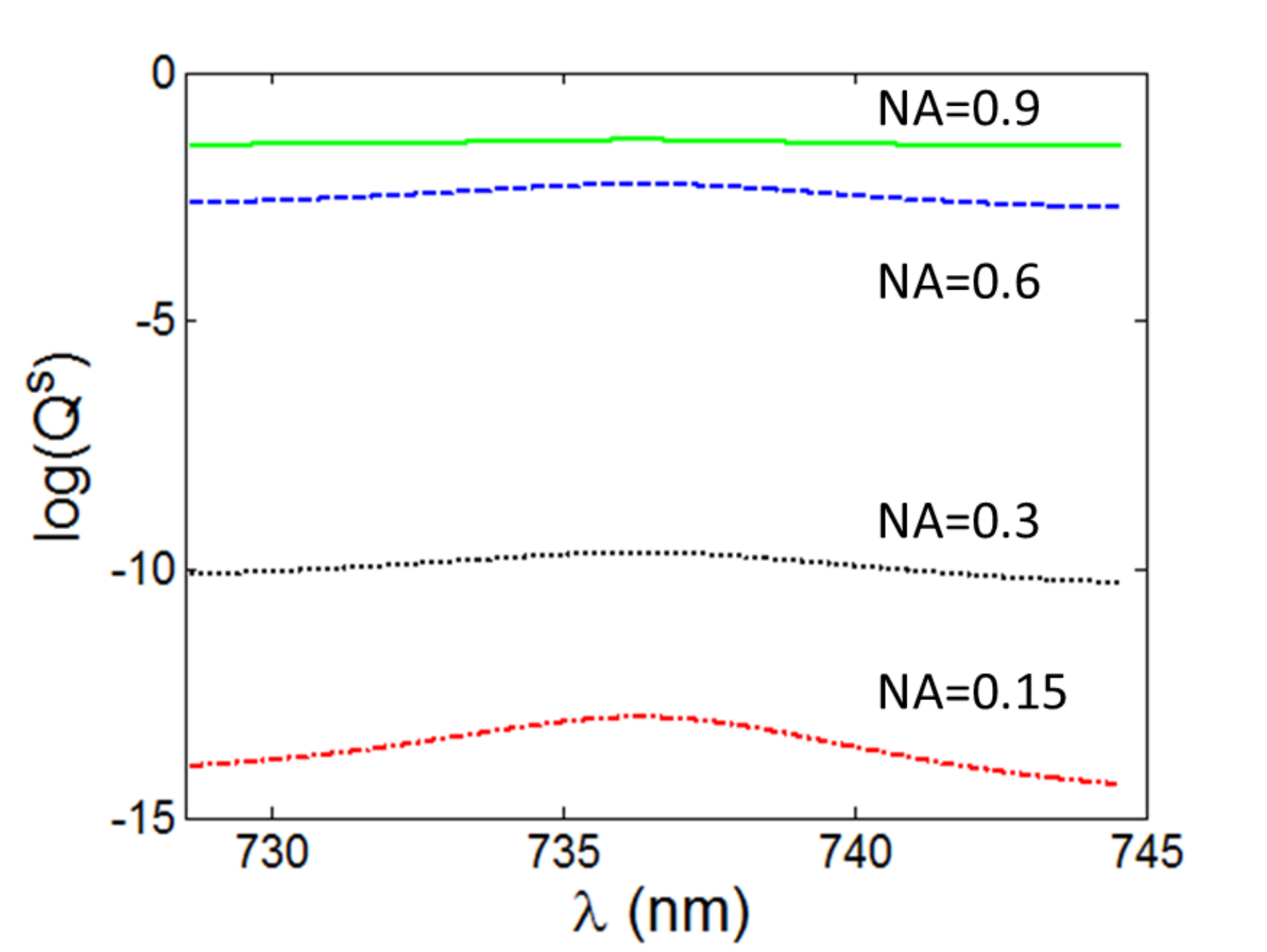}
\caption{Effect of the NA of the lens on the excitation of a single resonance at $\lambda=736.6$ nm. A LG$_{6,0}$ with helicity $p=1$ is used for all the cases, but different lenses are used to focus it down to the particle. The particle has a radius $R=1.3 \ \mu$m, and a relative index of refraction $n_r=1.5$. In all the cases, $Q_s$ is plotted in logarithmic units.}
\label{F_excitation}
\end{figure}
In order to make up for it, a higher NA lens can be used. This is shown in Figure \ref{F_excitation}, where the same resonance presented in Figure \ref{F_Enhancement} is excited with a LG$_{6,0}$ focused with different NAs. It can be observed that there is a difference in more than 10 orders of magnitude between the scattering produced by a paraxial beam (or NA$\approx 0.25$, see section \ref{Ch3_non}) and a highly non-paraxial one. Thus, highly non-paraxial beams with a well-defined AM and helicity are a good choice to excite single resonances, and in particular WGMs.\\\\ 
WGMs are widely used in physics. Their incredibly high Q factors make them very useful to probe any sort of disturbance in the environment \cite{Braginsky1989,Schiller1991,Knight1997,Oraevsky2002,Vahala2003}. Their definition can be found in a very good review done by Oraevsky \cite{Oraevsky2002}. He defines a WGM as a single multipolar mode, with $j$ large, $m_z=j$ and $\xi=1$, where $\xi$ is defined in the next lines\footnote{Oraevsky does not state anything about the parity of the multipolar mode, nor the frequency. Thus, a WGM can have both magnetic or electric parity, and can also be excited at any wavelength, as long as the other conditions are fulfilled. I will denote them as $\mathbf{A}_{jj}^{(y)}$.}. Even though defining something as `large' is not very mathematical, it makes sense in this context. The WGM resonances are computed as the roots of the Mie coefficients of order $j$. Each Mie coefficient has multiple roots, which are usually listed by the root number $\xi$. The Q factor of the resonance decreases with the root number $\xi=1,2,..$, therefore given a WGM mode (\textit{i.e.}, given $j$) the narrowest WGM resonance happens for $\xi=1$ \cite{Oraevsky2002}. Due to losses, it is experimentally very challenging to obtain Q factors larger than $10^9$, therefore a large $j$ can be understood as such that makes the Q factor of a resonance larger than $10^9$. There is not a quantitative answer to this, because the Q factor not only depend on $j$, but it also depends on the relative refractive index of the sphere with respect to the surrounding medium, $n_r$.\\\\
Now, using beams of light with a well-defined $J_z$ makes it straightforward to excite modes of light with a single $m_z^*=j$, where $j$ is large. The control of $m_z$ is done with the azimuthal number of a LG$_{l,q}$ beam and its helicity $p$: $m_z^*=l+p$.  However, following Oraevsky's definition, this is not enough to excite a WGM - only one $j$ needs to be excited, too.\\\\
To achieve that, the idea is to use the properties of the Mie coefficients for dielectric particles\footnote{Remember that a dielectric particle has a real index of refraction.}. In this case, the Mie coefficients, $a_j$ and $b_j$, that is the multipolar moments of the sphere, are complex and their absolute values are bounded between zero and one, $0\leq \left\lbrace \vert a_j\vert,\vert b_j\vert \right\rbrace \leq 1$. Both $a_j$ and $b_j$ are very close to zero for small $x = 2\pi R/ \lambda$, and they start to grow for a value of $x$ which is proportional to the order of the mode $j$. The proportionality value is a function $f(n_r)$ that depends on $n_r$. For example, when $n_r=1.5$, it can be computationally verified that the Mie coefficients approximately start to grow when $x \approx 4j/5$ (see Figure \ref{ajbj}) \cite{Zambrana2013OE}. Then, if $x < 4j/5$ both $a_j$ and $b_j$ are negligible. For $x > 4j/5$ their absolute values oscillate between 0 and 1. This is depicted in Figure \ref{ajbj}, where $a_j$ and $b_j$ are plotted for $j=1,10,20$. It can be observed that all the Mie coefficients follow the pattern described above: their absolute value is 0 until $x \approx 4j/5$, and for $x \geq 4j/5$ they oscillate between 0 and 1. 
\begin{figure}[tbp]
\centering\includegraphics[width=12cm]{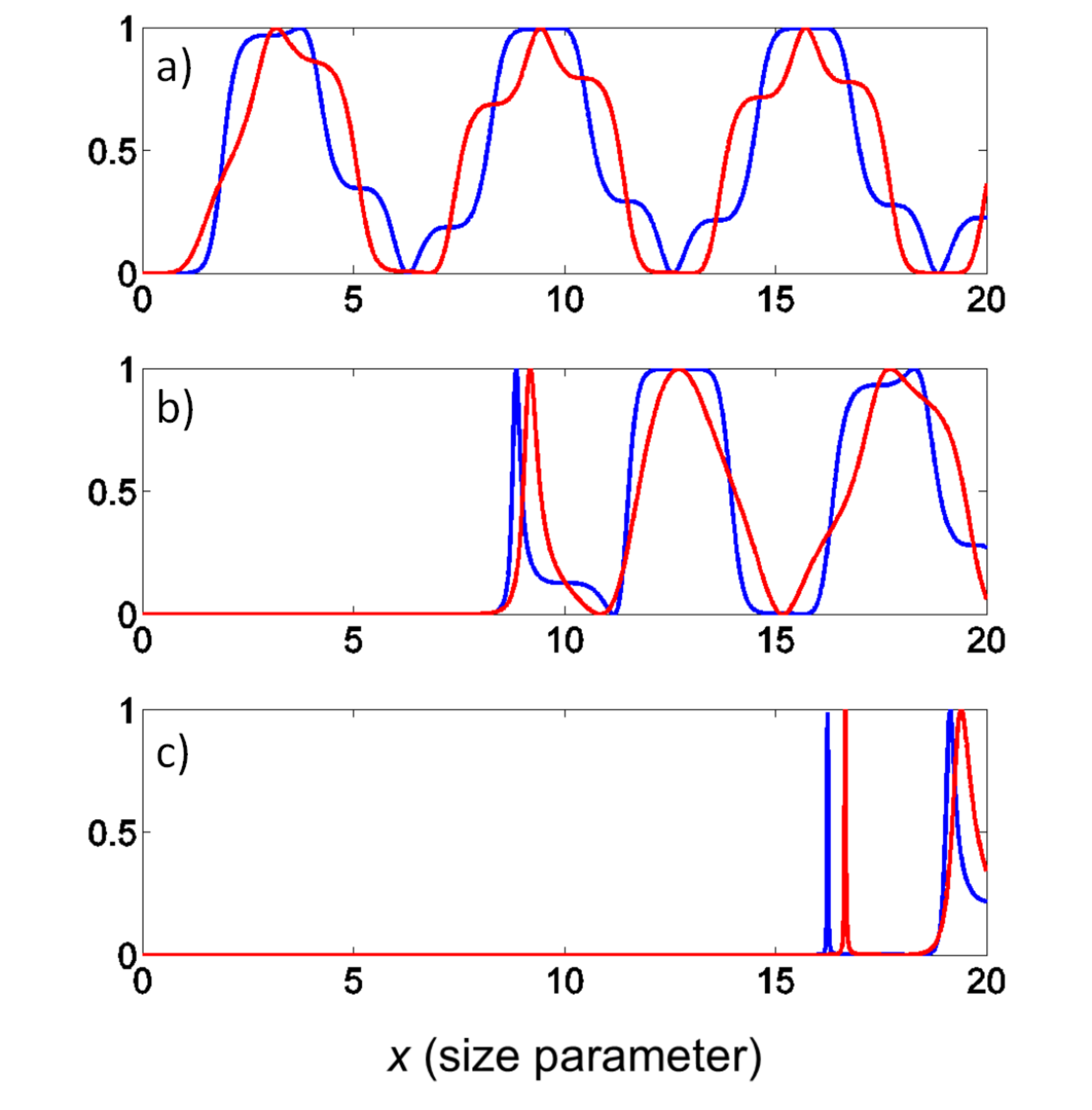}
\centering\caption{Norm of the Mie coefficients $\vert a_j \vert $ (in red) and $\vert b_j \vert$ (in blue) for $j=1,10,20$ in a), b) and c) respectively, as a function of the size parameter $x=2\pi r/\lambda$ . The relative index of refraction is $n_r=1.5$. It can be seen that all of them start being significantly different from zero when $x \approx 4j/5$. \label{ajbj}}
\end{figure}
Hence, it always exists an interval around $x \approx 4j/5 $ where $a_j$ and $b_j$ start growing and the higher Mie coefficients are approximately zero (as they start growing for $x \approx 4(j+1)/5$ ). Consequently, if the incident beam has a $J_z=m_z^*$\footnote{As mentioned in chapter \ref{Ch1}, this is an abuse of language. It means that the beam is an eigenstate of $J_z$ with eigenvalue $m_z^*$.} and we pick a particle whose size $R$ is such that $kR \approx f(n_r) \cdot m_z^*$, only modes with $j=m_z^*$ will be excited. This condition can be achieved by tuning the wavelength of the incident beam, $\lambda = 2\pi /k$. In summary, a single WGM with indexes $j$ and $m_z=j$ ($\mathbf{A}_{jj}^{(y)}$) can be excited by doing the following two steps process:
\begin{enumerate}
\item We create a cylindrically symmetric beam with a large value of $J_z = m_z^* = l +p$ and we select a MO with a large NA to tightly focus it. The wavelength of the laser is supposed fixed.
\item We choose a sphere whose radius $R$ is such that the following condition is fulfilled: $R \approx \lambda \cdot m_z^* \cdot f(n_r) /(2\pi) $
\end{enumerate}
Next, I will illustrate how precise and flexible this method of exciting WGMs is by giving three examples of excitation of WGMs. In the three examples, $\lambda$ will be fixed at $\lambda=532$ nm:
\begin{itemize}
\item \underline{\textsc{Example 1}}: I want to excite a WGM with $j=15$, $\mathbf{A}_{15,15}^{(y)}$. Hence, a beam of light with $\ms = 15$ needs to be used. That can be a LG$_{14,0}$ beam with a helicity $p=1$. Now, the size of the sphere needs to be selected so that only a single multipolar mode is excited. All the spheres are surrounded by air and made of a material with $n_1=n_r=1.5$. It can be computationally proved that $n_r=1.5 \rightarrow f(n_r)\approx 0.8$ \cite{Zambrana2013OE}. As I want to excite a mode with $j=15$, then I need that $kR / f(n_r) \approx 15 \rightarrow kR \approx 12$. Given the excitation wavelength of $\lambda=532$ nm, the condition is fulfilled by a particle with $R=1.0 \ \mu$m. It can be checked that this multipolar resonance has a magnetic parity and a Q factor of $Q=70$.
\item \underline{\textsc{Example 2}}: A WGM with $j=40$ needs to be excited and the available spheres and surrounding medium are still such that $n_r=1.5$. A  LG$_{39,0}$ beam with $p=1$ is created to excite the resonance. Then, the radius of the sphere needs to be such that $kR / f(n_r) \approx 40 \rightarrow kR \approx 31$, which is fulfilled by $R = 2.6 \ \mu$m. In this case, the Q factor is $Q=5\cdot 10^4$, and the parity is still magnetic.
\item \underline{\textsc{Example 3}}: The material which the sphere is made of is changed and now $n_r = 2$. It can be computationally checked that this implies that $f(n_r)\approx 0.6$. I want to excite the same multipolar modes as in the previous case, \textit{i.e.} $\mathbf{A}_{40,40}^{(m)}$. Therefore, I need to used a LG$_{39,0}$ mode with $p=1$. The condition on the radius of the sphere is the same one, but the result is different due to the change of material: $kR / f(n_r) \approx 40 \rightarrow kR \approx 23$, which yields $R = 1.9\ \mu$m and a Q factor $Q=1\cdot 10^6$.
\end{itemize}
WGMs or multipolar modes are normal modes of the sphere, and as such they are symmetric under parity transformations (see sections \ref{Ch1_multipoles} and \ref{Ch2_Symmetries}).
\begin{figure}[htbp]
\centering\includegraphics[width=\columnwidth]{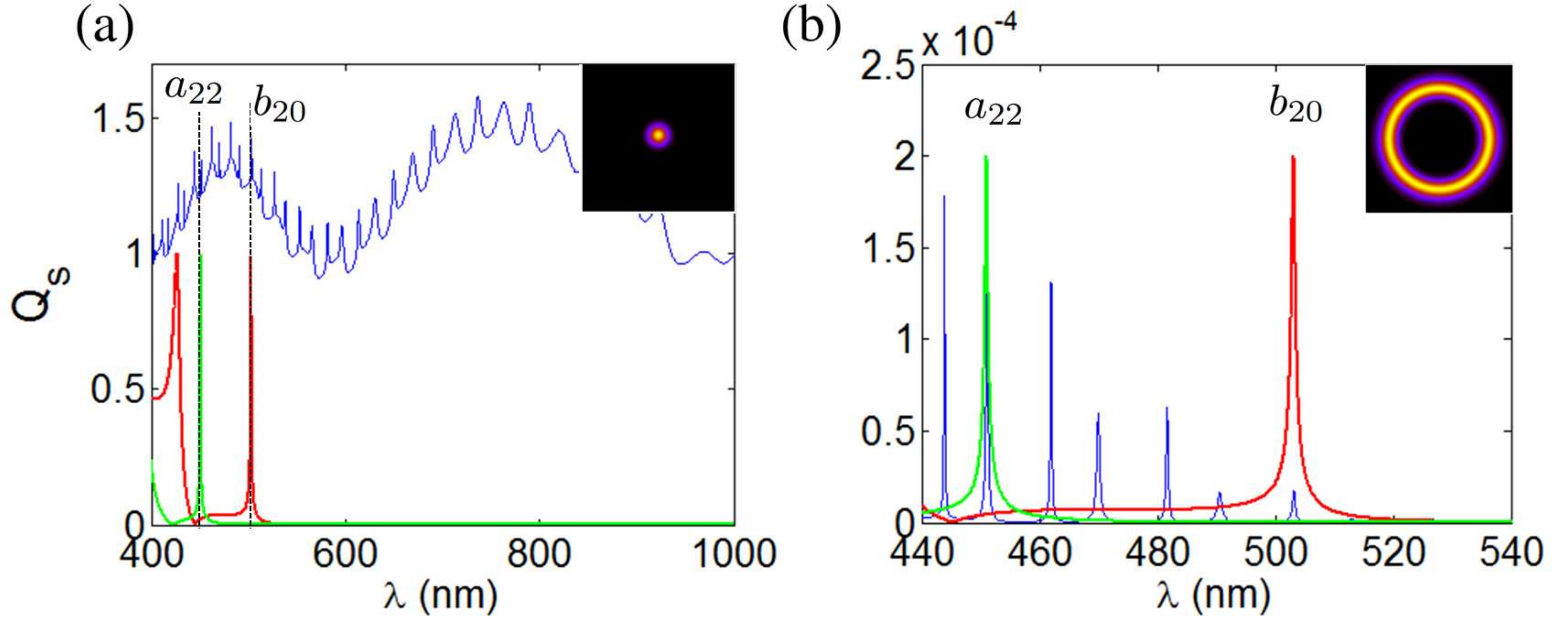}
\caption{Suppression of background in the scattering efficiency ($Q_{s}$). Input beams are a) LG$_{0,0}$ and b) LG$_{18,0}$, both with $p=1$. In a) an aplanatic lens with NA$=0.25$ is used, whereas in b) an aplanatic lens with a NA=0.9 is used. The rest of parameters are kept constant for both plots, \textit{i.e.} $R=1.3 \ \mu$m, $n_r= 1.5$. The insets indicate the typical profile of the beam used to excite the sphere. The scattering efficiency is represented with a blue continuous line. The Mie coefficients $b_{20}$ and $a_{22}$ are plotted with a red and a green line respectively. In a) we have indicated with a dashed line the position of two particular resonances for these modes. Note that the ordinate axis in a) and b) are different. The y-axis scale in both plots is given by $Q_s$. Then, $b_{20}$ and $a_{22}$ are re-scaled to match the values of $Q_s$.} 
\label{F_resonances}
\end{figure}
It is interesting to see what happens when we decompose the scattering into its two orthogonal helicity components. For an electric or magnetic resonance, the scattering is split equally into two orthogonal helicity components \cite{Ivan2012PRA,Zambrana2012}. This effect is put forward in Figures \ref{F_resonances}, \ref{F_intensity}.\\\\
First, Figure \ref{F_resonances} shows the background-suppression effect previously shown in Figures \ref{F_Qswavelength}, \ref{F_Enhancement}. That is, in Figure \ref{F_resonances}(a) I show the scattering efficiency produced by a Gaussian beam on a particle whose radius is $R=1.3 \  \mu \text{m}$ and whose relative index of refraction is $n_r=1.5$. Two resonances given by $a_{22}$ and $b_{20}$ are singled out. Their contribution to the scattering is seen to be a single ripple, due to the large background produced by all the other modes. However, in (\ref{F_resonances}(b)), where a focused LG$_{18,0}$ is used, the scattering efficiency is completely dominated by the ripple structure, as the background produced by the low order multipolar modes is not excited due to the large value of $J_z$. \\\\
Then, in Figure \ref{F_intensity}, the intensity of the electric field ($\vert \E \vert^2$) in all the space is plotted at the $b_{20}$ resonant condition, $\lambda=503$ nm, and $2$ nm out of resonances at $\lambda = 505$ nm. The particle and the surrounding medium are still the same, \textit{i.e.} $R=1.3 \  \mu \text{m}$, $n_r=1.5$. $\vert \E \vert^2$ is plotted for two excitation cases: $\Ei = \text{LG}_{0,0}\spphat$ and $\Ei = \text{LG}_{18,0}\spphat$, both of them with $p=1$. Then, $\Ei = \text{LG}_{0,0}\spphat$ is focused with a NA$=0.25$ and $\Ei = \text{LG}_{18,0}\spphat$ with a NA$=0.9$. The resonant behaviour of the scattered field is especially clear when the opposite component of the helicity $\Lambda^-$ is examined, since the contribution due to the incident field is 0. Figures \ref{F_intensity}(a)-(d) depict the fact that the influence of the $b_{20}$ resonance on the scattering is minimum, as the same behaviour is observed for on and off-resonance cases. However, in Figures \ref{F_intensity}(e)-(h) the situation is very different. There are two orders of magnitude of difference in intensity between the cross helicity component in Figure \ref{F_intensity}(f) and \ref{F_intensity}(h). Finally, it can be observed that unlike Figure \ref{F_intensity}(e-h), where the field can be described with a single mode, many multipoles are needed to describe the highly complex pattern seen in Figure \ref{F_intensity}(a-d). In fact, the intensity profile observed in Figure \ref{F_intensity}(a-d) has drawn a lot of attention since 2004, where it was first characterized as photonic nanojet \cite{Chen2004}. Since then, nanojets have been extensively studied and applied in many different fields \cite{Ferrand2008,Devilez2008,Gerard2008,Heifetz2009}.\\\\
Previous works on the excitation of metallic particles by LG beams were carried out in  \cite{vandeNes07}. There, the use of purely numerical techniques allowed the authors to distinguish certain properties of the scattered field. 
\begin{figure}[tbp]  
\begin{center}
\includegraphics[width=12.5cm]{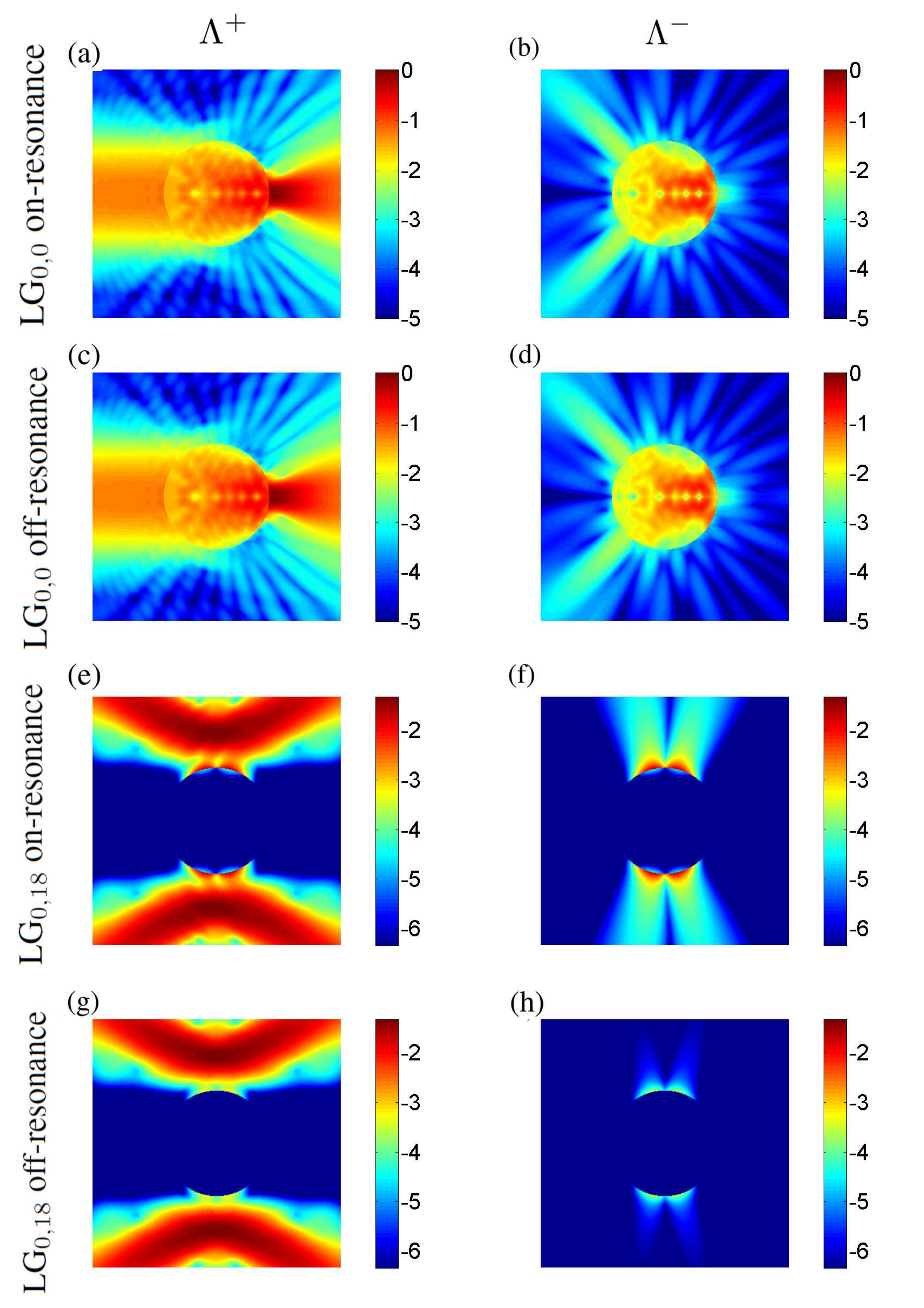}
\end{center}
\caption{Projection into the helicity basis of the intensity of the interior and total field from a sphere with parameters $\left\lbrace R=1.3 \ \mu \mathrm{m},n_r=1.5 \right\rbrace $. The helicity of the incident beam is always $p=1$. The intensity is plotted in logarithmic scale, where 0 corresponds to the maximum of intensity in a).  Images a), b), e) and f) have been simulated for a $\lambda_{\mathrm{resonance}}=503$ nm, whereas a $2$ nm shift has been introduced in the off-resonance wavelength for the others, \textit{i.e.} $\lambda_{\mathrm{off-resonance}} = 505$ nm. The simulations for the LG$_{0,0}$ beam have been executed with a NA=0.25 lens  in order to be able to excite the resonance of order 20. Note that the scattering is dominated by the off-resonance modes, since a change of 2nm in $\lambda$ leaves the field invariant. Calculations with higher numerical apertures give similar results. On the contrary, a lens with a NA=0.9 has been used for the excitation with a LG$_{0,18}$. In both cases, the entrance pupil of the lens was filled. Note that in contrast with b), the mode of order 20 is greatly enhanced when we tune the wavelength to the resonance.}
\label{F_intensity}
\end{figure}
However, due to the large absorption of silver, the ripple structure could not be enhanced in a similar fashion to what I have shown. On the contrary, the control of the AM of LG modes was used to achieve optical transparency in dielectric particles in \cite{Rury2012}. LG modes were also used in some recent scattering experiments with silica spheres \cite{Gabi2012OL}. The theory developed here allows for a better understanding of the experiments. In fact, the same calculations could also help to expand the computations of the optical forces when the incident beam is not a plane wave \cite{Stilgoe2008} and when the particle is not much smaller than the wavelength \cite{Nieto-Vesperinas2010,Novitsky2011,Dogariu2012}. Last but not least, the technique presented in the present chapter to control the multipolar content of cylindrically symmetric beams, as well as its application to excite single multipolar modes could open up new possibilities in the forefront research of magnetic resonances \cite{Garcia-Etxarri2011, Kuznetsov2012, Miroshnichenko2012, Filonov2012,preKuznetsov2013}.

\begin{savequote}[10cm] 
\sffamily
``Physics is like sex: sure, it may give some practical results, but that's not why we do it'' 
\qauthor{Richard P. Feynman}
\end{savequote}

\chapter{Control of the helicity content in scattering}
\graphicspath{{ch4/}} 
\label{Ch4}

\newcommand{\aljms}{\alpha_{jm_z^*}^p}
\newcommand{\aljm}{\alpha_{jm_z}^p}
\newcommand{\Ajmspp}{\mathbf{A}_{j\ms}^{p}}
\newcommand{\Ajmsmp}{\mathbf{A}_{j\ms}^{-p}}

\newcommand{\Ajmmp}{\mathbf{A}_{jm_z}^{-p}}
\newcommand{\Tmp}{T_{{m_z^*}p}(R,n_r)}
\newcommand{\Tp}{T_{p}(R,n_r)}
\newcommand\nal{n_{\mathrm{Al}_2\mathrm{O}_3}}
\newcommand\nsi{n_{\mathrm{SiO}_2}}

\section{GLMT problem in helicity basis}\label{Ch4_GLMT_helicity}
In the previous chapter, I have used the GLMT developed in chapter \ref{Ch2} to find the scattering off a sphere illuminated with a cylindrically symmetric beam with a well-defined helicity. As explained in chapter \ref{Ch2}, the problem is described with the multipolar fields ($\Am,\Ae$) since they are the normal modes of the system. Consequently, and as we have seen in section \ref{Ch3_WGM}, the resonances of the system have a well-defined parity (either electric or magnetic). In the current chapter, I will show that a wealth of information about the sample can be easily obtained by changing the basis in which the problem is described. The same information could be obtained with the expressions derived in chapter \ref{Ch2}, but the change of basis will unveil this information in a straightforward manner. I will describe the problem with the multipolar fields with a well-defined helicity (see section \ref{Ch1_multipoles}). Their relation with the multipolar fields (with well-defined parity)\footnote{Due to their common use, the multipolar fields with well-defined parity will be simply referred to as multipolar fields.} is the following one \cite{Lifshitz1982,Zambrana2013}:
\begin{equation}
\Ajmp = \dfrac{\Am + i \Ae} {\sqrt{2}} \qquad \Ajmm = \dfrac{\Am - i \Ae} {\sqrt{2}}
\label{E_MultiHel}
\end{equation}
It can be checked that these modes have a well-defined helicity, \textit{i.e.} they are eigenstates of the $\Lambda$ operator. In order to check that, the following relations between the multipolar fields need to be taken into account \cite{Rose1957,Zambrana2012}: 
\begin{equation}
\Lambda \Am= i \Ae \qquad \Lambda \Ae = -i \Am
\end{equation}
Then, it is straightforward to check that:
\begin{equation}
\Lambda \Ajmp = + \Ajmp \qquad \Lambda \Ajmm = - \Ajmm
\label{E_Ajmp}
\end{equation}
As discussed in chapter \ref{Ch2}, duality transformations are generated by the helicity operator \cite{Calkin1965,Zwanziger1968}. It is well-known that Maxwell equations are not symmetric with respect to electric and magnetic fields in the presence of charges \cite{Jackson1998}. This is due to the lack of magnetic monopoles in the universe \cite{Bendtz2013}. However, it was proven in \cite{Ivan2013} that the macroscopic Maxwell equations for isotropic and homogeneous media can be dual-symmetric if some conditions are fulfilled. Microscopically, duality symmetry is still broken, but the collective effect of all the charges and currents in the medium restores the symmetry in the macroscopic approximation. In \cite{Ivan2013}, the non-conservation of helicity was carefully quantified. In the same way as it happens with any other generator of a symmetry, if the helicity of a light beam is preserved upon interaction with a material medium, this necessarily implies that the system is symmetric under its associated duality symmetry. I will refer to these sort of media as `dual', \textit{i.e.} media that preserve the helicity content of the incident beam upon scattering. In order to study the dual behaviour of spheres, I will use beams of light with a well-defined helicity with value $p$. Its decomposition into multipolar modes of well-defined helicity will be the following one:
\begin{equation}
\Ei = \displaystyle\sum_{j,m_z}  \aljm \Ajmpp
\label{E_Ei_hel}
\end{equation}
where $\aljm$ are the amplitudes that determine the multipolar content of the incident beam\footnote{I will not call the coefficients $\aljm$ beam shape coefficients as I reserve that name for the coefficients $\gjm$ and $\gje$ that multiply the multipolar fields in the GLMT described in chapter \ref{Ch2}}. Then, the scattered and interior field can be obtained following the procedure shown in section \ref{Ch2_Scatt}. The following fields are obtained: 
\begin{equation}
\begin{array}{lll}
\Es &=& \displaystyle\sum_{j,m_z}  \aljm \left[ \dfrac{b_j+ a_j}{2} \Ajmpp + \dfrac{b_j- a_j}{2}\Ajmmp \right]  \\
\Eint &=& \displaystyle\sum_{j,m_z}  \aljm  \left[ \dfrac{c_j+ d_j}{2} \Ajmpp + \dfrac{c_j- d_j}{2}\Ajmmp \right]
\label{E_fields_hel}
\end{array}
\end{equation}
where $\left\lbrace a_j,b_j,c_j,d_j\right\rbrace$ are the Mie coefficients \cite{Bohren1983}. The rest of quantities defined for the GLMT in chapter \ref{Ch2} could be equally derived, but I will not compute them as I will not need them for my purposes.

\section{Kerker conditions}\label{Ch4_Kerker}
\begin{figure}[htb]
\centering\includegraphics[width=\columnwidth]{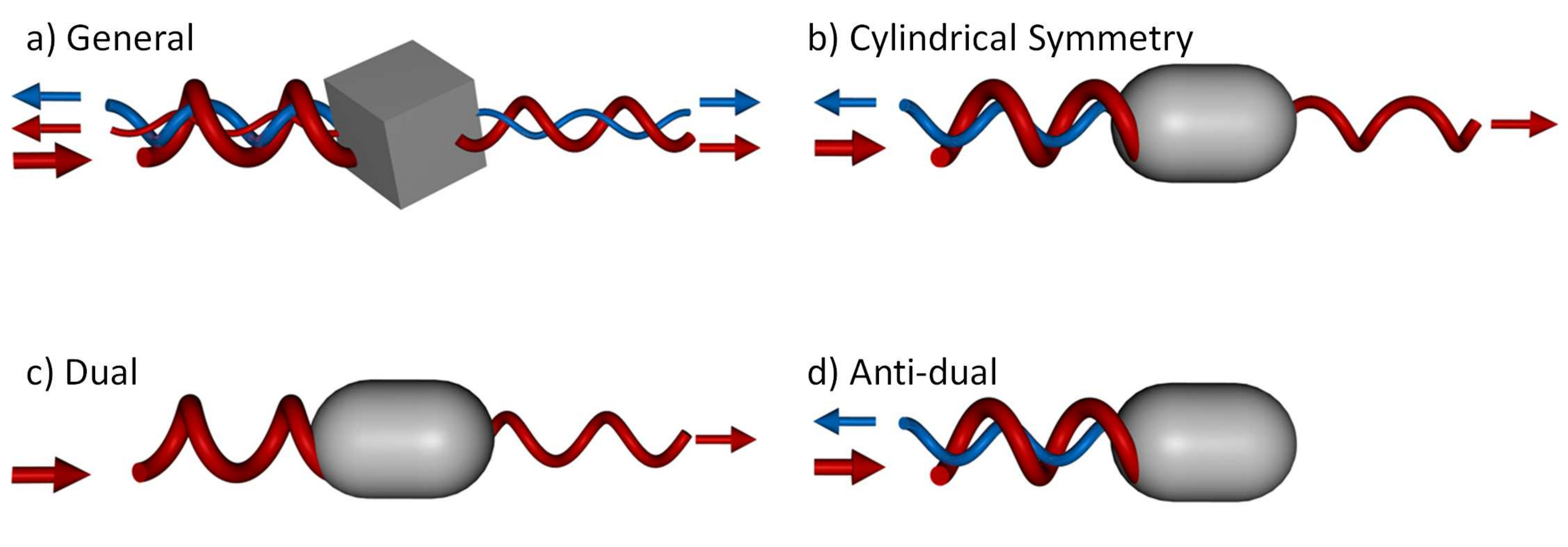}
\caption{Sketch of relations between Kerker conditions and cylindrical and Duality symmetry. The large red helix pointing the scatterer represents an incident plane wave with a well defined helicity. The other helices represent scattered plane waves. The small red helix has the same helicity as the incoming wave, while the small blue helix has the opposite helicity. a) A general scattering process with neither cylindrical nor dual symmetries is shown. In general, both back and forward scattering have contributions from both helicities. b) Scattering from a cylindrically symmetric object. In this case, symmetry imposes that forward scattering is of the same helicity than the incident field, and backward scattering of opposite helicity. c) A dual and cylindrically symmetric system. Only forward scattering is allowed with the same helicity. d) An anti-dual system with cylindrical symmetry. Only back scattering is allowed with flipped helicity.}
\label{F_kerker}
\end{figure}

In 1983, Kerker \textit{et al.} demonstrated that a sphere with $\mu \neq 1$  could have zero scattering in both back and forward directions \cite{Kerker1983}. These two anomalous conditions, nowadays known as first and second Kerker conditions, arguably passed unnoticed until 2006, when Mehta and co-workers published experimental evidence of zero forward and back-scattering by magnetic spheres \cite{Mehta2006,Mehta2006prb}. Since then, a lot of work has been done in this field, as researchers seek to independently control both the electric and magnetic resonances of different structures and control its directionality \cite{Garcia-Camara2008,Garcia-Camara2011,Garcia-Etxarri2011,Kuznetsov2012}. Recently, different research groups have managed to measure the two Kerker conditions in different regimes. Geffrin and co-workers measured these two conditions in the microwave regime \cite{Geffrin2012}. Whereas Person \textit{et al.} and Fu \textit{et al.}  measured the first Kerker condition in the dipolar approximation for optical wavelengths \cite{Person2013,Fu2013}. Nonetheless, the measurement of the Kerker conditions in the optical regime for arbitrary large particles still remains an unsolved problem.\\\\
In this chapter, I will extend the Kerker conditions to systems with cylindrical symmetry. We will see that this extension naturally arises from equations (\ref{E_fields_hel}). Indeed, looking at equations (\ref{E_fields_hel}), it can be seen that the scattered and interior fields do not generally preserve the helicity of the incident field (see Figure \ref{F_kerker}(a) and \ref{F_kerker}(b)). That is, in general, an incident beam with a well-defined helicity such as $\Ei$ impinging on an isotropic and homogeneous sphere will give rise to scattered and interior fields which will not be eigenvectors of the helicity operator $\Lambda$. This is evident from equations (\ref{E_fields_hel}), as multipolar modes with helicity $-p$ ($\Ajmmp$) are created upon scattering. The underlying reason for this is that the two pairs of Mie coefficients $\left\lbrace a_j,b_j \right\rbrace$ and $\left\lbrace c_j,d_j \right\rbrace$ are not generally equal and therefore the amplitudes modulating $\Ajmmp$ will be different from zero.\\\\ However, as it was proven by Kerker \textit{et al.}, the Mie coefficients can be equal for some cases, \textit{i.e.} $a_j(x)=b_j(x) \ \ \forall j,x$ when $\epsilon_r=\mu_r$ \cite{Kerker1983}. This is the so-called first Kerker condition (K1), and it implies zero-backscattering from the sphere in consideration. Actually, observing equations (\ref{E_fields_hel}), it can be seen that the K1 also implies that the sphere in consideration will preserve the helicity of the incident field $\Ei$ upon scattering and therefore be dual (see section \ref{Ch4_GLMT_helicity} to refresh the definition of a dual system). Indeed, for the scattered field, the amplitudes of the components with opposite helicity to the incident one are given by $\alpha_{j,m_z}^p (b_j- a_j)/2$. Hence, when $a_j=b_j$ they are all zero, granting the scattered field with the same helicity as the incident one. This feature has been overlooked in the past, and it will enable me to extend K1.\\\\
The K1 is a particular case of a more general condition that restores the EM duality symmetry in material media \cite{Ivan2013}. In general,
\begin{equation}
\dfrac{\epsilon_i}{\mu_i} =\text{const} \Longleftrightarrow  \Lambda\text{ conservation } 
\label{E_eps_mu}
\end{equation} 
for any medium made of an arbitrary number of isotropic and homogeneous sub-media $i$, $\epsilon_i$ and $\mu_i$ being its electric permittivity and magnetic permeability. That is, when an arbitrary light beam impinges on a medium such that the electric and magnetic fields behave symmetrically, i.e. condition (\ref{E_eps_mu}) is met, the helicity of this beam is not changed regardless of the geometry of this medium. Now, even though the zero-backscattering condition or K1 was introduced using spheres, it can be demonstrated that it can be relaxed to any system with cylindrical symmetry \cite{Zambrana2013}. Suppose we have a dual scatterer and impose that it must also be cylindrically symmetric around the $z$ axis. Then, besides helicity, the $J_z$ of the incident beam must also be preserved. Under these conditions, a plane wave incident along the symmetry axis will not backscatter. The proof comes from the definition of helicity $\Lambda={\mathbf{J}}\cdot{\mathbf{P}}/|{\mathbf{P}}|$ (see section \ref{Ch1_symm}). Imagine that a plane wave is directed in the $z$ axis (its linear momentum is $\mathbf{P}=P_z\mathbf{\hat{z}}$) and it has a well defined helicity $\Lambda = J_z\cdot P_z / \vert P_z \vert = J_z = 1$.
A plane wave with $\mathbf{P} = - P_z \mathbf{\hat{z}}$ must necessarily have $\Lambda=-J_z$. If the system preserves $J_z$ and $\Lambda$, such plane wave can never exist. Thus, it is very natural to extend K1 from a sphere such that $\epsilon_r=\mu_r$ to a dual system with cylindrical symmetry, \textit{i.e.} a system that preserves both helicity and the $z$ component of the AM (see Figure \ref{F_kerker}(c)), as both entities have zero backscattering \cite{Zambrana2013,Ivan2013}. \\\\
The second Kerker condition (K2) can also be extended in a similar fashion. K2 predicts zero forward scattering for dipolar magnetic particles. Here, I will relax its use to cylindrically symmetric systems of any size. In order to do this, I am going to define what an ``anti-dual'' scatterer is. It is a scatterer whose scattered light flips the helicity state of the incident beam. That is, if $\Ei=\sum \alpha_{jm_z} \mathbf{A}_{jm_z}^{p}$, then $\Es=\sum \beta_{jm_z} \mathbf{A}_{jm_z}^{-p}$. Looking at equations (\ref{E_fields_hel}), the conditions for an anti-dual sphere can be found. For $\Es$, the amplitudes of the components with the same helicity as the incident field are given by  $\alpha_{j,m_z}^p (b_j+ a_j)/2$. Thus, a necessary and sufficient condition for a sphere to be anti-dual is that:
\begin{equation}
a_j(x)=-b_j(x) \ \ \forall j,x
\label{E_K2}
\end{equation}
If this condition is expressed in the dipolar approximation, \textit{i.e} only terms with $j=1$ are considered, it yields $a_1=-b_1$. This is the so-called K2.\\\\
With this new perspective, the zero-forward scattering condition for spheres (K2) can be cast again as a particular case of a more general system, a system which is cylindrically symmetric and anti-dual. Once again, only considering the symmetries of such a system, it is possible to derive that the forward scattering must be zero. To prove that, a plane wave travelling in the positive $z$ axis, whose helicity is $\Lambda = J_z\cdot P_z / \vert P_z \vert = 1$ is used as $\Ei$. Now, as the system is anti-dual, the helicity in all the directions (in particular in forward) have to flip, giving $\Lambda=-1$. However, $J_z$ cannot change since the system is cylindrically symmetric, and $ P_z / \vert P_z \vert=1$ in the forward direction. Consequently, this plane wave cannot exist (see Figure \ref{F_kerker}(d)).\\\\
At this point, it is important to note that finding an exact anti-dual sphere is challenging. Some authors have noticed that dielectric, isotropic and homogeneous spheres cannot behave as anti-dual materials in the strict sense, as they would contradict the Optical Theorem\footnote{Their argument is only valid for plane wave excitation, though. As it is explained in \cite{GLMT_book}, the optical theorem only applies to Mie theory, and fails to provide meaningful results in GLMT, where more complicated beam of lights can be used.} \cite{Nieto-Vesperinas2011,Alu2010,Garcia-Camara2011}. But a possible alternative would be to use particles with gain, as the Mie coefficients can get values greater than 1 \cite{Garcia-Camara2011,Zambrana2013,Zambrana2013OE}.

\section{Helicity control in scattering}\label{Ch4_trans}
\begin{figure}[tbp]
\centering\includegraphics[width=\columnwidth]{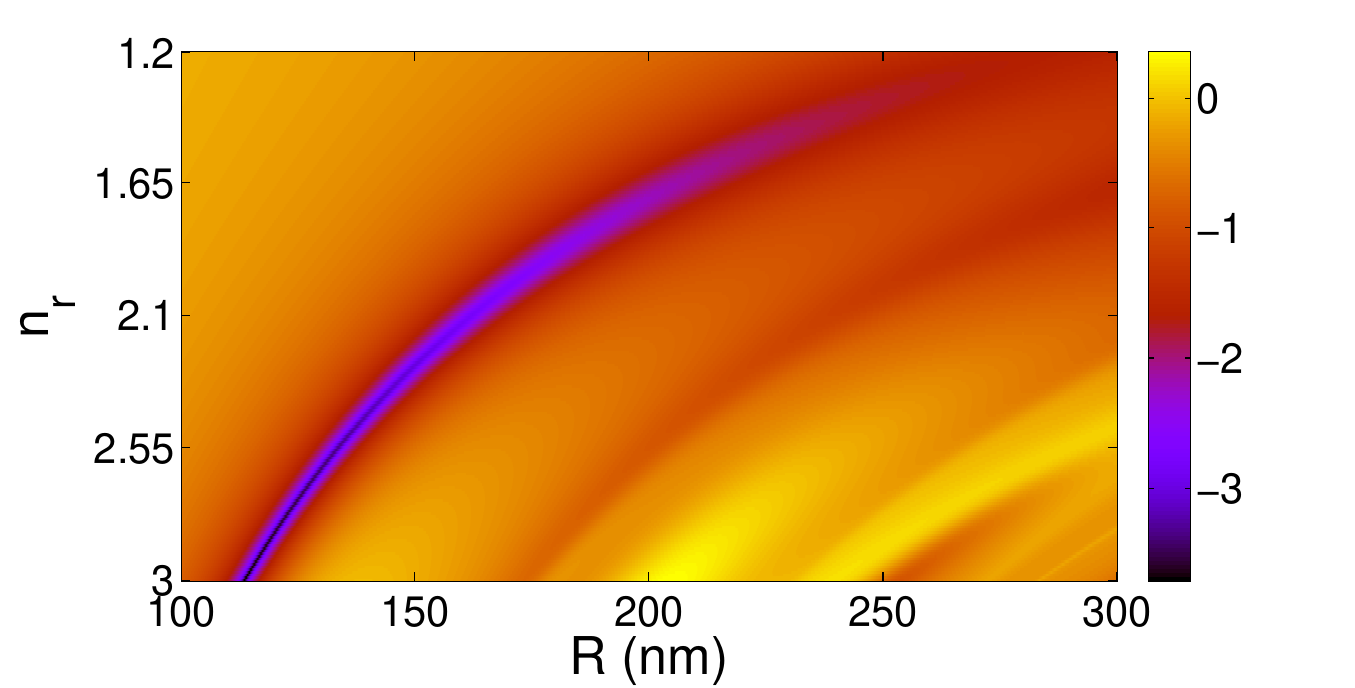}
\caption{ Plot of the $log$ of the Transfer function $\Tp$ for $p=1$, \textit{i.e.} $\log\left( T_{1}(R,n_r)\right) $, as a function of the radius $R$ of the particle (horizontal axis) and the relative index of refraction $n_r$ (vertical axis). The radius of the particle is varied from 100nm (left) to 300nm (right) and the relative index of refraction goes from $1.2 \leq n_r \leq 3$.}
\label{Tm}
\end{figure}

Equations (\ref{E_Ei_hel}, \ref{E_fields_hel}) give us the expression of the electric field in all space. Then, the magnetic field can be obtained using the Maxwell equations given by equations (\ref{E_Max_EH_mono}). The following expressions for the $\Hh$ field are obtained:
\begin{eqnarray}
\Hi & = & \dfrac{1}{\mu}\displaystyle\sum_{j,m_z}  p \cdot \aljm \Ajmpp \\ 
\Hs &=& \dfrac{1}{\mu} \displaystyle\sum_{j,m_z}  p \cdot \aljm \left[  \dfrac{b_j+ a_j}{2} \Ajmpp - \dfrac{b_j- a_j}{2}\Ajmmp \right]  \label{E_Hshel} \\ 
\Hint &=& \dfrac{1}{\mu_1} \displaystyle\sum_{j,m_z}  p \cdot \aljm  \left[ \dfrac{c_j+ d_j}{2} \Ajmpp - \dfrac{c_j- d_j}{2}\Ajmmp \right]
\end{eqnarray}
The energy density of the EM field can be computed as \cite{Jackson1998,preBliokh2013}:
\begin{equation}
\mathtt{w}=\dfrac{1}{16\pi} \left( \epsilon \vert \E \vert^2 + \mu \vert \Hh \vert ^2 \right)
\label{E_w}
\end{equation}
Then, if $\mathtt{w}$ is integrated on a surface, the energy of the EM field per unit of length is obtained:
\begin{equation}
\mathtt{W}= \int_{\Omega} \dfrac{1}{16\pi} \left( \epsilon \vert \E \vert^2 + \mu \vert \Hh \vert ^2 \right) d\Omega
\label{E_W}
\end{equation}
To find out more information about the scattering process, I will particularize equation (\ref{E_W}) for the scattered field, as it contains all the information about the scatterer. Given equations (\ref{E_fields_hel}, \ref{E_Hshel}), the orthogonality relations given in section \ref{Ch1_basis}, the definition of multipoles with well-defined helicity (\ref{E_MultiHel}), the following relations yield:
\begin{eqnarray}
\mathtt{W}^\mathrm{s} & = & \dfrac{1}{2} \displaystyle\sum_{j,m_z} \vert \aljm \vert^2 \left( \vert a_j \vert^2 + \vert b_j \vert^2\right) \\ 
\label{dual}
\mathtt{W}_p^\mathrm{s} & = & \dfrac{1}{4} \displaystyle\sum_{j,m_z} \vert \aljm  \vert^2 \vert b_j +   a_j \vert^2 \\
\mathtt{W}_{-p}^\mathrm{s} & = & \dfrac{1}{4} \displaystyle\sum_{j,m_z} \vert  \aljm  \vert^2 \vert b_j -   a_j \vert^2  
\label{antidual}
\end{eqnarray} 
$\mathtt{W}^\mathrm{s}$ represents the total energy of the scattered field. Due to the orthogonality of the multipolar fields, the scattered energy can be split into two orthogonal components, and these are $\mathtt{W}_p^\mathrm{s}$ and $\mathtt{W}_{-p}^\mathrm{s}$. They account for the energy scattered in modes with the same helicity as the incident field ($\mathtt{W}_p^\mathrm{s}$) and with the opposite helicity ($\mathtt{W}_{-p}^\mathrm{s}$).\\\\
As described in the previous section, note that in general the scattered field will always carry energy in modes with the opposite helicity, \textit{i.e.} $\mathtt{W}_{-p}^\mathrm{s} \neq 0$. However, if $a_j(x)=b_j(x) \  \forall j,x$, the particle only scatters energy with the same helicity as the incident beam, conserving the helicity of the EM field and therefore behaving as a dual medium. Note also that if $a_j(x)=-b_j(x) \ \ \forall j,x$, then $\mathtt{W}_{p}^\mathrm{s} = 0$. That is, the scattered field has the opposite helicity to the incident one and therefore the sphere behaves as an anti-dual scatterer. It can be proven that a spherical particle will only be dual if $\mu_r=\epsilon_r$ \cite{Kerker1983,Ivan2013}. Nonetheless, here I will relax this definition in order to address some experimentally relevant cases \cite{Ivan2013JCP,Ivan2013prb,Ivan2013OE}. I will consider that a particle of radius $R$ behaves in a dual or anti-dual manner at a given wavelength if all the scattered energy goes into the same helicity component or the opposite one at that wavelength.\\\\
These two facts have been experimentally verified in the dipolar regime \cite{Geffrin2012,Person2013,Fu2013}, experimentally achieving $a_1=b_1$ and $a_1 \approx -b_1$. I will show how to extend these conditions to other regimes and shapes. As a result, it will be clear that even a dielectric sphere can approximately fulfil the anti-dual condition or extended K2. In order to achieve these purposes, I will define a transfer function $T_{p}(R,n_r)$. Its definition depends on the helicity value of $\Ei$, the radius $R$ of the sphere in consideration and the relative refractive index $n_r$:
\begin{equation}
\Tp=  \dfrac{\mathtt{W}_{-p}^\mathrm{s}}{\mathtt{W}_p^\mathrm{s}} = \dfrac{\sum_{j,m_z} \vert \aljm \vert^2  \vert b_{j}  -  a_{j} \vert^2  } {\sum_{j,m_z}\vert \aljm \vert^2 \vert b_{j}  +  a_{j} \vert^2 } 
\label{Tj}
\end{equation}
$\Tp$ is the ratio between the scattered light going into modes with opposite helicity with respect to the incident light ($\mathtt{W}_{-p}^\mathrm{s}$), and the scattered light going into modes with the same helicity ($\mathtt{W}_{p}^\mathrm{s}$), for a given helicity ($p$) of the incident beam. Thus, $\Tp$ varies from 0 to infinity. When $\Tp$ tends to zero, the particle is dual and all the scattered light has helicity $p$: in other words, it fulfils the extended K1 \cite{Zambrana2013,Zambrana2013OE}. On the contrary, when $\Tp$ tends to infinity, the particle is anti-dual and all the scattering is transferred to the cross helicity, $-p$: it fulfils the extended K2 \cite{Zambrana2013,Zambrana2013OE}.\\\\
Given an incident field $\Ei$ and a sphere with radius $R$ and relative refractive index $n_r$, $\Tp$ can be computed and the combinations of $\{ \lambda, R, n_r \}$ that make the particle behave as dual or anti-dual can be studied \cite{Zambrana2013OE}. For example, Figure \ref{Tm} depicts the transfer function of sphere excited with a Gaussian beam at $\lambda=780$ nm. It is seen that the sphere can behave as dual for many resonant combinations of $\{R, n_r\}$. 
This way of looking at duality behaviour and Kerker conditions represents a considerable improvement over the methods used before, where only the first order Mie coefficiens were used \cite{Garcia-Camara2008,Garcia-Camara2011,Garcia-Etxarri2011}. As it can be observed in Figure \ref{Tm}, not all the points sitting on the curve $a_1=b_1$ are equally dual. Generally speaking, it is due to the fact that the particle has a size (or an index of refraction) for which the dipolar approximation does no longer apply. The transfer function defined in (\ref{Tj}) exactly quantifies this fact, as it accounts for the contributions of all the higher multipolar orders present in scattering.\\\\
It is also interesting to see how the induced duality strongly depends on the relative index of refraction. It is apparent from Figure \ref{Tm} that the dielectric sphere gets closer to the dual condition when $n_r$ gets larger. That means that the helicity of the incoming beam will be better preserved when particles with a high refractive index embedded in a low refractive index medium are used.\\\\
In addition to the duality considerations mentioned above, Figure \ref{Tm} also depicts the anti-dual behaviour of the particle. Two main features can be observed. First, as it can be deduced from the colorbar, the anti-dual condition is not achieved as finely as the dual is. That is, the extension of K2 is more difficult to achieve. Hence, the forward scattering is never reduced as much as the backward is. This feature had already been observed and analysed in \cite{Nieto-Vesperinas2011}. Moreover, it is consistent with the few experiments done until now \cite{Geffrin2012,Fu2013}. 
\begin{figure}[tbp]
\centering\includegraphics[width=\columnwidth]{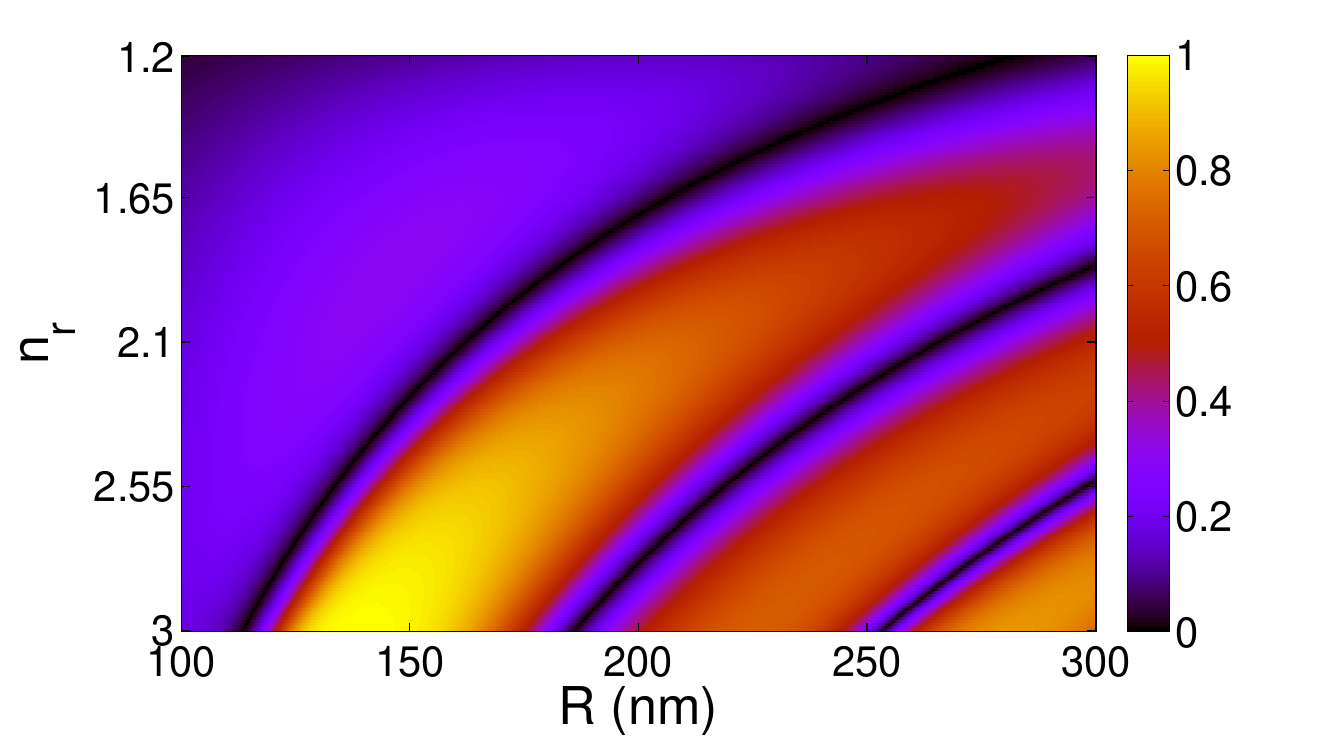}
\caption{Plot of $ \vert a_1-b_1 \vert $ as a function of the radius of the sphere $R$ (horizontal axis) and the relative index of refraction $n_r$ (vertical axis). It can be observed that there are three major regions where $ \vert a_1-b_1 \vert =0$. The wavelength is set to $\lambda=780$nm. \label{F_a1b1}}
\end{figure} 
Secondly, if the relative index of refraction is maintained, the anti-dual condition is held for larger particles than the dual one \cite{Gomez-Medina2011,Garcia-Etxarri2011}.\\\\ 
It is worth mentioning that the same method for characterizing the dual behaviour of a medium can be undertaken even though its geometry is not spherical. In that case, the analytical expression of the transfer function cannot be given, but the idea is the same one. The scattered energy can still be split into two orthogonal components with a well-defined helicity content. \textit{i.e.} $\mathtt{W}^\mathrm{s}= \mathtt{W}_{p}^\mathrm{s}+\mathtt{W}_{-p}^\mathrm{s}$. Then, the only difference lays in the calculation of $\mathtt{W}_{p}^\mathrm{s}$ and $\mathtt{W}_{-p}^\mathrm{s}$. Whereas this calculation for a sphere only depends on the Mie coefficients, for a general medium is achieved in the following way \cite{Ivan2013}:
\begin{equation}
\mathtt{W}_{p}^\mathrm{s} = \Vert \Es + i p \Hs \Vert ^2
\label{E_Tp}
\end{equation}
where the norm is computed as a surface integral. Therefore, the equation for the transfer function in general is:
\begin{equation}
\Tp = \dfrac{\Vert \Es - i p \Hs \Vert ^2}{\Vert \Es + i p \Hs \Vert ^2}
\label{E_Tp_general}
\end{equation}
It can be checked that equations (\ref{dual}, \ref{antidual}) are re-obtained if the expressions (\ref{E_fields_hel}, \ref{E_Hshel}) for $\Es$ and $\Hs$ are introduced in equation (\ref{E_Tp}). For a more general case though, numerical methods are needed to compute $\Es$ and $\Hs$. This method has been applied both in \cite{Ivan2013} and in \cite{Ivan2013JCP} to compute the helicity content of the scattered field. As mentioned in chapter \ref{Ch2}, the calculation of $\Es$ and $\Hs$ depends on $\Ei$, but an eigenmode calculation could retrieve calculations which are independent of the excitation beam.

\section{Angular momentum-induced helicity transfer}\label{Ch4_AM}
In this section, I will particularise the equations (\ref{E_Ei_hel}, \ref{E_fields_hel}) for the case of $\Ei$ being cylindrically symmetric (and with a well-defined helicity). As in the previous chapter, I will consider that its eigenvalue for $J_z$ is $m_z^* = l + p$, where $l$ is the topological charge of the beam in the paraxial approximation. Thus, equations (\ref{E_Ei_hel}, \ref{E_fields_hel}) will be simplified to:
\begin{eqnarray}
\Ei &=& \sum_{j=\vert \ms \vert}^{\infty} \aljms \Ajmspp \label{E_Eih} \\ 
\Es &=& \sum_{j=\vert \ms \vert}^{\infty} \aljms \left[ \dfrac{b_j+ a_j}{2} \Ajmspp + \dfrac{b_j- a_j}{2}\Ajmsmp \right] \label{E_Esh} \\
\Eint &=& \sum_{j=\vert \ms \vert}^{\infty}  \aljms  \left[ \dfrac{d_j+ c_j}{2} \Ajmspp + \dfrac{d_j- c_j}{2}\Ajmsmp \right] \label{E_Einth}
\end{eqnarray}
Then, the transfer function can be re-written as:
\begin{equation}
\Tmp = \dfrac{\mathtt{W}_{-p}^\mathrm{s}}{\mathtt{W}_p^\mathrm{s}} = \dfrac{\sum_{j=\vert \ms \vert}^{\infty} \vert \aljms \vert^2  \vert b_{j}  -  a_{j} \vert^2  } {\sum_{j=\vert \ms \vert}^{\infty} \vert \aljms \vert^2 \vert b_{j}  +  a_{j} \vert^2 } 
\label{E_Tmsp}
\end{equation}
Given an incident cylindrically symmetric beam, I will show how the dual and anti-dual properties of the particle previously depicted in Figure \ref{Tm} dramatically change when a higher angular momentum mode is used as $\Ei$. To exemplify this, a cylindrically symmetric beam with $\ms= 5$ and $p=1$ is used. In particular, the beam is a left circularly polarized LG$_{4,0}$ focused with a lens whose NA$=0.9$. Its decomposition into multipoles is given by Figure \ref{F_CJMP} in section \ref{Ch3_non}. 
\begin{figure}[tbp]
\centering
\includegraphics[width=\columnwidth]{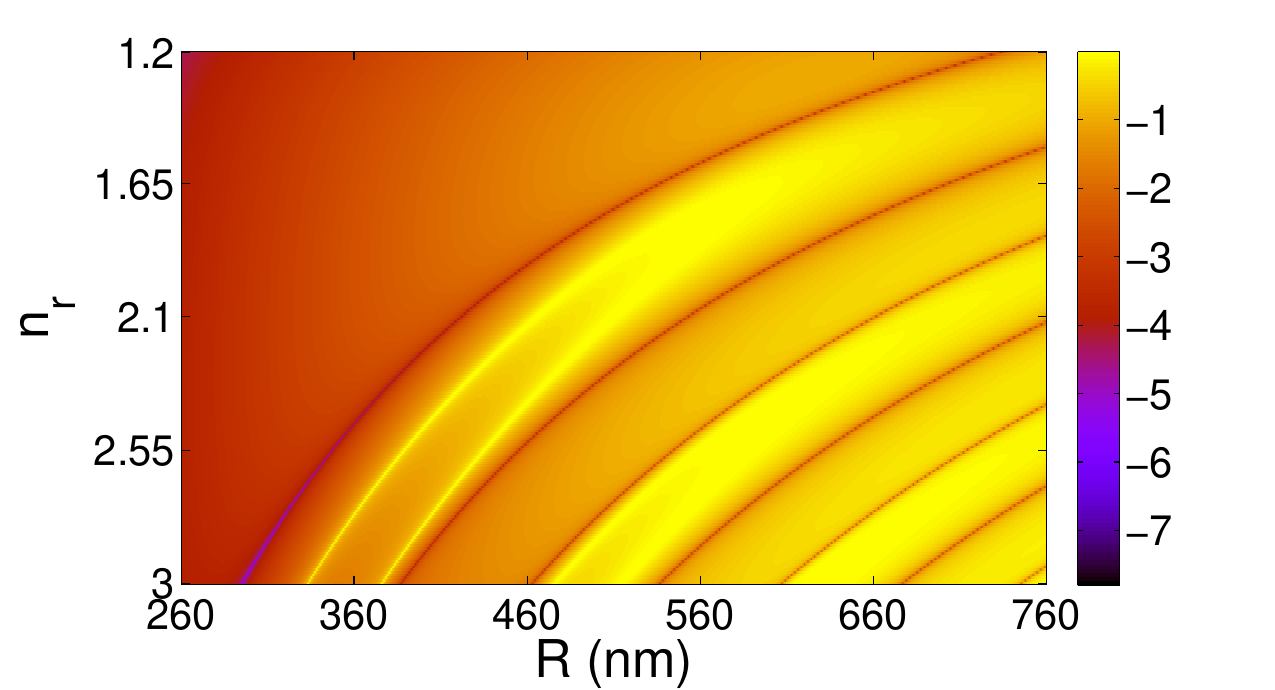} 
\caption{ Plot of $\log \left( \vert a_5-b_5 \vert \right) )$ as a function of the radius $R$ of the particle (horizontal axis) and the relative index of refraction $n_r$ (vertical axis).  \label{a5b5}}
\end{figure}
\begin{figure}[tbp]
\centering
\includegraphics[width=\columnwidth]{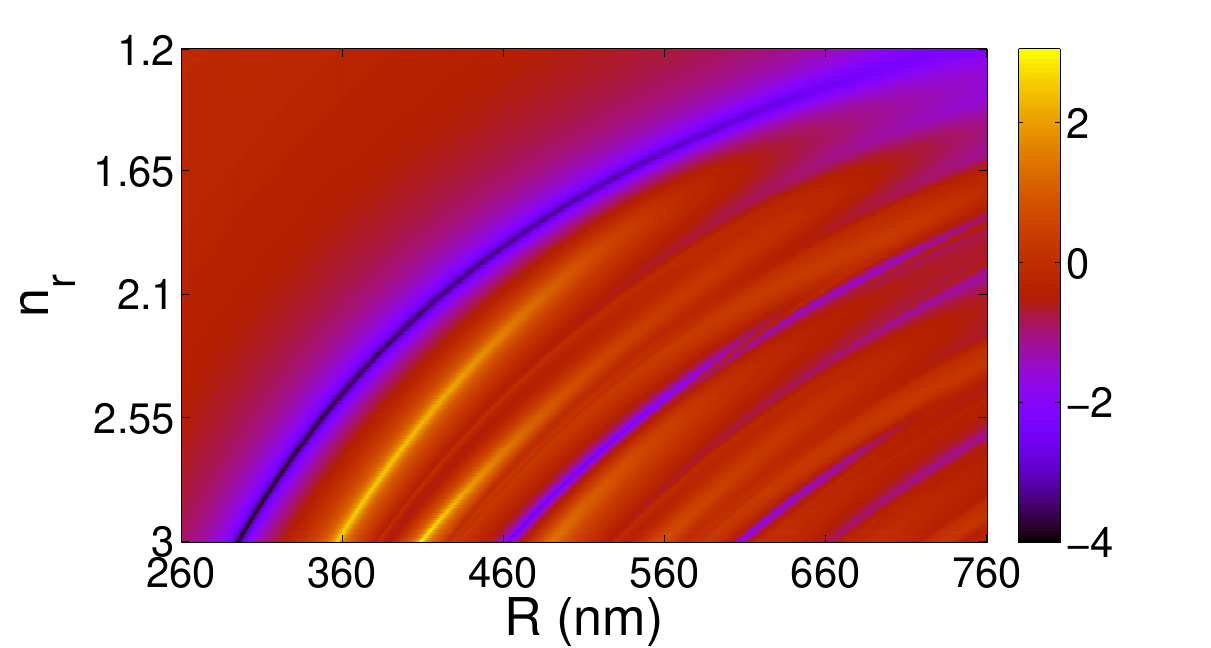} 
\caption{ Plot of the $log$ of the transfer function $T_{\ms p}(r,n_r)$ for $m_z=5$ and $p=1$, $\log \left( T_{51}(R,n_r) \right) $, as a function of $R$ and $n_r$.\label{Tmp_a5b5}}
\end{figure}
If we compare equations (\ref{E_Eih}) and (\ref{E_Eifinal}), it can be observed that the relation between the beam shape coefficients $\Cjmsp$ and $\aljms$ is $ i^j (2j+1)^{1/2} \Cjmsp \sqrt{2} = \aljms$. Then, equation (\ref{E_Tmsp}) can be used to compute the transfer function for this new excitation scenario.\\\\
In Figure \ref{a5b5}, the shape of $\log \left( \vert a_5-b_5 \vert \right)$ as a function of $R$ and $n_r$ is displayed ($\lambda =780$ nm), where the $\log$ function has been used to make the plot more readable. Although Figure \ref{a5b5} has a similar shape to Figure \ref{F_a1b1}, where $ \vert a_1-b_1 \vert $ was plotted, some differences can be found. The range of sizes for the particles to achieve the condition $a_5=b_5$ has increased for the same interval of $n_r=[1.2,3]$. Before, it spanned 200nm, whereas now it spans 500nm. Nonetheless, as previously discussed, this plot does not give any information regarding the higher multipolar orders. Therefore, I have plotted $\log \left( T_{51}(R,n_r) \right)$ in Figure \ref{Tmp_a5b5} to capture this behaviour.\\\\
As in the $j=1$ case, an increase in $n_r$ is linked to an increase of the dual properties of the sphere. Also, it can be inferred from a comparison between  Figure \ref{Tmp_a5b5} and Figure \ref{Tm}, that the dual and anti-dual conditions are fulfilled with a better approximation in this new occasion. Indeed, the minimum value of the colorbar drops almost an order of magnitude more, and the maximum value increases more than two orders of magnitude. That is, for some certain combinations of $R$ and $n_r$, the energy of the scattered field in the modes of opposite helicity is $10,000$ times smaller than the energy going to the original helicity of the incident field; whereas for some other certain conditions, the scattered field energy is dominated by modes with the opposite helicity with a ratio of 500 to 1. As in the incident Gaussian beam case, an increase in $n_r$ is linked to an increase of the dual and anti-dual properties of the sphere. To summarize, some very general conclusions can be reached after a careful look into these results:
\begin{itemize}
\item The larger the relative refraction index of the particle $n_r$ is, the more accurately the two extended Kerker conditions shown in \cite{Zambrana2013} can be achieved.
\item Fixing the index of refraction and size of the particle, we can always approximately induce the duality symmetry by choosing a right combination of AM and optical frequency of our laser, regardless of how big the sphere is with respect to the wavelength.
\end{itemize}
Even though I have not included other plots for other excitation beams, I have confirmed these conclusions by doing calculations similar to the ones presented with different sizes, wavelengths, refractive indexes, and excitation beams. The results have always been consistent with the conclusion above. Finally, note that in these simulations $n_r \leq 3$ and $\lambda = 780$nm. With these two conditions, the smallest a particle can be to induce duality symmetry is 120nm. This size could be reduced down to 81nm if a $\lambda=532$nm was used. 

\section{Possible experimental implementations}\label{Ch4_experiments}
In the previous section, I have compared different scenarios where the EM duality symmetry can be induced with dielectric spheres and arbitrarily high AM modes. I have shown that it is possible to induce dual and anti-dual behaviours for certain incident beams $\Ei$, regardless of the nature of the particle. In this section, two hypothetical experiments in the laboratory will be quantified where the duality condition can be achieved.\\\\
I will consider spheres made of Silica and Alumina. Their respective refractive indexes at $\lambda=780$nm are $\nsi=1.54$ \cite{Palik1985} and $\nal=1.76$ \cite{Bass1994}. I will suppose that they are embedded in water, therefore their respective relative refractive index will be $n_r^{\mathrm{SiO}_2}=1.16$ and $n_r^{\mathrm{Al}_2\mathrm{O}_3}=1.32$. The necessary equipment in order to induce this dual behaviour will be a tunable laser, a set of wave-plates and linear polarisers to create well-defined helicity states \cite{Nora2014}, an element to modify the spatial properties of light such as a Spatial Light Modulator, an imaging system (two microscope objectives), and a dielectric sphere embedded in a homogeneous, lossless and isotropic medium. Then, the way to proceed to induce helicity conservation will be the following. I consider the particle and its embedding medium as a given system. It will be clear that the helicity of light can be conserved regardless of the nature of the particle (size and index of refraction), as long as it is approximately spherical, the surrounding medium is homogeneous, lossless and isotropic, and considering a tunable laser with a broad enough wavelength modulation.\\\\
Once the two parameters defining the particle $R$ and $n_r$ are known, the range of size parameters $x$ that could be achieved with a tunable laser can always be computed. Supposing that the tunable laser can offer wavelengths spanning from 700nm to 1000nm (that would be the case of a Ti:Sapphire laser, for example), the achievable $x$ will belong to the interval $\left\lbrace 6.28\, r(\mu \mathrm{m}), \ 8.98\, r(\mu \mathrm{m}) \right\rbrace$. Now, as it has been proven in the previous section, there also exist a large number of radii $R^*$ (and consequently, $x^*$) for which the dual condition is achieved. This is a consequence of the fact that given $n_r$, there exist different radii of particles $R$ that make the particle dual provided adequate excitation beams with angular momentum $m_z$ are used. Hence it is highly probable that regardless of $R$ and $n_r$ the dual condition $\Tmp \approx 0$ can be achieved. In fact, this statement is even more true inasmuch as $R$ gets bigger. To be more specific, suppose that we have four different spheres. Two of them are made of Silica and the other two are made of Alumina. For each of the materials, suppose that the radii of the spheres are $R_1=325$nm and $R_2=700$nm. Now, given these sizes and their respective index of refraction, the $R$ dependence on the horizontal axis of Figure \ref{Tm} and Figure \ref{Tmp_a5b5} can be turned into a $x=2\pi R / \lambda$ dependence and obtain the wavelength for which the dual condition will be achieved.\\\\
The results are presented in Table \ref{T_experiment} for the different four combinations of materials and sizes $\left\lbrace R_1^{\mathrm{SiO}_2},R_2^{\mathrm{SiO}_2}, R_1^{\mathrm{Al}_2\mathrm{O}_3}, R_2^{\mathrm{Al}_2\mathrm{O}_3} \right\rbrace  $ and for the two different excitation beams considered in sections \ref{Ch4_trans} and \ref{Ch4_AM}, \textit{i.e.} $\ms = 1$ and $\ms =5$. 
\begin{table} 
\caption{Wavelengths at which the dual condition is achieved depending on the AM of the incident beam. The bold wavelengths are those at which the dual condition could be achieved with the range of wavelengths available in a Ti:Sap laser. The dual conditions are achieved with a minimum precision of $T_{m_zp}(r,n_r)=2\%$ for the four different cases.}
\renewcommand{\arraystretch}{1}
\label{T_experiment}  
\begin{center} 
\begin{tabular}{| l | l | l |}
\hline  $\lambda$ (nm)& $m_z=1$ & $m_z=5$ \\
\hline $R_1^{\mathrm{SiO}_2}$ & \textbf{859} & 330 \\
\hline  $R_2^{\mathrm{SiO}_2}$ & 1860 & \textbf{710} \\
\hline $R_1^{\mathrm{Al}_2\mathrm{O}_3}$ & \textbf{986} & 377 \\
\hline  $R_2^{\mathrm{Al}_2\mathrm{O}_3}$ & 2122 & \textbf{812} \\
\hline 
\end{tabular}
\end{center}
\end{table}
It can be observed that regardless of the size and the material, the duality condition can be achieved if a wavelength is chosen. Moreover, it is seen that when the particle is larger, higher order beams are needed to reach the duality condition, for a fixed $\lambda$ range. Finally, it can also be observed that when the relative refractive index $n_r$ is increased, the duality condition is pushed to longer wavelengths.\\\\
All this evidence, enables me to conclude that dual systems are easily realisable in the laboratory if an arbitrary dielectric sphere is properly matched with a proper light beam. The results for aggregates of particles or metamaterials are different, though. In those cases, each of the individual cells is typically smaller than the wavelength, therefore the results about high order AM modes do not apply there, as their behaviour is dipolar due to their size. However, Figure \ref{Tm} can still be used to engineer some dual individual cells \cite{Ivan2013prb,Xavi2013spie}.

\begin{savequote}[10cm] 
\sffamily
``It doesn't matter how beautiful your theory is, it doesn't matter how smart you are. If it doesn't agree with experiment, it's wrong.'' 
\qauthor{Richard P. Feynman}
\end{savequote}

\chapter{Experimental Techniques}
\graphicspath{{ch5/}} 
\label{Ch5}
\newcommand{\Uxy}{U(x,y)}
\newcommand{\I}{I(\mathbf{r})}
\newcommand{\pha}{\arg \left\lbrace  \U \right\rbrace}
\newcommand{\txy}{t(x,y)}
\newcommand{\vort}{e^{il\phi}}

\section{Holography principles} \label{Ch5_Hol}
\begin{figure}[tbp]
\centering
\includegraphics[width=\columnwidth]{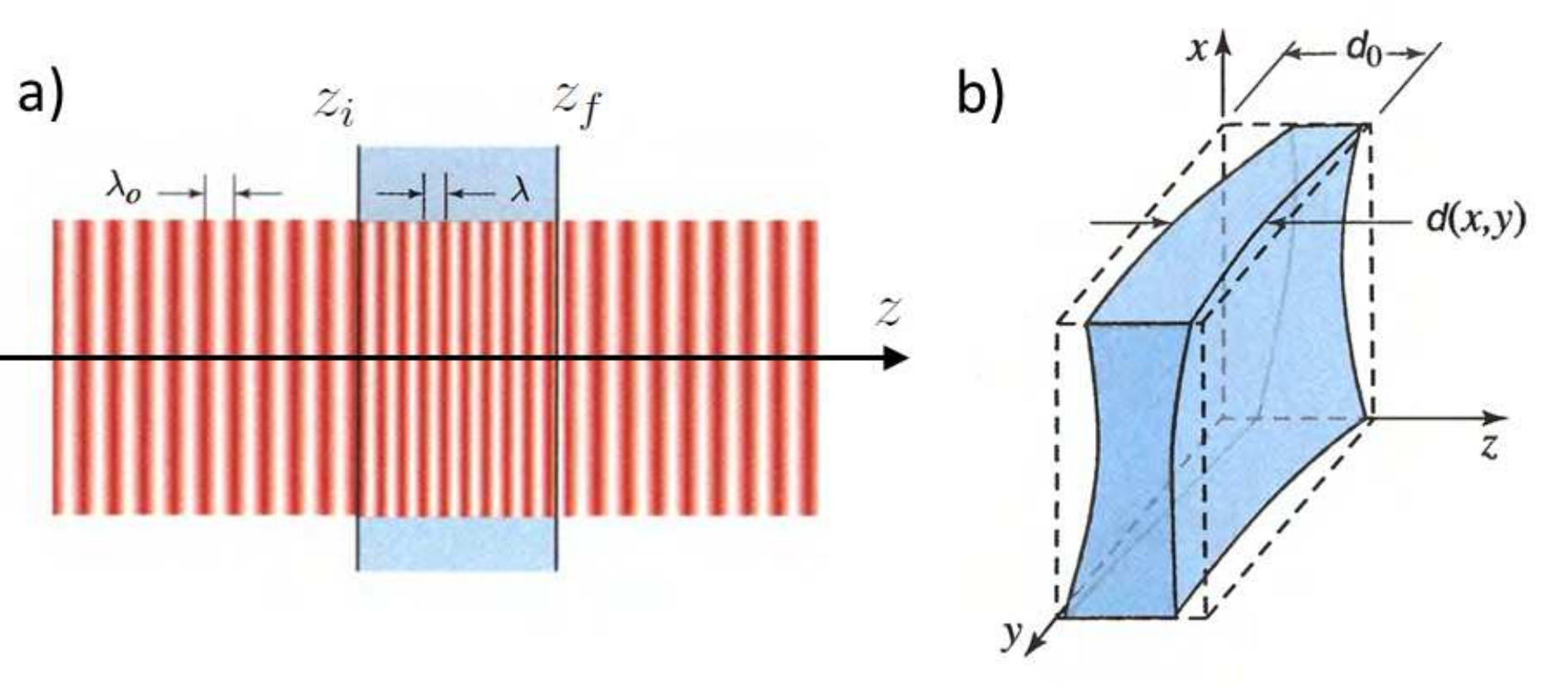} 
\caption{ Sketch of two different transmission plates. a) Thin layer of glass with constant thickness $z_f-z_i=d_0$. b) Thin layer of glass with variable thickness $d(x,y)$. Both figures have been copied from \cite{Saleh2007} with the permission of Prof. M.C. Teich.   \label{Fig61}}
\end{figure}

The aim of this chapter is to introduce the reader to different experimental techniques that will be used to carry out the experiments in chapters \ref{Ch6} and \ref{Ch7}. As shown in all the previous chapters, the precise characterization of both the spatial properties and polarization of light is crucial to control light-matter interactions at the nano-scale. In particular, a great emphasis has been put into using cylindrically symmetric beams, \textit{i.e} beams of light which are eigenstates of $J_z$. Here, it will be shown how to experimentally create these beams in the laboratory with two different techniques: Computer Generated Holograms (CGH) and Spatial Light Modulators (SLM). Both techniques are based on holography, whose basis are grounded on Fourier optics \cite{Saleh2007,Goodman2005}. In Fourier optics, the polarization does not play any role - the whole description of light phenomena can be done with a scalar magnitude named complex wave-function $\Uu$. Furthermore, as it has been done in all the previous chapters, a harmonic time dependence $\exp(-i\omega t)$ will be considered, therefore $\Uu=\U \exp(-i \omega t)$, where $\U$ is referred to as complex amplitude. Then, $\U$ can be retrieved with the intensity $\I=\vert \U\vert^2$ and the phase $\pha$ of the wave:
\begin{equation}
\U = \sqrt{\I} \exp \left( i \pha \right)
\end{equation}
Only paraxial waves will be considered here \cite{Lax1975}. Paraxial waves were described in section \ref{Ch1_symm}. For these waves, the complex amplitude can be expressed as $\U = A(\mathbf{r}) \exp (i k z)$, where $A(\mathbf{r})$ is the envelope of the wave which varies slowly with $z$. Then, a very important magnitude in wave optics is the complex amplitude transmittance of an optical element, $\txy$. This magnitude is defined as the ratio between the complex amplitude of the wave after and before the optical element. That is,
\begin{equation}
\txy = \dfrac{U(x,y,z_f)}{U(x,y,z_i)}
\end{equation}
where $z_i$ and $z_f$ are the two $z$ planes where the optical element begins and finishes. In Figure \ref{Fig61}(a), a sketch of a thin layer of glass is depicted, with $z_f-z_i=d_0$. The excitation wave is paraxial and propagating in a direction normal to the interface of the material. In this case, $\txy$ is naturally defined as $\U$ at the outer face of the layer over $\U$ at the inner face. Then, if the refractive index of the layer is $n$, 
\begin{equation}
U(x,y,z_f) \approx \exp \left[ -inkd_0 \right] U(x,y,z_i) \Longrightarrow t(x,y) \approx \exp \left[ -inkd_0 \right]
\label{E_txy_layer}
\end{equation}
where $\exp(-inkd_0)$ is the phase shift introduced by the optical element\footnote{Here, reflection at the interfaces and absorption in the material are neglected.}. However, when the optical element is not plane-parallel, the definition of $t(x,y)$ is a bit more complex. If the element lies between $z_i$ and $z_f$, we can define a function $d(x,y)$ which gives the thickness of the optical element as a function of its transverse coordinates $(x,y)$. The air occupies the rest of it, \textit{i.e.} the thickness of the air is $d_0 - d(x,y)$ (see Figure \ref{Fig61}(b)). Considering the result given by equation (\ref{E_txy_layer}), the transmittance of this variable-thickness optical element is:
\begin{equation}
t(x,y) \approx \exp\left[ -inkd(x,y) \right] \exp \left[  -ik (d_0 - d(x,y) )\right] \approx h_0 \exp \left[ -i (n-1) k d(x,y) \right]
\label{E_txy_general}
\end{equation}
with $h_0 = \exp \left[ -ikd_0  \right]$ being a constant phase factor.
Now, a \textbf{hologram} is an optical element (a transparency) that contains a coded record of an optical wave. That is, it contains the amplitude and the phase of the wave that has been recorded, allowing for a complete reconstruction of the field. For example, suppose that a transparency was made with $t(x,y)=U(x,y)$, with $U(x,y)$ being the wave to be reconstructed. If this transparency was illuminated with a wave $U(x,y,z_i)=1$, then the transmitted wave would be $U(x,y,z_f)=t(x,y) \cdot 1 = U(x,y)$. However, it is challenging to encode both amplitude and phase on a transparency. Next, a method to reconstruct $t(x,y)$ only modulating the intensity of the beam is explained. Then, in sections \ref{Ch5_CGH} and \ref{Ch5_SLM}, two different techniques to carry out phase modulation holograms will be discussed.

\subsection{Intensity modulation holograms} \label{Ch5_Imod}
\begin{figure}[tbp]
\centering
\includegraphics[width=\columnwidth]{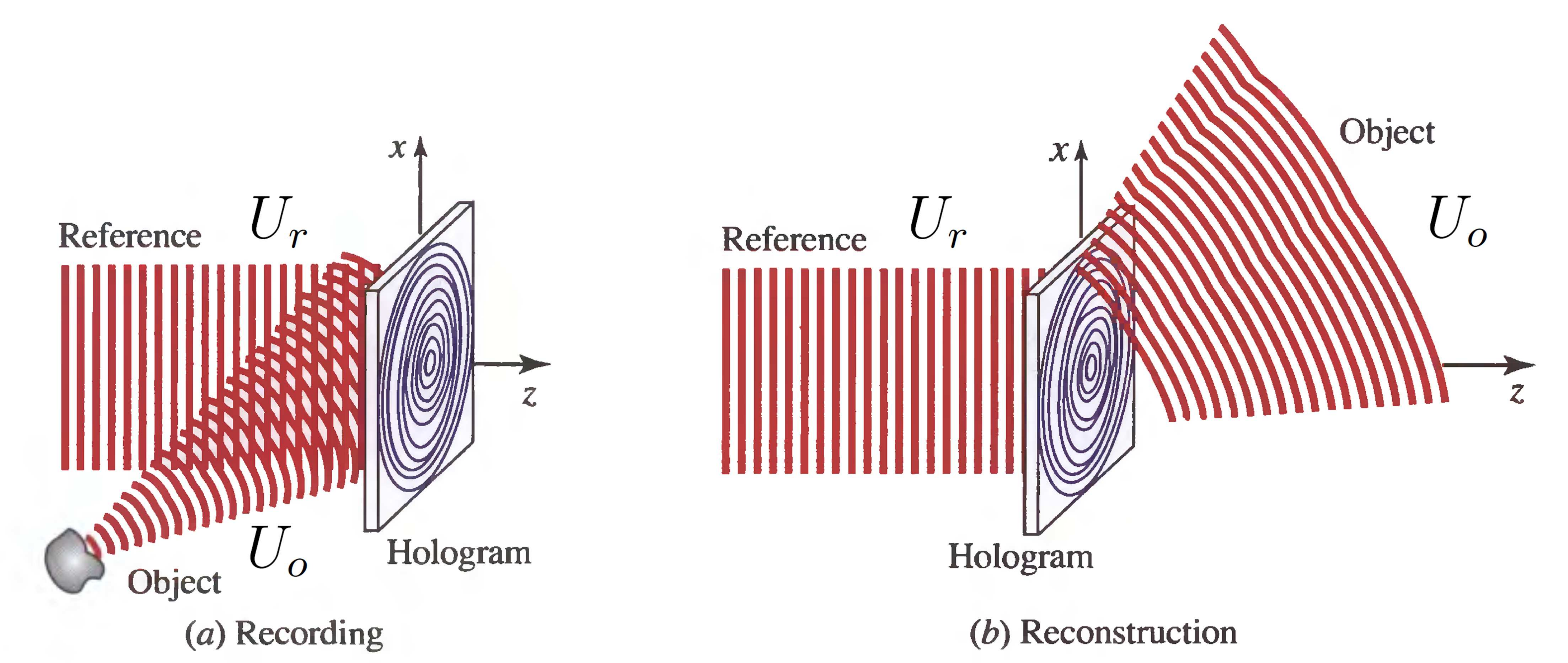} 
\caption{a) The intensity interference pattern of a reference wave $U_r$ and the object $U_o$ is recorded on a holographic plate. b) The object wave is recovered after the interference pattern is shone onto the plate. Both figures have been copied from \cite{Saleh2007} with the permission of Prof. M.C. Teich. \label{Fig62}}
\end{figure}
There are different techniques to make intensity (or amplitude) modulation holograms. Since the first holograms made by Gabor in 1948  \cite{Gabor1948}, a lot of work was done by many researchers \cite{Kasper1985,Hariharan2002,Saxby2010}. An intensity modulation hologram encodes the intensity and the phase as a form of intensity pattern\footnote{The first holograms were made of photographic plates, which indeed are only sensitive to the intensity of light shone on them.}. In order to do that, an interference pattern needs to be recorded so that the information of the phase of the wave is carried through. Then, if the wave that wants to be reconstructed is named object wave $U_o$ and the wave that will impinge on the hologram is called reference wave $U_r$, the transmittance needs to be proportional to their interference:
\begin{equation}
t \propto \vert U_o + U_r \vert^2 =  I_r  +  I_o  + U_r^* U_o + U_r U_o^*
\end{equation}
where $U_o$, $U_r$, and consequently all the rest of terms can take different values for different transverse coordinates $(x,y)$. When the reference wave illuminates the transparency, the following wave is obtained as a result:
\begin{equation}
U = t U_r \propto U_r ( I_r + I_o ) + I_r U_o + U_r^2 U_o^*
\label{E_tUr}
\end{equation}
Any wave can be chosen to illuminate the transparency. It is convenient to choose a wave such that $I_r$ is a constant, \textit{i.e.} $I_r(x,y)=I_r$. If that is the case, equation (\ref{E_tUr}) can be re-formulated as:
\begin{equation}
U(x,y) \propto I_r + I_o(x,y) + \sqrt{I_r} U_o(x,y) + \sqrt{I_r} U_o^*(x,y)
\label{E_Uxy}
\end{equation}
Now, to make sure that the four terms of the equations are properly separated in space, a oblique phase can be added to $U_o$. That is, if $U_o'(x,y)=U_o(x,y)\exp(ikx\sin\theta)$, then equation (\ref{E_Uxy}) can be expressed as:
\begin{equation}
U'(x,y) \propto I_r + I_o(x,y) + \sqrt{I_r} U_o(x,y) \exp(ikx\sin\theta) + \sqrt{I_r} U_o^*(x,y) \exp(-ikx\sin\theta)
\label{E_Uxy_sin}
\end{equation}
\begin{figure}[tbp]
\centering
\includegraphics[width=\columnwidth]{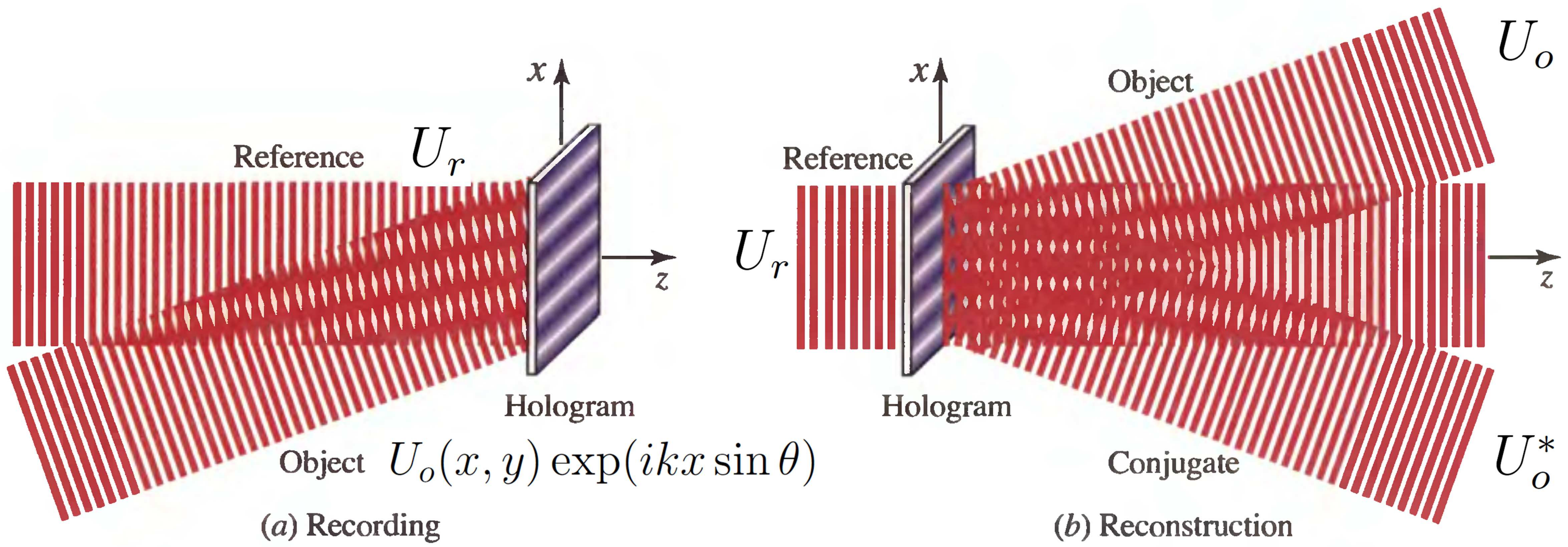} 
\caption{a) Recording of the hologram. b) Reconstruction of the hologram, giving rise to three components: a constant wave, $U_o\exp(ik\sin\theta x)$ and $U_o^*\exp(-ik\sin\theta x)$ Both figures have been copied from \cite{Saleh2007} with the permission of Prof. M.C. Teich. \label{Fig63}}
\end{figure}
The first two terms of equation (\ref{E_Uxy_sin}) are waves that travel in the $z$ axis. However, the third and fourth terms of the equations are waves travelling in an oblique manner, as shown in Figure \ref{Fig63}. The third term represents the object wave $U_o$ travelling at an oblique angle $\theta$ with respect to the $z$ axis and modulated by a constant factor $\sqrt{I_r}$. On the contrary, the fourth term represents the conjugate wave $U_o^*$ travelling at an angle $-\theta$ with respect to the $z$ axis. Hence, the addition of an oblique plane wave allows for the total reconstruction and isolation of the object wave, which is the aim of any hologram. In fact, the addition of an oblique phase is widely used in holography, and will also be considered in sections \ref{Ch5_CGH} and \ref{Ch5_SLM} when phase-only holograms are introduced.

\section{Computer Generated Holograms} \label{Ch5_CGH}
The meaning of the word hologram changed with the advent of modern computers. Before then, a real object wave needed to be encoded in the hologram. The invention of computers allowed for the creation of holograms representing optical waves coming from non-existing objects. Holography became an inverse scattering problem, where the object wave was given and the diffraction pattern was sought \cite{Brown1966}. The first CGHs made by Brown and Lohmann \cite{Brown1966} were a very specific kind of amplitude-only modulated holograms: they were binary, \textit{i.e.} a set of apertures on an opaque mask. More sophisticated methods to produce intensity-modulated CGH using gray-tones came up later, even though their fabrication remained challenging \cite{Burch1967,Lee1978}. A review of the history of intensity-only CGH can be found in \cite{Tricoles1987}. Then, in the 1970's, another type of holograms was discovered \cite{Kirk1971}. Using phase-only holograms (which were already known), Kirk managed to encode amplitude information in a phase-only hologram. However, their use went almost unnoticed until 1999, when Davis and co-workers used SLMs (instead of photographic films or transparencies) to encode amplitude and phase modulation in a phase-only hologram \cite{Davis1999}. Now, even though phase-only CGH did not start being used to encode amplitude modulation until the 2000's, they were very popular to change and encode phase shifts. The fabrication processes had been carefully characterized by Hariharan in the 1970's \cite{Hariharan1971,Hariharan1971AO,Hariharan1971OC,Hariharan1971IN,Hariharan1972,Hariharan1980,Hariharan2002}, and their use was extended in imaging \cite{Dammann1971}, Fourier optics \cite{Goodman1970} and even acoustics, where they were first conceived \cite{Metherell1967,Larmore1969}. Here, I will describe an applications of phase-only CGH in the field of singular optics, and in particular in the creation of beams of light with a well-defined AM. Light beams with phase singularities, also known as vortex or doughnut beams\footnote{See section \ref{Ch3_paraxial} to know a bit more about why they are called in this manner.}, boomed after the seminal paper by Allen and co-workers in 1992 \cite{Allen1992}. However, even when the connection between phase singularities and the AM of light had not been properly established, CGH had already been used to produce vortex beams \cite{Vasara1989,Bazhenov1990,Heckenberg1992,Leseberg1987,Bazhenov1992,Heckenberg1992OL}. In fact, they kept being mainly used to manipulate the AM of light up until the beginning of 2000, when the use of SLMs took over \cite{Dholakia2002,Grier2003,Leach2003,Tao2003}. Next, I will describe why CGH are useful to manipulate the AM of light, and how they achieve it.

\subsection{Phase singularities}
\begin{figure}[tbp]
\centering
\includegraphics[width=11cm]{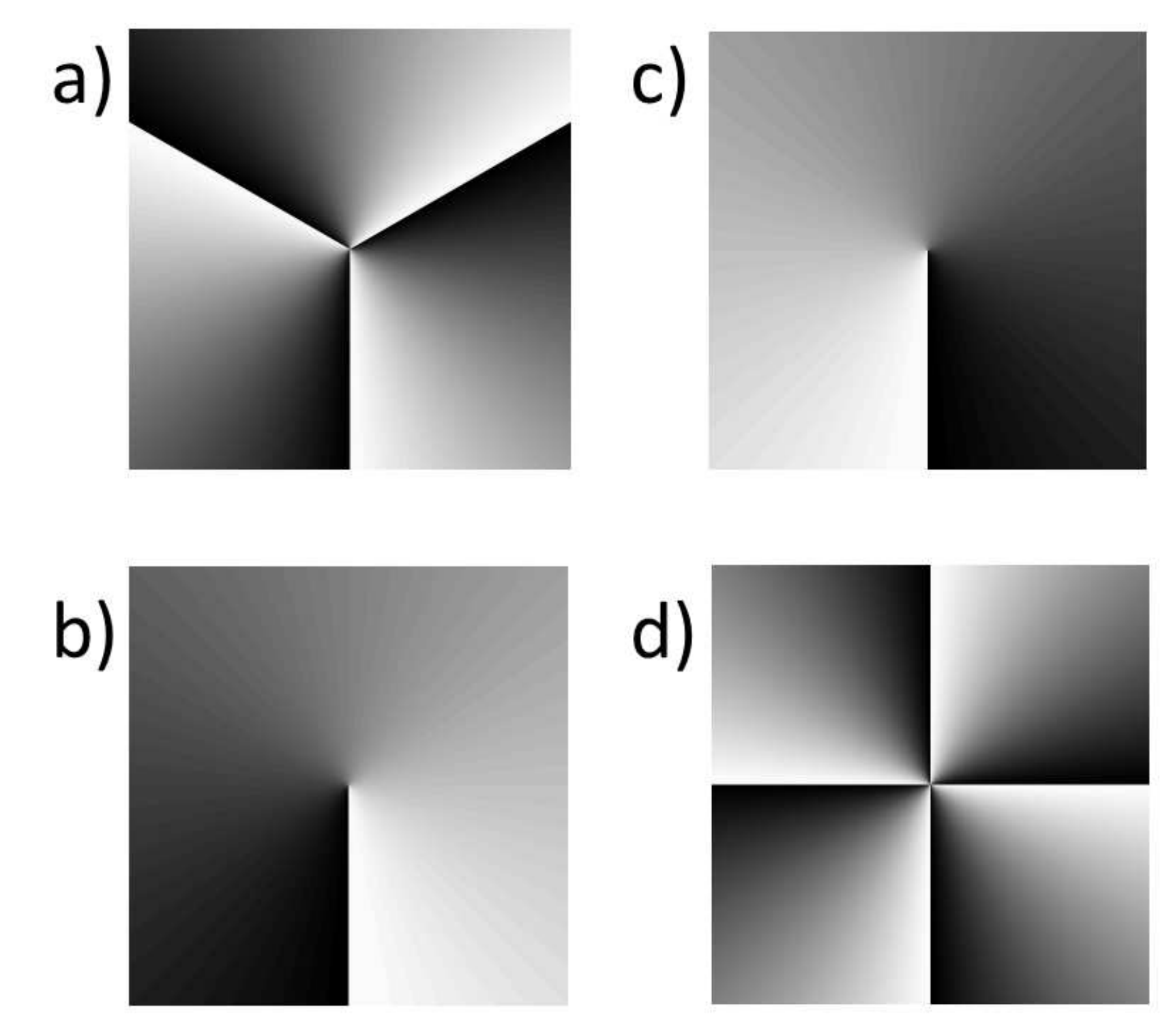} 
\caption{Phase profile of a paraxial vortex beam on a transverse plane $x-y$: $\arg \left\lbrace  U_l(x,y) \right\rbrace$ for a) $l=-3$, b) $l=-1$, c) $l=1$, d) $l=4$. Black means $\left[ U_l(x,y) \text{mod} 2\pi \right]=0$, and white $\left[ U_l(x,y) \text{mod} 2\pi \right]=2\pi$  \label{Fig64}}
\end{figure}

In chapter \ref{Ch1}, Bessel beams and multipolar fields have been introduced. Both of them are EM modes, \textit{i.e.} they are basis of solutions of the Maxwell equations (\ref{E_Max_EH_mono}). Also, they are both eigenstates of $J_z$. Therefore, following the nomenclature given in chapter \ref{Ch3}, they are cylindrically symmetric. Cylindrically symmetric beams have been used in chapters \ref{Ch3} and \ref{Ch4} to show many different effects - from excitation of WGMs, to inducing duality symmetry in non-dual materials. In chapters \ref{Ch3} and \ref{Ch4}, a cylindrically symmetric beam has been decomposed as a superposition of multipolar fields, as these are the normal modes of a sphere. Unfortunately, multipolar fields are experimentally difficult to produce, as they propagate in all directions of space.\\\\
Here, it will be considered that a cylindrically symmetric beam is a general superposition of Bessel beams (with either TE/TM character or well-defined helicity):
\begin{equation}
\Ei_m = \int_0^{\infty} \krho \dd \krho \ \left( a_{\krho} \BTE + b_{\krho} \BTM \right) = \int_0^{\infty} \krho \dd \krho \ \left( a'_{\krho} \Bmpp + b'_{\krho} \Bmpm \right)
\label{E_BBnonpar}
\end{equation}
Now, because of their definition as a basis of solution of Maxwell equations, Bessel beams include paraxial and non-paraxial waves. However, as mentioned in section \ref{Ch5_Hol}, only paraxial waves will be considered in this chapter. Therefore, equation (\ref{E_BBnonpar}) needs to be re-expressed in the paraxial approximation, \textit{i.e.} $\krho \ll k$. Using equations (\ref{E_BBplane}, \ref{E_BBTEM}, \ref{E_Bmp}, \ref{E_sphi}, \ref{E_ptheta}), it can be observed that the four different Bessel modes can be expressed in the paraxial limit as:
\begin{eqnarray}
\BTE & \approx & A_{\krho} e^{im\phi} e^{ik_z z} \left[ \rhohat J_{\epsilon}(\krho \rho) + i\phihat J_{\delta}(\krho \rho) \right] \label{E_BTEpar}\\
\BTM & \approx & B_{\krho} e^{im\phi} e^{ik_z z} \left[ \rhohat J_{\delta}(\krho \rho) + i\phihat J_{\epsilon}(\krho \rho) \right]\\
\Bmpp & \approx & C_{\krho} J_{m-1}(\krho \rho) e^{i(m-1)\phi} e^{ik_z z} \ \sphat \\
\Bmpm & \approx & D_{\krho} J_{m+1}(\krho \rho) e^{i(m+1)\phi} e^{ik_z z} \ \smhat \label{E_Bmpppar}
\end{eqnarray}
with $A_{\krho},B_{\krho},C_{\krho},D_{\krho}$ being functions of $\krho$, and $J_{\delta}(\krho \rho)=J_{m-1}(\krho \rho)-J_{m+1}(\krho \rho)$, $J_{\epsilon}(\krho \rho)=J_{m-1}(\krho \rho)+J_{m+1}(\krho \rho)$. Taking into account the asymptotic properties of Bessel functions \cite{Watson1995}, which yield $J_m (x) \propto x^m$ when $x \longrightarrow 0$, it can be seen that a cylindrically symmetric beam in the paraxial approximation can be done with waves whose complex amplitude, when $\rho \rightarrow 0$, is of the kind:
\begin{equation}
U_l(\rho,\phi,z) \propto \rho^l e^{il\phi} e^{ik_z z}  \label{E_Um}
\end{equation}
and whose polarization has to be properly matched depending on the value of $J_z$ that needs to be created. That is, if a value of $J_z=m^*$ is needed, then $U_{l}(\rho,\phi,z)$ can be used if the beam is radially or azimuthally polarized, whereas $U_{(l-1)}$ or $U_{(l+1)}$ ought to be used for left or right circular polarization respectively.\\\\
Polarization states such as $\sphat$ and $\smhat$ can be easily achieved using wave plates \cite{Saleh2007}. On the contrary, $\phihat$ and $\rhohat$ polarization states are more challenging to obtain, as typically interferometry was needed \cite{Tidwell1990,Tidwell1993}. Nowadays, with the invention of liquid-crystal polarization converters \cite{Stalder1996,Marrucci2006}, radial and azimuthal polarization states are not as challenging to prepare as before. For simplicity though, only circular polarization states will be considered in all the experiments of this thesis concerning AM states.\\\\
Independently of the polarization state though, it is clear looking at equations (\ref{E_BTEpar}-\ref{E_Bmpppar}) that the beam must possess a phase dependence of the kind $e^{il\phi}$ to be cylindrically symmetric beams. Figure \ref{Fig64} depicts the phase of $U_l(x,y) $ on a transverse plane for $l=-3,-1,1,4$. It can be observed that the phase winds around the center of the transverse plane $(x,y)=(0,0)$. In fact, $(x,y)=(0,0)$ is a singular point, as the angle $\phi$ is not well-defined. Any point whose phase is not well-defined is named \textbf{phase singularity}. A point which is singular in phase must have a zero intensity, so that the field still fulfils the Helmholtz equation. Then, the azimuthal number $l$ gives the order of the singularity.
Thus, the aim of the phase-only CGHs made for this project will be to create beams with a twisting phase such as $e^{il\phi}$. In fact, as mentioned in section \ref{Ch5_Imod}, an oblique phase of the kind $\exp(ik_x x)$ will be added to the spiral phase $e^{il\phi}$ so that the hologram separates the different diffraction orders coming out of the hologram. 

\subsection{Fabrication of phase-only CGH}
It is clear now that a hologram which imprints a $e^{il\phi}$ phase dependence onto the beam is needed. Now, if a paraxial wave of the kind $\U=A(\mathbf{r})\exp(ik_zz)$ goes through a hologram with $t(x,y)=e^{il\phi}$, the field after the hologram will be a vortex beam. Equation (\ref{E_txy_general}) gives a way to produce that kind of hologram. If a homogeneous and isotropic material is modified in a way that its thickness has a $\phi$ dependence, then $t(x,y)$ can be made equal to $e^{il\phi}$:
\renewcommand{\arraystretch}{1}
\begin{equation}
\left.
\begin{array}{lcc}
t(x,y) & \approx & h_0 \exp \left[ -i (n-1) k d(x,y) \right]\\
t(x,y) & = & \exp (il\phi)
\end{array} \right\rbrace \Longrightarrow d(x,y) = d(\phi) \label{E_tdphi}
\end{equation}
That is, a homogeneous material made of an index of refraction $n$ can be used to create phase singularities of order $l$ if its thickness varies according to the following expression:
\begin{equation}
d(\phi)= \dfrac{l \phi \lambda}{(n-1) 2\pi}
\label{E_dphi}
\end{equation}
Note that equation (\ref{E_dphi}) is wavelength-dependent, what means that in theory the thickness $d(\phi)$ to create a perfect phase dependence $\vort$ can only be achieved for a single wavelength. However, in practice, the efficiency of CGHs does not considerably vary for a broad range of wavelengths. The method to obtain $d(\phi)$ is based on the following idea. As it was explained in section \ref{Ch5_Imod}, photographic films are only sensitive to the intensity of light. In a conventional black and white negative developing process, a film transforms the intensity captured by the camera into a grey scale where black means high exposure and white low exposure \cite{Haist1979}. Then, the negative (which could be understood as an amplitude hologram) can be transformed into a phase hologram by a bleaching and fixing process. The bleaching and fixing processes remove all the silver salts and the unexposed gelatin, consequently giving different widths to the material depending on the exposure. More importantly, the efficiency of the hologram increases \cite{He1995}. The fabrication process is detailed in Appendix \ref{App2Fab}.

\subsection{Characterization} \label{Ch5_CGHchara}
The performance of CGH can be characterized by different parameters. One of the most relevant ones is their efficiency. Here, I define what the efficiency of a CGH is, and how to measure it. Then, in Appendix \ref{App2Cha}, the characterization of the CGH fabricated throughout my project and their dependence on some controllable parameters is detailed. The wavelength of the laser used to perform the characterisation of the holograms is $\lambda=632.8$nm.\\\\
Even though the efficiency is not the only parameter to characterize the performance of a hologram, it is certainly a very important one. The efficiency is a ratio between the power of the incident beam onto the hologram, and the power going to the first diffraction order of the hologram (that is, the object wave in Figure \ref{Fig63}). Note that if the hologram absorbs some power, then the efficiency drops.
\begin{figure}[tbp]
\centering
\includegraphics[width=14cm]{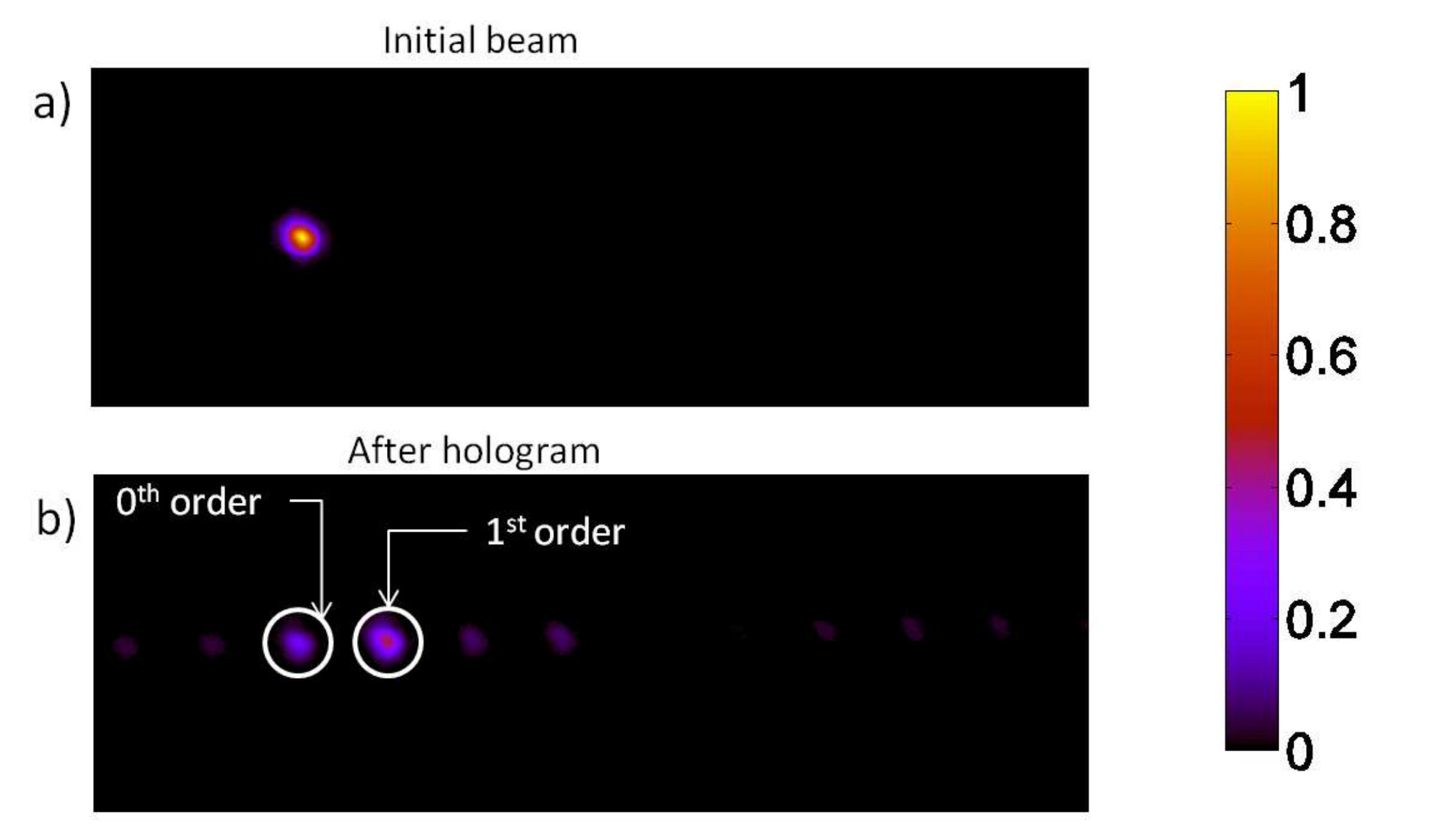} 
\caption{a) CCD image of the reference beam when no hologram is on the optical path. b) Typical diffraction pattern created by a CGH when the efficiency is over 30\%. The 0th order is at the same position as the reference beam, whereas the 1st order has been tilted. The colour bar indicates goes from 0 to 1, where 1 is taken as the absolute intensity maximum of the reference beam. \label{Fig610}}
\end{figure}
As depicted by Figure \ref{Fig610}, the intensity is measured with a CCD camera. That is, a snapshot is taken of the incident beam when the hologram is removed from the optical path (Figure \ref{Fig610}(a)). Next, the hologram is added in the optical path making sure that the beam almost fills the hologram dimensions. The hologram tilts the incident beam and its effect is again captured by the CCD camera (Figure \ref{Fig610}(b)). Due to the imperfections of the hologram, other diffraction orders arise giving rise to the pattern in Figure \ref{Fig610}(b). Note that in Figure \ref{Fig610}(b), there is light at the same position where the incident beam is in Figure \ref{Fig610}(a). This is the so-called 0th-order, \textit{i.e.} the incident beam going through the hologram without being modified by it. The tilt given by the hologram depends on the number of lines of the hologram, \textit{i.e.} the number of complete $2\pi$ phase jumps. In fact, the angle of diffraction given by the hologram is given by:
\begin{equation}
\Delta\theta_x = \dfrac{\lambda}{\Delta x} \label{E_kx}
\end{equation}
where $\Delta\theta_x$ is the angle that the 1st diffraction order forms with the direction of the initial beam (or the 0th order); $\Delta x$ is the width of a $2\pi$ phase ramp in the hologram (see Figure \ref{Fig65} in Appendix \ref{App2Fab}); and $x$ is the direction in which the phase jump is given. In Appendix \ref{App2SLM}, it will be determined how the efficiency critically depends on the oblique phase $\exp (ik_x x)$ added to the hologram, which is proportional to $(\Delta x)^{-1}$. Also, a detailed explanation of how to experimentally separate the different diffraction orders coming out of a hologram, \textit{i.e.} how to obtain Figures \ref{Fig610}, \ref{eff_SLM} and \ref{Fig619}, is given in Appendix \ref{App2Sep}.

\section{Spatial Light Modulators} \label{Ch5_SLM}
\begin{figure}[tbp]
\centering
\includegraphics[width=13cm]{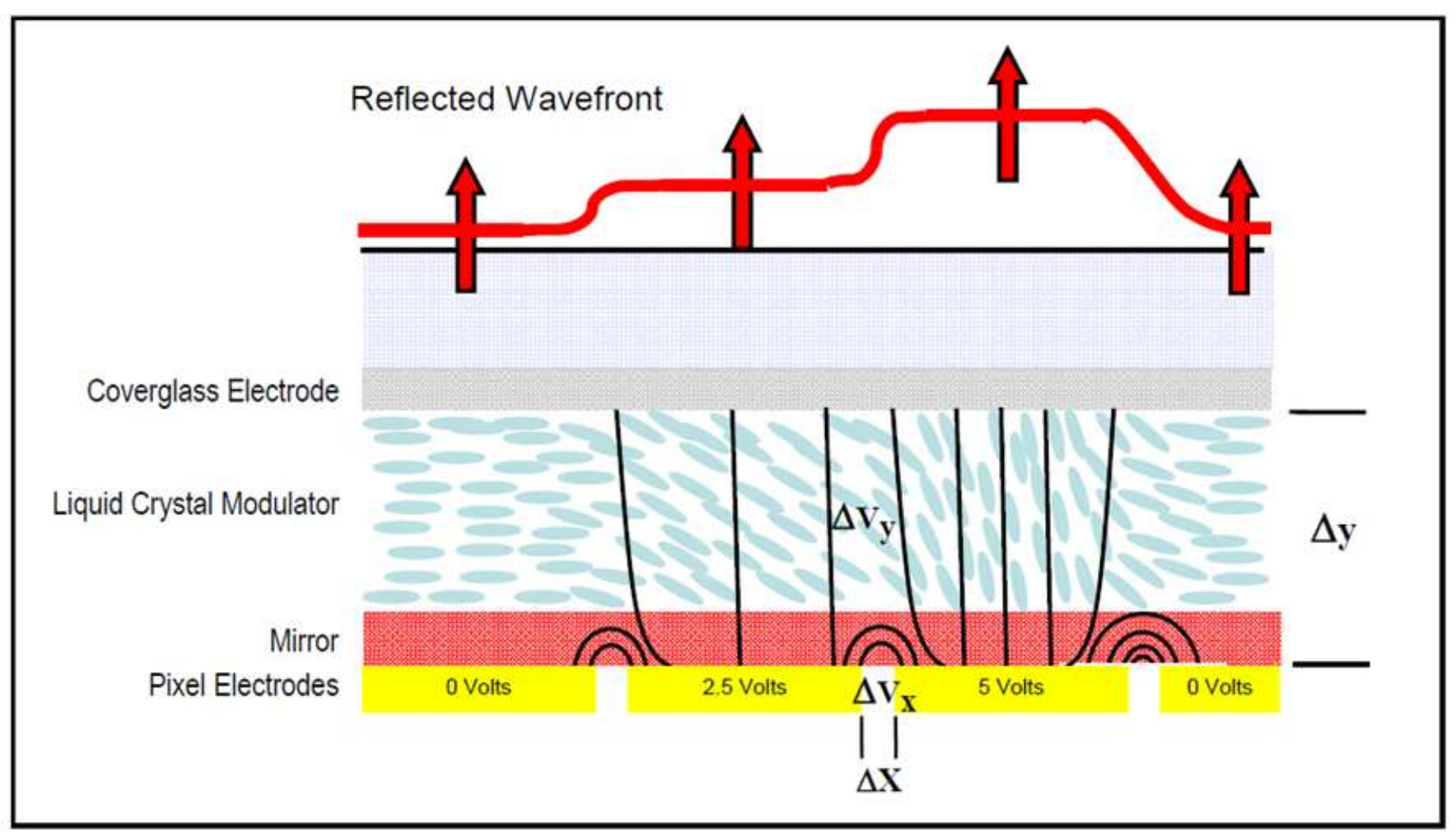} 
\caption{Schematics of a transverse section of an SLM. Two electrodes and a mirror enclose a volume filled with liquid crystals. The potential difference applied by the electrodes sets the orientation of the liquid crystal, thus achieving the desired transmittance function $t(x,y)$. Figure copied from \cite{BNS_pdf} with their permission. \label{Fig612}}
\end{figure}

The technology to digitally control nematic liquid crystals boomed at the end of the 1980's giving rise to SLMs that could modify the phase, the amplitude and the polarization of beams \cite{Konforti1988,Collings1989,Neff1990,Amako1991}. A sketch of a transverse section of an SLM can be observed in Figure \ref{Fig612}. Two electrodes close a volume that is filled with liquid crystals and a mirror. First, the incident light on the SLM is transmitted through the coverglass electrode. Afterwards, it goes through the volume enclosing the liquid crystals. Their orientation is determined by the difference of potential between the two electrodes, which is at the same time linked to the CGH displayed on the computer. Due to the anisotropy of the liquid crystals, their index of refraction will be different for two orthogonal incident linear polarizations. Then, controlled changes in their orientation will vary the index of refraction for one of polarization as a function of the transverse coordinates $(x,y)$, giving rise to a computer-controlled transmittance $t(x,y)$\footnote{Unlike the CGHs, because of the anisotropy of the liquid crystals, $t(x,y)$ is very polarisation-dependent.}. Then, the light hits a mirror and is reflected back off. The mirror sits on top of a grid of pixel electrodes, which set the potential difference on the liquid crystal volume in conjunction with the first electrode. Thus, in most of cases, the SLM technology works in reflection, in contrast to the CGH holograms described in section \ref{Ch5_CGHchara}, which work in transmission (see Figure \ref{Fig613}).
One decade after their boom, the ability of SLMs to interactively control holograms in real time made them very prolific in the field of optical micro-manipulation and optical tweezers \cite{Reicherter1999,Hayasaki1999,Liesener2000,Curtis2002,Grier2003,Dienerowitz2008}. Nowadays, the holographic optical tweezers are used in a myriad of fields and it is even possible to find portable commercial optical tweezers set-ups \cite{RichardThesis,Richard2014}. Now, as commented in section \ref{Ch5_CGH}, the use of SLMs to produce beams carrying phase singularities highly escalated after 2002 \cite{Dholakia2002,Grier2003,Leach2003,Tao2003}. Actually, the use of phase-only CGH for these purposes is practically obsolete now. Nonetheless, in this section I will characterize the performance of SLMs and compare it with the performance of CGH. It will be seen that even though CGHs are less efficient and certainly less flexible, they can still be useful for certain applications.

\subsection{SLM characterization} \label{Ch5_SLMcharac}
\begin{figure}[tbp]
\centering
\includegraphics[width=\columnwidth]{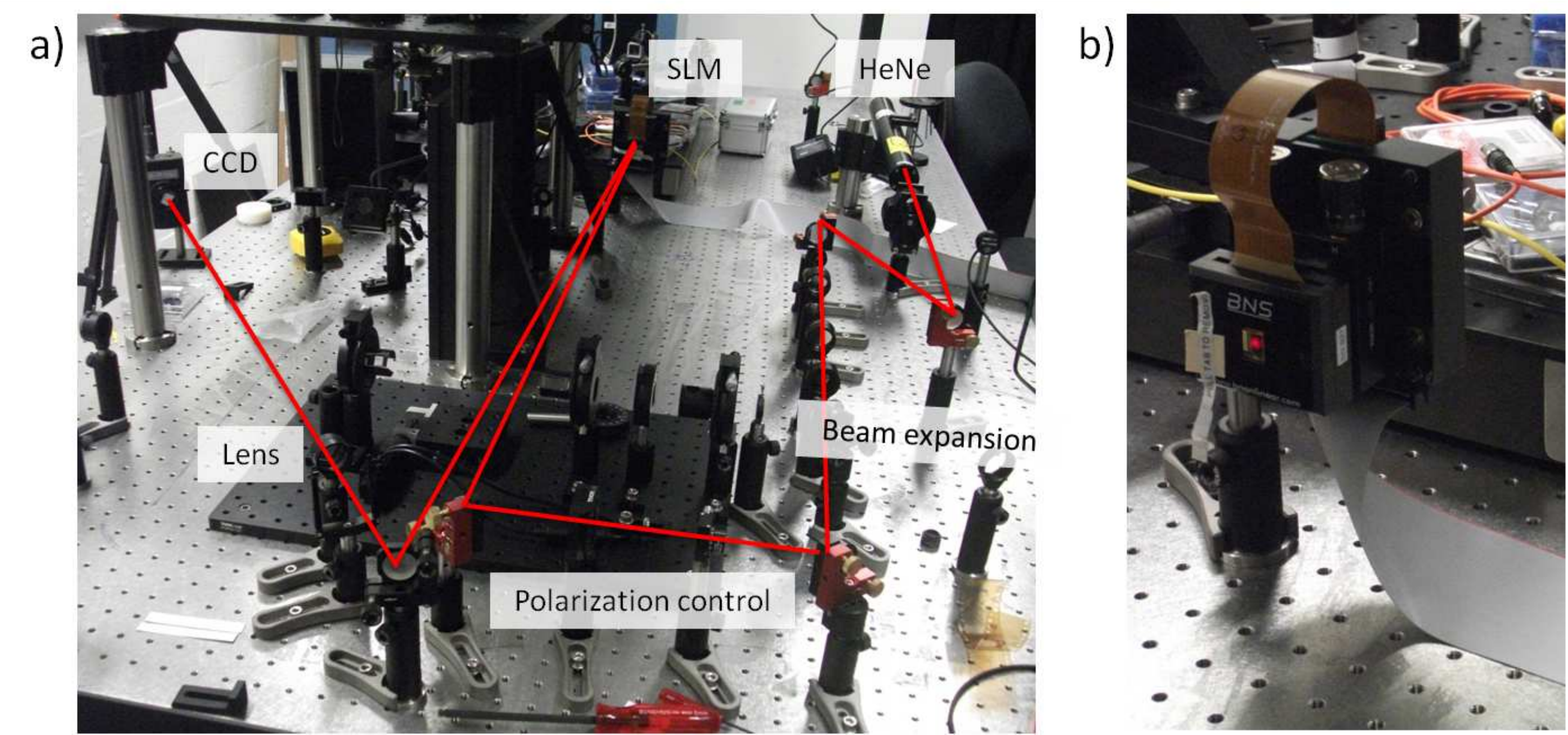} 
\caption{a) Picture of the experimental set-up to characterize the XY series SLM from BNS. A Gaussian beam from a He-Ne laser is expanded, then its polarization is controlled before it is sent over to the SLM. The SLM changes the phase and reflects off the beam, which is focused with a lens onto a CCD camera. b) Picture of the SLM. The incident beam almost fills in all the SLM chip. \label{Fig613}}
\end{figure}
\begin{figure}[tbp]
\centering
\includegraphics[width=14cm]{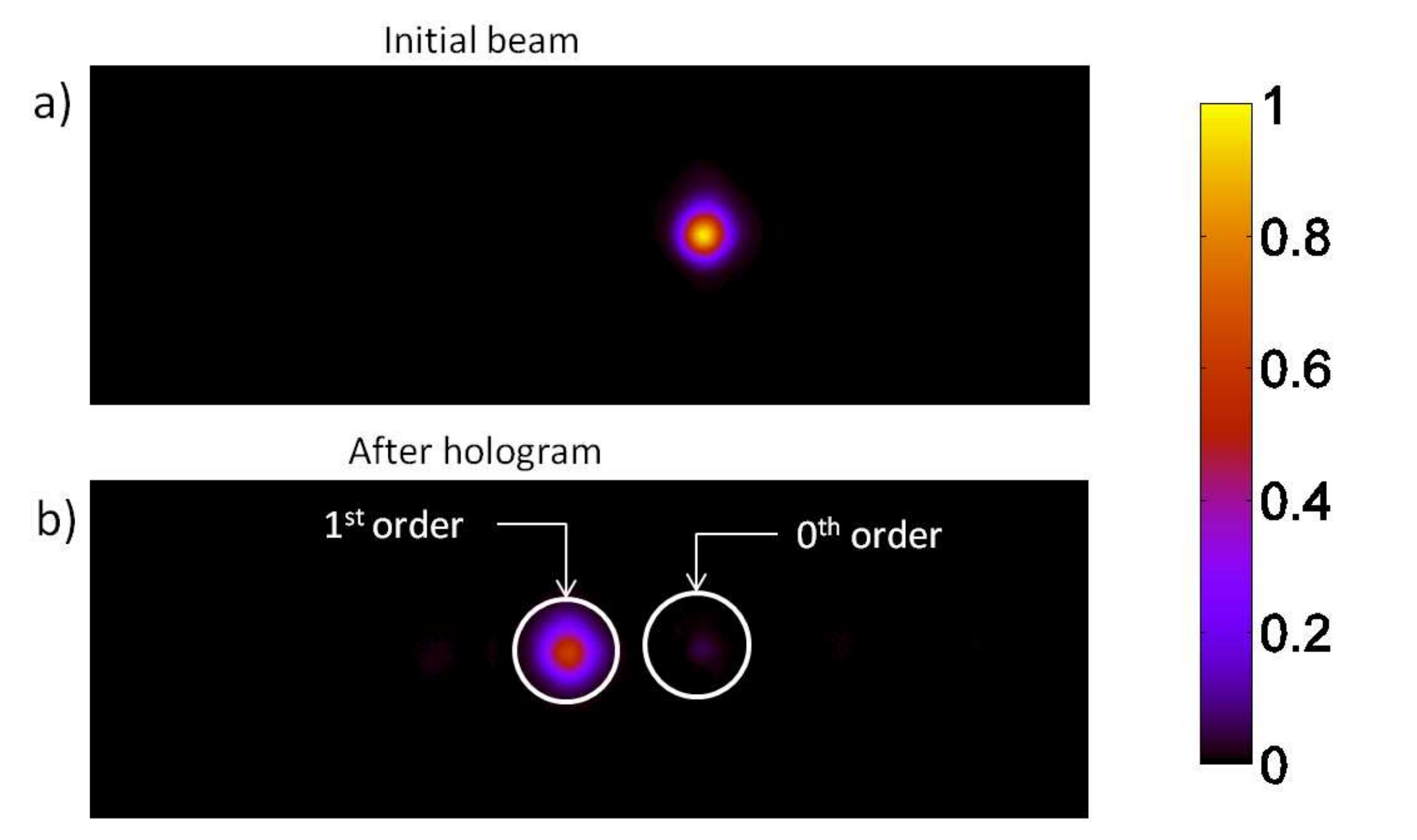} 
\caption{a) CCD image of the reference beam when the SLM works as a mirror. b) Typical diffraction pattern created by the SLM when the efficiency is over 80\%. The 0th order is at the same position as the reference beam, whereas the 1st order has been tilted. The colour bar indicates goes from 0 to 1, where 1 is taken as the absolute intensity maximum of the reference beam. \label{eff_SLM}}
\end{figure}

Here, similarly to what has been done in section \ref{Ch5_CGHchara}, the performance of an SLM is characterized. The SLM used in this chapter as well as in chapter \ref{Ch6} and \ref{Ch7} is a \hyperlink{http://www.bnonlinear.com/}{Boulder Nonlinear Systems}\footnote{http://www.bnonlinear.com/} (BNS) XYseries optimized for wavelengths ranging from $[630-800]$ nm. The SLM characterization is done again in terms of the SLM efficiency. Figure \ref{Fig613}(a) shows a real picture of the set-up to test the efficiency of the SLM. The laser beam is given by a He-Ne laser, whose $\lambda=632.8$nm. The beam is expanded to properly fill the SLM chip (see Figure \ref{Fig613}(b)). Then, its polarization is controlled by a linear polariser and a half-wave plate. Later, a mirror sends the beam to the SLM, which changes its phase. Finally, a lens Fourier-transforms the beam and a CCD camera records the image at the Fourier plane. Then, the efficiency is measured as the intensity of the laser beam hitting the SLM divided over the intensity going to the first diffraction order. Note that this efficiency measurement takes into account the losses of the SLM: imperfect reflection of the mirror, as well as diffraction losses due to the grid of pixels. Similarly to Figure \ref{Fig610}, a typical diffraction pattern is captured with a CCD camera in Figure \ref{eff_SLM}. Figure \ref{eff_SLM}(a) shows the position of the beam on the CCD camera when the SLM works as a mirror (the phase shift is 0). Then, Figure \ref{eff_SLM}(b) displays the diffraction pattern created by the SLM. Note that unlike the diffraction pattern created by the CGH shown in Figure \ref{Fig610}, most of the intensity goes to the 1st diffraction order.   
In Appendix \ref{App2SLM}, the efficiency of the SLM is characterized as a function of different controllable parameters in the laboratory.

\subsection{CGHs or SLMs? Pros and cons}
Due to their flexibility and interactive control, SLMs have taken over the role that CGHs used to play in the field of optical micro-manipulation, and in particular in the field of the AM of light. However, CGHs made of photographic films can still be useful for some applications. Thus, to conclude with this study of different methods to create phase-only holograms, the pros and cons of using each of the two methods will be discussed.
\begin{itemize}
\item \underline{\textsc{Price}}. Making CGH in the way described in section \ref{Ch5_CGH} is cheap. All the equipment and materials needed are easily available. On average, making 60 holograms (which corresponds to two photographic films) costs less than \$100. In contrast, the cheapest SLM in the market (from \hyperlink{www.cambridgecorrelators.com/}{Cambridge Correlators}\footnote{www.cambridgecorrelators.com/}) is valued \$1000. And then, the rest of them are in the range of \$15000-\$25000. 
\item \underline{\textsc{Efficiency}}. The CGH made and characterized in this thesis reached efficiencies values of 45\%. However, it is known that the efficiency can reach values of the order of 60\%. The main reason why those values were not reached is because the chemical process is not optimised and some silver has remained in the sample, thus producing some absorption. In contrast, the efficiencies measured with the SLM from BNS are constantly above 80\%. In order to visualize the difference of efficiency, Figure \ref{Fig619} displays the diffraction pattern from a CGH and the one from an SLM. It is seen that with the SLM, almost all the light goes into the 1st diffraction order, whilst with the CGH the 0th and 2nd order are also clearly seen. The cheap SLMs from Cambridge Correlators do not have such high efficiencies, though. Theirs is usually below 60\%. 
\begin{figure}[tbp]
\centering
\includegraphics[width=11cm]{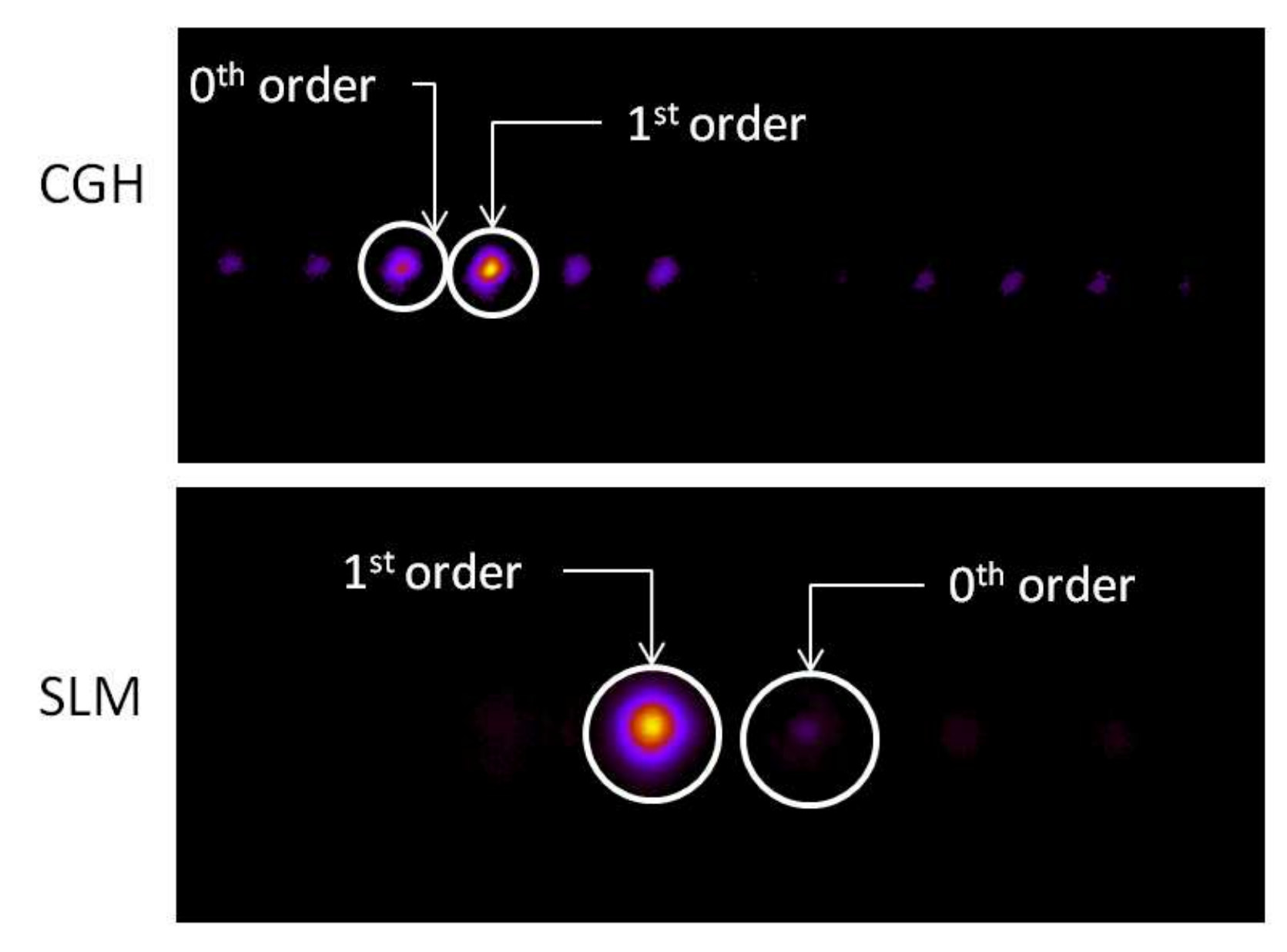} 
\caption{CCD image of a typical diffraction pattern created by a CGH and a SLM. It is seen that the SLM sends most of the light to the first diffraction order, whereas the light coming out of the CGH is split between many diffraction orders. \label{Fig619}}
\end{figure}
\item \underline{\textsc{Polarization}}. Polarization does not play any role for CGHs. That is, the transmittance $t(x,y)$ given by the hologram is completely independent of the polarization of the incoming beam. However, for SLMs, the polarization of the incident beam is crucial. Nematic liquid crystals form an anisotropic medium, therefore a change of polarization alters their response dramatically. In order to achieve the desired phase shift, the incident beam must be linearly polarized in a certain direction given by the axis of the liquid crystals \cite{Konforti1988}.
\item \underline{\textsc{Flexibility}}. CGH are optical elements that cannot be varied once they have been made. SLMs are dynamically controlled, therefore they can steer the beam interactively in real time. This makes them much more flexible.
\item \underline{\textsc{$\lambda$ dependence}}. The phase shift given by both CGHs and SLMs is wavelength dependent (see equation (\ref{E_txy_general})). However, for the CGH, the efficiency remains almost constant for a large range of wavelengths. In contrast, the efficiency of an SLM is very dependent on the range of $\lambda$s for which it has been designed. When the SLM is tested outside that range, the efficiency drop considerably. One of the main reasons for the drop in efficiency is the drop in reflectivity by the mirror standing between the liquid crystals and the pixel electrodes.
\item \underline{\textsc{Set-up}}. CGH work in transmission, whereas SLMs typically work in reflection. 
\end{itemize}
A summary of the pros and cons can be found in Table \ref{T_proscons}.
\renewcommand{\arraystretch}{1.3}
\renewcommand{\tabcolsep}{0.2cm}
\linespread{1}
\begin{table} 
\caption{Summary features CGHs and SLMs.}
\begin{center} 
\begin{tabular}{|c|p{6.2cm}|p{6.2cm}|}
\hline   & CGH & SLM \\ 
\hline Price & \$100 to make 60 holograms. & Cambridge Correlators: \$1500. Others: \$15000-\$25000. \\
\hline Efficiency & 40\% easily achievable. Chemistry needs to be refined to go above 50\%. & Cambridge Correlators: 55\%. Others: $>60$\%, 85\% with BNS.\\
\hline Polarization & Does not play any role. & Efficiency highly depends on polarization.\\
\hline Flexibility & Cannot be changed once it is made. & Controlled by computer. Can be changed interactively. \\
\hline $\lambda$ dependence & Phase shift depends on $\lambda$. Efficiency is not very affected by $\lambda$. & Performance depends a lot on $\lambda$.\\
\hline Set-up & Transmission. & Reflection.\\
\hline
\end{tabular}
\end{center}
\label{T_proscons}
\end{table}

\section{Fabrication of samples}
In order to carry out the experiments done in chapters \ref{Ch6} and \ref{Ch7}, different samples have been needed. In this section, both the samples and the methods to obtain them are broadly described. Notice that this is not the main topic and this thesis and consequently the level of detail of this section is lower than some others.
\subsection{Nano-apertures on metallic films} \label{Ch5_apertures}
\begin{figure}[tbp]
\centering
\includegraphics[width=\columnwidth]{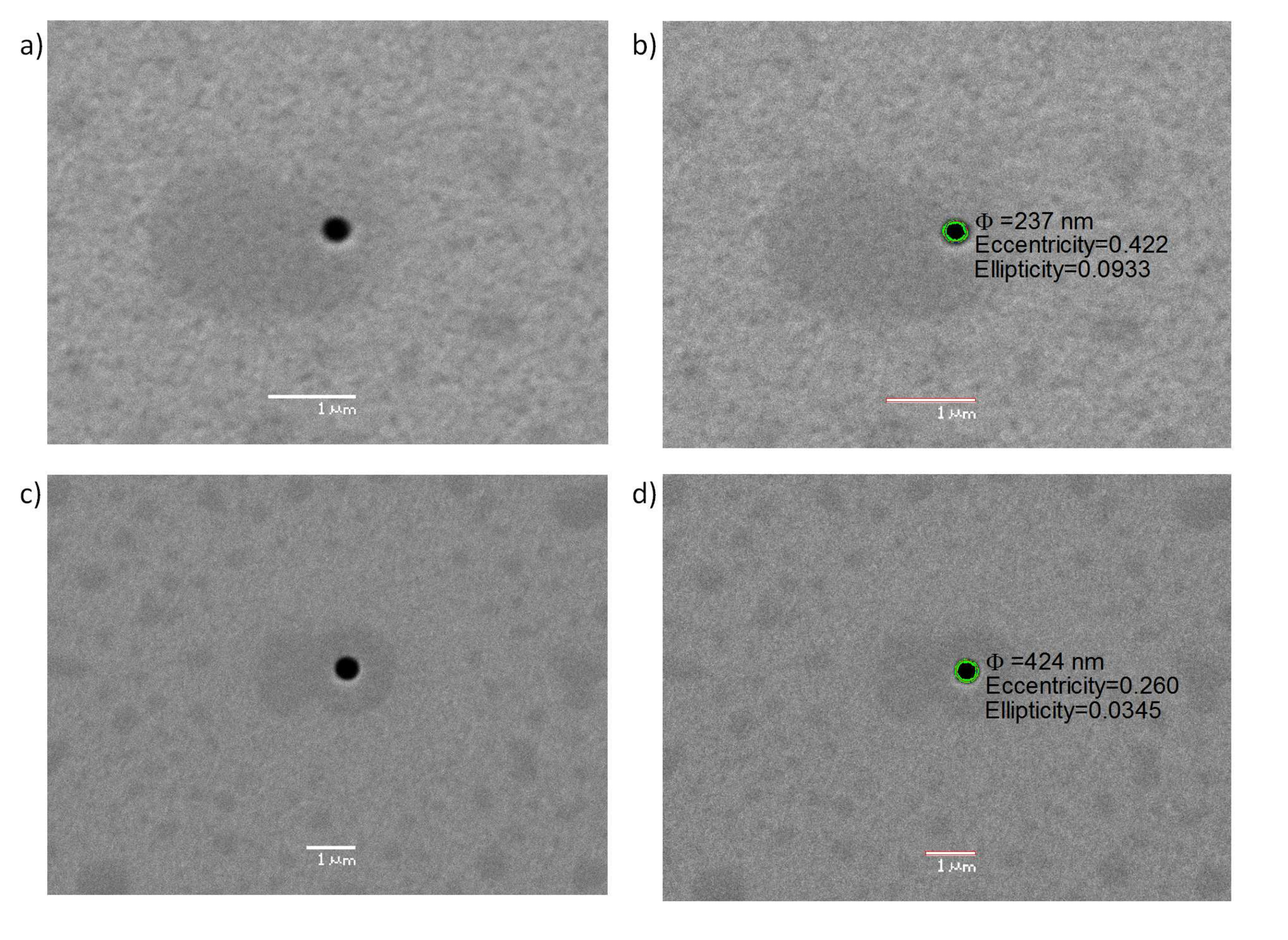} 
\caption{SEM images of two nano-apertures of diameter $\Phi =237$nm (a) and b)) and $\Phi = 424$nm (c) and d)). None of the images is post-processed. Images b) and d) have been used to retrieve the values of $\Phi$, Eccentricity and Ellipticity \label{Fig620}}
\end{figure}
In chapter \ref{Ch6}, some experiments using circular nano-apertures are presented. I assisted Alexander E. Minovich in the fabrication of these circular nano-apertures at the ACT Node of the Australian Nanofabrication Facility, in Canberra. First, a glass film of 1mm thickness was cleaned using a plasma etching technique. Afterwards, a gold layer of 200nm was deposited on top of the glass layer using thermal evaporation. Then, the gold layer on the glass film was milled using a \hyperlink{http://anff-act.anu.edu.au/HTM/detail/flagship_services.htm}{FEI Helios NanoLab Focused Ion Beam system}\footnote{http://anff-act.anu.edu.au/HTM/detail/flagship$\_$services.htm}. The circular nano-apertures have diameter sizes ranging from 200-450 nm, and they are all 50 $\mu$m apart, so that the coupling of surface plasmons launched from one nanohole to the closest one is avoided. The nano-apertures have been imaged using a secondary electron Scanning Electron Microscope (SEM, JEOL JSM-6480) operated at 10KeV, at the Microscopy Unit of Faculty of Science in Macquarie University. Then, the images have been analyzed with Matlab, where the boundaries of the nano-apertures were determined by selecting the pixels whose intensity was below the 10\% of the maximum. Figure \ref{Fig620}(a) depicts a SEM image of a 237 nm nano-aperture and Figure \ref{Fig620}(b) shows the same image with a boundary traced by a Matlab code to determine the diameter ($\Phi$), the eccentricity and the ellipticity of the hole. Likewise, Figure \ref{Fig620}(c) and Figure \ref{Fig620}(d) show the same features for a larger aperture of $\Phi = 424 $ nm. All the images on Figure \ref{Fig620} have not been post-processed. The results for the 12 milled nano-apertures are summarized in Table \ref{T_holes}.  
\renewcommand{\arraystretch}{1.3}
\renewcommand{\tabcolsep}{0.2cm}
\linespread{1}
\begin{table} 
\caption{Characterization of the circular nano-apertures images taken with the SEM.}
\begin{center} 
\begin{tabular}{|c|c|c|}
\hline  $\Phi$ (nm) & Eccentricity & Ellipticity \\ 
\hline 212 & 0.324 & 0.0538\\
\hline 214 & 0.414 & 0.0897\\
\hline 237 & 0.422 & 0.0933\\
\hline 253 & 0.462 & 0.113\\
\hline 317 & 0.295 & 0.0446\\
\hline 325 & 0.279 & 0.0396\\
\hline 333 & 0.118 & 0.00693\\
\hline 341 & 0.360 & 0.0672\\
\hline 424 & 0.260 & 0.0346\\
\hline 429 & 0.388 & 0.0783\\
\hline 432 & 0.273 & 0.0381\\
\hline 433 & 0.310 & 0.493\\
\hline
\end{tabular}
\end{center}
\label{T_holes}
\end{table}

\subsection{Single spheres on a substrate}\label{Ch5_spheres}
\begin{figure}[tbp]
\centering
\includegraphics[width=12cm]{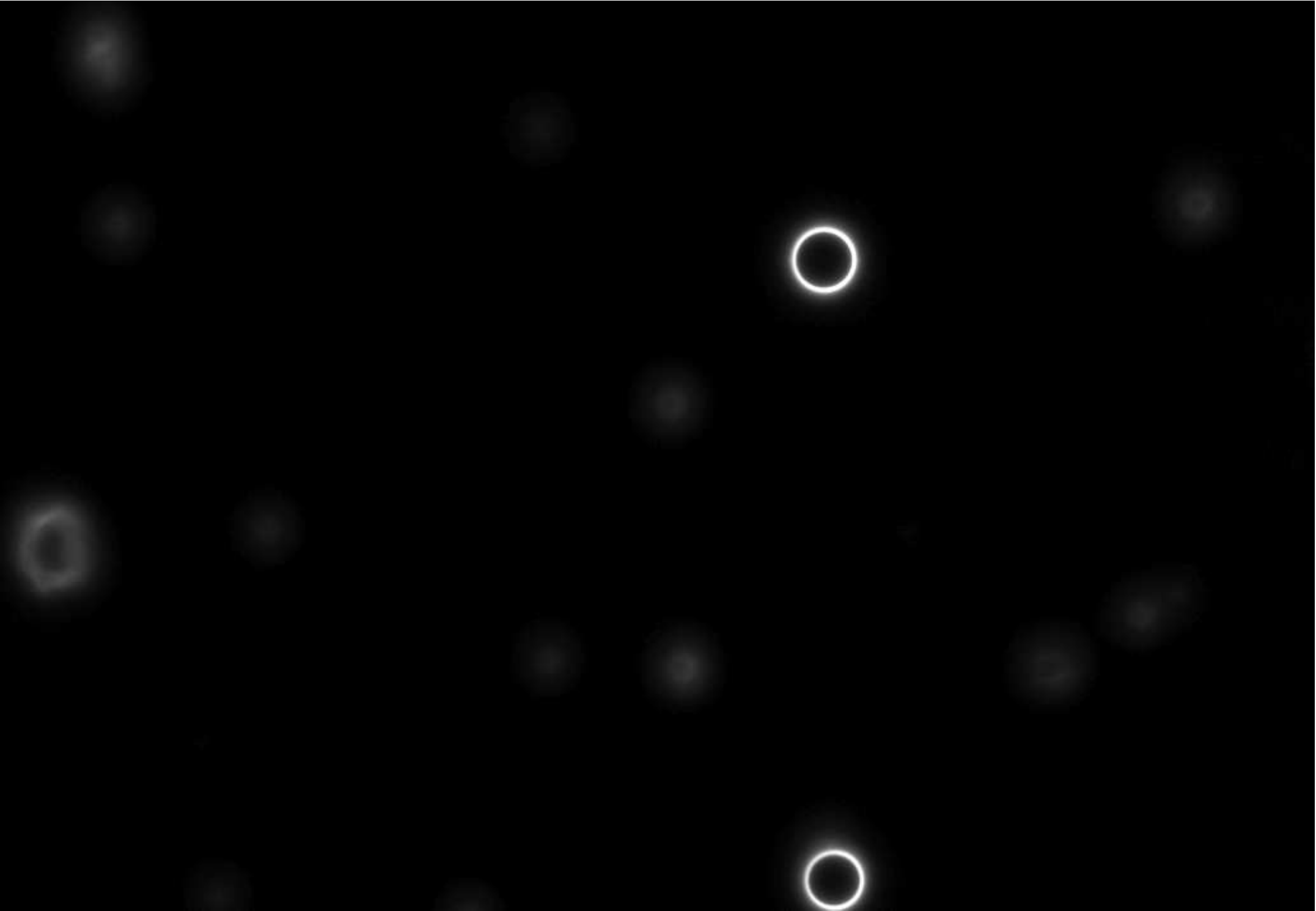} 
\caption{Dark-field image of $\Phi=950$ nm titania particles deposited on a coverslip after being spun. The images have been obtained from directly above the coverslip.\label{Fig641}}
\end{figure}

The theory developed in chapters \ref{Ch3} and \ref{Ch4} is based on the assumption that an isolated sphere is embedded in an isotropic, homogeneous medium. It is not trivial to achieve these conditions experimentally. One option would be levitating a single sphere. This was done already in the pioneering experiments of Ashkin \cite{Ashkin1971,Ashkin1974,Ashkin1975,Ashkin1976}. However, levitating particles of diameters smaller than 1 $\mu$m is experimentally hard. Even though objects of the order of 100 nm have been recently levitated \cite{Krishnan2010,Gieseler2012}, the experimental set-ups typically use vacuum chambers for cooling and other equipment that was not available in my current set-up. Now, due to the fact that we wanted to be able to use small particles to reach the dipolar regime to reproduce results from chapter \ref{Ch4}, another technique to have them embedded in a homogeneous medium was considered. This technique is spin-coating.\\\\
Spin-coating is a simple technique that is used to uniformly deposit a layer of material (usually a polymer or an organic material) on top of a flat surface. A good review of this process can be found in \cite{Spincoating}. The process starts by mixing particles with a resin. A droplet of the mix (with a controlled volume) is deposited on top of a planar surface, a microscope slide in this case. The microscope slide has been previously plasma-etched to remove any impurities that the glass surface could have. The microscope slide with a droplet on top is located on the spin-coater. Three parameters need to be entered to control the spin-coater: the duration time ($t_{sc}$), the maximum angular speed ($\omega_{sc}$), and the angular acceleration ($\alpha_{sc}$) to reach $\omega_{sc}$. Then, the spin-coater spins for $t_{sc}$ seconds following a pattern assigned by $\alpha_{sc}$ and $\omega_{sc}$. Once the process is finished, the sample is removed and placed under the illumination of a UV lamp, which joins the monomers of the resin to form a solid polymer. In this way, after the last step, the particles find themselves trapped in a homogeneous medium. If the concentration of particles is right (neither too low nor too high), it will be easy to find particles under the microscope and some of them will be isolated, \textit{i.e.} single particles whose closest particles are at a distance of 10 $\lambda$s.\\\\
As it has been explained in chapters \ref{Ch3} and \ref{Ch4}, the relative index of refraction between the single spheres and the surrounding medium is a crucial parameter to observe many phenomena. In fact, the example given at section \ref{Ch3_WGM} and Figure \ref{Tm} clearly show that high $n_r$ play an important role to either excite WGMs, or to preserve helicity. The refractive index of polymers varies from 1.3-1.7 \cite{Liu2009}. The most common ones for lithographic processes, such as PMMA or SU-8, have $n \approx 1.5$ \cite{Kasarova2007,SU-8}. That means that with these typical polymers, particles with $n>2.4 $ are needed in order to obtain $n_r > 1.6$. \\\\
Finding spheres with high refractive indexes is hard. Silicon is a very good candidate working at telecommunication wavelengths with $n>3$, but it absorbs too much in the optical region \cite{Bass1994}. In the experiments carried out in chapter \ref{Ch7}, particles made of Titania (TiO$_2$) have been used. Titania can be found in different phases, amorphous, anatase and rutile. The three phases have different refractive indexes. At $\lambda=500$ nm, rutile has $n_{ru} = 2.7$ \cite{Devore1951}, anatase has $n_{an}=2.3$ \cite{Demirors2011}, and the amorphous phase $n_{am} = 1.8$ \cite{Demirors2011}. Spheres made of these materials have been purchased from a Canadian company, \href{http://www.mknano.com/}{MKnano}\footnote{http://www.mknano.com/}.\\\\
In order to increase even further the contrast, a low-refractive index resin from an Israeli company called \href{www.mypolymers.com/}{MY Polymers Ldt.}\footnote{www.mypolymers.com/} was acquired. The refractive index is $n_{re}=1.33$. However, mixing titania particles with the resins acquired from MY Polymers has turned out to be challenging. The resins are highly hydrophobic, that is, they repel water. Whilst titania particles are given in aqueous solutions as they are hydrophilic. The way to fix this problem is to mix the aqueous solution in methanol, which is also hydrophobic. Xavier Vidal and I have seen that when the aqueous solution of titania particles is mixed with methanol in a 1/10 or 1/100 water/methanol manner, then the mix of titania and resin give acceptable results. That is, even though lots of particles still agglomerate, some of them can be isolated and surrounded by an homogeneous and isotropic medium. However, we have not been able to produce samples where the air-polymer interface is parallel to the coverslip surface. Since we are interested in symmetric samples, we have dismissed them due to these asymmetries.\\\\
Instead, we have done something else. We have deposited the aqueous solution of titania particles directly on top of a coverslip. The sample has been spun using the spin-coater to redistribute the particles across the sample and get some isolated single spheres. The result is displayed by Figure \ref{Fig641}, which is a dark-field image of a certain region of one of our samples. Even though these samples do not have spherical symmetry, they still keep the cylindrical one. Thus, they are still good candidates to test the theories developed in chapters \ref{Ch3} and \ref{Ch4}, as they must preserve $J_z$.

\begin{savequote}[10cm] 
\sffamily
``For a succesful technology, reality must take precedence over public relations, for nature cannot be fooled.'' 
\qauthor{Richard P. Feynman}
\end{savequote}

\chapter{Experiments with single nano-apertures}
\graphicspath{{ch6/}}
\label{Ch6} 

\newcommand{\Tt}{\mathbf{T}}
\newcommand{\Top}{\overline{\mathbf{T}}}
\newcommand{\Mz}{M_{\left\lbrace \zhat \right\rbrace }} 

\newcommand{\Dm}{\mathbf{D}_{mk_{\rho}}}
\newcommand{\Cm}{\mathbf{C}_{mk_{\rho}}}
\newcommand{\modk}{\vert \mathbf{k}_0 \vert }
\newcommand{\Rz}{R_z}
\newcommand{\Einp}{\mathbf{E}_{+}^{\mathbf{in}}}
\newcommand{\Einm}{\mathbf{E}_{-}^{\mathbf{in}}}
\newcommand{\Einpp}{\mathbf{E}_{p}^{\mathbf{in}}}
\newcommand{\Emp}{\mathbf{E}_{m,p}^{\mathbf{in}}}
\newcommand{\Esmp}{\mathbf{E}_{m,p}^{\mathbf{s}}}
\newcommand{\Esmpp}{\mathbf{E}_{m,p}^{\mathbf{s} \  p}}
\newcommand{\Esmpm}{\mathbf{E}_{m,p}^{\mathbf{s} \ -p}}
\newcommand{\Epq}{\mathbf{E}_{p,l}^{\mathbf{in}}}
\newcommand{\Eppq}{\mathbf{E}_{-p,-l}^{\mathbf{in}}}
\newcommand{\Etpq}{\mathbf{E}_{p,l}^{\mathbf{t}}}
\newcommand{\Etppq}{\mathbf{E}_{-p,-l}^{\mathbf{t}}}

\newcommand{\Atpqp}{A_{p,l}^{\mathbf{t}}(x,y)}
\newcommand{\Btpqpp}{B_{p,l}^{\mathbf{t}}(x,y)}

\section{Introduction} \label{Ch6_Intro}
In this chapter, a way of controlling the scattering of a given nano-structure with the incident field is shown. As it was explained in section \ref{Ch1_apla}, the conventional way to control the scattering response of a structure is by characterizing its scattering matrix $\Sconv(\mathbf{g},\mathbf{m})$, where $\mathbf{g}$ and $\mathbf{m}$ are general sets of variables describing the geometrical and material properties of the structure, respectively. Then, assuming a $\shat$ or $\phat$ plane wave-excitation, the geometrical and/or material properties of the structure are modified to control the optical response given the corresponding plane wave excitation. There is an alternative and less-explored way of controlling the response of structures, and that is by tuning the properties of the incident beam. That is, once the scattering matrix has been characterized, a certain response can be obtained as a weighted superposition of individual responses of different plane waves:
\begin{equation}
\Eout(\mathbf{r}) = \left( \int \dd^3 \mathbf{k_i'} \ \exp (i \mathbf{k_i'} \cdot \mathbf{r}) \Sconv(\mathbf{g},\mathbf{m}) *  \Ein (\mathbf{k_i'}) \right) (\mathbf{r})
\label{E_Eoutconv}
\end{equation}
where $\mathbf{k_i'}$ is each of the different plane waves that take part in the superposition, and $\Ein (\mathbf{k_i'})$ is the Fourier transform of the incident field, which modulates the plane wave decomposition. Both approaches are equally challenging and complementary. For example, once a structure is characterized, the knowledge of the response to different input beams can be used to control its scattering. Alternatively, once the structure is chosen but not characterized, the incoming beams used to probe it can be modified to maximize the information retrieved from the sample. In this sense, exploiting the symmetries of the sample allows for identifying correlations between input and output fields. \\\\
Here, the scattering of a non-chiral sample is controlled with vortex beams to experimentally induce a huge CD. These results are presented in section \ref{Ch6_CD}, but before that an analysis of the experiment in terms of symmetries is carried out.

\section{Symmetries} \label{Ch6_symm}
\begin{figure}[tbp]
\centering
\includegraphics[width=12cm]{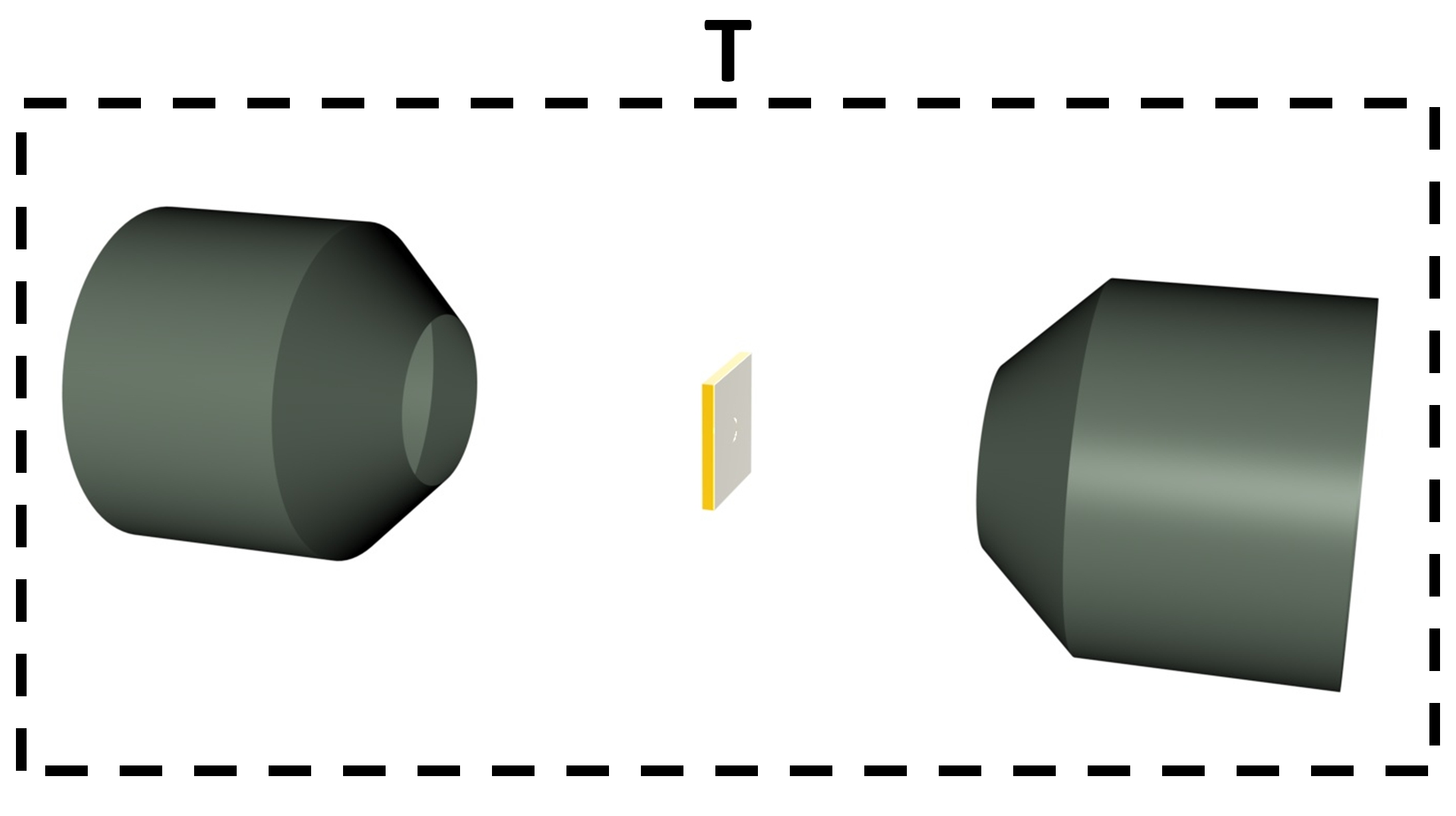} 
\caption{Schematics of the Target = MOs + Sample. MO$_1$ focus light onto the sample and the scattering is collected by MO$_2$. \label{Fig71}}
\end{figure}

The interaction of light with nano-apertures is a problem that has been carefully studied by many scientists. In particular, the work of Ebbesen \textit{et al.} \cite{Ebbesen1998} showing that nano-apertures could have an extraordinary transmission due to the coupling of light and SPPs opened up lots of possibilities among different fields in optics, micro-manipulation, biophysics and condensed-matter \cite{Mock2002,Mathieu2011,Mcdonnell2001,Haes2002}. Lots of these studies have focused on how the nano-apertures couple with SPPs depending on many different parameters \cite{Homola1999,Barnes2003,Zayats2005,Genet2007}. Some others have focused on the radiation diagram of these nano-apertures and have studied how the $\shat$ and $\phat$ polarization states are transmitted through the apertures both theoretically and numerically \cite{Aigouy2007,Lalanne2009,Sol2011,Ivan2011,Sol2012,Yi2012}. Finally,  the excitation of surface plasmon with subwavelength nano-apertures has also been used to induce spin-orbit interactions \cite{Bliokh2008,Yuri2008,Vuong2010,Bliokh2010,Rodriguez-Herrera2010,Yuri2013}.\\\\
Here, it will be shown that a symmetry characterization of both samples and excitation beams can allow for a predictive description of light-matter interactions at the nano-scale. First, the symmetries of the nano-apertures shown in section \ref{Ch5_apertures} are characterized. As described in section \ref{Ch5_apertures}, the sample is a 1mm glass film coated with a gold layer of 200nm. The sizes of the nano-apertures as well as their deviation from a circle are given by Table \ref{T_holes}. 
Then, with a very good approximation, the circular nano-apertures are symmetric under rotations along the $z$ axis, where the $z$ axis stands for the normal direction of the sample. With the same assumption, they are also symmetric under any mirror transformation that contains the $z$ axis. Last but not least, since the samples behave as linear media (see section \ref{Ch2_Symmetries}), they are also symmetric under time translations and therefore the frequency of light is preserved in the interaction. In contrast, spatial translations and duality symmetry are broken. All these symmetry considerations of the samples are inherited by the scattering matrix  $\Sconv(\mathbf{g},\mathbf{m}) $. Thus, the scattering matrix commutes with the following three operators\footnote{See section \ref{Ch1_symm} to see the definition of these operators}:
\begin{equation}
\left[ \Sconv, R_z \right]=\left[ \Sconv, \Mz \right]=\left[ \Sconv, T_{\Delta t} \right] = 0 \label{E_comm}
\end{equation}
Now, even though the experiment will be done with a single nanohole, the CD study will be carried out with a more general system formed by two MOs and the sample (see Figure \ref{Fig71}). This more general system will be referred to as target ($\Tt$). The idea is that the microscope objectives are symmetric under the same transformations as the sample: rotations around the $z$ axis, mirror symmetries containing the $z$ axis and temporal translations (see section \ref{Ch1_apla}). Thus, target and sample are equivalent in terms of symmetries. This fact will be exploited in the study of CD in section \ref{Ch6_CD}. Moreover, it will be considered that the action of the two microscope objectives and the sample can be characterized by a linear integro-differential operator that inherits all the properties of the target. I will denote this operator as $\Top$. Similarly to the scattering matrix, $\Top$ fulfils the following commutation rules:
\begin{equation}
\left[ \Top, R_z \right]=\left[ \Top, \Mz \right]=\left[ \Top, T_{\Delta t} \right] = 0 \label{E_Tcomm}
\end{equation}
 
\subsection{Duality symmetry} \label{Ch6_duality}
\begin{figure}[tbp]
\centering
\includegraphics[width=\columnwidth]{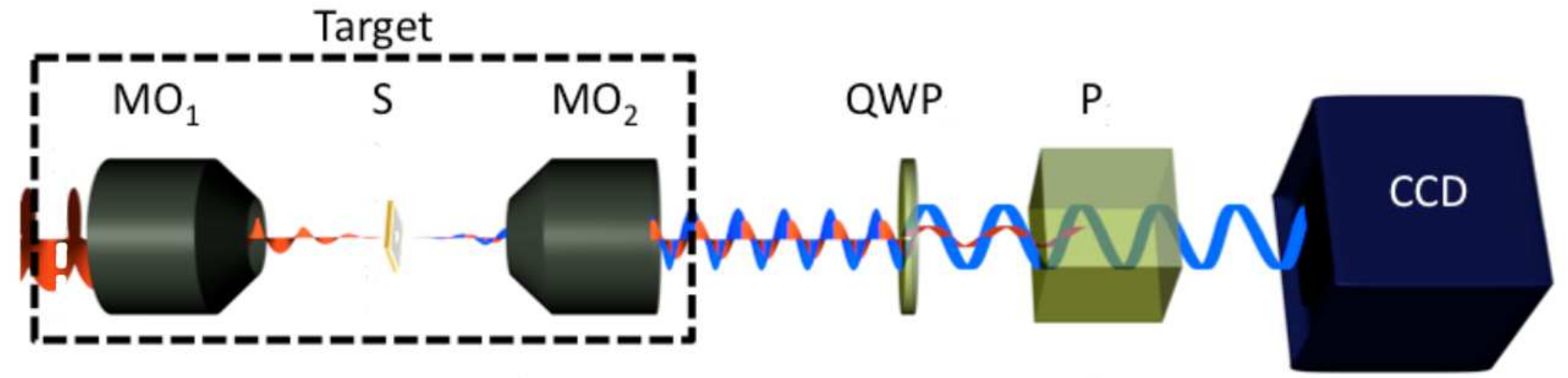} 
\caption{Schematics of a helicity projection scheme. A beam with a well-defined helicity is focused down to a sample that breaks duality symmetry. As a consequence, an additional helicity component is created. A quarter-wave plate and a linear polariser are used to separate the two helicity states.  \label{Fig72}}
\end{figure}

One main difference between the MOs and the sample is that the MOs preserve helicity (see section \ref{Ch1_apla}), whereas the sample does not. In this subsection, the mechanism by which duality symmetry is highly broken in the sample is explained. As it has been explained in previous chapters, the helicity operator $\Lambda$ is the generator of duality transformations. Therefore, the helicity of light is not preserved upon interaction when the sample is not dual, and vice versa.\\\\
It is known that the microscopic Maxwell equations for material media always break duality symmetry \cite{Jackson1998,Ivan2013}. However, it was proven in \cite{Ivan2013} that duality symmetry can be restored in the macroscopic Maxwell equations if there is an impedance matching between all the media where the studied EM interaction takes place. That is, 
\begin{equation}
\dfrac{\epsilon_i}{\mu_i} =\text{const} \Longleftrightarrow  \Lambda\text{ conservation } 
\label{E_duality_Iv}
\end{equation} 
for any medium made of an arbitrary number of isotropic and homogeneous sub-media $i$, $\epsilon_i$ and $\mu_i$ being its electric permittivity and magnetic permeability. This is not the case for the sample, made of gold and glass. Clearly, 
\begin{equation}
\dfrac{\epsilon_{\text{glass}}}{\mu_{\text{glass}}} \neq \dfrac{\epsilon_{\text{air}}}{\mu_{\text{air}}} \neq \dfrac{\epsilon_{\text{gold}}}{\mu_{\text{gold}}}
\end{equation}
therefore duality symmetry must be broken by the sample.\\\\
In contrast, as shown in section \ref{Ch1_apla}, MOs are made in a way such that helicity is preserved\footnote{As a consequence, the action of two MOs preserves the polarisation state. This statement has been experimentally verified in the laboratory using a \hyperlink{http://www.thorlabs.hk/newgrouppage9.cfm?objectgroup_id=1564}{Thorlabs polarimeter}. The value of the third Stokes parameter $S_3$ remained the same when it was measured with or without the MOs.}. 
That is, consider a paraxial beam of light with a well-defined helicity $p$. Because the beam is paraxial, the polarization has to be circular ($\spphat$). Now, it is well-known that when a beam is tightly focused, the polarization changes so that the transversality of the field is maintained \cite{Novotny2006,Ivan2012PRA,Bliokh2011}. However, because MOs preserve helicity, the helicity of the beam at the focus is still $p$.\\\\
Then, the interaction with the sample takes place. The sample is not dual-symmetric, therefore helicity is not preserved. That is, an opposite helicity component $-p$ is created upon scattering.\\\\
The scattered light is collected by the second MO, which also preserves helicity. If the collecting MO is placed at a focal distance of the sample, then the beam coming out of the objective is a paraxial field again. Hence, one component of the helicity is left circularly polarized (LCP) and the other one is right circularly polarized (RCP). LCP corresponds to $\sphat$ and RCP to $\smhat$. Then, by projecting the states back to a linear basis, the two different helicity components can be analysed separately.\\\\
The method is depicted in Figure \ref{Fig72}. 
Between MO$_1$ and QWP, the red helix represents a beam with a helicity $p$, and the blue helix represents a beam with a helicity $-p$. It can be seen that the beam has helicity $p$ before it interacts with the sample. Once the interaction has been taken place, the beam is a combination of red and blue, as a $-p$ helicity component has been created. The quarter-wave plate (QWP) changes the polarization state from LCP and RCP to horizontal and vertical, so that the linear polariser (P) can filter out one of the components. Note that the $-p$ helicity component has an exceptional signal ratio. That is, all the light analysed with the $-p$ helicity component exclusively comes from the sample and not from the incident field, except for the experimental errors introduced by the QWP and LP.\\\\
The fact that a nano-aperture scatters light in the crossed helicity component is not surprising. Any macroscopic material which does not fulfil equation (\ref{E_duality_Iv}) breaks duality symmetry and creates light with the crossed helicity component $-p$. As I showed in chapter \ref{Ch4}, some very specific conditions need to be fulfilled so that dielectric spheres behave in a dual way. However, the creation of modes with helicity $-p$ with a circular nano-aperture is interesting from different perspectives.\\\\
Firstly, due to the sub-wavelength nature of the nano-apertures, duality symmetry is highly broken. To analyse it, it is useful to define a ratio of light scattered into the crossed helicity component divided over the direct one, $\gamma$. The definition of $\gamma$ is very similar to the transfer function $\Tp$ defined in section \ref{Ch4_trans}, except for a subtlety: $\Tp$ takes into account the intensity which is backscattered, whereas $\gamma$ is only defined in transmission. If the $\gamma$ ratio is computed for the glass + gold multilayer system, it turns out to be of the order of $10^{-3}$ \cite{Nora2014}. However, the nano-aperture highly breaks duality symmetry and then $\gamma$ is of the order of 1 \cite{Nora2014}. This incredibly high helicity conversion ratio is due to the coupling of the scattered field to surface modes of the multilayer system, whose character is TE/TM \cite{Ivan2011,Nora2014}. Thus, the study of the $-p$ helicity component as well as its intensity ratio with respect to the direct one can allow for the characterization of the coupling of light to surface modes.\\\\
Secondly, the creation of modes with helicity $-p$ can also be very useful to link geometrical and material symmetries of the sample to spin-orbit interactions. In the literature, spin-orbit couplings are used to explain phenomena where the polarization (spin) and spatial properties of light (orbital) are linked. Due to the non-paraxial nature of the helicity formalism, spin-orbit interactions are naturally included. Thus, in section \ref{Ch6_AM}, it will be shown that the two helicity components give rise to different spatial modes with different spatial properties and polarizations. Last but not least, the spatial properties of the crossed helicity component have been also used in \cite{preNora2014} to develop a highly stable nanopositioning method to center a sample in only three steps.

\section{Circular Dichroism} \label{Ch6_CD}
In this section, the second method to control the scattering of a structure explained in section \ref{Ch6_Intro} is used to induce giant CD on a sample. Since its discovery in the 19th century, CD has been widely used in science. Defined as the differential absorption LCP and RCP \cite{Barron2004}, its uses are as diverse as protein spectroscopy, DNA studies and characterization of the electronic structure of samples \cite{Kelly2005}. In the advent of nano-photonic circuitry, a lot of work has been put recently into characterizing plasmonic components and optical metamaterials in terms of CD \cite{Decker2007,Kwon2008,Zhang2009,Wang2009,Fan2010,Guerrero-Martinez2011,Hendry2012,Sersic2012,Kuzyk2012}. The approach in all these works is the same one. CD wants to be induced with a structure that is illuminated under a plane wave excitation. In order to achieve it, structures with a non-trivial geometry made of a smart combination of materials are engineered. In fact, in all the cases cited before, the structures engineered are \textbf{chiral}. A chiral structure is such that none of its mirror symmetric images can be superimposed the sample itself \cite{Barron2004,Bishop2012}. One of the mains reasons why the design of the structures is chiral is an important theorem known in molecular physics:\\
\begin{quote}
\sffamily
``We know today, in \textit{molecular} terms, that the one necessary and sufficient condition for a substance to exhibit optical activity is that its molecular structure be such that it cannot be superimposed on its image obtained by reflection in a mirror.''
\qauthor{David M. Bishop, \textit{Group theory and chemistry}, (2012) page 20}
\end{quote}
The approach that it is taken here to achieve CD is completely different. The nano-structure that will be used to induce CD is a circular nano-apertures. Typically it was thought that a non-chiral sample could not induce CD. However, recent experiments with non-chiral plasmonic planar structures have shown that a non-zero CD can be obtained with a sample that lacks mirror symmetry with respect to the incident input beam, but it is non-chiral on its own \cite{Plum2009,Cao2012,Maoz2012,Ren2012}. Other attempts to create CD with a non-chiral sample have been carried out surrounding the sample with a chiral medium \cite{Abdulrahman2012,Xavi2013}. Here, I will go a step further and induce giant CD in a non-chiral sample such as a circular nano-aperture using light at normal incidence. In contrast to the previous approaches, the mirror symmetry of the system is broken with an internal degree of freedom of the input beam: its angular momentum. The results presented in this section show that CD can be induced in a symmetric sample if the beam used to make the CD measurement is a vortex beam. The reasons behind this interesting phenomenon are two.
\begin{figure}[htbp]
\centering\includegraphics[width=13cm]{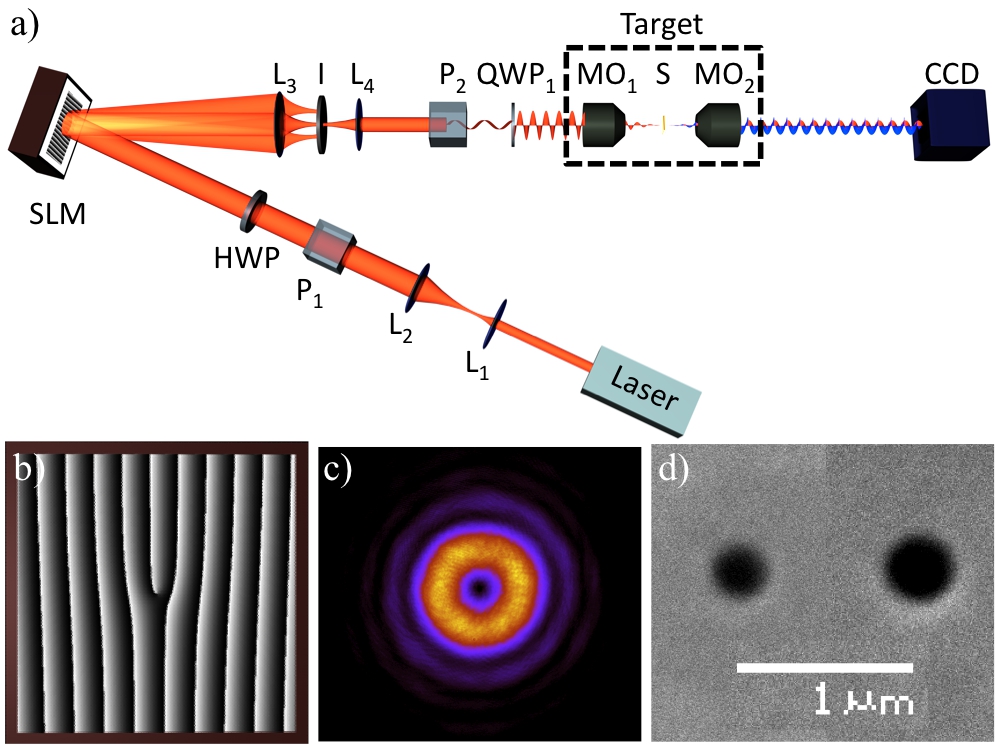}
\caption{(a) Schematic view of the optical set-up in consideration. LCP and RCP vortex beams go through a sub-wavelength circular aperture and their transmissivities are recorded with a CCD camera. (b) A pitchfork-like hologram is used to create a vortex beam. (c) Intensity profile of one of the vortex beams used in the experiment. (d) SEM images of two isolated circular nano-apertures.}
\label{set-up}
\end{figure}
Firstly, the two LCP and RCP beams used to carry out the CD measurement are not mirror symmetric. And secondly, the LCP and RCP vortex beams have an AM that differs in two units. \\\\
A sketch of the experimental set-up used in the experiment is depicted in Figure \ref{set-up}. It can be divided into three parts: preparation of states, non-paraxial interaction with the sample, and measurement. For the state preparation, a HeNe laser working at $\lambda=632.8$nm was used. It produces a collimated, linearly polarized Gaussian beam. The beam is expanded with a beam expander (lenses L$_1$-L$_2$) to match the dimensions of the chip of the SLM. Both lenses are achromatic with an anti-reflection coating for $650-1050$nm. Their focal lenses are $f_1=200$nm and $f_2=300$nm respectively, making the magnitude of the beam expander equal to 1.5. Then, the polarization state of the beam is modified with a linear polariser (P$_1$) and a half-wave plate (HWP) to maximize the efficiency of the SLM. The linear polariser is a Glan-Taylor Calcite Polarizer with a $650-1050$nm coating. The HWP is an achromatic zero-order wave plate from Newport (10RP02-28). Then, the SLM creates a vortex beam by displaying an optimized pitchfork hologram \cite{Richard2011} (see Figure \ref{set-up}(b)). Proper control of the pitchfork hologram allows for the creation of a phase singularity of order $l$ in the center of the beam, \textit{i.e.} the phase of the beam twists around its center from $0$ to $2 \pi l$ radians in one revolution. Note that when $l=0$, the SLM behaves simply as a mirror. As explained in section \ref{Ch5_CGHchara} and appendix \ref{Appendix2}, the SLM produces different diffraction orders. In order to filter the non-desired orders of diffraction, a modified 4-f filtering system with the lenses L$_3$-L$_4$ and an iris (I) in the middle is used. Lens L$_3$ (achromatic AC254-150-B-ML from Thorlabs) Fourier-transforms the beam, and the iris selects the 1st diffraction order and filters out the rest. Then, lens L$_4$ (achromatic AC254-125-B-ML from Thorlabs) is used to match the size of the back-aperture of the microscope objective that will be used to focus the beam down to the sample. But before focusing, the preparation of the input beam is finished by setting its polarization to either LCP or RCP. This is done with a linear polariser (P$_2$), which is also a GT10-B like P$_1$, and a quarter-wave plate (QWP$_1$). The QWP$_1$ is a 10RP04-28 zero-order quartz wave-plate from Newport. Since the beam is collimated, this change of polarization does not appreciably affect the spatial shape of the input beam.\\\\
After this initial preparation, the light is focused down to interact with the sample (S), \textit{i.e.} one of the single circular nano-apertures listed in Table \ref{T_holes}. The focusing MO$_1$ is a water-immersion \hyperlink{http://www.olympusamerica.com/seg_section/uis2/seg_uis2.asp}{Olympus LUMFLN60XW}\footnote{http://www.olympusamerica.com/seg\_section/uis2/seg\_uis2.asp} with NA=1.1 and a long working distance (WD=1.5mm). The nano-apertures are centred with respect to the incident beam with a nano-positioning stage. The sample is mounted on a piezo electric transducer on closed-loop with resolution below 0.5nm (translation range 300 $\mu$m with 20-bit USB interface and noise floor of tens of picometers). Then, the centring is done maximizing the intensity going through the aperture when a Gaussian beam (vortex beam with charge $l=0$) is used. When a vortex with $l\neq 0$ is used, the central position is found by minimizing the intensity going through the nano-aperture. The interaction of the light and the centred nano-aperture occurs in the non-paraxial regime. Typically, the nano-aperture only allows a small part of the incoming beam to be transmitted. The transmitted light is scattered in all directions. This facilitates the coupling of light to superficial modes of light, as it has been described theoretically and experimentally by many authors \cite{Solthesis,Ivan2011,Yi2012}. The light transmitted through the sample is collected by another microscope objective (MO$_2$). MO$_2$ is an \hyperlink{http://www.olympus-ims.com/en/microscope/mplfln/}{Olympus MPLFLN100x}\footnote{http://www.olympus-ims.com/en/microscope/mplfln/}, whose NA=0.9 and WD=1mm. MO$_2$ collects the transmitted light through the aperture and images it on a charged-couple device (CCD) camera, which sits more than 1m apart of MO$_2$. The CCD camera is used to capture the transmitted intensity.\\\\
The procedure to measure CD is the following: First, a vortex beam of topological charge $l$ is created with the SLM. Secondly, the QWP$_1$ is rotated to create a LCP state. Then the sample is centred with respect to the beam at the focal plane of MO$_1$ in the manner explained before. The transmitted intensity $I_l^L$ is measured with the CCD camera, where $L$ stands for the polarization of the beam (LCP) and $l$ for its topological charge. Later, the QWP$_1$ is rotated again to change the polarization state to RCP. The sample is re-centred. Finally, the transmitted intensity $I^R_l$ is measured. Given the two measured values $I^L_l$ and $I^R_l$, the CD measurement given an incident vortex beam with charge $l$ can be computed as:
\begin{equation}
\text{CD}_l(\%)=\dfrac{I_l^L-I_l^R}{I_l^L+I_l^R} \cdot 100
\label{CD}
\end{equation} 
\begin{table}
\caption{\label{TabCD}Measurements of CD($\%$) for three different phase singularities $l$ as a function of the diameter of the nano-aperture. CD is computed using equation (\ref{CD}).}
\begin{center}
\begin{tabular}{|c|c|c|c|}
\hline $\Phi$(nm)& $l=-1$& $l=0$ & $l=1$\\
\hline 
$d_1=237$  & $-78 \pm 6$ & $-0.130 \pm 0.003 $ & $68 \pm 6$  \\
$d_2=212$  &  $-91 \pm 6$ & $-1.843 \pm 0.004 $ & $84 \pm 6$ \\
\hline 
$d_3=325$ & $-10 \pm 4$ & $0.596 \pm 0.004 $ & $9.8 \pm 1.1$ \\
$d_4=317$  & $-7 \pm 5$ & $1.258 \pm 0.005 $ & $12 \pm 4$ \\
$d_5=333$  & $-9 \pm 4$ & $0.841 \pm 0.005 $ & $13 \pm 2$ \\
\hline
$d_6=432$  & $-22.3 \pm 0.8$ & $0.596 \pm 0.003 $ & $27.1 \pm 0.8$ \\
$d_7=429$ & $-34 \pm 2$ & $-1.055 \pm 0.007 $ & $37.5 \pm 1.3$ \\
$d_{8}=433$  & $-46.8 \pm 1.3$ & $0.629 \pm 0.004 $ & $45.9 \pm 0.8$\\ 
\hline
\end{tabular}
\end{center}
\end{table}
\begin{figure}[htbp]
\centering\includegraphics[width=15.5cm]{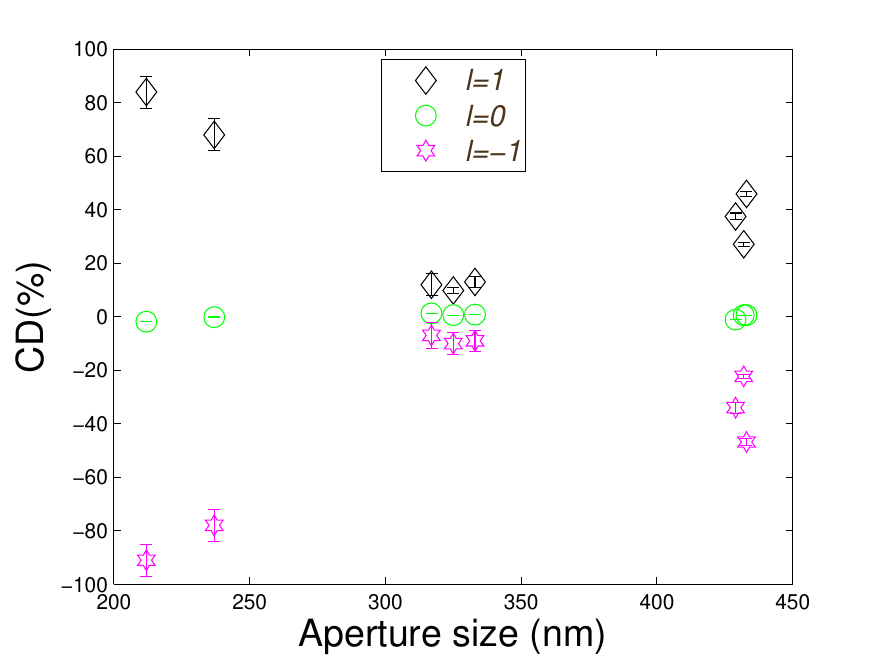}
\caption{Measurements of CD($\%$) for three different phase singularities $l$ as a function of the diameter of the nano-aperture. CD is computed using equation (\ref{CD}). Black diamonds are used to represent the CD obtained with $l=1$; green circles for $l=0$ and magenta hexagon stars for $l=-1$.}
\label{CD_fig}
\end{figure}
The results of the experiment are presented in Table \ref{TabCD} and Figure \ref{CD_fig}. The first column of Table \ref{TabCD} shows the size of the nano-aperture in consideration. The rest of columns show the measured CD using beams with different phase singularities of order $l=-1,0,1,$ respectively. As expected, due to the fact that the circular nano-aperture is mirror symmetric and the incidence is normal, the CD is very close to zero when the incident beam is Gaussian ($l=0$). The residual CD can be attributed to the small asymmetries of the sample (characterized in section \ref{Ch5_apertures}) or the incoming beam. When vortex beams with $l=-1,1$ are used the situation is very different. A non-null CD is obtained. This value is of the order of $90\%$ for some nano-apertures, even though the incidence is normal and the nano-aperture is still mirror symmetric. Because the theoretical maximum value for CD given the formula (\ref{CD}) is 100\%, it is fair to say that for certain sizes a giant value of CD has been obtained. Furthermore, it can be observed that there is an underlying symmetry relating the value of CD$_1$ and CD$_{-1}$. Indeed, CD$_1 \simeq -$CD$_{-1}$. Hereafter the results are discussed and the mechanisms by which CD$_1 \simeq -$CD$_{-1} \neq 0$ are unveiled. It will be seen that symmetries considerations are almost enough to predict these results. However, it will be shown that not only the symmetries of the sample need to be taken into account, but also the symmetries of light. The symmetries of the target have been analysed in section \ref{Ch6_symm}. $\Tt$ is symmetric under time translations, rotation around the $z$ axis, and mirror symmetries containing off a plane containing the $z$ axis. Thus, $\Top$ fulfils the commutation relations given by equation (\ref{E_Tcomm}). Now we will turn our attention to the symmetries of the light beams. The electric field of the light beam incident on $\Tt$ can be described within the paraxial approximation with the complex vector:
\begin{equation}
\Epq= A\rho^l e^{\left( il\phi + i k z \right)} e^{(-\rho^2/w_0)} \spphat
\label{Epq}
\end{equation}
where $\spphat = (\xhat + i p \yhat) / \sqrt{2}$, $\xhat$ and $\yhat$ being the horizontal and vertical polarization vectors, $A$ is a normalisation constant, $k=2 \pi / \lambda$ with $\lambda$ the wavelength in consideration, and $\rho$, $\phi$, $z$ are the cylindrical coordinates. An implicit harmonic $\exp(-i \omega t)$ dependence is assumed, where $\omega = 2\pi c /\lambda$ is the angular frequency of light, and $c$ is the speed of light in vacuum. Note that $p=1$ refers to LCP and $p=-1$ to RCP. Now, if we apply a rotation around the $z$ axis by an angle $\theta$ to this beam, the transformed field acquires a constant phase: 
\begin{equation}
\Rz(\theta) \Epq =\exp(-i (p+l) \theta) \Epq
\end{equation}
That is, $\Epq$ is a cylindrically symmetric beam (see section \ref{Ch3_cylsymm}). Thus, $\Epq$ is also an eigenstates of the generator of rotations around the $z$ axis, \textit{i.e.} the $z$ component of the total AM $J_z$:
\begin{equation}
J_z \Epq = (p+l) \Epq
\end{equation}
The $J_z$ eigenvalue is $m=p+l$. Also note that a beam with $p \neq \pm 1$ would be elliptically polarized (linearly when $p=0$) and would no longer be an eigenstate of $J_z$, since a rotation of the field would not leave it invariant except for a phase (see Appendix \ref{Appendix} for more information about rotations of EM fields). The helicity operator $\Lambda$\footnote{See section \ref{Ch1_symm} to find its differential expression.} can also be applied to $\Epq$. If only paraxial terms are considered, then the following result yields:
\begin{equation}
\Lambda \Epq \approx p \Epq
\end{equation}
Hence, in the paraxial approximation, a beam $\Epq$ with $p=1$ is both LCP and also an eigenvector of $\Lambda$ with value $1$. In the same way, a beam $\Epq$ with $p=-1$ is RCP and its helicity equals to $-1$. It is important to emphasize that this simple relation between helicity and polarization is only valid because the light beam is being described in the paraxial regime. When the paraxial approximation does not hold, the polarization of the field and helicity can no longer be so simply related with helicity, as it is shown in chapter \ref{Ch1} and \cite{Ivan2012PRA}. Now, if a mirror transformation is applied to the incident beam, the following result is obtained:
\begin{equation}
\Mz \Epq = \exp(i \alpha) \Eppq. 
\label{-Epq}
\end{equation}
where $\alpha$ is a phase given by the specific mirror transformation chosen\footnote{Remember that $\Mz$ is any mirror symmetry that contains the $z$ axis. Therefore, a rotation angle is needed to define a particular $\Mz$. If $\Mz$ is written as $M_{\nhat}$, the unitary normal vector to the plane is $\nhat = \cos \alpha \xhat + \sin \alpha \yhat$.}. That is, a beam with flipped values of $J_z$ and $\Lambda$ is obtained. As shown in section \ref{Ch1_symm}, this is a consequence of the fact that both $J_z$ and $\Lambda$ anti-commute with mirror symmetry transformations $\Mz$ \cite{Messiah1999}: 
\begin{eqnarray}
M^{\dagger}_{\left\{\zhat\right\}} J_z \Mz & = & - J_z \\
M^{\dagger}_{\left\{\zhat\right\}} \Lambda \Mz & = & - \Lambda \label{Eq_Hel_mirror}
\end{eqnarray}
In order to explain the experimental results given by Table \ref{TabCD} and Figure \ref{CD_fig}, the field transmitted through the nano-apertures and collimated by MO$_2$ $\Etpq$ is classified with the parameters $p$ and $l$ from the incident field: $p=-1,1$ and $l=-1,0,1$. Remember that for the incident field $\Epq$, $p$ is modified with QWP$_1$ and $l$ with the SLM (see Figure \ref{set-up}). Mathematically, the field $\Etpq$ can be obtained through the use of the target operator $\Top$, which can be found using the Green dyadic formalism and contains all the relevant information about target $\Tt$ \cite{Martin1998,Ivan2011}. That is, $\Etpq=\Top \{\Epq\}$, where the action of $\Top$ on the incident field will in general be in the form of a convolution. As $\Top$ is simply the mathematical description of $\Tt$, $\Top$ inherits the symmetries of $\Tt$. Thus, due to the cylindrical and mirror symmetries of the target and given an incident field $\Epq$, the following two statements hold: First, the transmitted field $\Etpq$ will also be an eigenstate of $J_z$ with the same eigenvalue of $m=p+l$. Second, two incident beams which are mirror images of each other will produce two transmitted fields which will be mirror images. The proof of both statements is given next. Due to the bijective properties of the exponential function, the commutation relation between $\Top$ and $R_z$ can be extended to the generator of rotations $J_z$: 
\begin{equation}
J^{\dagger}_z \Top J_z =  \Top
\end{equation}
Then, given that $\Epq$ is an eigenvector of $J_z$, $\Etpq$ must also be an eigenvector of $J_z$ with the same eigenvalue:
\begin{equation}
J_z \left[ \Etpq \right] =J_z \left[ \Top \{\Epq\} \right]=\Top \{ J_z \left[ \Epq \right] \} = (p+l) \Top \{\Epq\} = (p+l) \Etpq
\end{equation}
Then, as explained in section \ref{Ch6_symm}, given a sample which is symmetric under mirror transformations, its scattering matrix must commute with $\Mz$, \textit{i.e.} $\left[ \Top, \Mz \right]=0$. Now, given two mirror symmetric beams such as $\Epq$ and $\Eppq$ (see equation (\ref{-Epq})), it can be checked that their respective transmitted fields ($\Etpq$ and $\Etppq$) are related with a mirror symmetry:
\begin{equation}
\begin{array}{ll}
\Etpq & =\Top \{ \Epq\}  = \Top \{ \exp(i\alpha) \Mz \Eppq \} = \Mz \Top \{ \exp (i \alpha)\Eppq \} \\
  &= \exp(i \alpha) \Mz \Etppq
\end{array}
\label{mirror} 
\end{equation}
That is, given two mirror symmetric beams such as $\Epq$ and $\Eppq$ (see equation (\ref{-Epq})), their transmitted beams $\Etpq=\Top \{\Epq\}$ and $\Etppq=\Top \{\Eppq\}$ are connected via a mirror symmetry: $\Etpq=\exp(i\alpha)\Mz\mathbf{E}_{-p,-l}^{\mathbf{t}}$. This last result is one of the key points to understand the CD measurements with vortex beams presented in Table \ref{TabCD} and Figure \ref{CD_fig}. Indeed, in equation (\ref{CD}), the intensities $I_l^L$ and $I_l^R$ can be obtained from the transmitted electric field as:
\begin{equation}
I_l^{L/R}=\int_{-\infty}^{\infty} \int_{-\infty}^{\infty} |\mathbf{E}_{+1/-1,l}^{\mathbf{t}}|^2 {dx} {dy},
\label{inten}
\end{equation}
where the integral is taken on the plane of the detector (in our case CCD chip of the camera). Then, for a mirror symmetric sample, it can be proven that:
\begin{equation}
I_l^{L/R}=I_{-l}^{R/L}
\label{Iq}
\end{equation}
The proof of equation (\ref{Iq}) is done in Appendix \ref{Appendix3}. Next, equation (\ref{Iq}) is applied to prove that CD$_0=0$ and that CD$_l=-\text{CD}_{-l}$. When $l=0$, it is easy to see that $I_0^L=I_{0}^R$. Substituting this in the definition of CD, \textit{i.e.} equation (\ref{CD}), yields CD$_0=0$. However, when $l\neq0$, equation (\ref{Iq}) leads to $I_l^L-I_l^R = I_{-l}^R - I_{-l}^L = - (I_{-l}^L-I_{-l}^R)$, which implies that CD$_l=\text{CD}_{-l}$, in very good agreement with our measurements\footnote{Note that all the experimental values are consistent with $\text{CD}_l=\text{CD}_{-l}$, within the experimental error bars.}. \\\\
Powerful as these symmetry considerations are, they still cannot explain the measured quantitative results of CD, nor their variation with the diameter of the nano-aperture. They only indicate that when $l\neq0$ the two opposite circular polarizations are not the mirror image of each other and then the associated CD$_l$ does not have to be zero. Here, the AM of light plays a crucial role again. In general, it can be observed that CD measurements compare the differential ratio of electromagnetic fields with opposite circular polarization and a difference of AM of 2 units. For example, CD$_{l=1}$ relates $\vert \mathbf{E}_{1,1}^{\mathbf{t}} \vert^2$ and $\vert \mathbf{E}_{-1,1}^{\mathbf{t}} \vert^2 $, whose respective AM values are $m_{p=1}=1+1=2$ and $m_{p=-1}=1-1=0$. It is then interesting to observe that even though CD is usually defined as the differential absorption of circular polarization states, it also measures the differential absorption of AM states. In the case of the nano-aperture, this is the most probable cause of the giant value of CD obtained in the experiments. Even though the sample is cylindrically symmetric, thus preserving the AM of field, input beams with different values of AM have very different scattering amplitudes\footnote{If the sample was a sphere, the scattering amplitudes would be the Mie coefficients}. This was shown and used in chapter \ref{Ch3} to enhance the ripple structure and excite WGMs.
\begin{figure}[htbp]
\centering\includegraphics[width=\columnwidth]{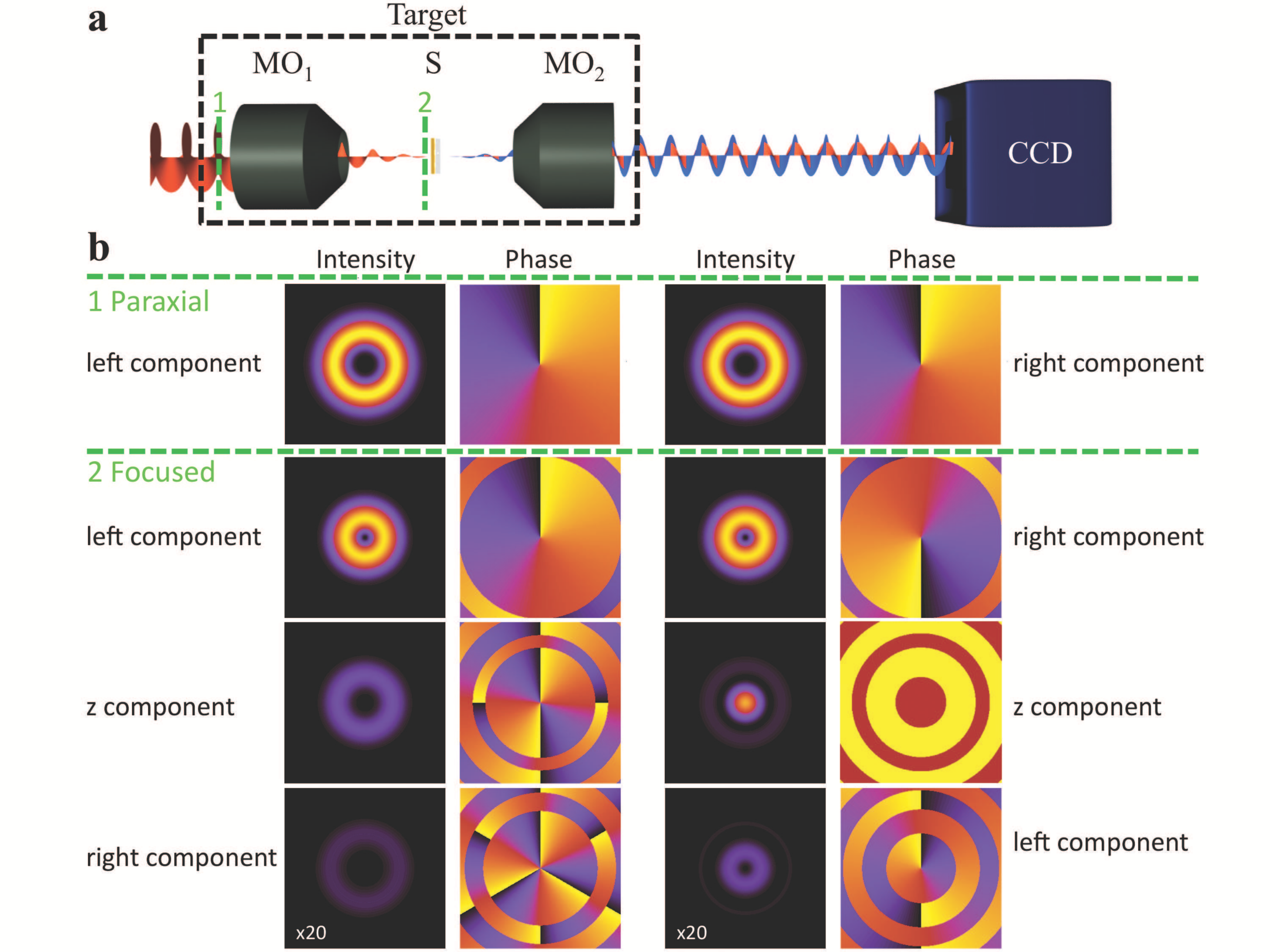}
\caption{\textbf{a)} Sketch of the addition to the set-up in order to perform a projective measurement of the circular polarisation states. The projective measurement is done in the following manner. Firstly, the metallic sample is removed from the nano-positioning system. Secondly, a LCP beam $\mathbf{E}_{1,l}^{\mathbf{in}}$ is focused with MO$_1$ and collimated with MO$_2$. Then, a second quarter-wave plate (QWP$_2$) and a linear polarizer (P$_3$) are used to select either the LCP (direct component) or the RCP (crossed component) of the transmitted beam. At last, the sample is put back on the nano-positioning device, which places it at the focal plane of MO$_1$, and re-centres it with the incident beam. \textbf{b)} Intensity and phase plots of the two beams used to carry out the measurement of CD$_{l=1}$, \textit{i.e.} $\mathbf{E}_{1,1}^{\mathbf{in}}$ and $\mathbf{E}_{-1,1}^{\mathbf{in}}$. In the upper row, the intensity and phase of the beams are shown on the back-aperture plane of the MO$_1$. Both the intensity and the phase can be described with equation (\ref{Epq}). In contrast, the three rows below show the intensity and the phase of the same beams ($\mathbf{E}_{1,1}^{\mathbf{in}}$ and $\mathbf{E}_{-1,1}^{\mathbf{in}}$) at the focal plane of MO$_1$. As it can be observed, even though their paraxial intensities and phases are analogous, their structure is completely different at the focal plane. This is a direct consequence of the fact that the AM of both beams differ in two units.}
\label{Projection}
\end{figure}
\\\\Furthermore, 
using the model of the aplanatic lens \cite{WolfII1959,Novotny2006}, one can check that the fields at the focal plane of MO$_1$ produced by $\mathbf{E}_{-1,l}^{\mathbf{in}}$ and $\mathbf{E}_{1,l}^{\mathbf{in}}$ are completely different. Indeed, due to their different AM content, $\mathbf{E}_{1,l}^{\mathbf{in}}$ has a vortex beam in the $\zhat$ polarization, whereas $\mathbf{E}_{-1,l}^{\mathbf{in}}$ does not (see Figure \ref{Projection}(b))\footnote{The fields shown in Figure \ref{Projection}(b) can be obtained using the expression for the Bessel beams given by equation (\ref{E_Bmp}).}. Consequently, their multipolar decompositions are also very different (see Figure \ref{F_CJMP}), surely giving rise to very different coupling scenarios.\\\\
This fact can be analytically proven for spheres, where different couplings can occur not only depending on the geometry of the particles but also on the beams used to excite these particles (see chapter \ref{Ch3}). But actually, very similar effects have been experimentally observed for other plasmonic structures \cite{Yuri2008,Banzer2010SRR,Banzer2010,Rodriguez-Herrera2010}.\\\\
In conclusion, a new method to induce CD has been demonstrated. Instead of tweaking the scattering matrix $\Sconv$ of the structure by to changing the geometry or material properties, vortex beams have been used carry out the measurement. As a result, it has been proven that a giant CD can be induced on a single non-chiral sub-wavelength sample, at normal incidence. To do so, the transmission of beams with three different phase singularities of order $l=-1,0,1$ has been studied. It has been seen that the results can be conveyed from a symmetry perspective. In particular, it has been proven that even for mirror symmetric systems, CD can be induced if the LCP and RCP beams are not connected via a mirror symmetry. In addition, it has been shown that the information carried by CD measurements has two different contributions: the differential scattering of different circular polarization states but also the differential scattering of different angular momentum states. Thus, it is expected that vortex beam-induced CD will be able to unveil properties of the sample which are hidden to standard CD measurements. Finally, it is interesting to see that in other related phenomena, such as molecular optical activity, the interplay between the two symmetries associated to helicity and AM (electromagnetic duality and rotational symmetry) are also essential to fully understand the problem from first principles \cite{Ivan2013JCP}. 

\section{Angular momentum preservation}\label{Ch6_AM}
\begin{figure}[htbp]
\centering\includegraphics[width=12cm]{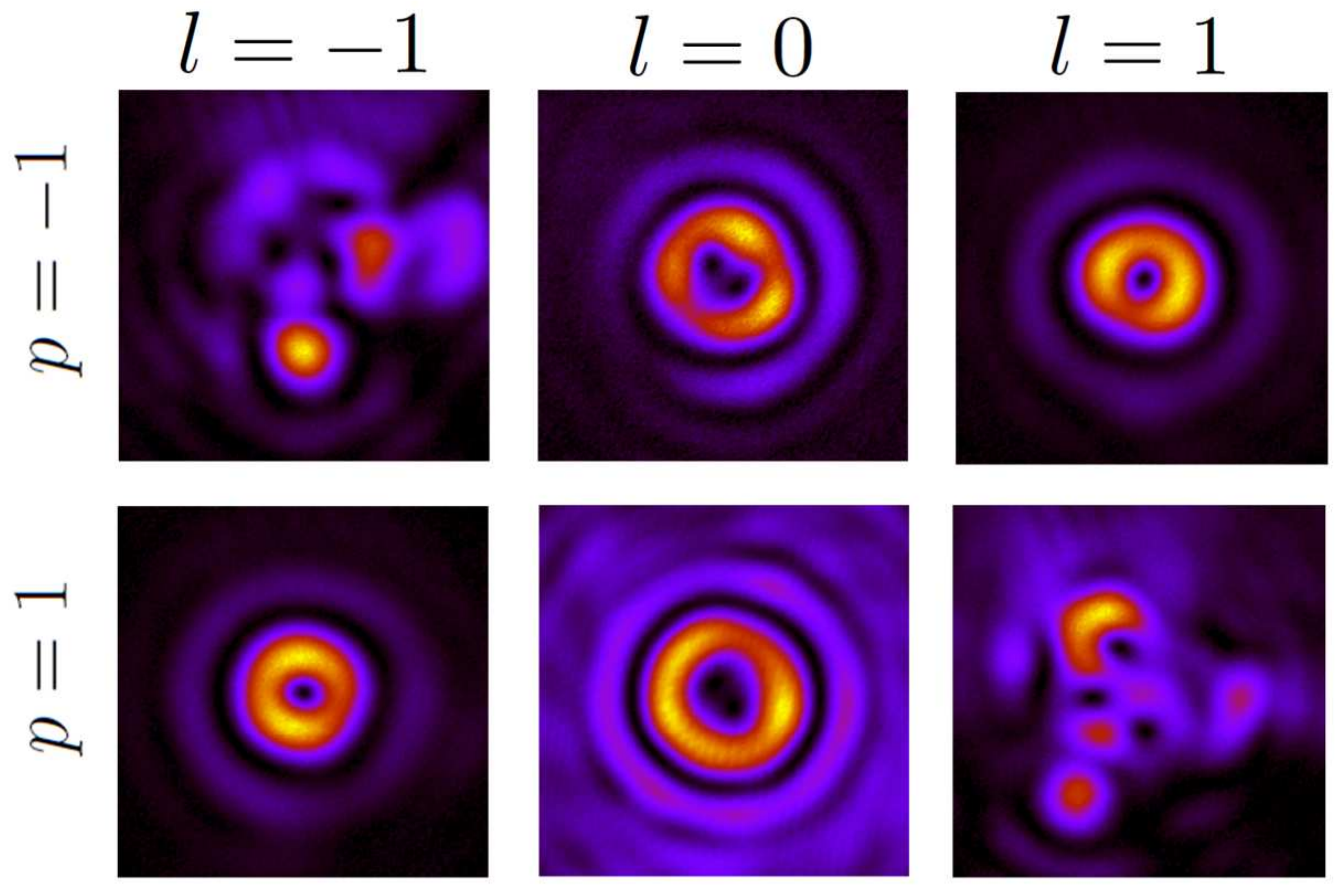}
\caption{Crossed component $\vert \Btpqpp \vert ^2$ for $p=-1,1$ and $l=-1,0,1$. The values of $p$ and $l$ are the ones carried by the incident beam $\Epq$. Given a row $p$ and a column $l$, the image represents $\vert \Btpqpp \vert ^2$, which is a mode of light with polarization $-p$ and a phase singularity $l+2p$. The six far-field images have been taken using the nano-aperture $d_8$}
\label{I_beam}
\end{figure}

In this section, in order to further explore the AM scattering from nano-apertures, an experimental characterization of the modes transmitted through the nano-apertures has been carried out. As mentioned earlier, due to the cylindrical symmetry of the nano-apertures, both $\Epq$ and $\Etpq$ are eigenstates of $J_z$ with value $(p+l)$. Nevertheless, because the sample is non-dual, helicity is not preserved upon interaction, therefore $p$ and $l$ may not be preserved individually. This phenomenon has been detailed in subsection \ref{Ch6_duality}: it is a consequence of the duality symmetry being highly broken by the nano-aperture and the multilayer system \cite{Ivan2012PRA,Ivan2013,Zambrana2013OE,Bliokh2013,Nora2014}. In order to measure it, two elements have been added to the set-up to perform a projective measurement of the two helicity components, and at the same time determine the topological charge $l$ of the corresponding mode (See Figure \ref{Projection}(a)). The two elements added in the set-up are a quarter-wave plate and a linear polarizer (QWP$_2$ and P$_3$ in Figure \ref{Projection}(a)). First, QWP$_2$ transforms the circular polarization into linear. Then, due to the fact that the two different helicity components are orthogonal, if one gets transformed into $\xhat$, the other one goes to $\yhat$. Therefore, the linear polariser P$_3$ serves as an analyser: it selects one of the two helicity components, and removes the other one. As mentioned earlier, due to the fact that the duality symmetry is highly broken by the sample, the helicity of the incident beam $\Epq$ is not preserved. Thus, the field $\Etpq$ comprises two helicity components, one of polarization $\spphat$ and another one of polarization $\sppmhat$: 
\begin{equation}
\Etpq = \Atpqp\spphat  + \Btpqpp\sppmhat
\label{decomp}
\end{equation}
where $\Atpqp$ and $\Btpqpp$ are the complex amplitudes of the two polarizations at the plane of the camera. $\Atpqp$ is denoted as the direct component, as it maintains the polarization state $\spphat$ and consequently the topological charge $l$, as AM needs to be conserved. The other orthogonal component is $\Btpqpp$, and it will be denoted as crossed component. The crossed component has a polarization state $\sppmhat$ when the incident state is $\spphat$. Due to the cylindrical symmetry, the value of the AM along the $z$ axis must be preserved. Therefore, when $p$ changes to $-p$, $l$ goes to $l+2p$. That is, the crossed component $\Btpqpp$ is a vortex beam whose topological charge differs in two units with respect to the incident beam $\Epq$. This is depicted in Figure \ref{I_beam}. The six images in Figure \ref{I_beam} are the CCD recorded images for $\Btpqpp$ when the values of $p$ and $l$ of the incident beam are $p=-1,1$ and $l=-1,0,1$. Hence, the image taken with $(p=1,l=1)$ corresponds to $B_{1,1}^{\mathbf{t}}(x,y)$, and consequently its vortex has an topological charge $l'=l+2p=1+2=3$. 
However, instead of observing a vortex of charge $l'=3$, three singularities of charge $l=1$ are observed. This occurs because higher order phase singularities are very unstable and prone to split into first order singularities \cite{Gabi2001OL,Ricci2012,Kumar2011,Rich2014}. Thus, in the current scenario, a phase singularity of order $l'$ will split into $\vert l' \vert$ singularities of order $sign(l')$. This is the principal feature that is captured in the snapshots of Figure \ref{I_beam}. They can be experimentally used to verify that the AM along the $z$ axis is conserved within the experimental errors. 
\begin{figure}[htbp]
\centering\includegraphics[width=13cm]{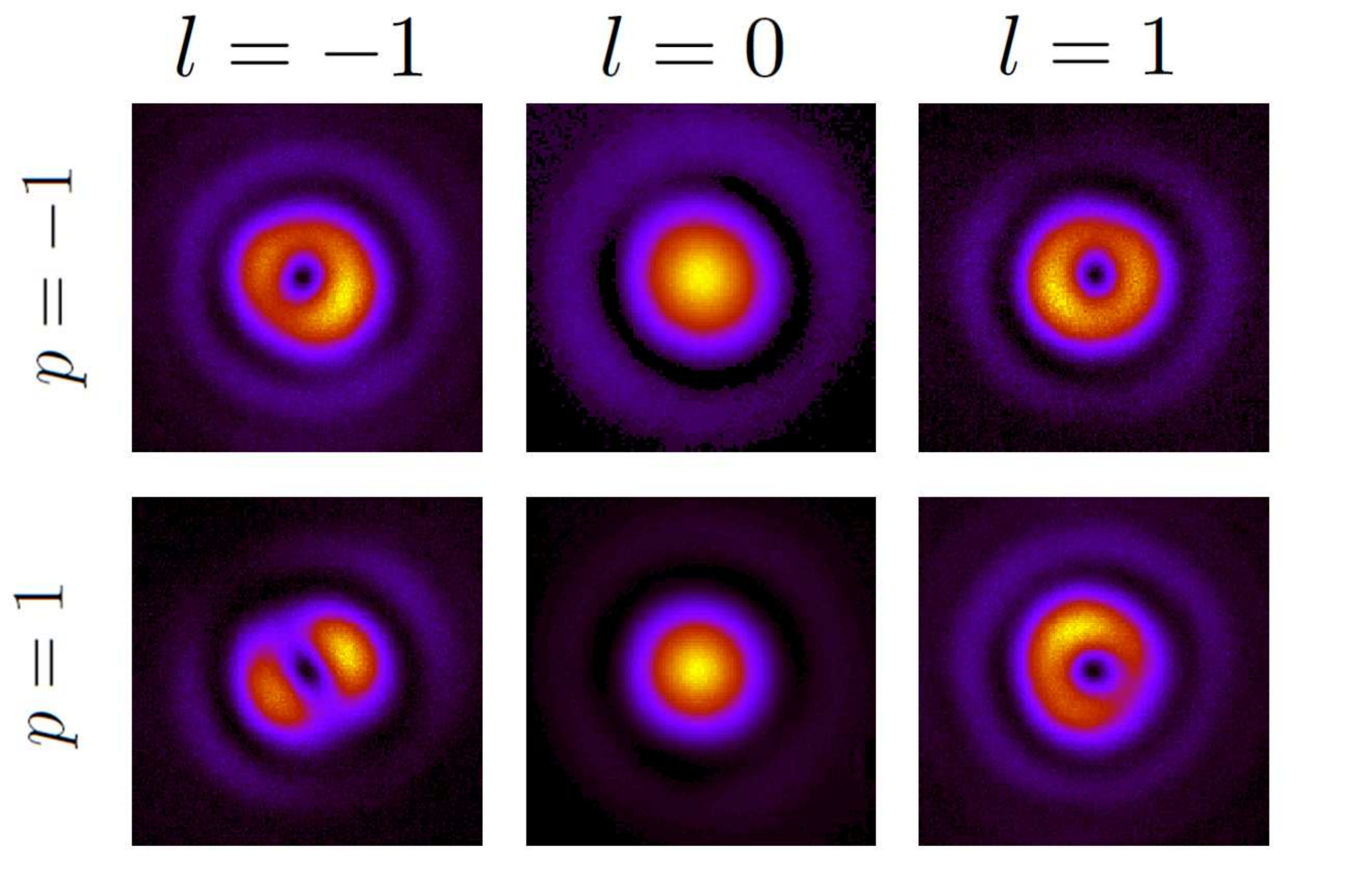}
\caption{Direct component $\vert \Atpqp \vert ^2$ for $p=-1,1$ and $l=-1,0,1$. The values of $p$ and $l$ are the ones carried by the incident beam $\Epq$. In contrast to Figure \ref{I_beam}, $p$ and $l$ for these images coincide with the values of $p$ and $l$ for the incident beam. All the snapshots have been taken in the far-field using the nano-aperture $d_6$.}
\label{Idirect}
\end{figure}
The preservation of AM in the $z$ direction is also observed with the direct transmitted component $\Atpqp$. This is depicted in Figure \ref{Idirect} for the same six incident beams. The intensity profiles are not as surprising as the ones shown in Figure \ref{I_beam}, as their intensity distributions are qualitatively the same ones as the incident beams $\Epq$, whose expression is given by equation (\ref{Epq}). The preservation of AM in a cylindrically symmetric system has also been reported by some other authors. In a very recent experiment, Gorodetski \textit{et al.} observed similar phenomena to the ones reported here \cite{Yuri2013}, even though their samples were not always cylindrically symmetric. Instead, they managed to transfer AM from their sub-wavelength samples to the EM field and measure it in the far field. Also very recently, Chimento and co-workers observed the preservation of the $z$ component of the AM using an incident Gaussian beam and a 20 $\mu$m circular aperture \cite{Chimento2012}. Finally, Vuong \textit{et al.} described how phase singularities can arise and change with the polarization of the incident field, in the near-field of sub-wavelength apertures \cite{Vuong2010}. However, to the best of my knowledge, this is the first time that the transmission of high AM through circular sub-wavelength apertures is measured in the far field.

\begin{savequote}[10cm] 
\sffamily
``We are trying to prove ourselves wrong as quickly as possible, because only in that way can we find progress'' 
\qauthor{Richard P. Feynman}
\end{savequote}

\chapter{Experiments with single spherical particles}
\graphicspath{{ch7/}} 
\label{Ch7}

\newcommand{\Ino}{I^{\text{norm}}}

\section{Scope}
The use of light to manipulate small particles has widely spread since the seminal works of Ashkin in the 1970's \cite{Ashkin1971,Ashkin1974,Ashkin1975,Ashkin1976}. Ashkin studied the levitation of particles of the order of 10 $\mu$m. He showed that the radiation pressure exerted on single particles of that size is enough to beat gravity and Brownian motion. His seminal works led to the development of what nowadays is known as optical tweezers \cite{Richard2013}. Besides, optical levitation was also used to measure scattering (or absorption) cross sections as a function of the size parameter $x=2\pi R/ \lambda$ from a single scatterer. Indeed, a monochromatic laser can be used to trap and levitate a single large droplet of water, \textit{e.g.} $\Phi = 100 \mu$m. Then, the power of the laser can be increased and as a result the droplet starts to evaporate. This allows for measuring the variation of scattering cross section as a function of $x$ \cite{Bohren1983,Bohren1983AO}. With the advance of laser technologies, other methods have been used by different authors to measure scattering efficiencies. Xifr\'{e}-P\'{e}rez \textit{et al.} used Fourier Transform Infra-red Spectrometer coupled to a microscope to measure the scattering cross section of Silicon particles in the infra-red \cite{Xifre-Perez2011}. Similarly, other authors have combined the use of continuum laser sources with spectral filters to make $\lambda$ scans, while keeping the size of the single structure constant \cite{Foot1989,Arnold1990,Hammer1998}. In this chapter, as it was pointed out in section \ref{Ch2_eff}, I use cylindrically symmetric beams to measure backward scattering efficiencies for single spheres. As it has been observed in chapter \ref{Ch3}, cylindrically symmetric beams can be used to control the scattering of spheres. They can enhance their scattering ripple structure, and in case high enough order modes are used, they can also excite WGMs. Furthermore, the scattering cross section of spheres excited by vortex beams has, up to my knowledge, never been measured as a function of $\lambda$ yet. The only similar experimental realisations have been done at one fixed wavelength and only using LG modes of order $l=-1,0,1$ \cite{Garbin2009,Petrov2012}. Here, not only $\lambda$ scans will be carried out, but also high AM beams such as $m=6$ will be used to study the interaction of symmetric light with these spheres. Even though the study is not fully complete, the results presented in this chapter ratify the fact that high AM can unveil some scattering resonances that are hidden under a plane-wave or Gaussian excitation.

\section{Experimental set-up}
\begin{figure}[tbp]
\centering
\includegraphics[width=14cm]{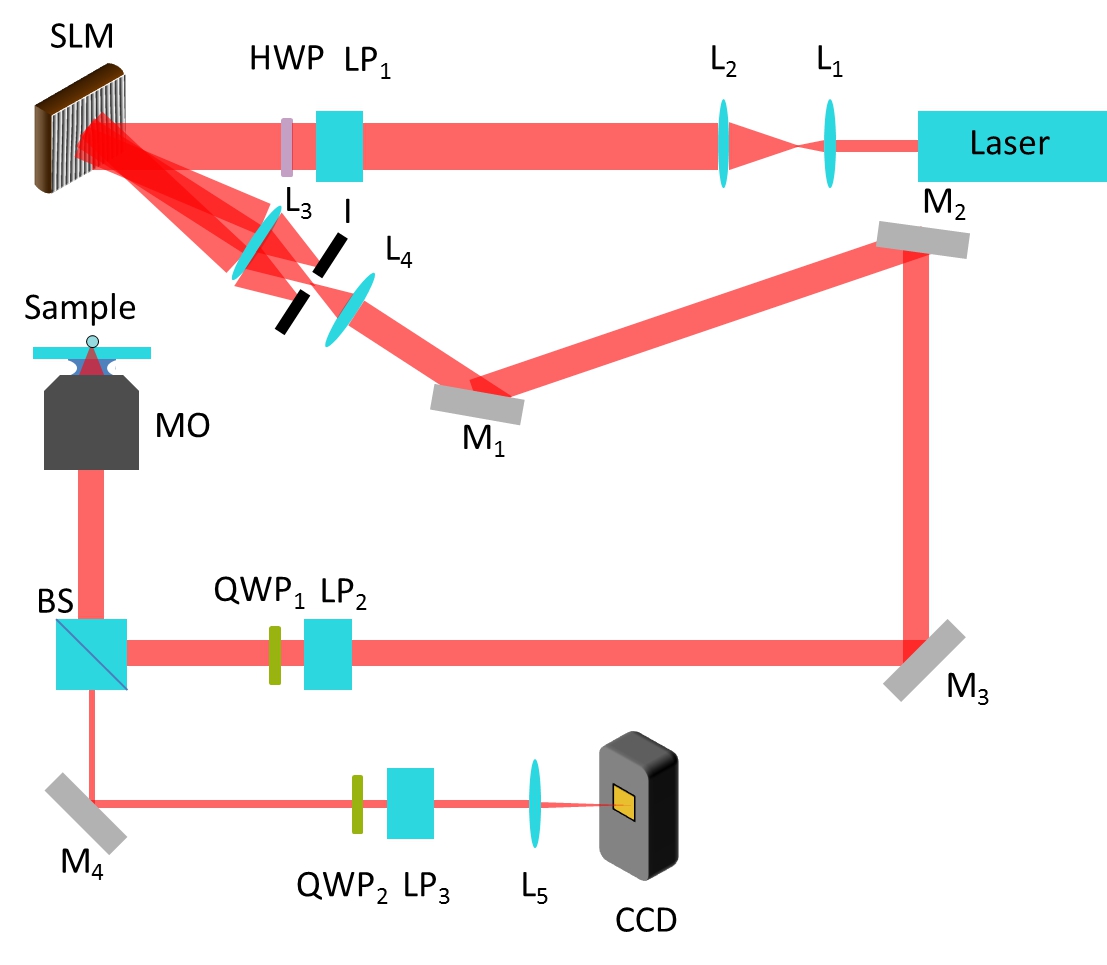} 
\caption{Schematics of the experimental set-up. A laser beam of variable wavelength $\lambda=[760-810]$nm is expanded with two lenses (L$_1$, L$_2$) and polarized with a linear polariser (P$_1$) and a half-wave plate (HWP) to match the requirements of the SLM. The SLM gives a spiral phase to the beam and with the use of two lenses (L$_3$, L$_4$) the first diffraction order is selected by an iris (I). Later, the polarisation of the beam is modified with a linear polariser (P$_2$) and a quarter-wave plate (QWP$_1$). \label{Fig81}}
\end{figure}
\begin{figure}[tbp]
\centering
\includegraphics[width=10cm]{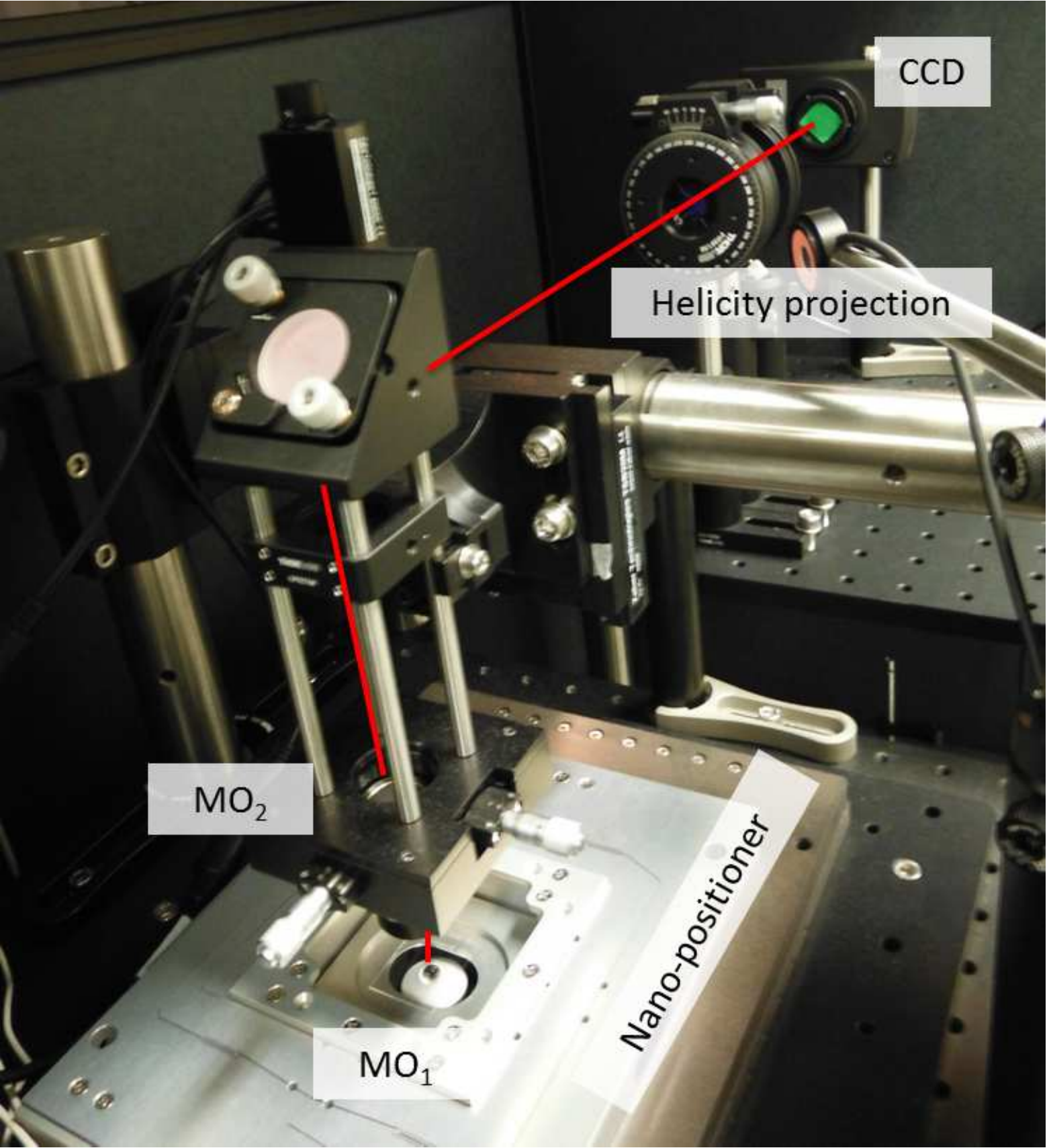} 
\caption{Picture of the forward scattering measuring set-up. MO$_1$ focus light on the sample, which lies on top of a sample-holder. The sample-holder is clamped onto a nano-positioning device. The forward scattering is collected by MO$_2$, which sends the light up to a mirror that reflects the light towards a helicity projector. The helicity projector is formed by a quarter-wave plate and a linear polariser. The projected state is imaged by a CCD camera.  \label{Fig82}}
\end{figure}
\begin{figure}[tbp]
\centering
\includegraphics[width=14cm]{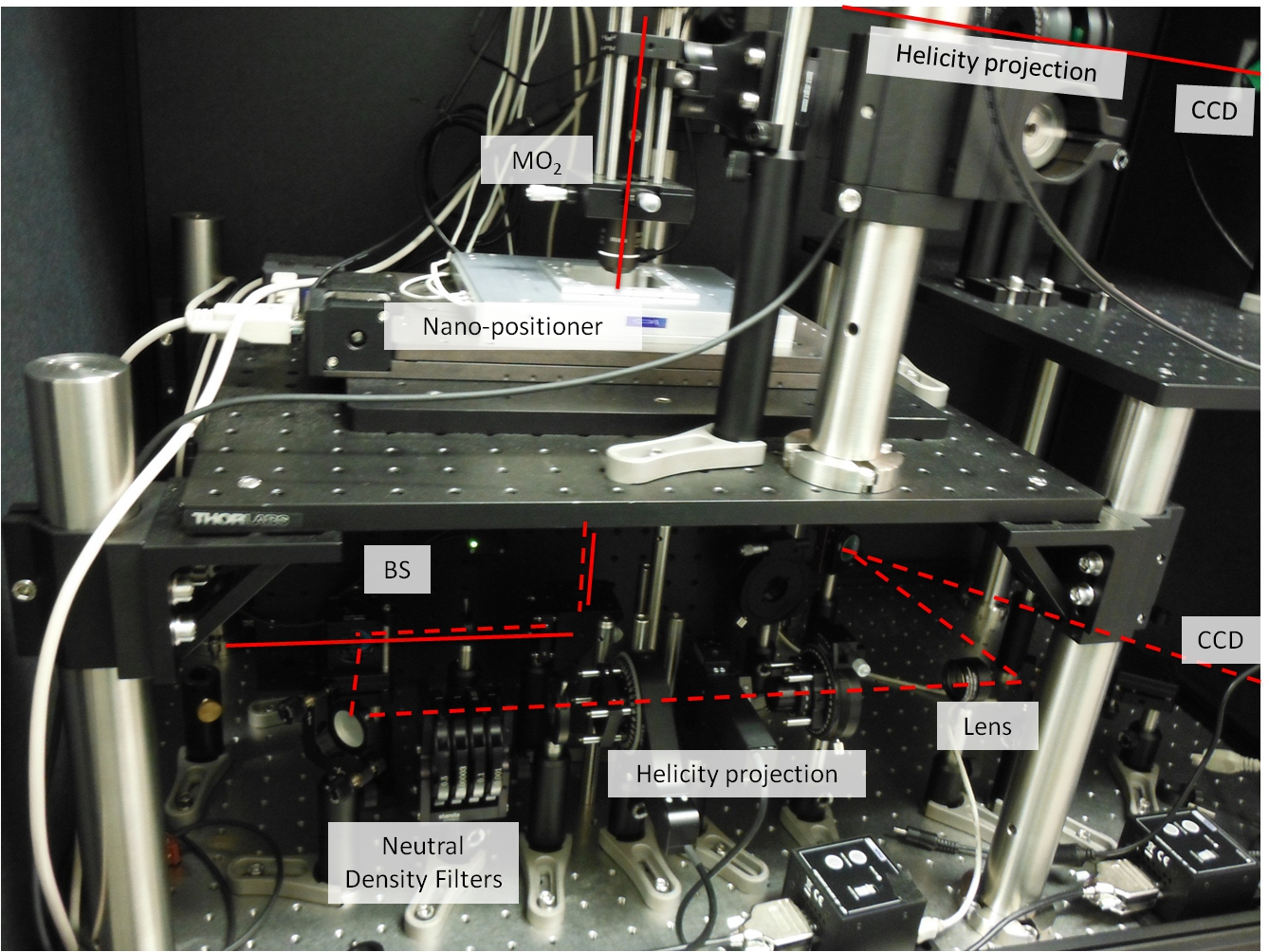} 
\caption{Picture of the backward scattering acqusition set-up. The light scattered off the particle in the backward semi-space is collimated by MO$_1$, which sends the beam downwards. A 45$^{\circ}$ mirror steers the light back to a horizontal plane. Then the scattered light is separated from the path of the incident light by a beam-splitter (BS). Afterwards, the light goes through some neutral dentisty filters and the helicity state is transformed using a quarter-wave plate and a linear polariser. Finally, a lens images the beam onto a CCD camera. \label{Fig83}}
\end{figure}

The experimental set-up is schematically displayed in Figure \ref{Fig81}. Note that the measurements presented in this chapter are done in backscattering. The experimental set-up allows for measuring forward scattering too, but it has not been used. The preparation of states is very similar to the one carried out in the previous chapter. That is, a monochromatic Gaussian mode is expanded using two lenses (L$_1$ and L$_2$) to match the chip size of the SLM; then, using a linear polariser and a half-wave plate (P$_1$ and HWP), its polarization is set to optimize the efficiency of the SLM; afterwards, the SLM modifies the phase of the beam to create a vortex beam of order $l$; the first diffraction order is selected using a pair of lenses and an iris (L$_3$, L$_4$ and I); then, its polarization is modified to create a state of well-defined helicity with a linear polariser and a quarter-wave plate (P$_2$ and QWP$_1$). Later, the light beam hits the back-aperture lens of MO$_1$ and is focused down to the sample.\\\\
Here, a subtlety needs to be commented. In the experiments with nano-apertures shown in chapter \ref{Ch6}, the wavelength of the laser is kept constant, therefore once the optical path has been aligned, the SLM is not changed. In the current case, the aim of the experiments makes this part a little bit more complicated. The laser used for this experiment is a \hyperlink{http://www.toptica.com/products/research\_grade\_diode\_lasers/tunable\_diode\_lasers.html}{Toptica DLpro tunable diode laser}\footnote{http://www.toptica.com/products/research\_grade\_diode\_lasers/tunable\_diode\_lasers.html}. Its wavelength range goes from 760-810 nm, and its linewidth is extremely narrow ($2\cdot 10^{-6}$ nm). The output power is over 50mW at all the different wavelengths, peaking at 785nm, where it gives 100mW. Due to the 50nm range of the laser, the hologram displayed by the SLM produces different diffraction angles at the two boundaries of the wavelength range. That is, $\Delta\theta_x^{760} \neq \Delta\theta_x^{810}$, where $\Delta\theta_x = \lambda / \Delta x$ (see equation (\ref{E_kx})). This is specially obvious at the focal plane of MO$_1$. Because of the different diffraction angles, the beam hits the back-aperture of MO$_1$ in a slightly different position, which creates a considerable displacement of the focus spot in the focal plane\footnote{In fact, this is the basic principle of holographic optical tweezers \cite{RichardThesis}}. Now, the SLM can correct for this displacement by displaying more or less $2\pi$ phase jumps. The formula to give the same diffraction angle for all wavelengths is:
\begin{equation}
\Delta x^{\lambda}=\dfrac{\lambda}{\lambda_{\mathrm{ref}}}\Delta x^{\lambda_{\mathrm{ref}}}
\label{Dx_lambda}
\end{equation}
where $\lambda_{\mathrm{ref}}$ is the reference wavelength for which the alignment has been done, and $\Delta x^{\lambda_{\mathrm{ref}}}$ is the total number of $2\pi$ phase ramps given by the hologram at $\lambda=\lambda_{\mathrm{ref}}$ divided by width of the SLM chip. I have chosen $\lambda_{\mathrm{ref}}$ to be the wavelength for which the provider of the SLM has made the LUT, which in this case is $\lambda=785$nm. Then, the alignment is done at $\lambda=785$nm with a certain $\Delta x^{\lambda_{\mathrm{ref}}}$ so that the different orders separate and the efficiency is still high.\\\\
The interaction between the incoming beam and the sample then takes place. The sample, as explained in section \ref{Ch5_spheres}, is a coverslip which has been spin-coated with a solution of TiO$_2$ particles on top. The slide is placed on a sample holder which is attached to a nano-positioning device. The nano-positioning device is also clamped onto a micro-positioning device. The micro-positioner is used to look for single spheres. Then the single sphere is centred with respect to the beam using the nano-positioning device, whose precision is 0.5nm (see section \ref{Ch6_CD}). Note that MO$_1$ is a water immersion objective, therefore the incoming beam must go through the coverslip first, and then is scattered off the particle.\\\\ 
The scattering is collected both in the forward and backward semi-spaces. Both of them analyse the scattering in the same fashion. Figure \ref{Fig82} shows the analyser in the forward direction. It can be observed that MO$_2$ (100x, NA$=0.9$) collects the scattered light and images it on the CCD camera. Before that, a quarter-wave plate projects the two helicity components onto two different linear polarisations so that a linear polariser can select one of the two and remove the other (see section \ref{Ch6_AM} and Figure \ref{Projection}). The backward scattering acquisition set-up is shown in Figure \ref{Fig83} and schematically in Figure \ref{Fig81}. The same MO$_1$ used to focus the light on the sample is used to collect the scattering. Then, a beam-splitter (BS) is used to separate the scattered light from incident beam. The scattered beam is separated into its two orthogonal helicity components with a quarter-wave plate and then another linear polariser is used to select one of the two components. Afterwards, the beam goes through a lens that images the sample onto a CCD camera.  

\section{Measuring methods}
The set-up explained above allows for measuring the direct and crossed helicity components both in forward and backward scattering. However, there are a few fundamental differences between the acquisition of data in forward or backward scattering. Forward scattering encompasses the scattering of the spheres in the forward semi-plane as well as all the direct light which does not interact with the sphere. That is, in Mie Theory notation, the measured electric field is the total field $\Et$, which is the addition of the scattered field $\Es$, and the incident one $\Ei$ (see section \ref{Ch2_Scatt}). Due to the size of the particles used in this experiment ($\Phi=950$nm), the contribution of $\Es$ to the scattering is much smaller than the direct light $\Ei$. Thus, an intensity recording of $\Et$ with a CCD camera does not yield a very good signal-to-noise ratio of the scattered field, since the latter is blurred by the incident beam. Two possible ways of overcoming this problem are explained next.\\\\
The first one uses the fact that the incident beams of light used in this experiment have a well-defined helicity. Due to this fact, all the incident light is vertically polarised after it has been collected with MO$_2$ and transformed with a QWP. Now, the scattered field has two helicity components, as the sphere is not dual, \textit{i.e.} $\Es = \Es_p + \Es_{-p}$. The direct component $\Es_p$\footnote{In the same way as I did in the previous chapter, I denote `direct component' to the scattering component with the same helicity as the incident light. The term `crossed component' is used to refer to the scattered light whose helicity is the opposite to the incident field.} is transformed to a vertical linear polarised state. In contrast, the crossed component is horizontal linearly polarised after the QWP. Thus, the linear polariser can filter out the light with helicity $p$ and leave exclusively $\Es_{-p}$. This method, even though is elegant and easy to implement, does not always perform as well as one would wish. If we define $t_r$\footnote{$t_r$ is defined here as the ratio between the unwanted component and the wanted one. Thus, it can take values between 0 and 1.} as the extinction ratio of a linear polariser, then the intensity of the incident beam $I^i$ must be smaller than $ I^s_{-p} \cdot t_r^{-1}$ for this projection to work. This method has been tried for the set-up shown in Figures \ref{Fig82}, \ref{Fig83} and $I^i \approx I^s_{-p} \cdot  t_r^{-1} $, therefore the experiment has not been done in this way.\\\\
The second method is based on dark-field microscopy principles \cite{Kuwata2003,Van2005,Kuznetsov2012}. MOs manufacturers make most of the objectives so that they create an image at infinity. Certainly, these are the ones used in this experiment. Then, their back-aperture plane contains the Fourier decomposition of the image field. Thus, a 4-f filter system (see Appendix \ref{App2Sep}) can be used to sort the Fourier content of the image field. The idea is that most of the incident light travels at small $\krho$'s, whereas the sphere scatters light for all angles, in particular those $\krho \sim k$. Hence, if the light at the center of the Fourier plane is blocked, only the spatial frequencies with large $\krho$'s will be imaged by the CCD. Those Fourier components mostly correspond to the scattering of the particle. Even though this method can potentially give good signal to noise ratios, we opted for the method presented in the next paragraph because it was more versatile.\\\\
The scattering can also be measured in the backward semi-space. And this is the way the measurements presented in this chapter have been carried out. 
\begin{figure}[tbp]
\centering
\includegraphics[width=13cm]{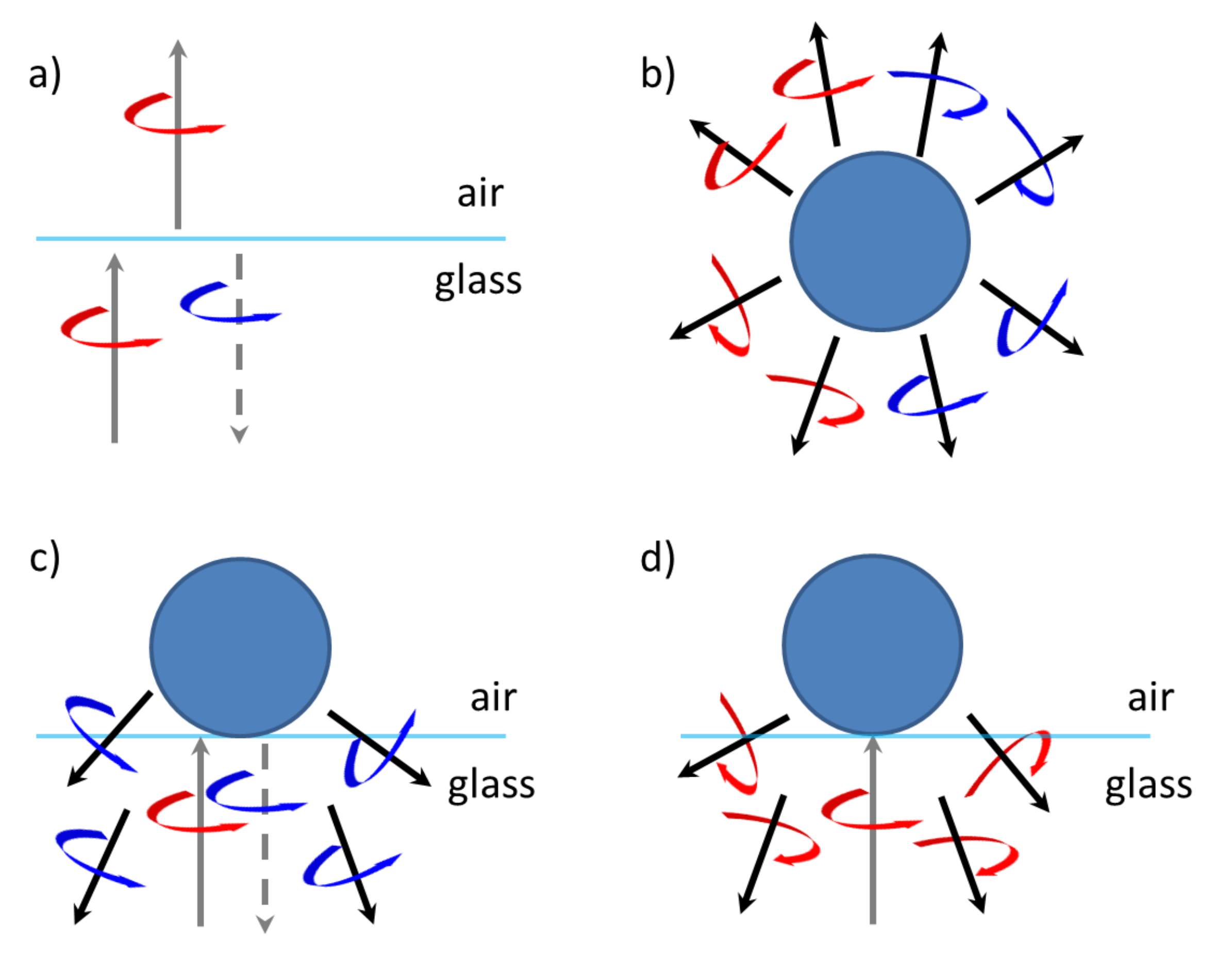} 
\caption{Schematics of the helicity content in different scattering events. a) A plane wave with a well-defined helicity (red) hits on an air-glass interface. The transmitted light preserves the helicity, whilst the reflected light flips the helicity value (blue). b) A non-dual sphere scatters intensity in both helicity components (red and blue). The division between left and write semi-spaces is merely aesthetic: both helicities are scattered in all directions. c) Backward crossed helicity (blue) component of the scattering of a sphere sitting on top an air-glass interface. The scattering encompasses the crossed helicity component in the backward semi-space as well as the reflection of the incident beam on the interface. d) Backward direct helicity (red) component of the scattering of a sphere sitting on top an air-glass interface. The scattering only encompasses the direct helicity component. \label{Fig84}}
\end{figure}
An intuitive model behind the measurements in backward scattering is presented in Figure \ref{Fig84}. Remember that the sample that is going to be used to carry out $\lambda$ scans with different optical vortices is the one depicted in Figure \ref{Fig641} - single TiO$_2$ particles with $\Phi=950$nm on a glass coverslip. Then, Figure \ref{Fig84}(a) sketches the interaction between an incoming plane wave (grey straight arrow) with a well-defined helicity (red spinning arrow) and a glass-air interface. It can be seen that a small fraction of light (generally it is about 4\% of the incident intensity) represented by a dashed arrow is reflected back. The rest of the 96\% of light goes through and maintains its helicity value\footnote{Even though the glass is not dual, the helicity must be preserved in the forward direction due to the cylindrical symmetry of the interface (see section \ref{Ch4_Kerker} for a more detailed explanation).}. Now, note that the 4\% of the light that is reflected changes its helicity (blue spinning arrow). The interface acts as a mirror. Without loss of generality, the $z$ axis can be placed in the direction normal to the interface. Then, the $z$ component of the incident $\mathbf{k}$ is flipped, and $k_x$, $k_y$ maintained. The same mirror symmetry applies to the polarisation vectors. That is, $\zhat \rightarrow - \zhat$, and $\xhat \rightarrow \xhat$, $\yhat \rightarrow \yhat$. That implies that the polarisation in the real space (computed with respect to some external reference frame in the lab) is maintained, but its projection to the linear momentum direction yields an opposite helicity \cite{Ivan2013OE}. The same conclusion can be reached using the anti-commutation relations between $\Lambda$ and $M_{\nhat}$ (see section \ref{Ch1_symm} and equation (\ref{Eq_Hel_mirror})).\\\\
Note that the previous explanation is only valid if the incident field is a plane wave with a well-defined helicity propagating in the direction normal to the interface. In the experiment, the field that impinges the coverslip is a fully vectorial focused beam with many different plane waves components, and therefore these considerations do not fully apply. In particular, the reflected light also has some components with the same helicity value as the incident beam, as the symmetry arguments why the helicity value must flip only apply for plane waves travelling at normal incidence to the interface (see section \ref{Ch4_Kerker}.) Then, the reflected beam is computed applying the reflection Fresnel coefficients to each of the $\shat$ and $\phat$ waves of the incident field and then adding them all together \cite{Born1999}.\\\\
Figure \ref{Fig84}(b) shows the fact that a non-dual sphere scatters light in both helicities. The division between left and write semi-spaces is purely aesthetic. Both helicities are scattered in all directions in space. The red spinning arrows represent the direct helicity component, while the blue ones represent the crossed component. It can be observed that the polarisation of a plane wave component scattered in the forward direction with helicity $p$ is the same one as the polarisation of a plane wave with helicity $-p$ scattered in the backward direction. This property is the key why the acquisition of data has been done in backward scattering.\\\\
Figure \ref{Fig84}(c) shows the excitation of a sphere with a beam with helicity $p$ (red), its reflection on the glass-air interface (dashed grey arrow), and the scattering crossed helicity component of the sphere in the backward semi-space. Now, because both the reflection from the interface and the scattering crossed component have the same helicity, they all acquire the same horizontal linear polarisation after they are projected to a linear basis by the QWP in the backward set-up (see Figure \ref{Fig83}). Similarly, the direct component of the scattering also gets projected onto the linear basis. However, its state becomes vertically polarised. Note that in this approximate model, all the light reflected off the glass is in the crossed helicity component. Therefore, if the linear polariser filters out the crossed component (or horizontal polarisation), all the remaining light will exclusively be from the direct scattering component of the sphere. This is depicted in Figure \ref{Fig84}(d), where it is shown that all the light with helicity $p$ on the backward semi-space comes from the scattering of the sphere. This is the way the cross section curves as a function of $\lambda$ have been measured. That is, instead of measuring all the scattered intensity, only the direct component of scattering in the backward semi-space has been recorded. The used measuring protocol is the following one:
\begin{enumerate}
\item \underline{\textsc{Alignment at $\lambda = 785$}nm}. As mentioned previously, the system is aligned at $\lambda = 785$nm. That is, the SLM is aligned with the two MOs and the camera so that all the different orders that will be used in the experiment ($l=-5,..,5$) are properly focused and collimated.

\item \underline{\textsc{Finding a suitable single sphere}}. The sample is moved with the nano-positioning device to single out a TiO$_2$ sphere on the glass substrate. In order to see that in screen of the computer, the focusing objective (MO$_1$) is moved out of focus so that the illumination is constant in the field of view of MO$_2$. Then, the position of the particles can be tracked with the CCD camera in the forward semi-space. It is important that the selected sphere is isolated enough, so that no multi-scattering phenomena take place. Figure \ref{Fig85} shows two CCD snapshots of single particles. Clearly, the single sphere on the left is not far enough from other spheres, therefore it has not been used. In contrast, the single sphere on the right snapshot is isolated enough. This is the particle that has been used for the experiments presented in this section.
\begin{figure}[tbp]
\centering
\includegraphics[width=7.6 cm]{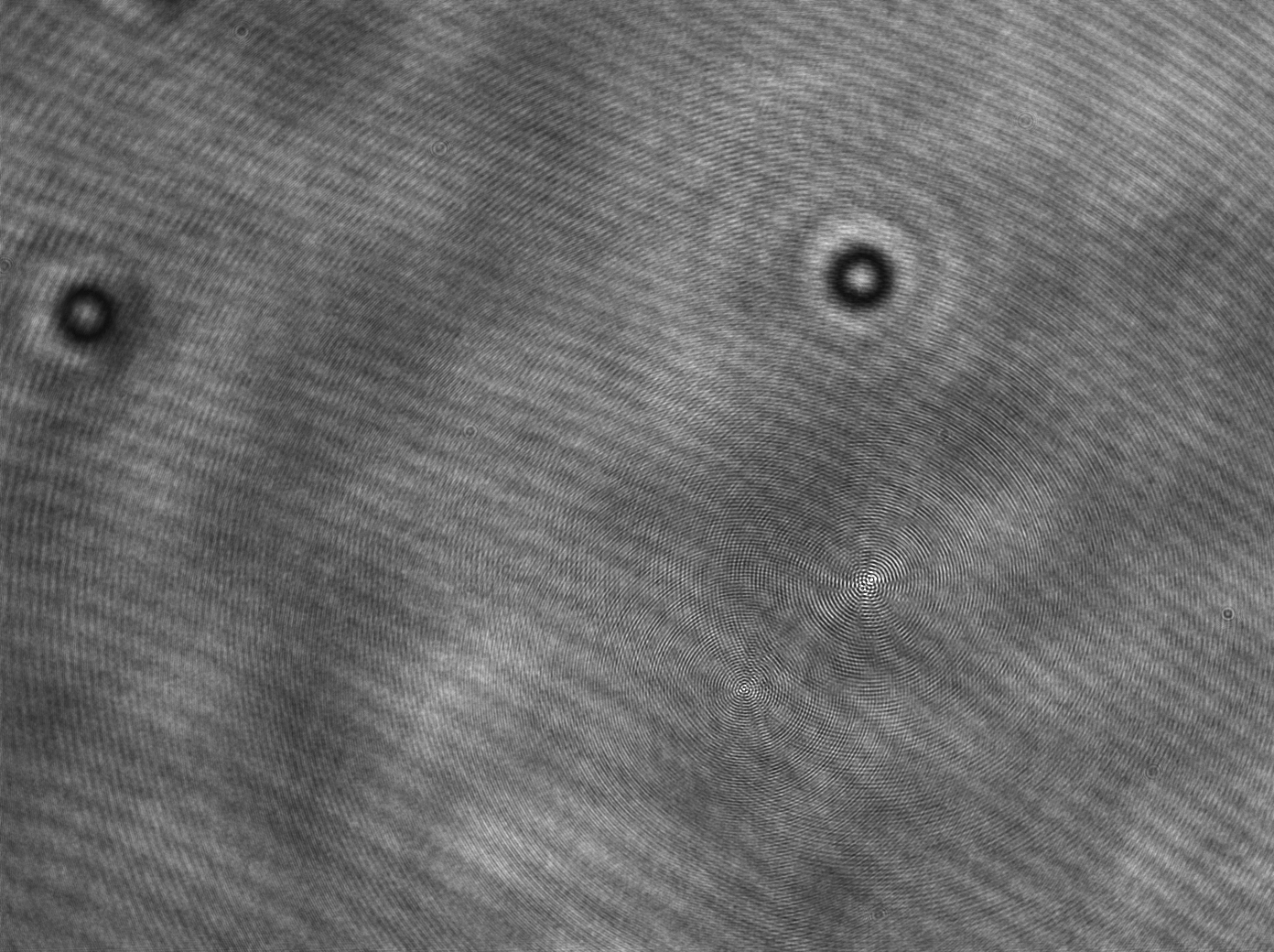} \includegraphics[width=7.6cm]{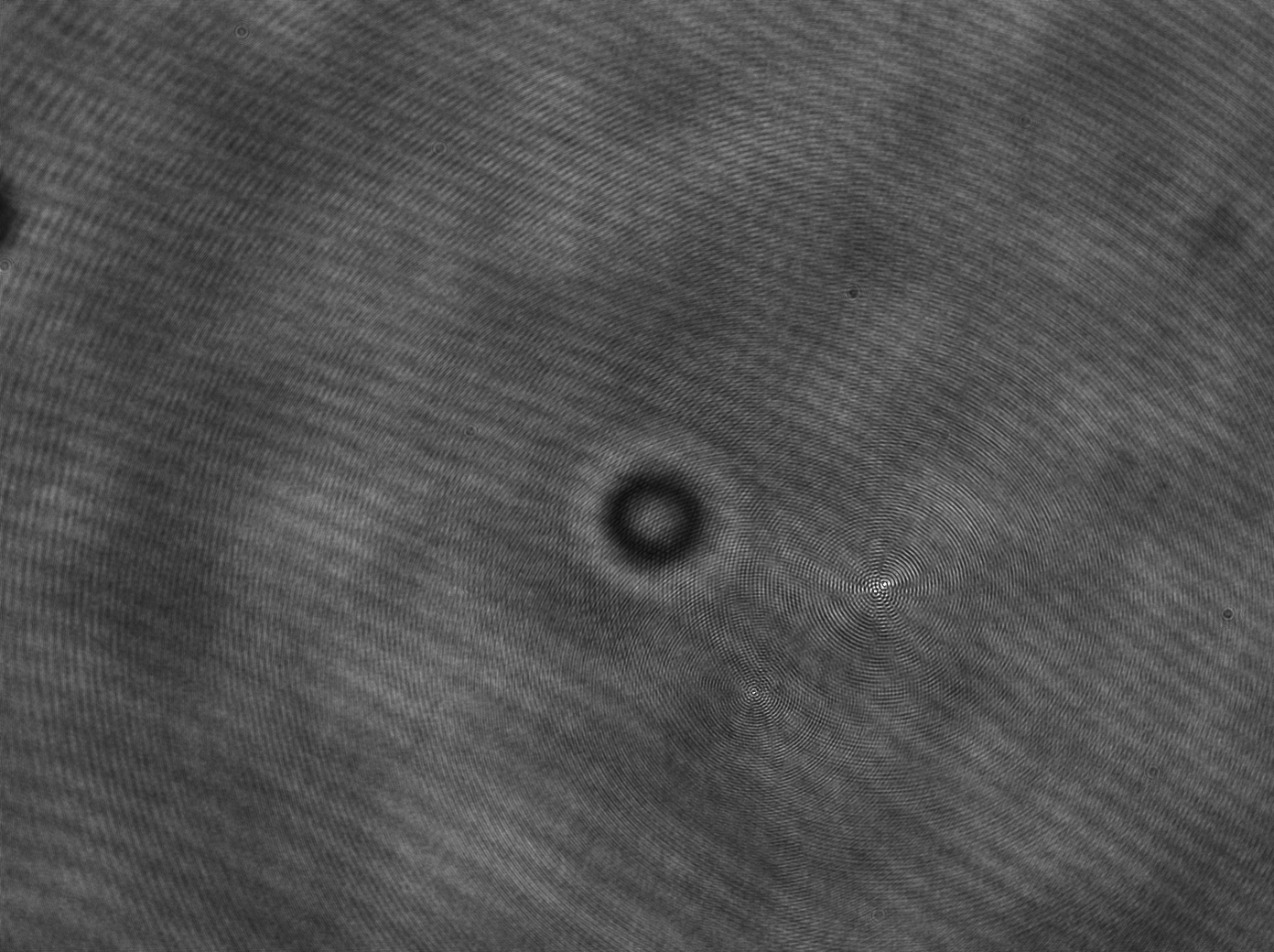} 
\caption{Forward imaging of the sample. Left: snapshot of a particle which is not isolated enough to perform single scattering experiments with it. Right: snapshot of a single sphere. The experiments have been carried out with this particle. There are no particles within 20 $\mu$m. All particles are made of TiO$_2$ and their $\Phi=950$nm. \label{Fig85}}
\end{figure}

\item \underline{\textsc{Background measurement at every $\lambda$}}. Once a sphere has been singled out, MO$_1$ is positioned to focus light on the particle. At the same time, MO$_2$ is moved mm's away from the focus, to avoid reflections from its lenses. Then, the particle is laterally moved tens of microns away from the position where the beam is focused. At this point, a snapshot of the beam is taken. This snapshot will be referred to as the background. The idea is that, this reflected light cannot be removed from the scattering measurements\footnote{As explained in Figure \ref{Fig84}(d), the background would be null if the incident field was a plane wave with a well-defined helicity propagating in the direction normal to the glass interface.}. Thus, it will be used to normalise the scattering of the particle. The background measurement depends on the incident mode, consequently a background snapshot will be taken for each mode $\Epq$, with $p=-1,1$ and $l=-5,..,5$. The process is repeated at every different $\lambda$.  

\item \underline{\textsc{Scattering measurement at every $\lambda$}}. The sphere is placed back at the centre of the beam. Now, centring the sphere with respect to the beam is not an easy task to do. In fact, it gets more and more cumbersome as the topological charge of the incident beam gets larger. The main reason why high optical vortices are hard to centre when they are focused was already mentioned in section \ref{Ch6_AM}. An optical vortex of charge $l$ breaks and it gives rise to $\vert l \vert$ vortices of charge $sign(l)$ \cite{Gabi2001OL,Ricci2012,Kumar2011,Rich2014}. Then, $\vert l \vert$ singular points can be found in the optical plane, which slightly break the cylindrical symmetry. In order to minimise this symmetry-breaking for every $\lambda$, $\mathbf{E}_{1,-3}^{\mathbf{in}}$ (for $p=1$ helicity modes) and $\mathbf{E}_{-1,3}^{\mathbf{in}}$ (for $p=-1$ modes) have been used to centre the particle with respect to the beam. It can be observed (see Appendix \ref{Appendix4}) that a multipolar field $\mathbf{A}_{jm_z}^{p}$ in the backward semi-space is identical to the multipolar field $\mathbf{A}_{jm_z}^{-p}$ in the forward semi-space. That is,
\begin{equation}
\mathbf{A}_{jm_z}^{p} \left( z<0 \right) = \mathbf{A}_{jm_z}^{-p}\left( z>0 \right)
\label{E_Ajmsemi}
\end{equation}
Now, when $\mathbf{A}_{jm}^{-p}$ is collimated and made paraxial, it yields an optical vortex of charge $l'=m+p$, where $m=p+l$ for beams of the kind $\Epq$, giving $l'=2p+l$. Then, in the direct component of backscattering, $\mathbf{E}_{1,-3}^{\mathbf{in}}$ and $\mathbf{E}_{-1,3}^{\mathbf{in}}$ yield vortices of charge $l'=1$ and $l'=-1$ respectively. Hence, if the CCD image obtained in backward scattering is a cylindrically symmetric beam with $\vert l \vert =1$\footnote{See Figures \ref{I_beam}($l=-1,p=1$) or \ref{Idirect}($l=1,p=-1$) to see an example of a scattered vortex beam with $\vert l \vert =1$.}, it means that the incident beam ($\mathbf{E}_{1,-3}^{\mathbf{in}}$ or $\mathbf{E}_{-1,3}^{\mathbf{in}}$) is properly centred with respect to the spherical particle. Then, once the particle has been centred, ten snapshots are taken of the scattering of the sphere for each combination of $p=-1,1$, $l=-5,..,5$.   

\item \underline{\textsc{CCD noise recording}}. To compensate for the inherent noise of the CCD camera due to electronic noise and others, thirty snapshots without light on the camera have been taken for each of the exposures used to record the background and scattering of the particle. In this way, an image of background noise is obtained, and it will be subtracted to all the rest of snapshots (background and particle).  
\end{enumerate}

\section{Wavelength scans}\label{Ch7_res}
\begin{figure}[tbp]
\centering
\includegraphics[width=\columnwidth]{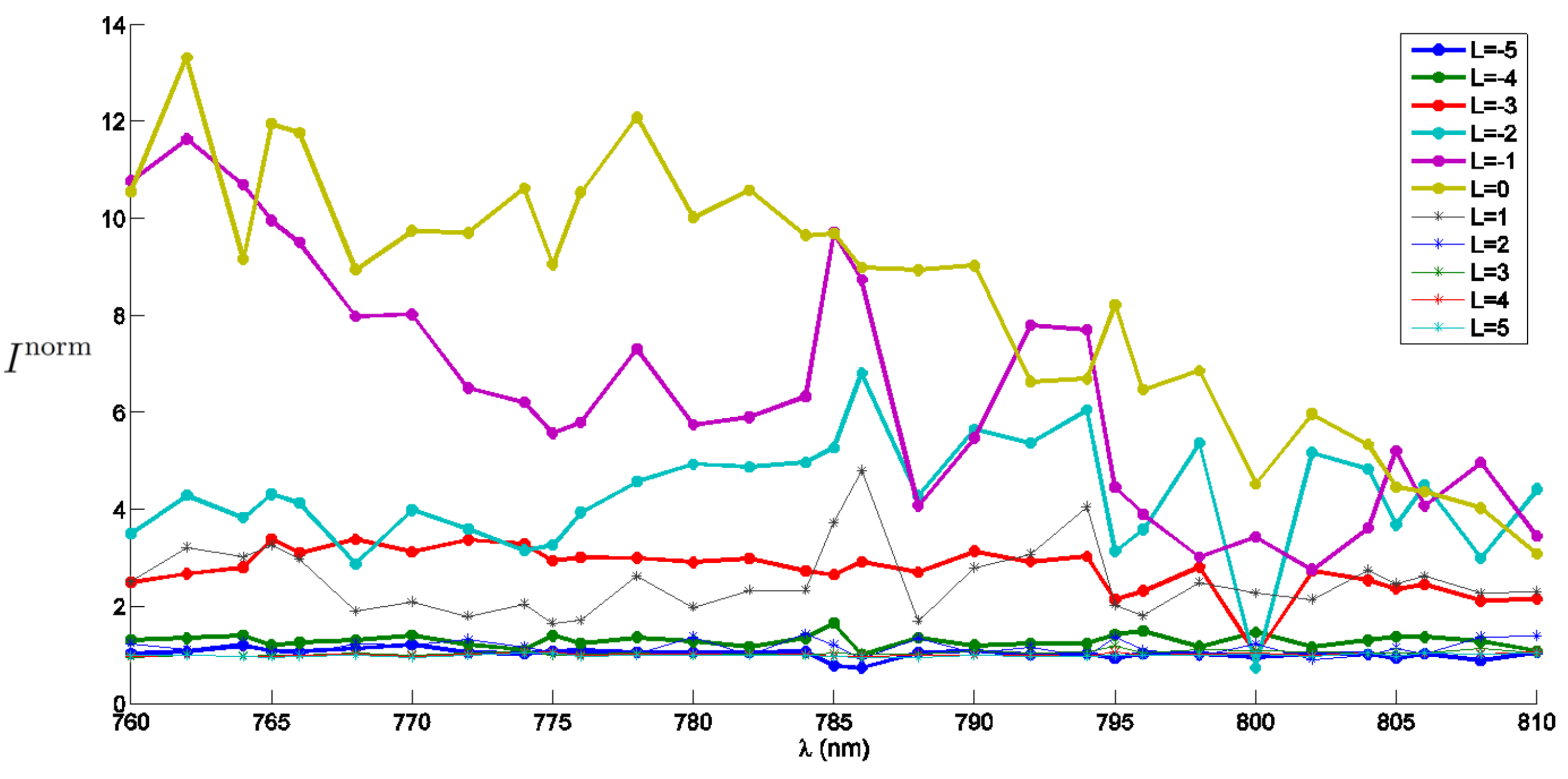}  
\caption{$\Ino$ as a function of $\lambda$. The incoming beams are of the kind $\mathbf{E}_{1,l}^{\mathbf{in}}$, where $l=-5,..,5$. Note the resonant behaviour of the modes with $l=\pm1$ and $l=2$ at 785nm. $\Ino$ is defined as the back-scattered intensity when the sphere is centred with respect to the incident beam over the intensity of the background. Both intensities are recorded with a CCD camera. \label{Fig86}}
\end{figure}

In the previous section, I have described the protocol to measure the backscattering of single spheres projected into their direct helicity component, given an incoming beam $\Epq$ and a wavelength $\lambda$. In this section, the collected data are shown and interpreted. First, $\lambda$ scans for $\Epq$ with $p=1$ and $l=-5,..,5$ are presented in Figure \ref{Fig86}. The L values in the legend correspond to the $l$ value of the incident mode used to record that plot. 
The horizontal axis shows the excitation wavelength $\lambda$, and the vertical axis a normalised intensity $\Ino$. $\Ino$ is computed as the scattered intensity divided over the background. The scattered intensity is computed when the single sphere that we want to study is centred with respect to the incident beam. In contrast, the background is computed to normalise the scattered intensity, \textit{i.e.} it is the reflection of the beam on the glass when the sphere is tens of microns away from the beam (see previous section). Consequently, the value of $\Ino$ is always $\Ino>1$. 
\begin{figure}[tbp]
\centering
\includegraphics[width=\columnwidth]{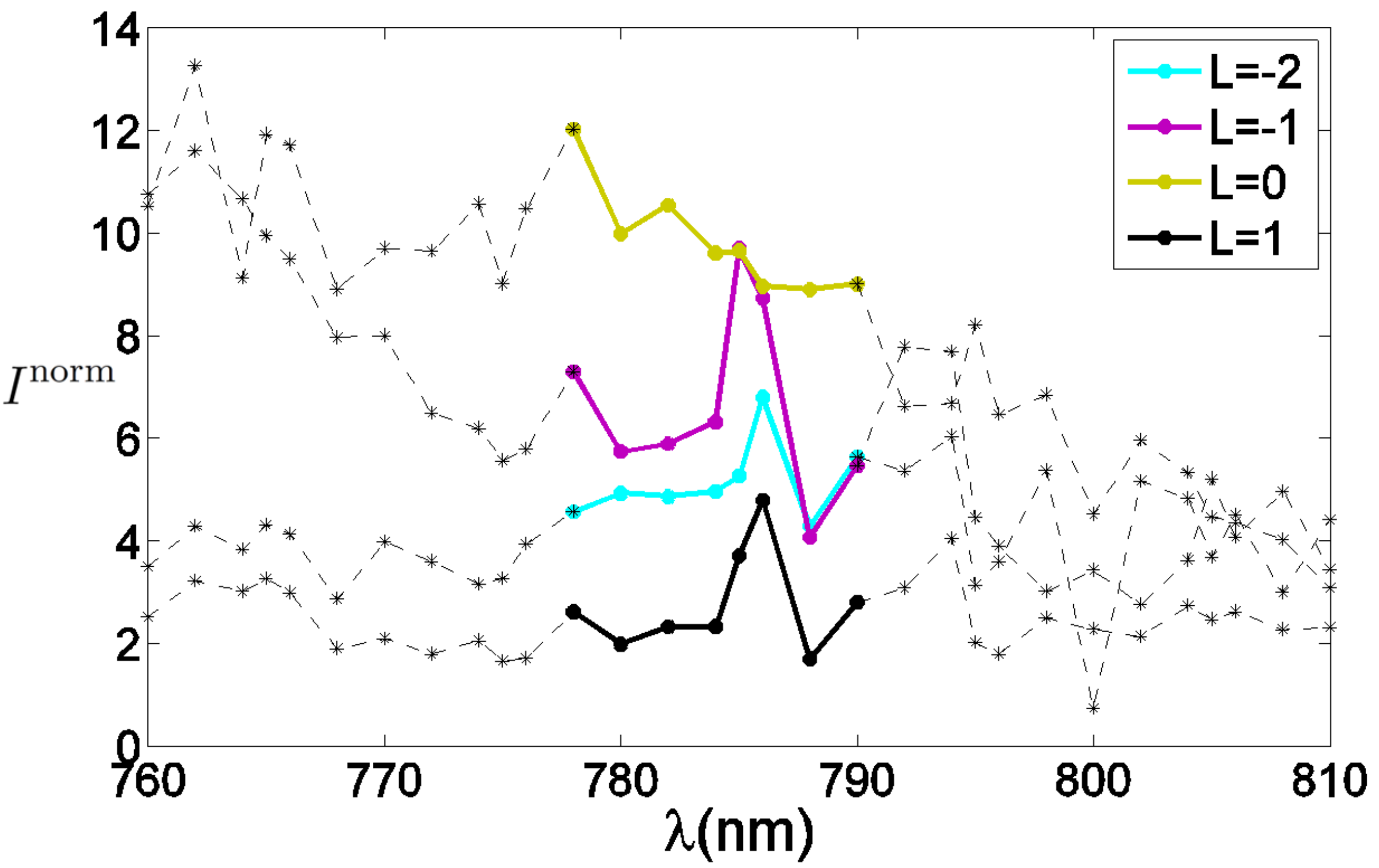}  
\caption{Resonant behaviour of $\Ino$ as a function of $\lambda$. The beams depicted in the figure are $\mathbf{E}_{1,l}^{\mathbf{in}}$, where $l=-2,-1,0,1$. The coloured lines highlight the resonant behaviour of the particle around $\lambda = 785$nm. The colors are the same ones as the ones used in Figure \ref{Fig86}. The rest of points, in a black dashed line, are left in the plot for completeness. It is clearly observed that the resonant behaviour of the particle is not observed with a Gaussian ($l=0$) excitation. The use of vortex beams is crucial to unveil the resonance. \label{Newres}}
\end{figure}
Because 10 pictures were taken for each of the scattering events, an average picture is computed\footnote{As mentioned before, the noise of the CCD camera is subtracted at every single snapshot (background and scattering).}. Then, a region of interest in the average picture is defined. Finally, the scattered intensity is computed summing all the individual values of the pixels in this region of interest of the average picture. The same region of interest is defined for the background picture and the values of the pixels in that region are summed up. That gives rise to the value of the background intensity. Then, as explained before, $\Ino$ is computed as the ratio between the scattered intensity over the background. Hence, every single point in Figure \ref{Fig86} is a dimensionless number computed as the ratio of two intensities, \textit{i.e.} $\Ino$ is an efficiency factor (see section \ref{Ch2_eff}).\\\\ 
\begin{figure}[tbp]
\centering
\includegraphics[width=\columnwidth]{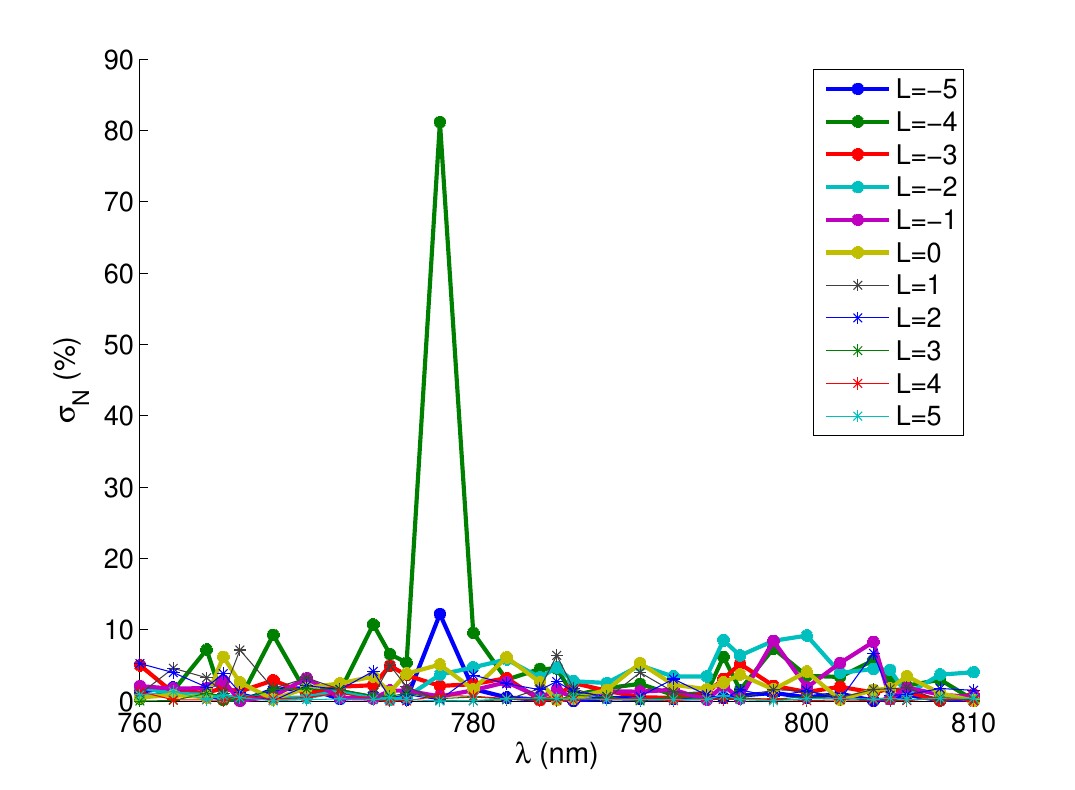}  
\caption{Standard deviation $\sigma_N(\%)$ of the recorded scattering as a function of the mode and the excitation $\lambda$. The incoming beams have helicity $p=-1$.\label{Fig87}}
\end{figure} 
These 10 different pictures are also used to compute statistical errors on the measurements. These are shown in Figure \ref{Fig87}. Except for one value giving a standard deviation of $\sigma_N(\%)\approx 80$, all the rest of errors are of the order of $\sigma_N(\%) \approx 10$. In order not to make the intensity plots even more complicated, these errors have not been displayed as error bars in Figure \ref{Fig86}.\\\\
Figures \ref{Fig86} and \ref{Newres} show very interesting results. Obviously, the $\lambda$ scans with the different modes of light yield different curves. Then, it can be observed that the sphere behaves very resonantly around 785nm.

\begin{figure}[tbp]
\centering
\includegraphics[width=12cm]{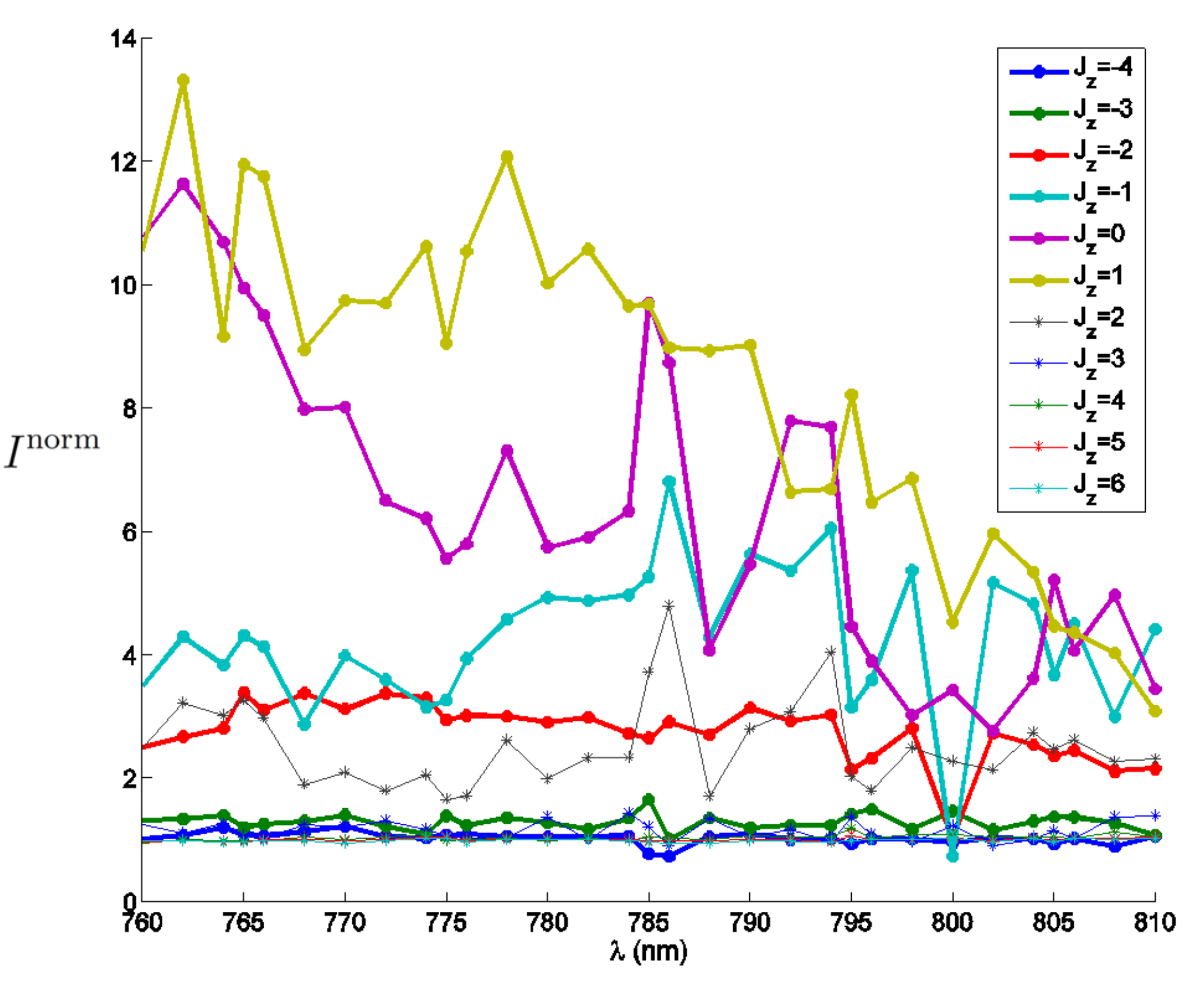} 
\caption{$\Ino$ as a function of $\lambda$. The incoming beams are labelled as a function of their AM content, $m=l+p$, where $p=1$ in this case. \label{Fig88}}
\end{figure}

\begin{figure}[tbp]
\centering
\includegraphics[width=14cm]{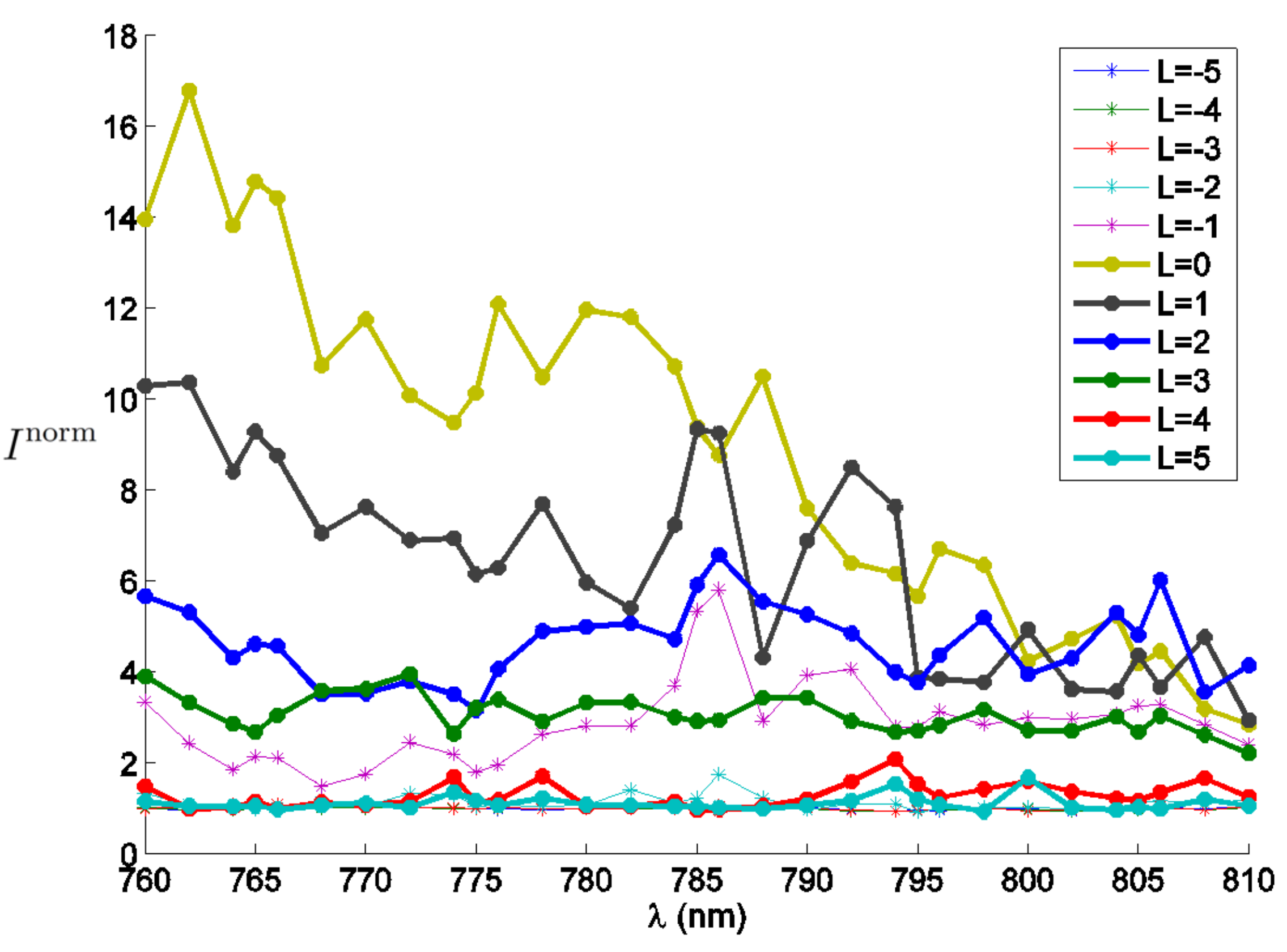} 
\caption{$\Ino$ as a function of $\lambda$. The incoming beams are of the kind $\mathbf{E}_{-1,l}^{\mathbf{in}}$, where $l=-5,..,5$. \label{Fig89}}
\end{figure}

\begin{figure}[tbp]
\centering
\includegraphics[width=14cm]{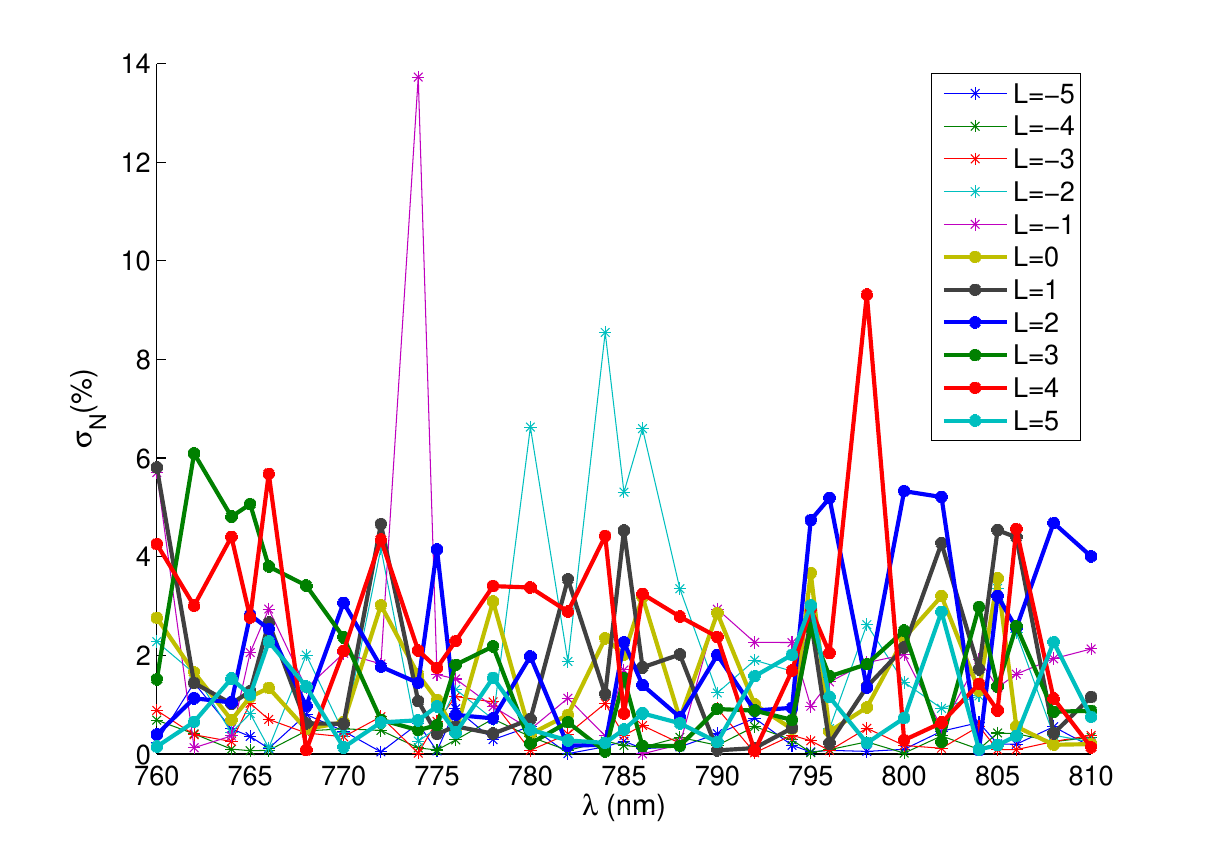} 
\caption{Standard deviation $\sigma_N(\%)$ of the recorded scattering as a function of the mode and the excitation $\lambda$. The incoming beams have helicity $p=-1$. \label{Fig810}}
\end{figure}

\begin{figure}[tbp]
\centering
\includegraphics[width=\columnwidth]{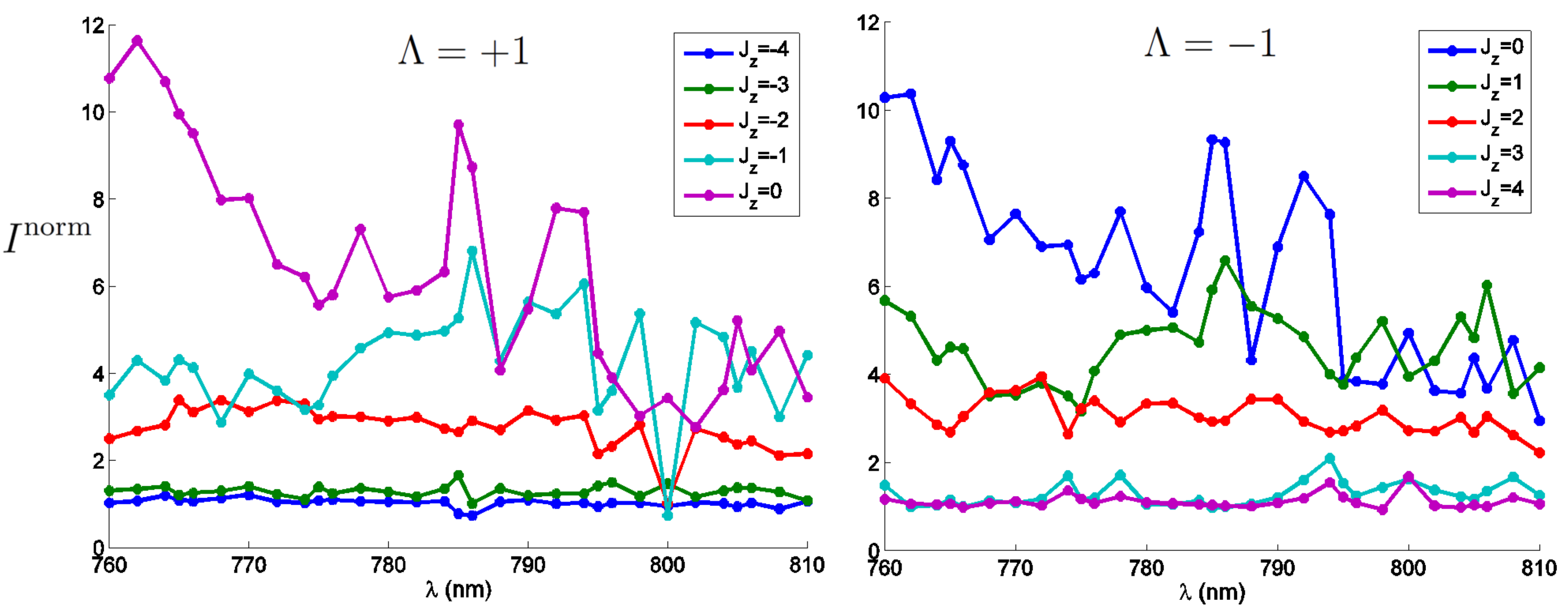} 
\caption{Comparative scattering of mirror symemtric beams. Left: the incident beams have $p=1$ and $m=l+p \leq 0$. Right: incident beams have $p=-1$ and $m=l+p \geq 0$.  \label{Fig811}}
\end{figure}

\begin{figure}[tbp]
\centering
\includegraphics[width=\columnwidth]{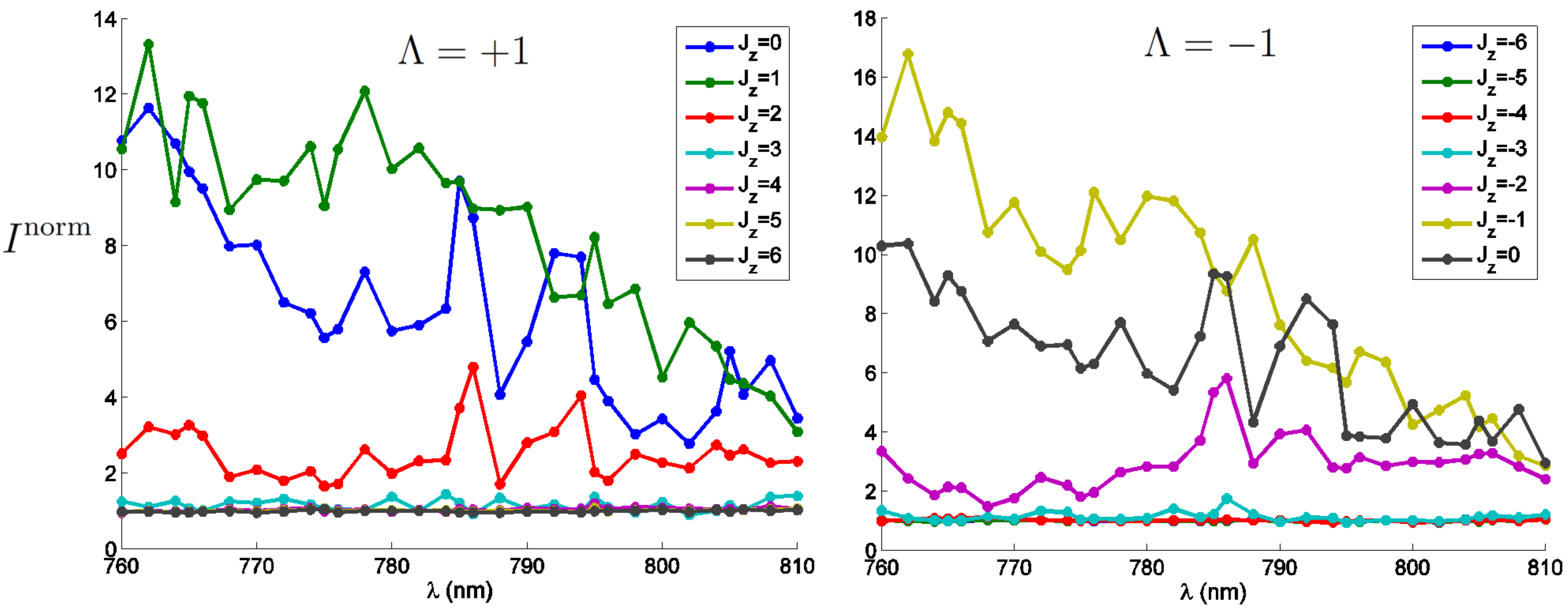} 
\caption{Comparative scattering of mirror symemtric beams. Left: the incident beams have $p=1$ and $m=l+p \geq 0$. Right: incident beams have $p=-1$ and $m=l+p \leq 0$. \label{Fig812}}
\end{figure}

In fact, the resonance is hidden under the Gaussian excitation ($l=0$), and the use of vortex beams is crucial to unveil it, as predicted by the theory in chapter \ref{Ch3} (see Figure \ref{F_Enhancement}). Also, as expected, the Gaussian mode ($l=0$) is the one that yields a larger intensity. However, the mode with $l=-1$ also gives a comparable amount of scattering. It is important to note the huge asymmetry between the modes with $l=1$ and $l=-1$. The mode with $l=-1$, scatters much more light than the one with $l=1$. This is due to their different AM content, $m=l+p$. Now, because a beam of $J_z=m$ can only excite Mie resonances of order $j \geq m$, it is convenient to re-label the incident beams in terms of their AM content. This is displayed in Figure \ref{Fig88}.\\\\
The $\lambda$ scans obtained for $\Epq$ with $p=-1$ and $l=-5,..,5$ are presented in Figure \ref{Fig89}. Again, their error bars are not plotted along with the average measurements. Instead, they are depicted in Figure \ref{Fig810}, yielding standard deviations smaller than $14\%$. A close look at Figure \ref{Fig89} shows that the results are very similar to the ones presented in Figure \ref{Fig86}. In fact, it is seen that there is an underlying symmetry relating the results for $\Epq$ and $\Eppq$. 
The underlying symmetry relates the scattering of vortex beams with an topological charge $l$ and a helicity $p$ with the scattering of modes with topological charge $-l$ and a helicity $-p$. This is the same symmetry relation that yielded CD$_q = - \text{CD}_{-q}$ in section \ref{Ch6_CD}: mirror symmetry. In fact, the sample used in this experiment (glass substrate with a single sphere on top) has exactly the same symmetries as the sample used in chapter \ref{Ch6} (circular nano-aperture in a metallic film). Those symmetries are mirror symmetry with respect to any plane containing the $z$ axis, rotations along the $z$ axis, and time translations (see section \ref{Ch6_symm} for a detailed explanation of the symmetries). The mirror symmetric scattering of the particle is highlighted in Figures \ref{Fig811}, \ref{Fig812}. In Figure \ref{Fig811}, the scattering produced by modes with $m \leq 0$ for beams with $p=1$, and $m \geq 0$ with $p=-1$ is compared. The resemblance of $I_{m,p}^{\text{norm}}(\lambda)$ and $I_{-m,-p}^{\text{norm}}(\lambda)$ is apparent. Similarly, Figure \ref{Fig812}, portraying the comparative scattering of modes with $m \geq 0$ for $p=1$ and modes with $m \leq 0$ for $p = -1$, also shows a great match between the results using the two different families of modes.  
Actually, one of the few significant differences is that in the plots for $p=1$, there is a drastic decrease of intensity for the modes with $l=0,-2,-3$ at $\lambda=800$nm. The fact that this decrease was not observed in its mirror symmetric scattering counterpart leads to believe that it is an experimental human error. \\\\
It is clear that the AM of light plays a crucial role in GLMT. As it was predicted in \cite{Zambrana2012,Zambrana2013JQSRT}, the AM of light can be used to unveil some hidden resonances in the Gaussian (or plane wave) excitation. Furthermore, it is clear that symmetry considerations allows for qualitative predictions of the results. In this case, the mirror symmetry of the sample allows for a prediction of the scattering intensity of beams $\Eppq$ once the scattering of $\Epq$ modes has been found. As future work, I would like to apply the same technique to structures with spherical symmetry, instead of spheres in contact with an interface. In order to do that, the method to embed particles in a polymer needs to be improved so that the faces of the sample are rather parallel and the scattering from the polymer is null. Another option would be to trap the particles with an optical tweezers set-up, which would be easy to implement in my current set-up. In a similar direction, I would like to carry out similar experiments with particles of different sizes. In particular, I would like to excite WGMs with direct light using the technique described in section \ref{Ch3_WGM}. For that matter, particles of large sizes as well as higher order optical vortices need to be used.

\section{Characterization of crossed helicity modes}
To conclude with this chapter, some of the CCD images that have been used to compute the intensity plots in Figures \ref{Fig86}-\ref{Fig812} are shown. Because the experiments have been done with the particles centred with respect to the incident beam, the images used to compute the background normalisation and the ones recording the scattering of the particles share the same features. That is a consequence of both system having the same symmetries. The basic difference between the images is the magnitude of the intensity. Next, the background CCD images are shown. As explained before, each of the snapshots has been taken using an incident beam of the kind $\Epq$. Therefore, the images are classified in terms of the charge of the optical vortex created by the SLM ($l$) and the helicity of the incident beam $p$. Note that, because of equation \ref{E_Ajmsemi}, the images show the spatial shape of the crossed helicity component in the forward direction. That is, given an incident beam of $J_z=m=l+p$, the captured image has $\vert l + 2p\vert $ optical vortices of charge $sign (l+2p)$. This fact is easily observable from the pictures when $\vert l + 2p\vert \leq 3$, yet it is difficult to appreciate when the number of singularities exceeds 3. Finally, note the mirror symmetry between the image profiles. That is, the images obtained for $\Epq$ are practically identical to the ones obtained with $\Eppq$. To the best of my knowledge, the experimental characterisation of crossed components of vortex beams with a well-defined helicity has not been published anywhere.
\begin{figure}[tbp]
\centering
\includegraphics[width=\columnwidth]{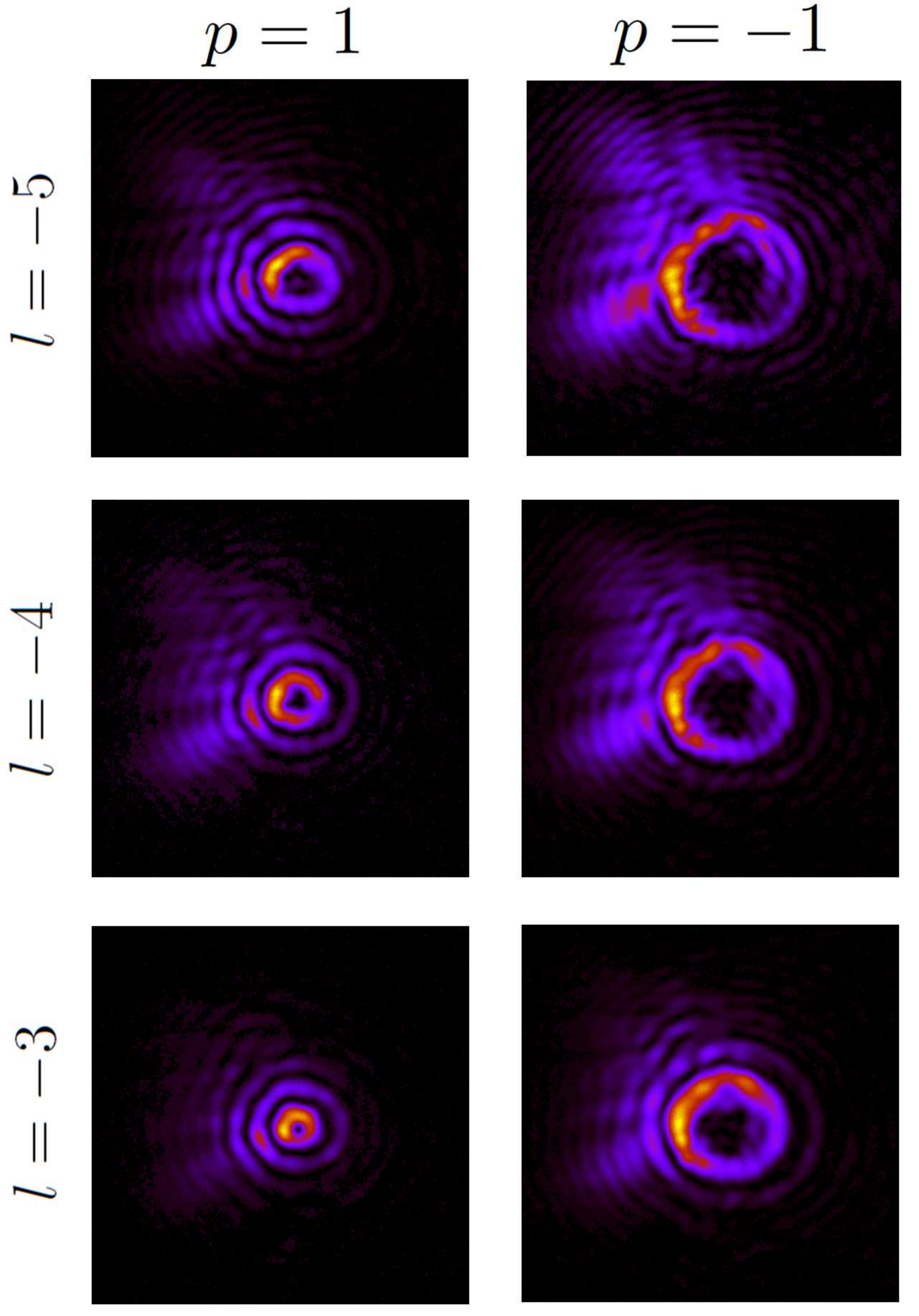} 
\caption{Background scattering intensity given an incident beam $\Epq$ with $p=\pm1$ and $l=-5,-4,-3$. The sphere is about $20\mu$m away.  \label{Fig814}}
\end{figure}
\begin{figure}[tbp]
\centering
\includegraphics[width=\columnwidth]{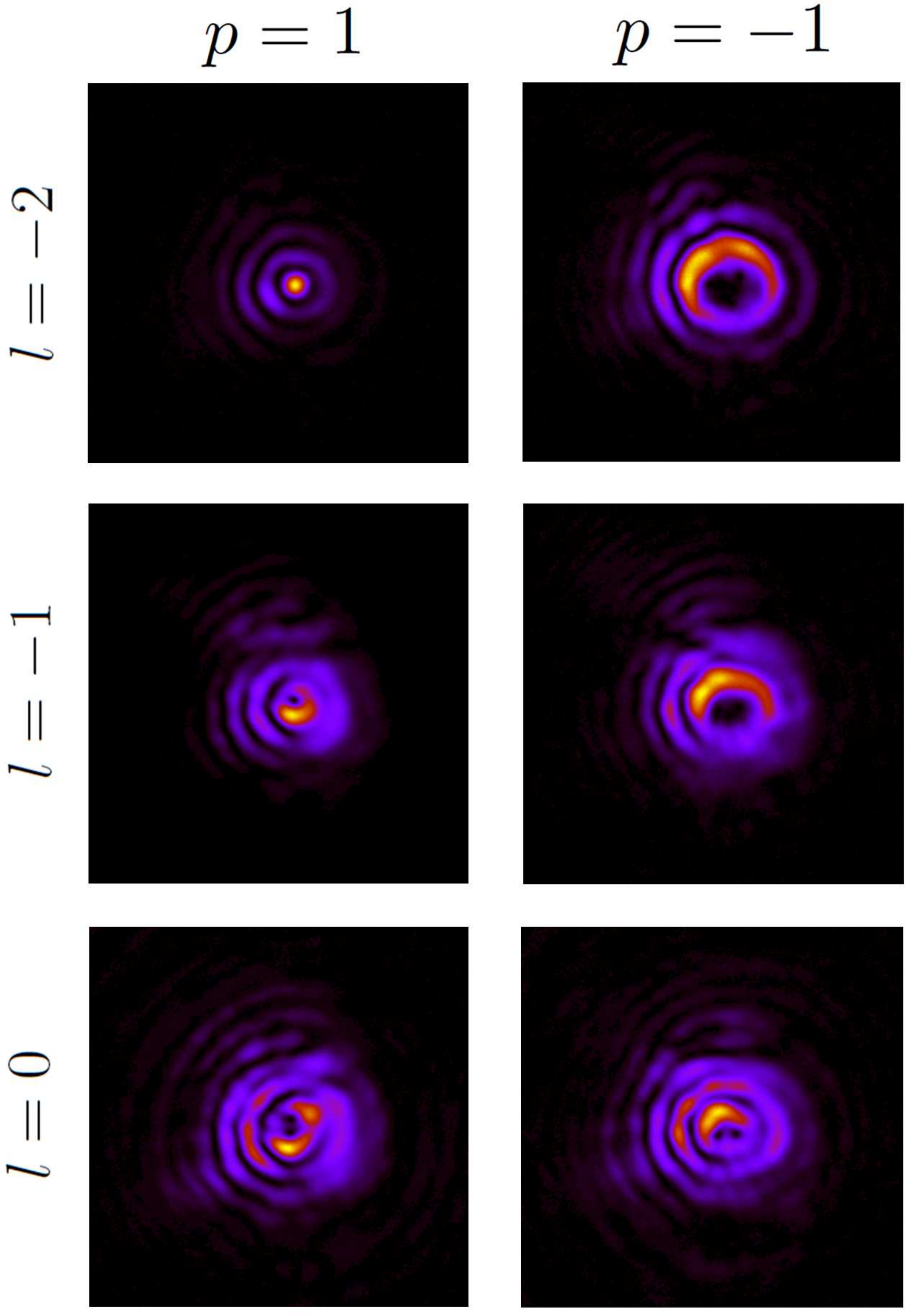} 
\caption{Background scattering intensity given an incident beam $\Epq$ with $p=\pm1$ and $l=-2,-1,0$. The sphere is about $20\mu$m away.  \label{Fig815}}
\end{figure}
\begin{figure}[tbp]
\centering
\includegraphics[width=\columnwidth]{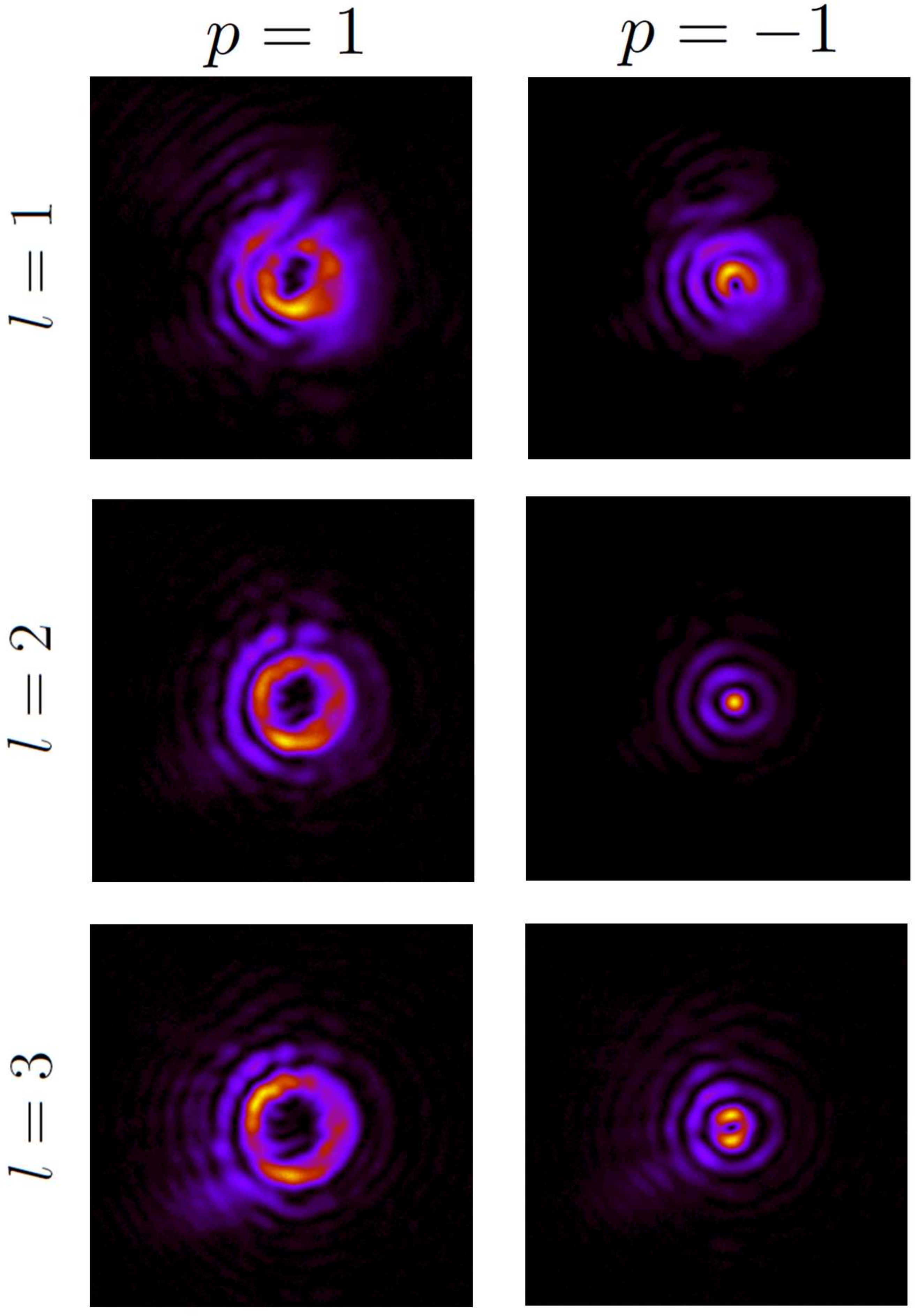} 
\caption{Background scattering intensity given an incident beam $\Epq$ with $p=\pm1$ and $l=1,2,3$. The sphere is about $20\mu$m away. \label{Fig816}}
\end{figure}
\begin{figure}[tbp]
\centering
\includegraphics[width=\columnwidth]{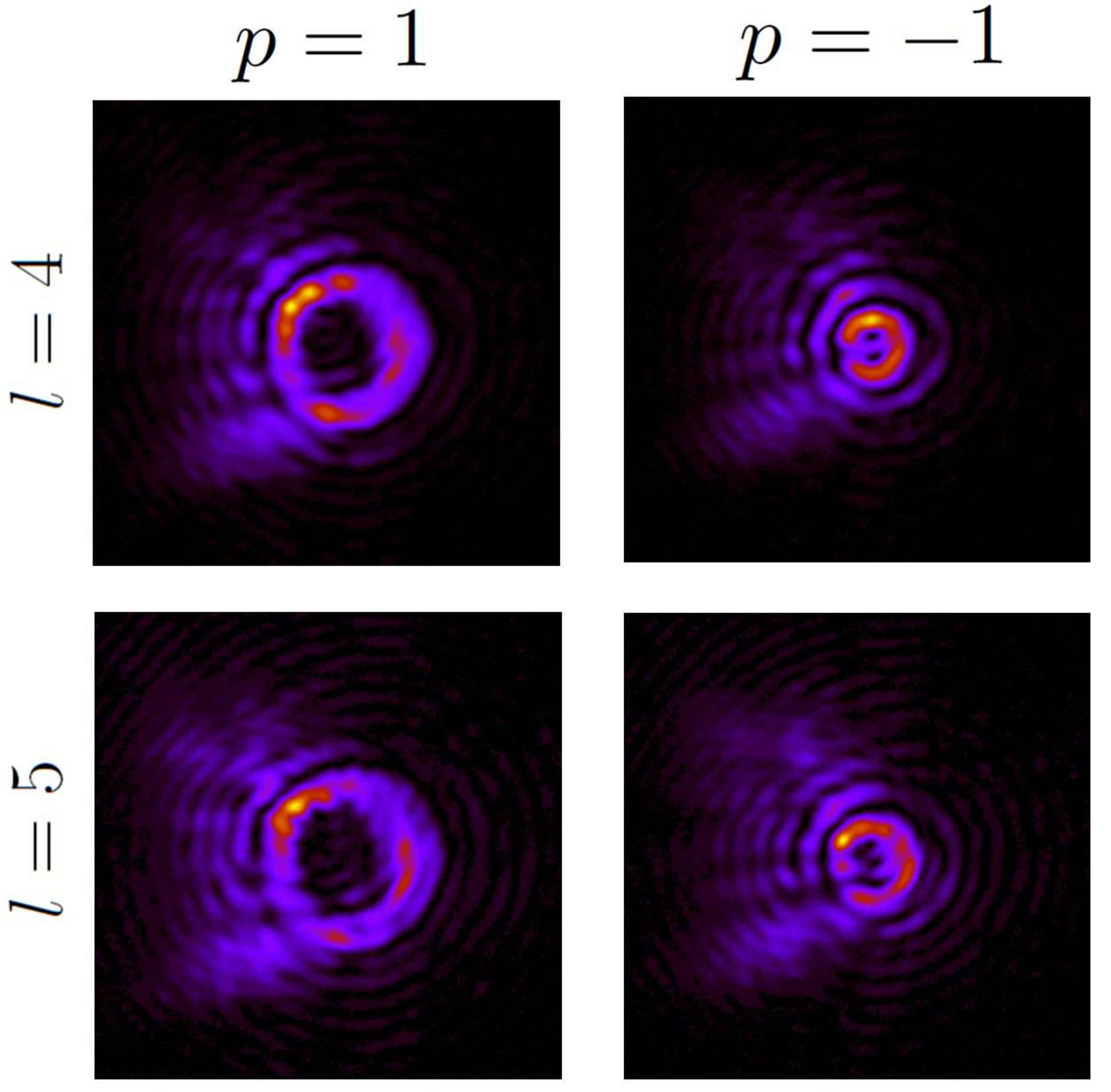} 
\caption{Background scattering intensity given an incident beam $\Epq$ with $p=\pm1$ and $l=4,5$. The sphere is about $20\mu$m away. \label{Fig817}}
\end{figure}

\begin{savequote}[10cm] 
\sffamily
``Education must provide the opportunities for self-fulfilment; it can at best provide a rich and challenging environment for the individual to explore, in his own way'' 
\qauthor{Noam Chomsky}
\end{savequote}

\chapter{Conclusions}
As technology advances, science is challenged with new problems and vice-versa. In particular, the development of dielectric photonic technologies in the 1980's made it clear that photonics could compete with electronics to encode, transmit and retrieve information. However, the current drive to shrink down our devices has enabled electronics to clearly outperform photonics. Nanophotonics is a young field, and there is still a lot of room for improvement. In this sense, the work presented in this thesis is done from a different approach to the mainstream research in nanophotonics. Most of efforts in overcoming the diffraction limit of light have been done in the field of plasmonics, where the materials and the geometry the nano-structures are made of are crucial to control light-matter interactions at the nano-scale. Here, light-matter interactions are described in terms of their symmetries. The fact that light can be experimentally modified so that it is symmetric under many different transformations is crucial in the process. Within this framework, the main findings of this work are the following ones: \\ \\
\textbf{Systematic study of EM modes in terms of their symmetries}. In chapter \ref{Ch1}, a systematic study of six different basis of transverse solutions of Maxwell equations is done. Moreover, the relations between the different basis are also given. Then, in chapters \ref{Ch3},\ref{Ch4},\ref{Ch6},\ref{Ch7}, different problems are faced. Depending on the symmetries of the problem, one of the basis is chosen over the others to describe the light-matter interaction. These studies have been partially published as part of \cite{Nora2012,Ivan2012PRA}.\\ \\
\textbf{Effect of the aplanatic lens on the multipolar content of a beam}. The aplanatic transformation of a lens, which is described in section \ref{Ch1_apla}, is used to modify the beam shape coefficients of a beam. That is, the NA of an aplanatic lens is used to control the multipolar content of a beam. This effect is part of the findings presented in \cite{Zambrana2012}.\\ \\
\textbf{Enhancement of ripple structure}. In Mie Theory, the scattering efficiency is computed as an infinite summation of Mie coefficients:
\begin{equation}
Q_s^{Mie} = \sum_{j=1}^{\infty} \dfrac{ 2j+1}{x} \left( \vert a_j \vert ^2 + \vert b_j \vert^2 \right)
\end{equation}
Due to the infinite summation, the resonances of the high order Mie coefficients are only perceived as little ripples in the cross section. These ripples have a very low Q factor ($\delta$ parameter in section \ref{Ch3_WGM}). In chapter \ref{Ch3}, it is proven that the use of LG beams (and in general any cylindrically symmetric beam) allows for the enhancement of the ripples. In fact, their Q factors can be increased by two orders of magnitude even using relatively low-order LG modes such as LG$_{6,0}$. The downside of it, there is much less scattering, as all the contributions of the low orders modes are suppressed from the scattering. These results were published in \cite{Zambrana2013JQSRT}. \\ \\
\textbf{Excitation of single multipolar modes}. In chapter \ref{Ch3}, a technique to excite single multipolar resonances regardless of their size and index of refraction is explained. In particular, the technique can be used to excite WGMs. The excitation of single modes is achieved in two steps. First, a cylindrically symmetric beam with a large value of $J_z=l+p$ is paraxially created, with $p$ being the helicity of the beam and $l$ the order of the phase singularity that carries. To increase the cross section, the paraxial beam is focused onto the sphere with a MO with a large NA. Then, the wavelength of the laser $\lambda$ and the radius of the sphere $R$ are chosen so that the following expression holds: $R \approx \lambda  (l+p)  f(n_r)/(2\pi)$, where $f(n_r)$ is a function of the relative refractive index that can be computationally checked. For example, when $n_r=1.5$, $f(n_r) \approx 0.8 $. This technique was explained in detail in \cite{Zambrana2012}. \\ \\
\textbf{Generalization of Kerker Conditions}. The Kerker conditions were demonstrated by Kerker and co-workers in 1983. The first Kerker condition (K1) states that a sphere with $\epsilon = \mu$ has zero backward scattering. Then, K2 states that a small particle with $\mu \neq 1$ can have zero forward scattering if $a_1=-b_1$. In section \ref{Ch4_Kerker}, it is demonstrated that Kerker conditions can be generalized to cylindrical objects of arbitrary size. In fact, it is proven that K1 is a specific case of a system symmetric under duality transformations and rotations along the $z$ axis. Then, in a very similar fashion, K2 is generalized to anti-dual cylindrically symmetric systems. That is, if a scatterer is anti-dual and symmetric under rotations along the $z$ axis, then its forward scattering must be 0. A detailed proof of this generalization was published in \cite{Zambrana2013}. \\ \\
\textbf{Inducing dual or anti-dual behaviours in dielectric spheres}. It is very challenging to find dual materials in nature. However, some materials can behave as such under certain conditions. In chapter \ref{Ch4}, an analytical method to induce duality in dielectric spheres is shown. The method can be applied to any geometry and material, but it becomes particularly efficient for spheres, as it can be computed analytically. In fact, using the same formalism, the anti-dual behaviour of spheres can also be probed. Then, the influence of the AM of light on this process is studied. It is demonstrated that the AM of light adds another degree of freedom to the previously discussed method. Thus, in principle, arbitrarily large spheres can behave as dual or anti-dual when high order AM modes are used. This method as well as some practical applications were shown in \cite{Zambrana2013OE}. \\ \\
\textbf{CD with non-chiral sample}. CD is a widely used technique in biology and chemistry. Recently, it has gained a lot of interest among the metamaterials and plasmonics community. In chapter \ref{Ch6}, a symmetry-based method to induce CD in a non-chiral sample is shown. It is observed that a circular nano-aperture can induce a giant CD when it is excited with vortex beams. The key reason why this phenomenon arises is that the two vortex beams used in the measurement are not mirror symmetric with respect to the other. Furthermore, it is explained that CD not only compares the differential absorption of left and right circular polarisation, but also the differential absorption of AM momenta states differing in two units. These results are presented in \cite{preZambrana2014}.  \\ \\
\textbf{Transmission of optical vortices through sub-wavelength nano-apertures}. In section \ref{Ch6_AM}, the transmission of optical vortices through a sub-wavelength circular nano-aperture is demonstrated. The decomposition of the transmitted light into its two helicity  components shows that optical singularities arise in scattering processes due to the fact that duality symmetry is broken. Similar experimental effects have been presented in \cite{preNora2014,Nora2014}. \\ \\
\textbf{Experimental excitation of scattering resonances with optical vortices}. Some of the predicted effects in chapter \ref{Ch3} have been experimentally verified in chapter \ref{Ch7}. In particular, it has been seen that vortex beams can unveil scattering resonances for dielectric spheres. That is, it has been observed that a relatively flat scattering cross section (as a function of the size parameter $x$) can be turned into a resonant one when the appropriate vortex beam is used to excite the single sphere. The effect has been corroborated with two sets of beams with different helicities. As a result, the same mirror symmetric properties observed in chapter \ref{Ch6} have been also measured. \\ \\
As it has been mentioned before, the findings of this thesis advance the control of light-matter interactions at the nano-scale using a new symmetry-based formalism. For example, thanks to the method put forward in section \ref{Ch3_WGM}, Q factors of the order of $Q=1\cdot 10^6$ are accessible with spheres of radius $R=1.9 \mu$m. Also, as shown in chapter \ref{Ch6}, a new wealth of information from nano-samples can be extracted carrying out a CD measurement with vortex beams. Figure \ref{CD_fig} in section \ref{Ch6_CD} depicts samples whose diameter sizes differ in only 100 nm retrieve CD values that differ in more than 70\%. Another effect with direct consequences in nano-science is the induced duality described in chapter \ref{Ch4}. The design of new metamaterials requires a great control of the building cells. In that sense, Figure \ref{Tm} can be used as a design map for new highly directional metamaterials, as dual spheres have zero-backscattering. The applications described so far are direct implications of the results presented in the previous chapters. Nevertheless, the findings in this thesis open up new possibilities in other fields. For instance, there is a growing interest in controlling the magnetic component of light. Different groups have recently fabricated structures that can enhance the magnetic component of light in certain regions. The semi-analytical techniques described in chapter \ref{Ch3} allow for scattering control of spheres. In particular, the technique can be applied so that the scattering is purely magnetic. Then, the analysis of the dual behaviour of a scatterer could result to be very useful to achieve very interesting regimes of optical forces. The fact that an anti-dual scatterer does not scatter light in forward and a dual scatterer does not scatter light in the backward direction could be linked to pushing and pulling forces. Also, the excitation of single high order multipolar modes could be used to carry out side-band cooling in quantum optomechanics experiments. The idea is that the excitation of WGMs with direct light can be easily added in an optical levitation set-up, which is a very optimum way of cooling down the centre of motion of a particle.

\appendix

\chapter{Rotation of electromagnetic fields}
\label{Appendix}
\newcommand{\Rr}{\mathbf{R}}
\newcommand{\Mpt}{\mathbf{M}(\phik,\thetak)}
\renewcommand{\arraystretch}{1}

In order to compute the rotation of EM fields, the following formulae are necessary. Given a EM field $\psixt$, a rotation of magnitude $\varphi$ along an axis $\nhat$ is computed using equation (\ref{E_rotpsi}):
\begin{equation}
 R_{\nhat}(\varphi) \left[ \psixt \right] = \mathbf{M}_{\nhat}(\varphi) \cdot \psi(\mathbf{M}_{\nhat}^{-1}(\varphi) \cdot \mathbf{r},t)
\label{A_rot} 
\end{equation}
where $\mathbf{M}_{\nhat}(\varphi)$ is given by the following expression
\begin{equation}
\mathbf{M}_{\nhat}(\varphi) = \begin{pmatrix} \cos \varphi & - \sin \varphi & 0 \\ \sin \varphi & \cos \varphi & 0 \\ 0 & 0 & 1  \end{pmatrix}
\end{equation}
if the cartesian axis are rotated in a way such that $\nhat = \zhat$ and $\varphi > 0$ corresponds to a dextrogyrate rotation. Now, this is not the only way of computing rotation matrices. The most usual way is using Euler angles. That is, the reference frame is considered clumped in the origin and then any rotation is expressed as the product of three rotations \cite{Tung1985}:
\begin{equation}
\mathbf{M}(\alpha,\beta,\gamma)= \mathbf{M_z}(\alpha)\mathbf{M_y}(\beta)\mathbf{M_z}(\gamma)
\end{equation}
where $\mathbf{M_z}(\alpha)$ and $\mathbf{M_y}(\beta)$ are given by:
\begin{equation}
\mathbf{M_z}(\alpha) =  \begin{pmatrix} \cos \alpha & - \sin \alpha & 0 \\ \sin \alpha & \cos \alpha & 0 \\ 0 & 0 & 1 \end{pmatrix} \qquad 
\mathbf{M_y}(\beta) = \begin{pmatrix}  \cos \beta & 0 & \sin \beta \\ 0 & 1 & 0 \\ - \sin \beta & 0 & \cos \beta \end{pmatrix}
\end{equation}
In this thesis, especially in chapter \ref{Ch1}, the operator of rotations $\rotation$ has been used. This operator rotates EM fields following equation (\ref{A_rot}). Then, the rotation matrix can be obteined as $\mathbf{M}(\alpha,\beta,\gamma)= \mathbf{M_z}(\phik)\mathbf{M_y}(\thetak)\mathbf{M_z}(0)$, which yields the following two rotation matrix:
\begin{equation}
\mathbf{M}(\phik,\thetak) =  \begin{pmatrix} \ct \cp & - \spp & \st \cp \\ \ct \spp & \cp & \st \spp  \\ -\st & 0 & \ct  \end{pmatrix}
\end{equation}
\begin{equation} 
\mathbf{M^{-1}}(\phik,\thetak) = \begin{pmatrix}  \ct \cp  &  \spp \ct & -\st \\ - \spp & \cp & 0 \\ \st \cp & \spp \st & \ct \end{pmatrix}
\end{equation}
As an example, the rotation of a plane wave is shown. The equation (\ref{E_p-x}) will be proven step by step:
\begin{equation}
\rotation \xhat \exp(ikz) = \Mpt  \xhat \exp \left( i k\zhat \cdot \mathbf{M^{-1}}(\phik,\thetak)  \mathbf{r} \right)
\end{equation}
\begin{equation}
i k\zhat \cdot \mathbf{M^{-1}}(\phik,\thetak)  \mathbf{r} = ik \left(  \st \cp x + \st \spp y + \ct z \right)= i \mathbf{k} \cdot \mathbf{r}
\end{equation}
\begin{equation}
\Mpt \xhat = \left( \begin{array}{c}
\ct \cp \\ \ct \spp \\ - \st
\end{array} \right) = - \phat
\end{equation}


\chapter{Holography}
\graphicspath{{ch_appendix2/}} 
\label{Appendix2}

\newcommand{\Dtx}{\Delta \theta_x}

This Appendix gives supplementary information to fully understand the holography techniques used in this thesis.
\section{Fabrication of phase-only CGH}\label{App2Fab}
\begin{figure}[tbp]
\centering
\includegraphics[width=\columnwidth]{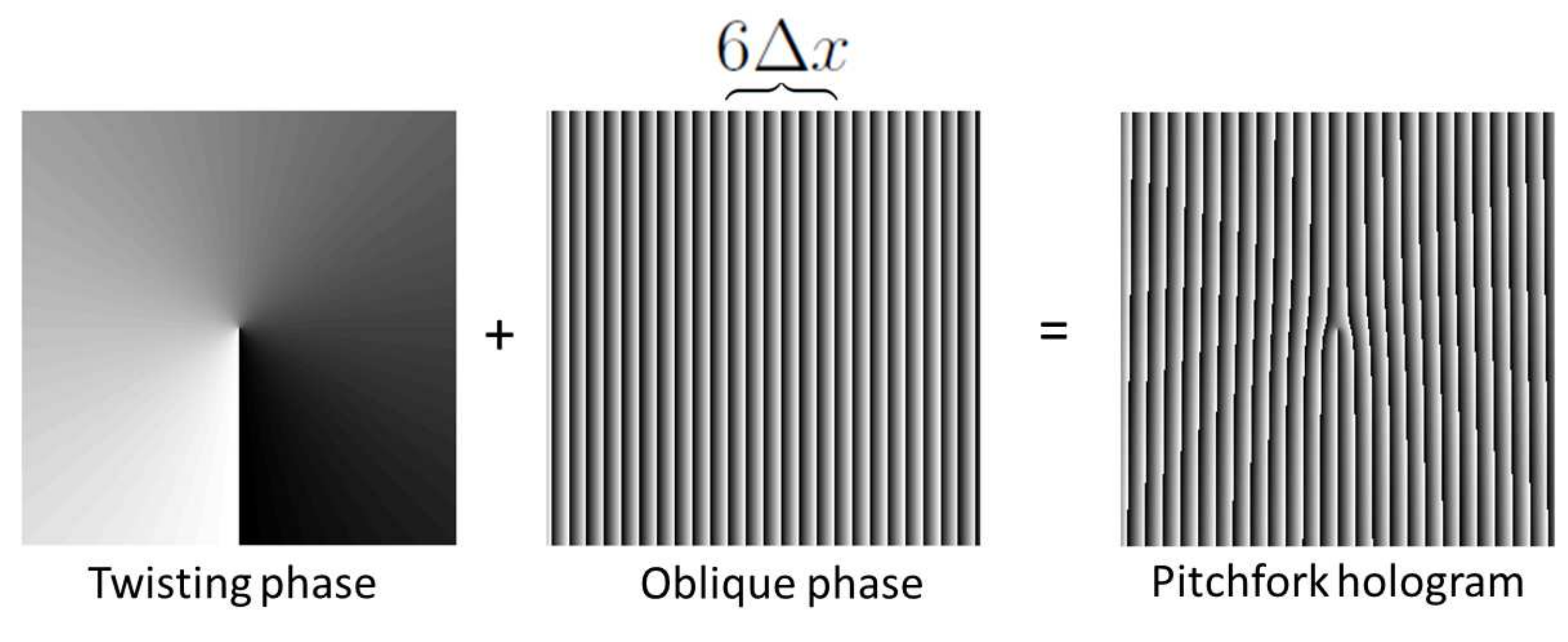} 
\caption{A pitchfork hologram is obtained adding a twisting and an oblique phase. The width of six $2\pi$ phase ramps is highlighted. \label{Fig65}}
\end{figure}
\subsubsection{Hologram design}
Designing the phase hologram is as simple as computing the desired phase as a function of the transverse coordinates $x,y$. In general, the computed phases will be additions of phase singularities of order $m$ and oblique phases. An oblique phase $e^{ik_x x}$ represents the tilt that a prism would imprint on the beam, where $k_x $ is related to the apex of the prism \cite{Saleh2007,Richard2011}. The larger $k_x$ is, the more the beam will be tilted. In fact, $k_x = 2\pi / \Delta x$, where $\Delta x$ is the width of a $2\pi$ phase ramp (see Figure \ref{Fig65}). Then, the superposition of an oblique phase and a phase singularity yields a so-called pitchfork hologram (see Figure \ref{Fig65}), which is used to create tilted vortex beams \cite{Richard2011,Karimi2009}.
\subsubsection{Hologram print}
Once the hologram has been digitally created, it has to be printed. In order to have a good efficiency, CGHs need to have a smooth gray-scale to go from black ($2\pi$ phase shift) to white ($0$ phase shift). This fact depends on many factors, and the first of them is the printer. The chosen printer needs to provide a good gray-scale. Generally, inkjet printers not only have better gray-scales than laser printers, but they are also cheaper \cite{Roser2008}. Thus, all the holograms made during my thesis were done with an inkjet printer \hyperlink{http://www.canon-europe.com/For_Home/Product_Finder/Printers/Inkjet/PIXMA_iP4850/}{Canon iP4850 Ink}. Also, printers can print both in black and white or color mode. The difference between the two modes is that when printing in black and white, the gray tones are done by printing black dots. If the gray tone is darker, then the black dots are more packed; otherwise, they are more separated. Therefore, the fact that this is seen as gray is an optic effect, but it is not truly gray. Colour printing goes through all the gray tones by mixing colors in a RGB (red green blue) basis. I printed holograms in both configurations to carry out the characterization presented in section \ref{Ch5_CGHchara}, but their efficiency was not seen to clearly depend on the printing technique.
\subsubsection{Photography of the hologram}
\begin{figure}[tbp]
\centering
\includegraphics[width=13cm]{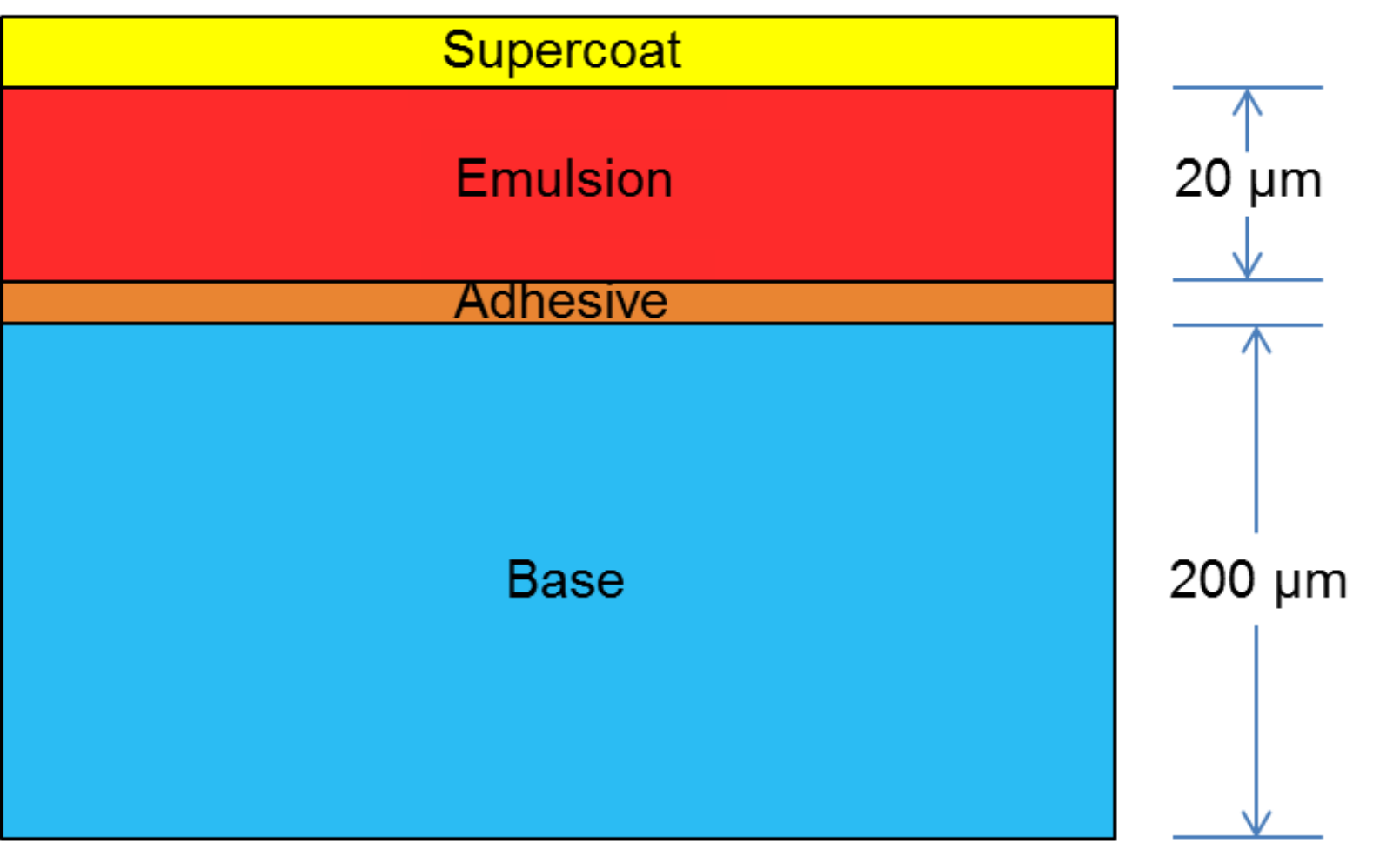} 
\caption{Sketch of a transverse section of a photographic film. The emulsion is sandwitched between the supercoat and adhesive layer, which separates the emulsion from the transparent base. The thickness of the emulsion is about 20 $\mu$m, and the thickness of the base is about 200 $\mu$m. \label{Fig66}}
\end{figure}
\begin{figure}[tbp]
\centering
\includegraphics[width=\columnwidth]{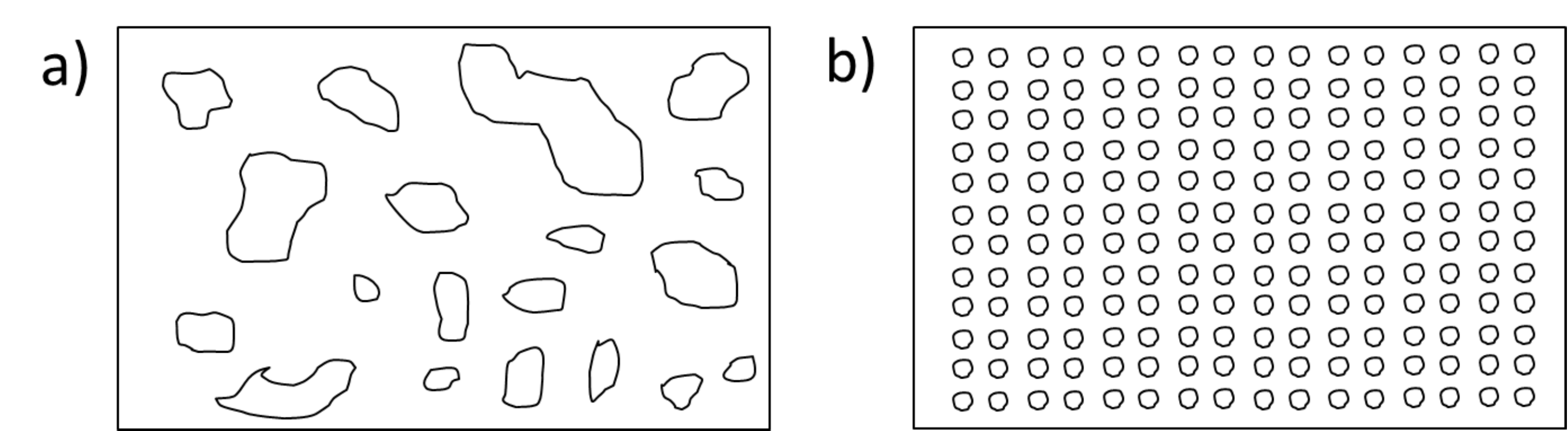} 
\caption{Schematics of a photographic emulsion for a) Generic photographic film with 100 lp/mm b) Gigabitfilm with 900 lp/mm. The generic one contains randomly located silver halide grains of different sizes, whilst the Gigabitfilm keeps both the size and the position of the grains in a much ordered manner \label{Fig67}}
\end{figure}

Once the hologram has been printed out, a picture needs to be taken. At this stage, several factors can play a very determinant role to obtain a good spiral thickness $d(\phi)$. The first one is the photographic film. A sketch of a photographic film can be found in Figure \ref{Fig66}. From top to bottom, it is formed by a supercoat, which protects the emulsion; an emulsion made of a photo-reactive material, which is generally composed of silver halide (AgBr) grains dispersed in a gelatine substrate; an adhesive layer so that the emulsion does not flow off the film; and a transparent base, which supports the rest of the film. Now, the key parameter for photographic films is the lines pairs per millimetre (lp/mm). One line pair is a pair of black and white lines next to each other. The larger the number of lp/mm is, the finer details the picture can resolve. There is a myriad of photographic films in the market, but the vast majority of them have a resolution of 100 lp/mm. In contrast, a German company called \hyperlink{http://www.gigabitfilm.com/}{Gigabitfilm}\footnote{http://www.gigabitfilm.com/} produces high resolution films with up to 900 lp/mm (see Figure \ref{Fig67}). These are the ones that have been used, granting a higher efficiency for the holograms. 
Another parameter that plays a very significant role is the camera. In order to make holograms, a \hyperlink{http://camerapedia.wikia.com/wiki/Minolta_SR-T_101}{Minolta manual reflex SR-T 101 camera} was used. It is important that the camera is reflex, so that what is seen by the viewfinder of the camera is seen in the same way by the objective (and film). Then, two parameters of the camera system need to be adjusted: shutter speed and aperture. The shutter speed, usually given in fractions of seconds, is the amount of time that the camera is permitting light from the lens to hit the film or digital sensor. A fast shutter speed is good for ``freezing" the action of a subject in motion. A slow shutter speed is good for purposely blurring the motion of a subject. After carrying out some batches of holograms, I have seen that holograms generally turn out better for relatively large exposition times (see section \ref{Ch5_CGHchara}). The aperture (or f-stop) refers to the size of the opening in the lens of the camera when a picture is taken. With aperture, smaller numbers (ex. f/4) mean a larger opening. Large numbers (ex. f/16) mean a small opening. Any time the amount of light is increased or decreased by a factor of two, the exposure changes by ``1 stop". Table \ref{T_apertures} shows the most common apertures. 
\renewcommand{\arraystretch}{1}
\begin{table} 
\footnotesize
\caption{Common lens apertures.}
\begin{center} 
\begin{tabular}{|c|c|c|c|c|c|c|c|c|}
\hline   larger aperture / more light & 1.7 & 2.8 & 4 & 5.6 & 8 & 11 & 16 & smaller aperture / less light \\ 
\hline
\end{tabular}
\end{center}
\label{T_apertures}
\end{table}
The aperture size of the lens is also related to the depth of field. A large aperture causes a small depth of field, and a small one causes a large depth of field. The depth of field of an image is related to the distance between the nearest and farthest objects in the scene that appear acceptably sharp in the image. In section \ref{Ch5_CGHchara}, the relation between f-stop, exposure time and hologram efficiency is shown. Also, the picture needs to be taken free of vibrations. In order to achieve that, the camera needs to be clumped on a stable tripod. Moreover, in order to even reduce more the vibrations, a trigger has been used to shoot the pictures. The last parameter that play a role on this step is the illumination of the printed CGH. Even though its dependence has not been numerically quantified, a uniform white illumination on the hologram is necessary to obtain a good photography.  
\begin{figure}[tbp]
\centering
\includegraphics[width=13cm]{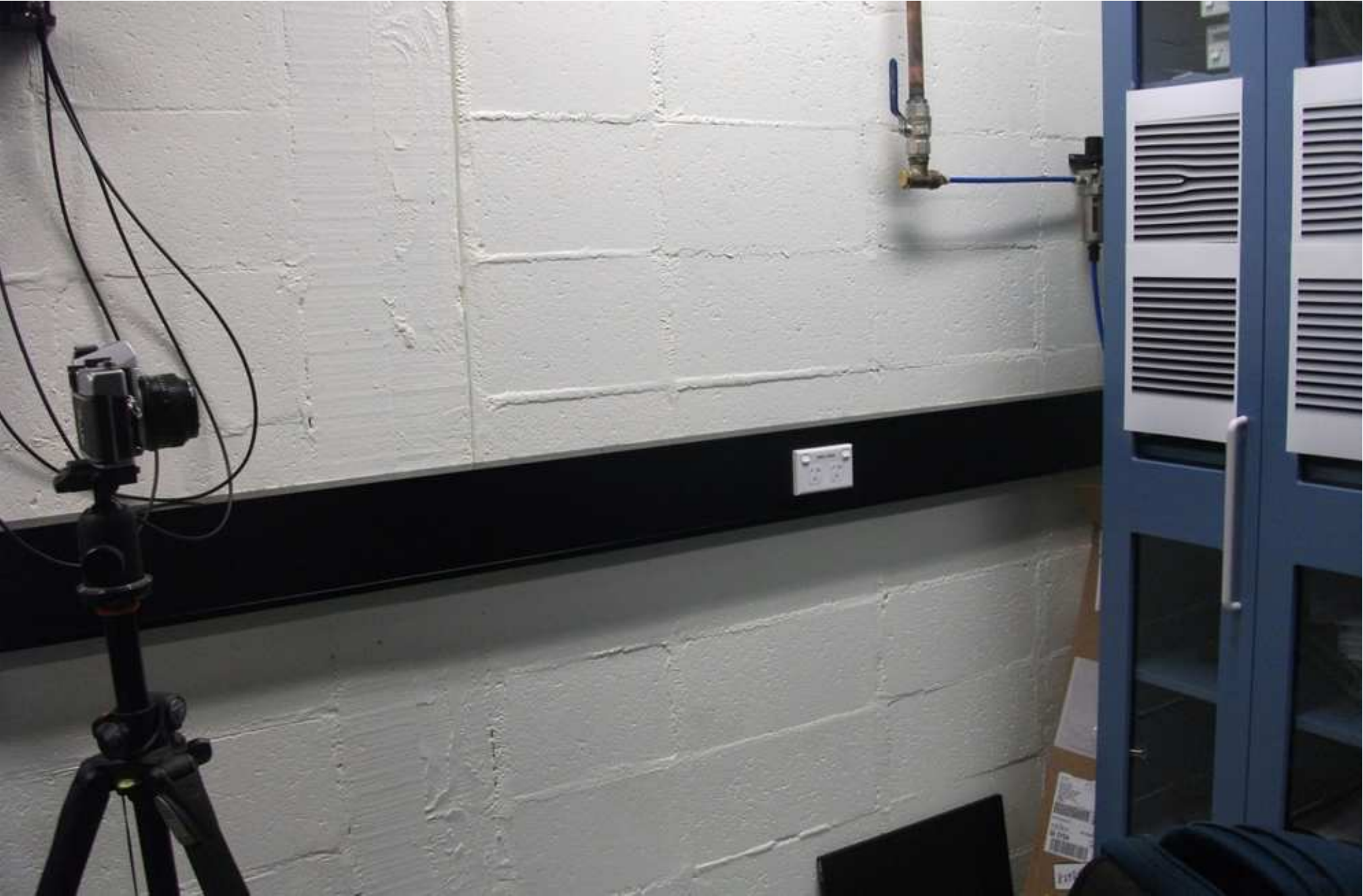} 
\caption{Photography set-up. The camera is mounted on a tripod at a certain distance of the printed holograms, which are stuck to a wall (or cupboard in this case). The illumination is white and uniform. \label{Fig68}}
\end{figure}
\subsubsection{Developing process} 
The development is the most important step in order to obtain efficient holograms. First, the film needs to be removed from the camera and placed in a developing tank, where the film is protected from light. This process needs to be done in a dark chamber, or somewhere where light is over 690 nm. Then, the whole developing process can be started, which is divided into the following steps:
\begin{enumerate}
\item \underline{\textsc{Development}}. The developer transforms the latent image\footnote{The latent image is an invisible image produced by the exposure to light of a photographic film. The chemical structure of the silver halides forming the emulsion changes, but those changes are not visible.} to metallic silver. Lots of different developers are available in the market, but I made my own one, as both the films and the application that they are used for (phase holograms) are not conventional. The main chemical needed for a developer is dihydroxybenzene or Catechol. This is mixed with NaOH and pure water using the following quantities:
\begin{itemize}
\item 0.6g catechol in 5ml of pure water.
\item 0.32g NaOH in 8ml of pure water
\item 300ml of pure water
\end{itemize}
To keep its developing properties, the mixing has to be made before being poured into the developing tank. It is important to continuously stir the developer while it is in contact with the film. This stage lasts 6min.
\item \underline{\textsc{Stop bath}}. Once the negative has been developed, the tanks needs to be quickly emptied and the stop bath needs to be poured. The stop bath is an acid mixture that halts the action of the developer by drastically changing the pH of the medium surrounding the film. The developer is a basic solution, and the stop bath is acid. It only needs to be stirred in for 30s. It is made of:
\begin{itemize}
\item 10ml acetic acid
\item 320 ml pure water
\end{itemize} 
\item \underline{\textsc{Bleaching}}. Bleaching is a crucial step in this fabrication process. Together with the fixing, it transforms the film from an amplitude to a phase-only hologram. There are different kinds of bleaching. Hariharan studied the efficiency of the different kinds of bleaching and reached the conclusion that the rehalogenising bleach allows for the greatest efficiency of phase-holograms \cite{Hariharan2002,Hariharan1971,Hariharan1971AO}. The principal difference between the rehalogenising bleaching process and the others is the addition of a rehalogenising agent, which in Hariharan experiments as well as mine is potassium bromide (KBr). The products used to make the rehalogenising bleach are the following ones:
\begin{itemize}
\item 0.3g potassium dichromate
\item 5g potassium bromide
\item 0.63ml sulfuric acid
\item 300ml pure water
\end{itemize}
A rehalogenising bleach converts the developed silver image into silver halides and modifies the gelatin matrix where the silver image has been formed. The new-created silver halides will be removed a posteriori in the fixing process, but the modified gelatin will stay. The sulphuric acid turns the medium acid so that the different chemical reactions can happen. The potassium dichromate dissolves the metallic silver created in the development and the potassium bromide re-converts the dissolved silver into silver halides. The solution needs to be stirred for 3min.
\item \underline{\textsc{Fixing}}. The fixer removes the silver salts, the unmodified gelatin (not exposed) and the supercoat. Instead, it leaves the transparent base and the modified gelatin whose thickness is directly proportional to the exposition - the most exposed areas have less modified gelatin removed. Usually, before the fixer is poured into the tank, the film is rinsed with pure water to remove all the substances left by the bleaching process. Then, the fixer is stirred in the developing tank for 1-2min. Two different fixers have been used yielding similar results:
\begin{enumerate}
\item 30ml Kodak T-Max fixer + 300ml pure water
\item 45g sodium thiosulfate + 325ml pure water
\end{enumerate}
After the fixing is done, the film is completely transparent and light-resistant. It is important to rinse the film for 2min after the fixer has been applied. Any residual fixer could corrode hologram. Then, it only needs to be dried and cut.
\end{enumerate}
\begin{figure}[tbp]
\centering
\includegraphics[width=\columnwidth]{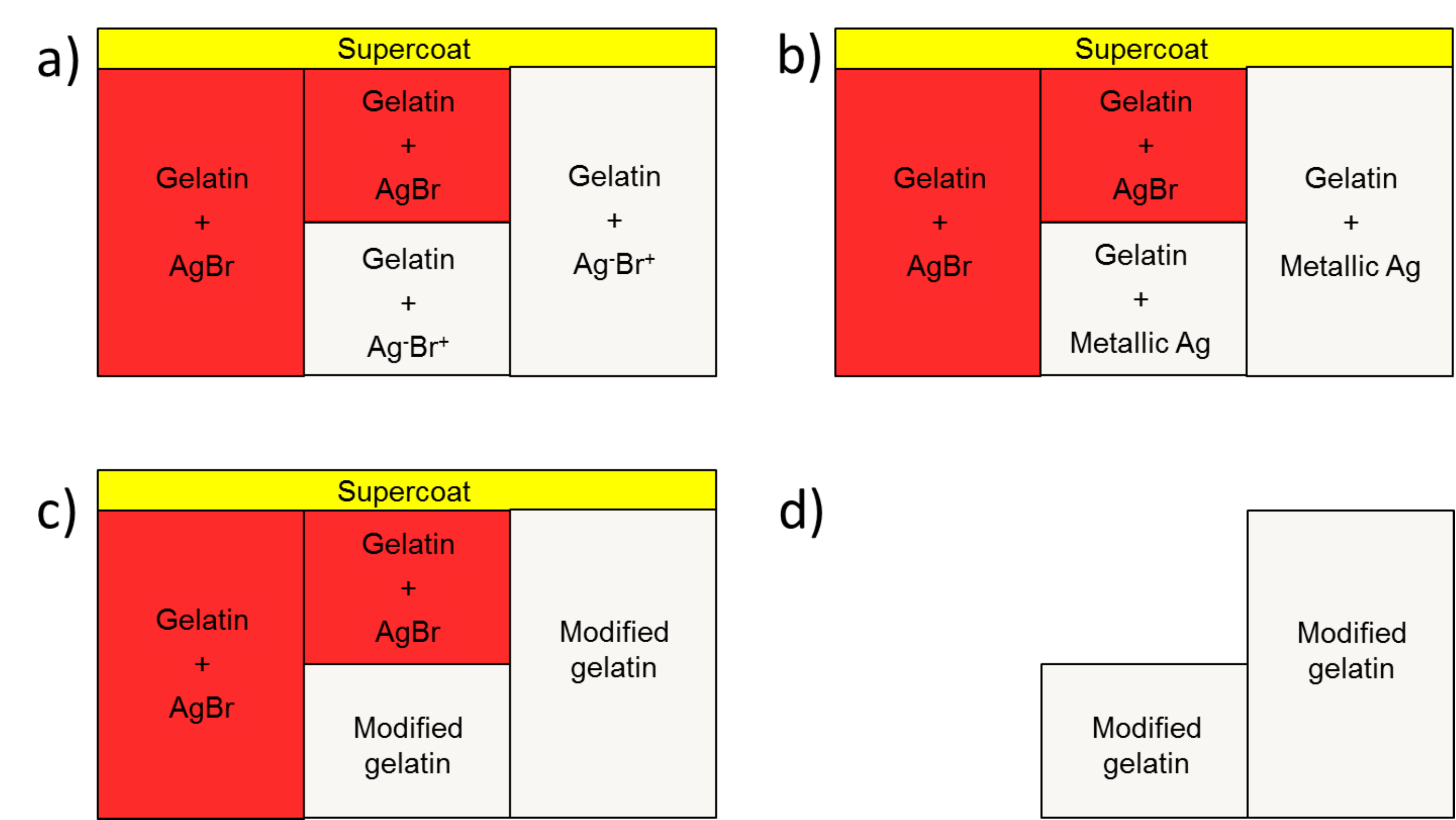} 
\caption{Phase-only CGH fabrication process. Grey (red) indicates that the film has (not) been exposed. a) Formation of latent image - silver halides become ionized. b) Developer converts ionized silver halides into metallic silver. c) Bleaching modifies the gelatin and converts the metallic silver back into silver halides. d) Fixing removes the supercoat and the unexposed emulsion. \label{Fig69}}
\end{figure}
Figure \ref{Fig69} summarizes the whole developing process explained in this section. In Figure \ref{Fig69}(a), the latent image is formed after exposure (Ag$^-$Br$^+$). Figure \ref{Fig69}(b) shows that the developer converts the ionized silver (Ag$^-$) into solid metallic Ag. Then, Figure \ref{Fig69}(c) displays the action of the stopping bath + bleaching. The metallic Ag is precipitated and there are cross-links in the gelatin molecules. The modified gelatin will remain after applying the fixing bath. Finally, Figure \ref{Fig69}(d) shows the action of the fixing bath, \textit{i.e.} removing the supercoat and the unexposed emulsion, while leaving the modified gelatine and the transparent base. A pattern of thickness given by the original picture is obtained as a result.

\section{Characterization of CGHs}\label{App2Cha}
\begin{figure}[tbp]
\centering
\includegraphics[width=13cm]{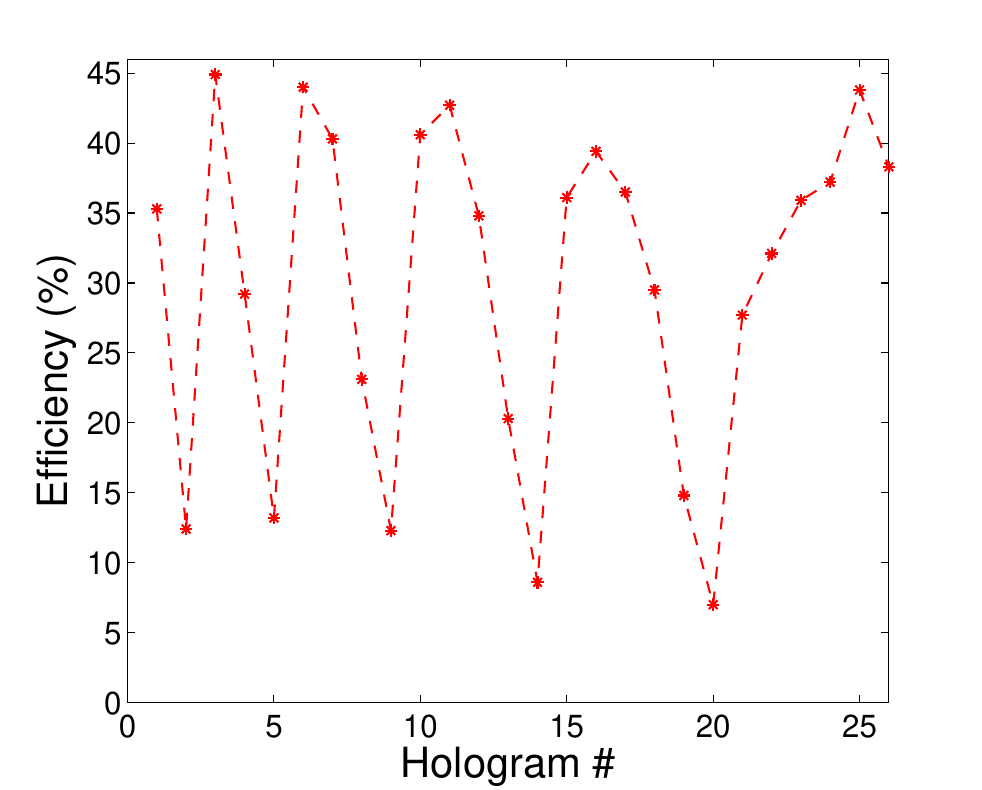} 
\caption{Efficiency of a batch of holograms as a function of shutter speed and aperture size. The holograms are the oblique phase given at Figure \ref{Fig65}. The dashed line is a guide of the eye. \label{Fig611}}
\end{figure}
The performance of holograms has been characterized in terms of the shutter speed and aperture size of the camera. All the rest of variables, \textit{i.e.} printing method, photographic film, illumination and chemical process have been previously optimized, so they will remain constant in this study. All the holograms characterized in this section have been imprinted with the oblique phase shown in Figure \ref{Fig65} has been used. Then, the efficiency has been computed for many different holograms. Every single hologram has been characterized by the shutter time and the f-stop used to illuminate it. The results are shown in Table \ref{T_CGH_charact} and Figure \ref{Fig611}. As it is observed in Figure \ref{Fig611}, for every single shutter speed, there is an optimum aperture size that maximizes the efficiency. For very slow shutter speeds, there is a very strong efficiency dependence on the aperture of the camera, whilst this dependence is much more smooth for short expositions. The results also suggest that slow shutter speed can produce higher efficiencies. Furthermore, it is seen that the maximum efficiencies are above $40\%$, in accordance to the typical results in the field \cite{He1995}. Nonetheless, it is worth commenting that higher efficiencies could be achieved if the chemical process was more accurate. Indeed, even though the phase-holograms are meant to be transparent, the ones used to carry out this characterization have absorption. If the chemical process was improved and the absorption was removed (maintaining the same thickness profile), then the efficiencies could reach values of $60 \%$. In fact, this is the result that is obtained when the efficiency is computed as a ratio between the intensity going to the first diffraction order and the intensity that is transmitted through the CGH.
\begin{table} 
\caption{Efficiency as a function of the aperture and the shutter speed. An aperture size of the kind 2.8-4 means that the aperture was set at an intermediate position between 2.8 and 4.}
\begin{center} 
\begin{tabular}{|c|c|c|c|}
\hline Hologram $\#$ & Shutter speed (s) & Aperture size & Efficiency ($\%$)\\ 
\hline 1 & 30 & 1.7 & 35.3\\
\hline 2 & 30 & 2.8 & 12.4\\
\hline 3 & 15 & 1.7 & 44.9\\
\hline 4 & 15 & 2.8 & 29.2\\
\hline 5 & 15 & 2.8-4 & 13.2\\
\hline 6 & 8 & 2.8 & 44.0\\
\hline 7 & 8 & 2.8-4 & 40.3\\
\hline 8 & 8 & 4 & 23.1\\
\hline 9 & 8 & 4-5.6 & 12.3\\
\hline 10 & 4 & 2.8-4 & 40.6\\
\hline 11 & 4 & 4 & 43.2\\
\hline 12 & 4 & 4-5.6 & 34.8\\
\hline 13 & 4 & 5.6 & 20.3\\
\hline 14 & 4 & 5.6-8 & 8.6\\
\hline 15 & 2 & 4 & 36.1 \\
\hline 16 & 2 & 4-5.6 & 39.4\\
\hline 17 & 2 & 5.6 & 36.5\\
\hline 18 & 2 & 5.6-8 & 29.5\\
\hline 19 & 2 & 8 & 14.8\\
\hline 20 & 2 & 8-11 & 7.0 \\
\hline 21 & 1 & 4 & 27.7\\
\hline 22 & 1 & 4-5.6 & 32.0\\
\hline 23 & 1 & 5.6 & 35.9\\
\hline 24 & 1 & 5.6-8 & 37.2\\
\hline 25 & 1 & 8 & 43.1\\
\hline 26 & 1 & 8-11 & 38.3\\
\hline
\end{tabular}
\end{center}
\label{T_CGH_charact}
\end{table}

\section{Holograms - separation of diffraction orders}\label{App2Sep}
\begin{figure}[tbp]
\centering
\includegraphics[width=\columnwidth]{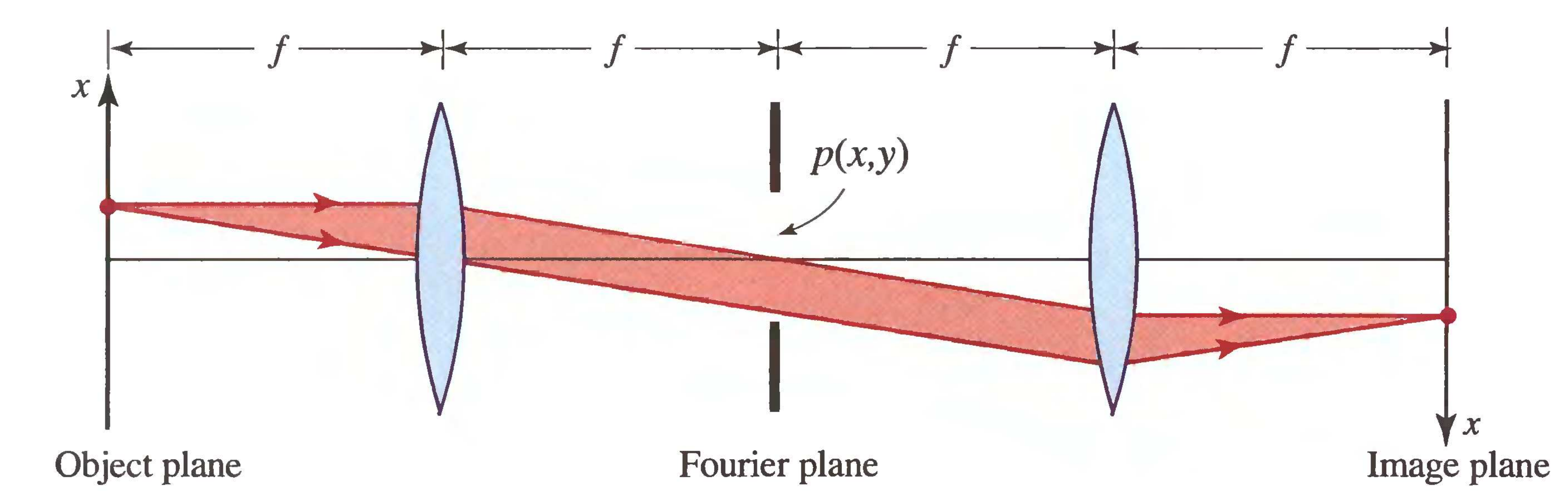} 
\caption{Sketch of a 4-f filtering system. The field at the object plane is Fourier-transformed by the first lens. At the Fourier plane, the diffraction orders split and a filter (either an iris or a pinhole) selects the 1st diffraction order. This order is re-imaged by the second lens. The figures has been copied from \cite{Saleh2007} with the permission of Prof. M.C. Teich\label{F_B1}}
\end{figure}

Here, the experimental basics to separate and isolate the 1st diffraction order of a hologram are explained\footnote{Note that all the equations given here are valid for both CGHs and SLMs.}. In section \ref{Ch5_Imod}, a method to separate the different contributions from a hologram has been explained. Indeed, if an oblique phase $\exp (ik_x x)$ is added to the hologram expression, the different diffraction orders of a hologram are separated by an angle $\Delta \theta_x = \lambda / \Delta x $, where $\Delta x = 2\pi/k_x$. Now, this implies that at a distance $z$ from the hologram the different diffraction orders are separated a transverse distance:
\begin{equation}
x_0= z \tan (\Delta \theta_x) \approx \dfrac{\lambda z}{\Delta x} \label{E_x}
\end{equation}
Nevertheless, the width of the diffraction orders also grows following Gaussian optics. That is, if the beam waist at the hologram plane is $w_0$, then the waist of every single diffraction order at a distance $z$ from the hologram is 
\begin{equation}
w=w_0\sqrt{l\left( 1+\left( \dfrac{z}{z_0} \right)^2 \right)} \approx \dfrac{\lambda z\sqrt{l}}{\pi w_0} \label{E_w}
\end{equation}
where $l$ is the order of the phase singularity given by the hologram. It is clear that the condition that assures that two diffraction orders are separated is 
\begin{equation}
x_0 \gg w \Longrightarrow \dfrac{\pi w_0}{\sqrt{l}} \gg  \Delta x \label{E_x0-w}
\end{equation}
Note that equations (\ref{E_x})-(\ref{E_x0-w}) are valid at distances such that a free-space Fourier Transform of the field at the plane of the hologram is carried out. In this case, the complex amplitude of the plane wave components at the plane of the hologram are related to different positions in the Fourier plane. Experimentally, a free-space Fourier Transform is carried out when the distance to the hologram $z$ is
\begin{equation}
z \gg  \dfrac{2\pi w^2}{\lambda \ \sqrt{ \left( \dfrac{2 \pi w}{\Delta x} \right)^2 - 4}}  \label{E_orders}
\end{equation}
Due to lack of space or need to make set-ups compact, it is not always possible to propagate beams. However, there is another way of doing a Fourier Transform, and that is using a lens. A widely used set-up to separate the diffraction orders of a hologram is a 4-f filtering system. A sketch of a 4-f filtering system is shown in Figure \ref{F_B1}. 
On one hand, the diffraction orders are separated a distance $x_0$ at the Fourier plane:
\begin{equation}
x_0 = \dfrac{k_x \lambda f}{2 \pi} = \dfrac{\lambda f}{\Delta x}
\end{equation}
where $f$ is the focal distance of the lens. On the other hand, the waist of a diffraction order at the focus of a lens is given by Gaussian optics:
\begin{equation}
w = \dfrac{f\lambda \sqrt{l}}{\pi w_0}
\end{equation}
Similarly to the free-space case, the condition that needs to be fulfilled is $x_0 \gg w $, which yields the exact same condition given by equation (\ref{E_orders}). A rule of thumb to separate the diffraction orders is:
\begin{equation}
\Delta x < \dfrac{w_0}{10}
\end{equation}

\section{SLM characterization}\label{App2SLM}
Here, the efficiency of the SLM is characterized as a function of some controllable parameters in the laboratory, such as the polarization of the incoming beam, the angle formed between the incident beam and the normal to the SLM, and the number of $2\pi$ phase jumps displayed at the hologram (see equation (\ref{E_kx})). Contrary to CGHs made of photographic films, the performance of the SLM is very dependent on all these parameters.
\begin{figure}[tbp]
\centering
\includegraphics[width=13cm]{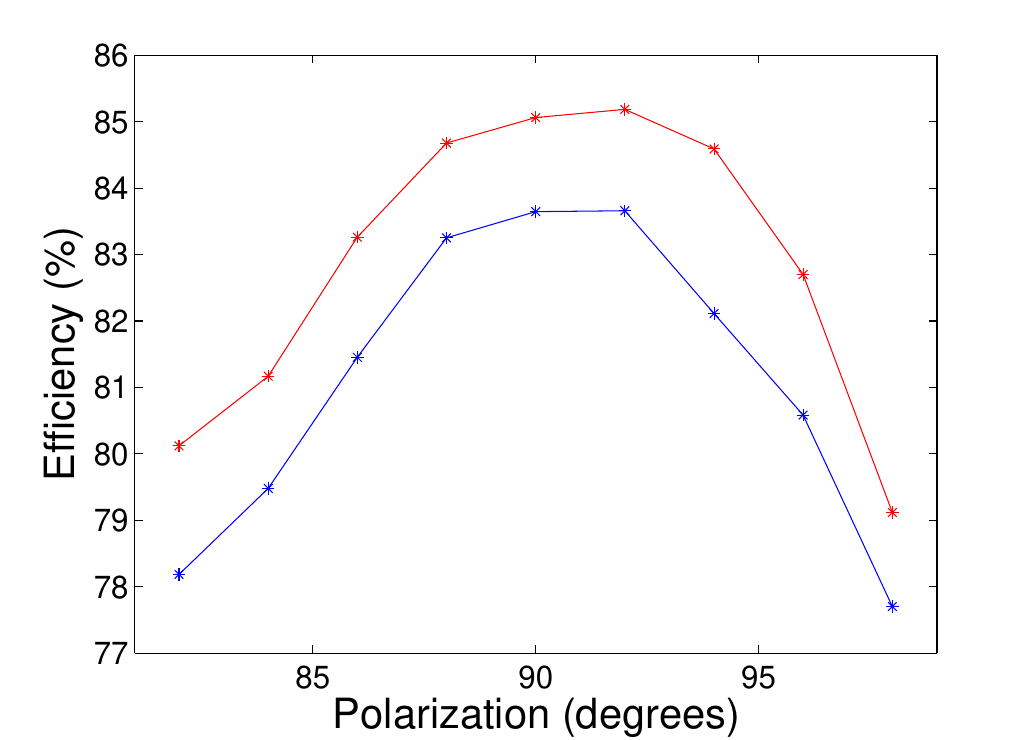}  
\caption{Efficiency dependence on the polarization angle. The angle formed by the normal of the incidence and the beam axis is 8'. The red markers display the efficiency in the Y axis, whereas the blue markers display the efficiency in the X axis. \label{Fig614}}
\end{figure}
\begin{figure}[tbp]
\centering
\includegraphics[width=13cm]{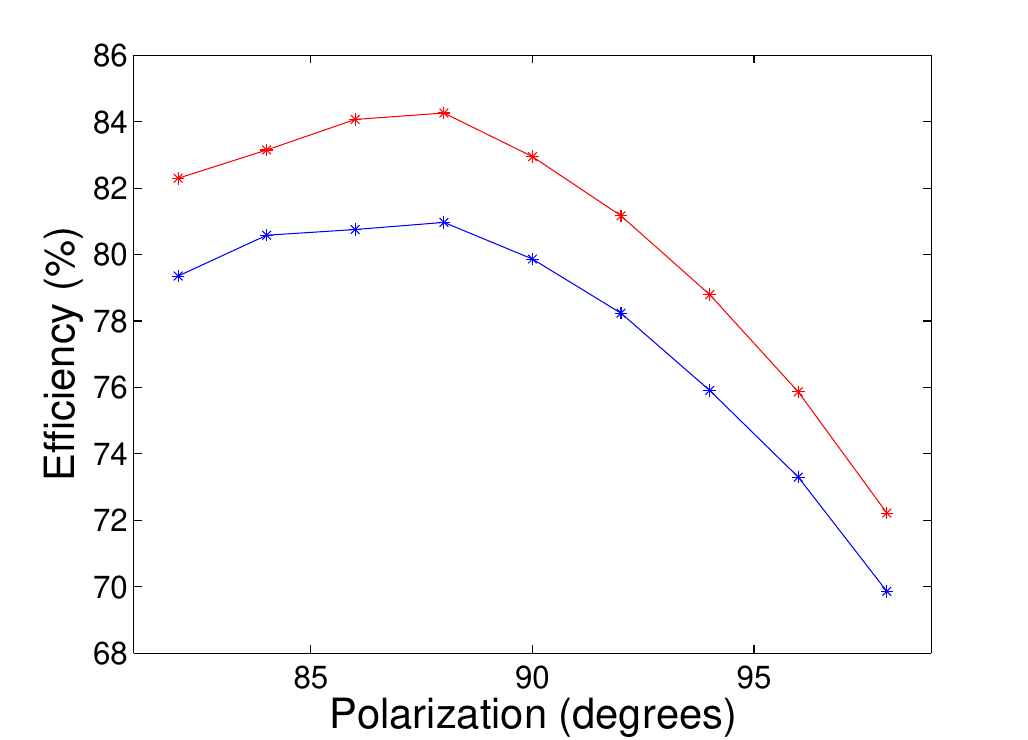} 
\caption{Efficiency dependence on the polarization angle. The angle formed by the normal of the incidence and the beam axis is 5 degrees. The red markers display the efficiency in the Y axis, whereas the blue markers display the efficiency in the X axis.\label{Fig615}}
\end{figure}
\begin{figure}[tbp]
\centering
\includegraphics[width=13cm]{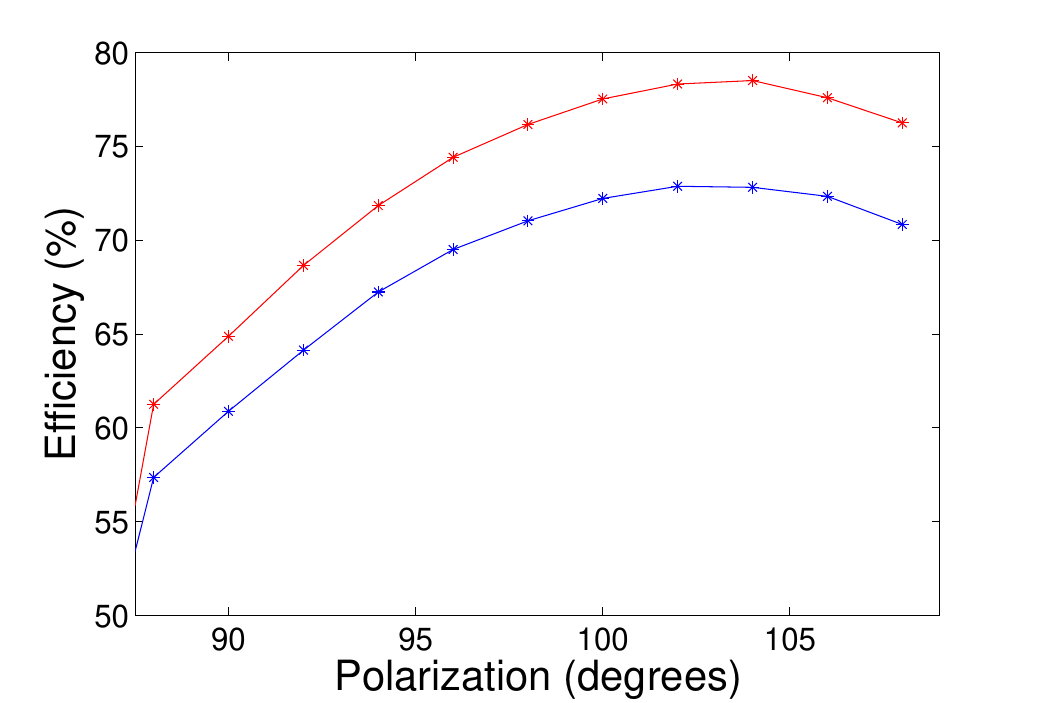} 
\caption{Efficiency dependence on the polarization angle. The angle formed by the normal of the incidence and the beam axis is $-$14 degrees. The red markers display the efficiency in the Y axis, whereas the blue markers display the efficiency in the X axis.\label{Fig616}}
\end{figure}
Figures \ref{Fig614} and \ref{Fig615} show the influence of the polarization and the angle of incidence on the efficiency\footnote{The hologram used in both figure gives twenty-four $2\pi$ phase jumps.}. In fact, in each of the Figures, two lines of data points are plotted. The red markers correspond the efficiency measured when the phase hologram is $t(x,y)=\exp (iky)$, while the blue markers are obtained with $t(x,y)=\exp (ikx)$. That is, the first (red) one measures the efficiency steering the beam in the $x$ axis, and the second one in the $y$ axis (the $y$ axis is perpendicular to the plane of the optical table). For all the cases, the vertical axis displays the efficiency of the SLM measured in the way it has been described above; whilst the horizontal axis displays the polarization angle. This polarization angle is a direct readout of the angular position of a rotation-mount where a half-wave plate is hold. Hence, the value is arbitrary, as a rotation of the SLM would make it vary. However, the number is given so that it can be checked how the maximum is displaced when the SLM is tilted. The first conclusion that can be drown from Figures \ref{Fig614} and \ref{Fig615} is that the efficiency significantly depends on the polarization of the incident beam. Second, it is observed that when the angle between the normal to the SLM and the beam axis increases, the efficiency decreases more rapidly for a same range of variation of polarization degrees. That is, Figure \ref{Fig614}, which corresponds to almost normal incidence, shows that a 10 degree variation in polarization is matched with a 5$\%$ variation in efficiency. Figure \ref{Fig615}, which depicts the efficiency data obtained when the angle between the incident beam and the normal to the SLM is 5 degrees, shows that a 10 degree variation in polarization is matched with a 12$\%$ variation in efficiency. Furthermore the maximum value for the efficiency drops a little and changes its position. The position of the maximum varies 5 degrees approximately, which is approximately the angle that the SLM has been rotated. In Figure \ref{Fig616}, where the results obtained when the angle between the normal to the SLM and the incident beam is $-$14 degrees are displayed, the same trends commented previously are maintained. That is, the efficiency of the maximum decreases; the angle for which the maximum of efficiency is moved as much as the angle between the normal of the SLM and the incident beam; and the gap between the efficiency in the $x$ and $y$ axis increases. The pixels are squares, so one would expect that the efficiencies in both axis were the same. However, a tilt of the SLM breaks the symmetry between the two axis and accounts for the measured difference.\\\\
Next, the importance of the hologram displayed on the SLM is characterized. In Figures \ref{Fig614}-\ref{Fig616} the same hologram was used. The hologram is the oblique phase in Figure \ref{Fig65} for the efficiency in the $y$ axis, and a $\pi/2$ rotation of it for the $x$ axis. This hologram contains twenty-four $2\pi$ phase jumps. In Figure \ref{Fig617} the efficiency is plotted as a function of the number of $2\pi$ phase jumps given by the hologram. The angle of incidence is set at 8' (same as Figure \ref{Fig614}) and the optimum position of the half-wave plate angle is set to maximize the efficiency. Obviously, as given by equation (\ref{E_kx}), the larger number of $2\pi$ phase jumps the hologram gives, the larger is the angle between the 0th and the 1st diffraction order. Then, it can be observed that there is an almost linear relation between the efficiency of the SLM and the number of $2\pi$ phase jumps in for this range of values. The larger the number of $2\pi$ phase jumps given by the hologram, the lesser pixels per ramp are used, and consequently the efficiency is reduced. 
\begin{figure}[tbp]
\centering
\includegraphics[width=13cm]{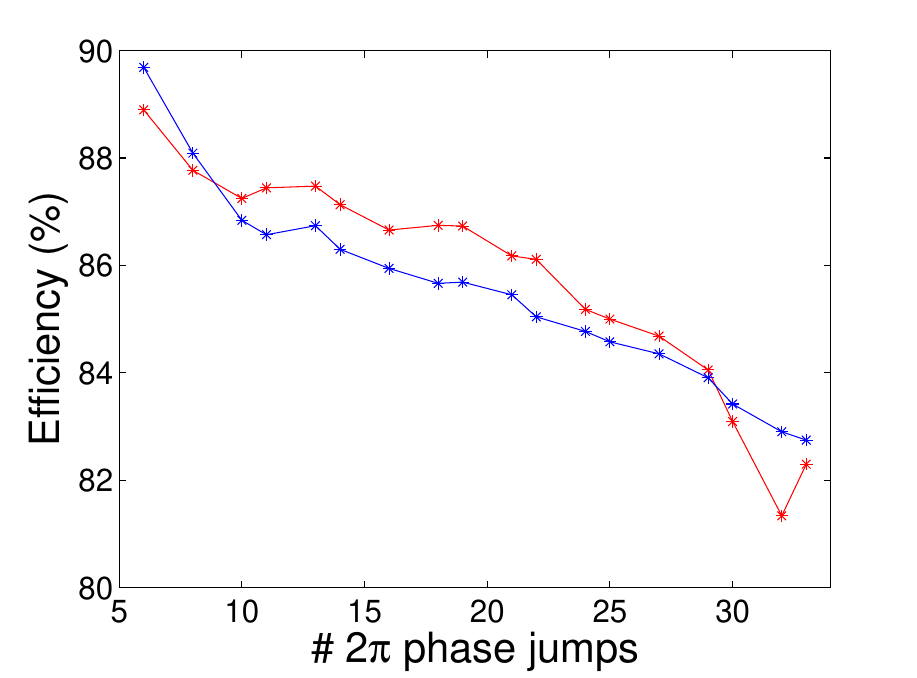} 
\caption{Efficiency dependence on the number of $2\pi$ phase jumps displayed on the SLM. The angle formed by the normal to the SLM and the beam axis is 8'. The polarization is set to optimize the efficiency. The red markers display the efficiency in the $y$ axis, whereas the blue markers display the efficiency in the $x$ axis.\label{Fig617}}
\end{figure}
To finalize, the Lookup Table (LUT) provided by the producer of the SLM (BNS) was tested. All the data points collected to plot Figures \ref{Fig614}-\ref{Fig617} were done at $\lambda=632.8$ nm, using a He-Ne laser (see Figure \ref{Fig613}). The phase shift given by an SLM highly depends on $\lambda$. The transmittance function created by the liquid crystals depends on $\lambda$ as shown in equation (\ref{E_txy_general}). This means that the efficiency of the SLM can vary a lot depending on $\lambda$. In order to account for this dependence and some other imperfections of the SLM such as non-linear behaviour of the pixels, LUTs change the theoretical value of pixels to give them an effective one which increases the performance of the hologram as a whole. In \cite{Richard2011}, a linear method to correct the imperfections of the SLM was developed. The idea is to multiply the phase of the phase-hologram $t(x,y)$ by a contrast constant $C$. Then, when the new $t'(x,y)=C t(x,y)$ is such that $t'(x,y)<-\pi$ or $t'(x,y)>\pi$, those phases are re-set to $-\pi$ or $\pi$. Using this method, Bowman \textit{et al.} managed to greatly increase the efficiency of an SLM. Using a hologram with ten $2\pi$ phase ramps and maximizing the polarization angle at an angle of incidence of 8', I tried to use the same method to improve the efficiency of the SLM. The results are shown in Figure \ref{Fig618}. The $y$ axis shows the efficiency and $x$ axis the contrast parameter $C$. Leaving $C=1$ means not doing any changes to the LUT. It can be observed that the highest efficiency is achieved for $C=1.02$. Hence, the LUT had been properly optimized at the frequency of the HeNe laser. I believe that whenever a new SLM or wavelength are used, this is a sanity check that should be done. In fact, some producers do not send LUTs along with the SLM. In those cases, it is highly recommended to apply the method described in \cite{Richard2011} to improve the performance of the SLM.
\begin{figure}[tbp]
\centering
\includegraphics[width=13cm]{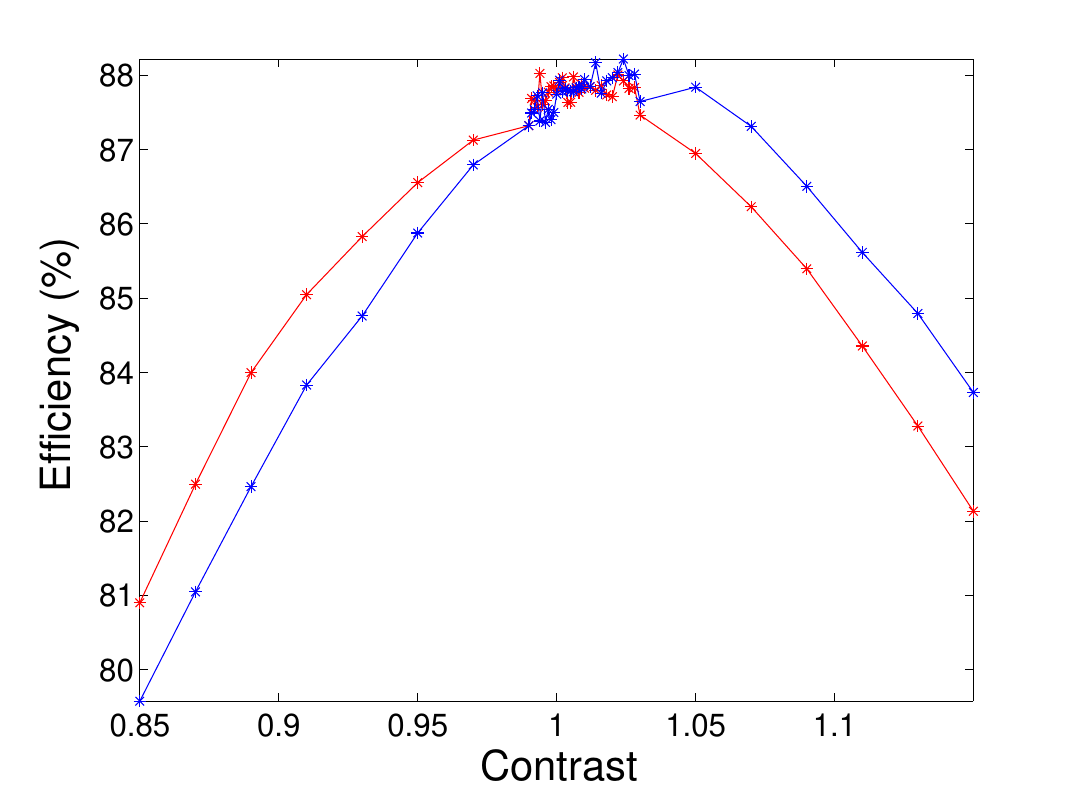} 
\caption{Efficiency dependence on the contrast. The hologram in consideration displays ten $2\pi$ phase ramps. The angle of incidence is 8', and the polarization is optimized to maximize the efficiency. \label{Fig618}}
\end{figure}

\chapter{Proof of $I_l^{L/R}=I_{-l}^{R/L}$}
\graphicspath{{ch_appendix3/}} 
\label{Appendix3}

In this appendix, the equation (\ref{Iq}) is demonstrated. In order to do that, the following relations given in chapter \ref{Ch6} are used: 
\renewcommand{\arraystretch}{1.5}
\begin{eqnarray}
I_l^{L/R} & = &\int_{-\infty}^{\infty} \int_{-\infty}^{\infty} |\mathbf{E}_{+1/-1,l}^{\mathbf{t}}|^2 {dx} {dy} \\
\Etpq  & = & \Atpqp\spphat  + \Btpqpp\sppmhat \\
\spphat^* \cdot \sppmhat & = & 0
\end{eqnarray}
Using all of them, it can be obtained that $I_l^{L/R}$ can be expressed as:
\renewcommand{\arraystretch}{1}
\begin{equation}
I_l^{L/R} = \int_{-\infty}^{\infty} \int_{-\infty}^{\infty} |A_{+1/-1,l}^{\mathbf{t}}(x,y)|^2 + |B_{+1/-1,l}^{\mathbf{t}}(x,y)|^2 {dx} {dy}
\label{iq}
\end{equation}
Following an identical procedure, the following equation yields for $I_{-l}^{R/L}$:
\begin{equation}
I_{-l}^{R/L} = \int_{-\infty}^{\infty} \int_{-\infty}^{\infty} |A_{-1/+1,-l}^{\mathbf{t}}(x,y)|^2 + |B_{-1/+1,-l}^{\mathbf{t}}(x,y)|^2 {dx} {dy}
\label{i-q}
\end{equation}
Choosing the mirror symmetric operator to be $\Mz = M_{\{x \rightarrow -x \}}$, then equation (\ref{-Epq}) can be used to get a relation between the coefficients in equations (\ref{iq},\ref{i-q}):
\begin{equation}
\begin{array}{lll}
\Etpq  & = & \Atpqp\spphat  + \Btpqpp\sppmhat  \\ 
= \Mz \Etppq &=&  \Mz \left( A_{-p,-l}^{\mathbf{t}}(x,y) \sppmhat + B_{-p,-l}^{\mathbf{t}}(x,y) \spphat  \right) \\
& = &  - A_{-p,-l}^{\mathbf{t}}(-x,y) \spphat - B_{-p,-l}^{\mathbf{t}}(-x,y) \sppmhat 
\end{array}
\end{equation} 
which implies that $\Atpqp = - A_{-p,-l}^{\mathbf{t}}(-x,y)$ and $\Btpqpp = B_{-p,-l}^{\mathbf{t}}(-x,y)  $, due to the orthogonality of $\spphat$ and $\sppmhat$. Consequently, it follows that:
\renewcommand{\arraystretch}{1.5}
\begin{equation}\begin{array}{ll}
I_l^{L/R} & = \displaystyle \int_{-\infty}^{\infty} \int_{-\infty}^{\infty} |A_{+1/-1,l}^{\mathbf{t}}(x,y)|^2 + |B_{+1/-1,l}^{\mathbf{t}}(x,y)|^2 {dx} {dy}  \\
& = \displaystyle\int_{-\infty}^{\infty} \int_{-\infty}^{\infty} |- A_{-1/+1,l}^{\mathbf{t}}(-x,y)|^2 + | - B_{-1/+1,l}^{\mathbf{t}}(-x,y)|^2 {dx} {dy} \\ 
& = \displaystyle\int_{-\infty}^{\infty} \int_{-\infty}^{\infty} |A_{-1/+1,-l}^{\mathbf{t}}(x^\prime,y)|^2 + |B_{-1/+1,-l}^{\mathbf{t}}(x^\prime,y)|^2 {dx^\prime} {dy} = I_{-l}^{R/L} 
\end{array}\end{equation}
as the integrations limits remain the same under the change $x \rightarrow -x^\prime$. \\

\chapter{Intensity plots of multipolar fields}
\graphicspath{{ch_appendix4/}} 
\label{Appendix4}

\begin{figure}[tbp]
\centering
\includegraphics[width=10.5cm]{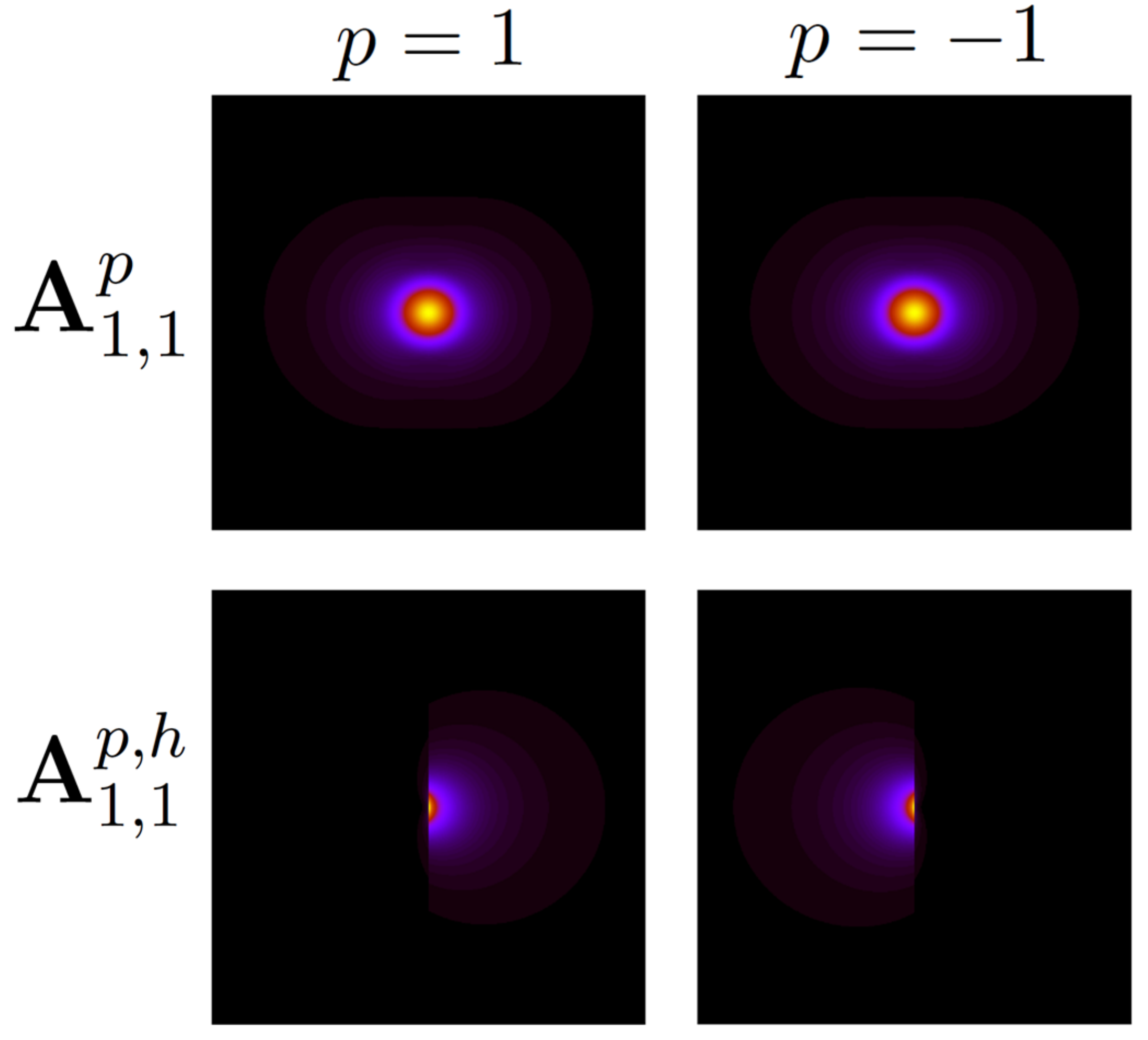} 
\caption{$\mathbf{A}_{1,1}^{p}$ and $\mathbf{A}_{1,1}^{p,h}$ for $p=\pm 1$. The plot area is a square of side $s=12\pi \lambda$. A contour of $0.5\pi \lambda$ around the origin has been removed to plot $\mathbf{A}_{1,1}^{p,h}$.  \label{D1}}
\end{figure}

In the current appendix, intensity distributions of multipolar fields are displayed. Bearing their spatial shape in mind is of special importance to understand the excitation of WGMs (section \ref{Ch3_WGM}), the suppression of backscattering (\ref{Ch4_Kerker}), and the intensity profiles recorded by the CCD camera in the experiments done in chapter \ref{Ch7}. Furthermore, it is fairly common to forget about the fact that there are four different kinds of multipolar fields depending on the radial function used: $j_j(kr)$, $n_j(kr)$, $h^{(I)}_j(kr)$, and $h^{(II)}_j(kr)$ . Also, as it has been shown across this whole thesis, the multipolar fields can be classified either in terms of parity or helicity. Hereafter, some frequently used multipolar fields are displayed. They are displayed for both parities, helicities, and typically used radial functions ($j_j(kr)$ and $h^{(I)}_j(kr)$). The plots are done in the $z-x$ plane, $z$ being the horizontal axis and $x$ the vertical one. Then, because they are cylindrically symmetric around $z$, the expression in the other planes can be deduced by applying a $2\pi$ revolution symmetry. The axis are normalised to the wavelength, so they are dimensionless. All the units are normalized to $\lambda$. Also, due to the properties of the Wigner D-matrices $\Djmp$ \cite{Tung1985}, the plots have only been done for $m_z>0$, as it can be observed that the following properties hold:
\begin{eqnarray}
\vert \mathbf{A}_{jm_z}^{p} \vert ^2 = \vert \mathbf{A}_{j(-m_z)}^{-p} \vert^2 \\
\vert \mathbf{A}_{jm_z}^{(y)} \vert ^2 = \vert \mathbf{A}_{j(-m_z)}^{(y)} \vert^2
\end{eqnarray}
First, the dipolar orders are displayed. It is apparent from Figure \ref{D1} (and it can be re-checked with all the rest of Figures displaying multipoles with well-defined helicity) that equation (\ref{E_Ajmsemi}) holds. In fact, equation (\ref{E_Ajmsemi}) is a consequence of the fact that
\begin{equation}
M_{\zhat} \vert \mathbf{A}_{jm_z}^{p} \vert^2 = \vert \mathbf{A}_{jm_z}^{-p} \vert^2
\end{equation}
Second, it can also be observed that $\Ajmpp$ has the same behaviour for the two $z<0$ and $z>0$ semi-spaces, whilst $\mathbf{A}_{jm_z}^{p,h}$ does not. This property is the basis of the measurements carried out in chapter \ref{Ch7}. Third, the stationary wave behaviour of the Bessel function is only observed for $ \mathbf{A}_{jm_z}^{(y)}$, but not for $\Ajmpp$. Remember that a spherical Bessel function $j_j(kr)$ is a superposition of two spherical Hankel functions $j_j(kr)=h_j^{(I)}(kr)+h_j^{(2)}(kr)$, where the spherical Hankel functions have a respective outwards and inwards radiative behaviour (see equations (\ref{E_jj}, \ref{E_hj}) in section \ref{Ch1_multipoles}). In addition, looking at all the captions of the different Figures, it is observed that the plot area does not remain constant. This has been done to visualize the spatial shape of the multipolar fields. Indeed, high order Hankel multipolar fields are highly evanescent, therefore the plot area has been reduced. In contrast, high order Bessel multipoles require a larger plot area. Also note that a central region surrounding the origin of coordinates has been removed from the plots of Hankel multipoles. The reason for this is that Hankel functions are singular at the origin. Finally, Figures \ref{D11}-\ref{D16} depict multipolar beams with $j=15$. Note the similarities between Figure \ref{D11} and Figure \ref{F_intensity} in section \ref{Ch3_WGM}, where almost a single mode was excited. And last but not least, Figure \ref{D16}, which depicts $\mathbf{A}_{15,1}^{(y)}$, shows an intensity plot which highly resembles the typical plots for WGMs field distribution. It is interesting to see that those plots are achieved when $m_z=1$ and $j$ is large, in contrast to Oraevsky's definition of WGM, which states that a WGM is a multipolar field with very large $j$ and $m_z=j$ \cite{Oraevsky2002}. That, of course, does not affect the Q factor of the sphere, as the Q factor is only given by the linewidth of a Mie resonance, usually the first one for WGMs \cite{Oraevsky2002}. And in Mie theory the $m_z$ value is degenerated, therefore a multipolar field $\mathbf{A}_{15,1}^{(y)}$ has the same Q-factor as $\mathbf{A}_{15,15}^{(y)}$.
\begin{figure}[tbp]
\centering
\includegraphics[width=10.5cm]{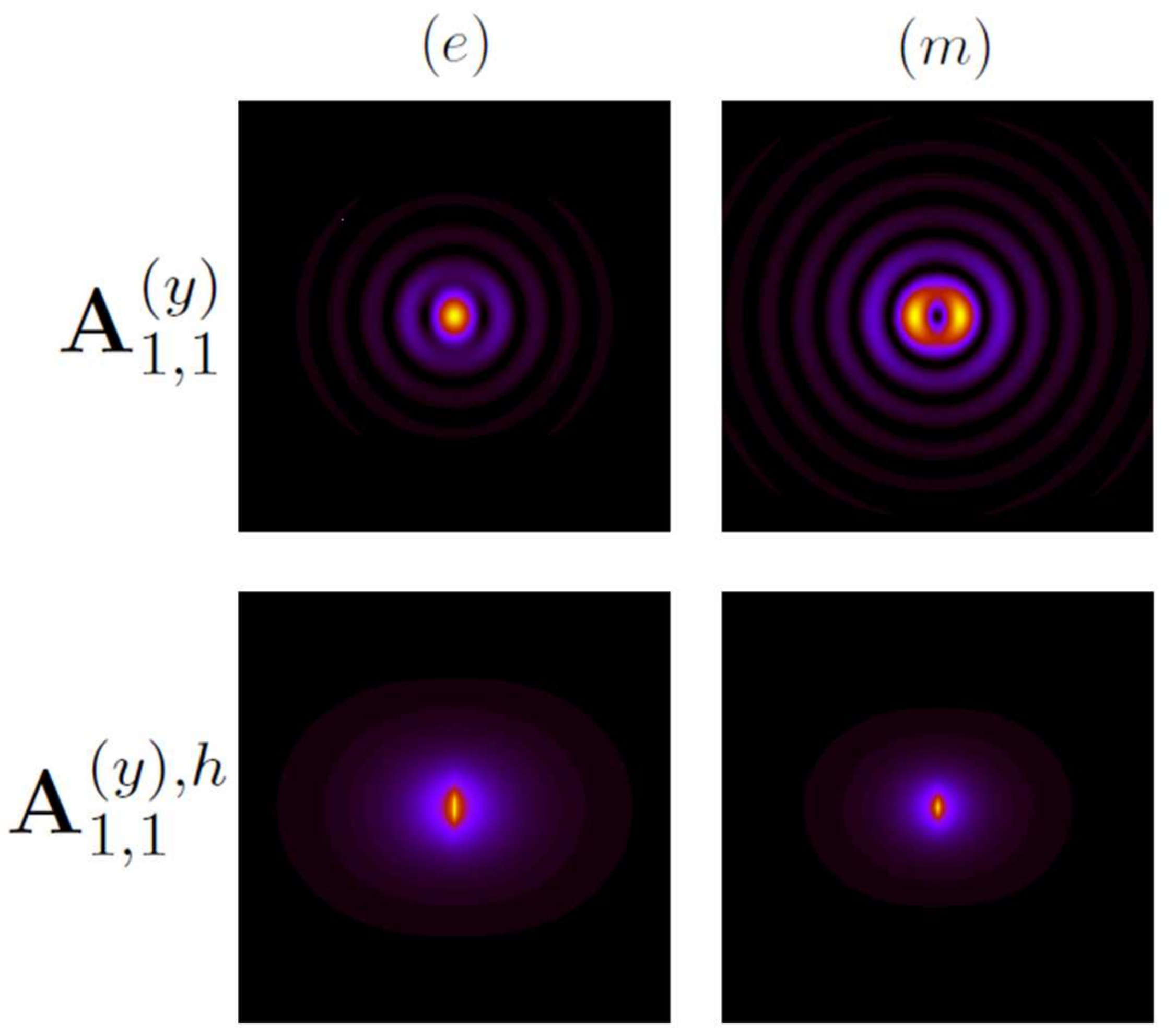} 
\caption{$\mathbf{A}_{1,1}^{(y)}$ and $\mathbf{A}_{1,1}^{(y),h}$ for $(y)=(e),(m)$. The plot area is a square of side $s=12\pi \lambda$. A contour of $0.5\pi \lambda$ around the origin has been removed to plot $\mathbf{A}_{1,1}^{(y),h}$. \label{D2}}
\end{figure}
\begin{figure}[tbp]
\centering
\includegraphics[width=10.5cm]{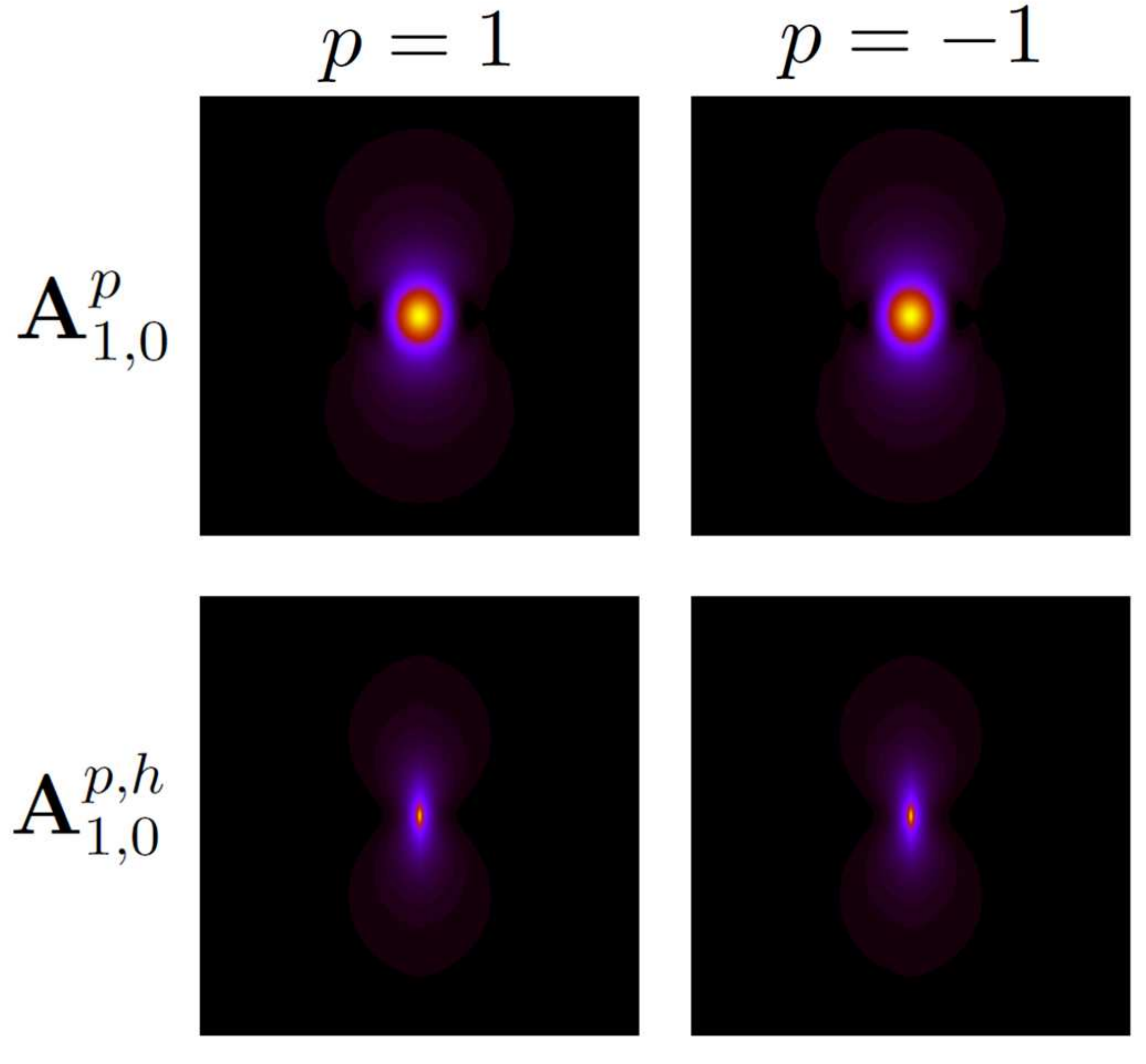} 
\caption{$\mathbf{A}_{1,0}^{p}$ and $\mathbf{A}_{1,0}^{p,h}$ for $p=\pm 1$. The plot area is a square of side $s=12\pi \lambda$. A contour of $0.5\pi \lambda$ around the origin has been removed to plot $\mathbf{A}_{1,0}^{p,h}$. \label{D3}}
\end{figure}
\begin{figure}[tbp]
\centering
\includegraphics[width=10.5cm]{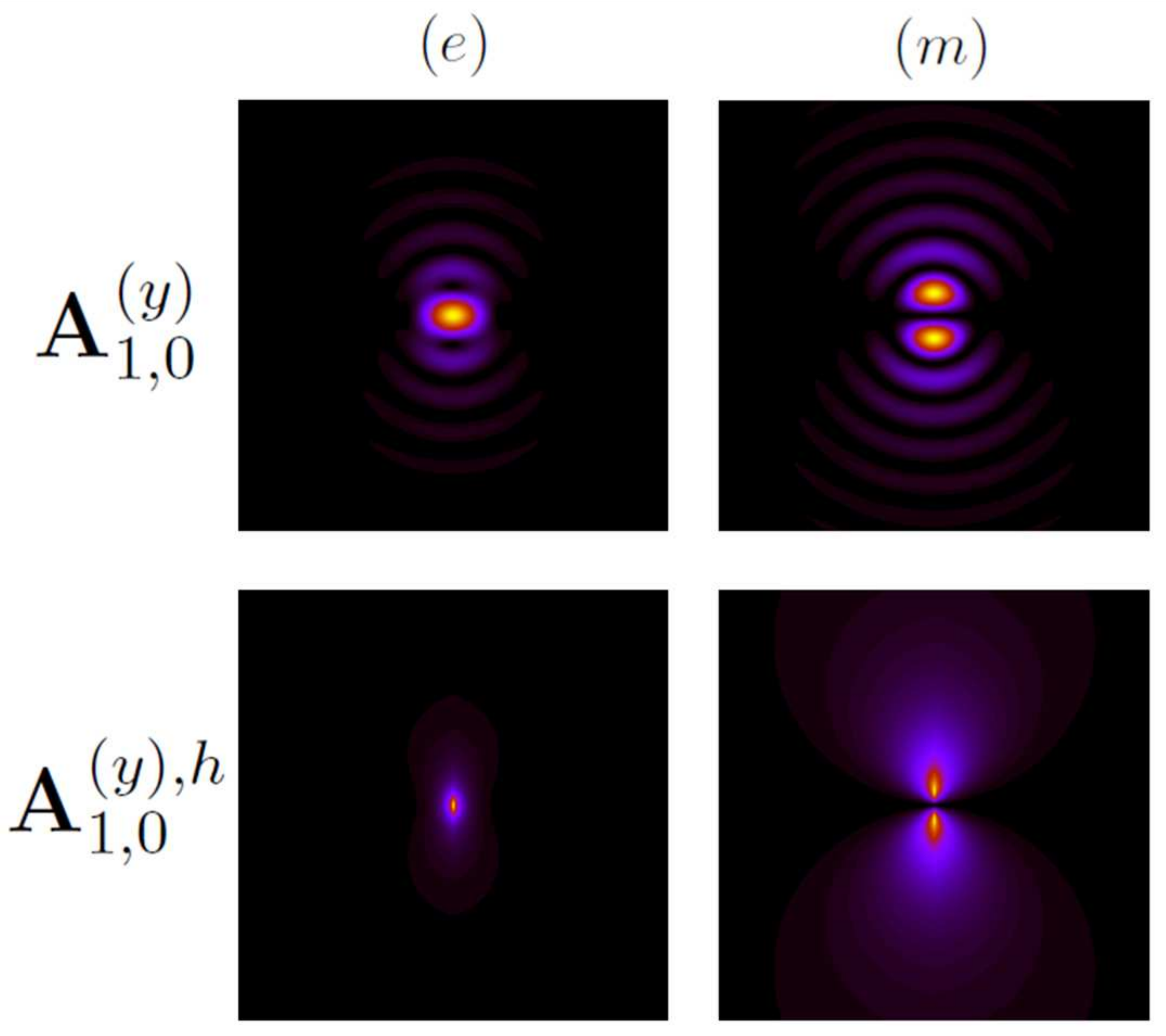} 
\caption{$\mathbf{A}_{1,0}^{(y)}$ and $\mathbf{A}_{1,0}^{(y),h}$ for $(y)=(e),(m)$. The plot area is a square of side $s=12\pi \lambda$. A contour of $0.5\pi \lambda$ around the origin has been removed to plot $\mathbf{A}_{1,0}^{(y),h}$. \label{D4}}
\end{figure}
\begin{figure}[tbp]
\centering
\includegraphics[width=10.5cm]{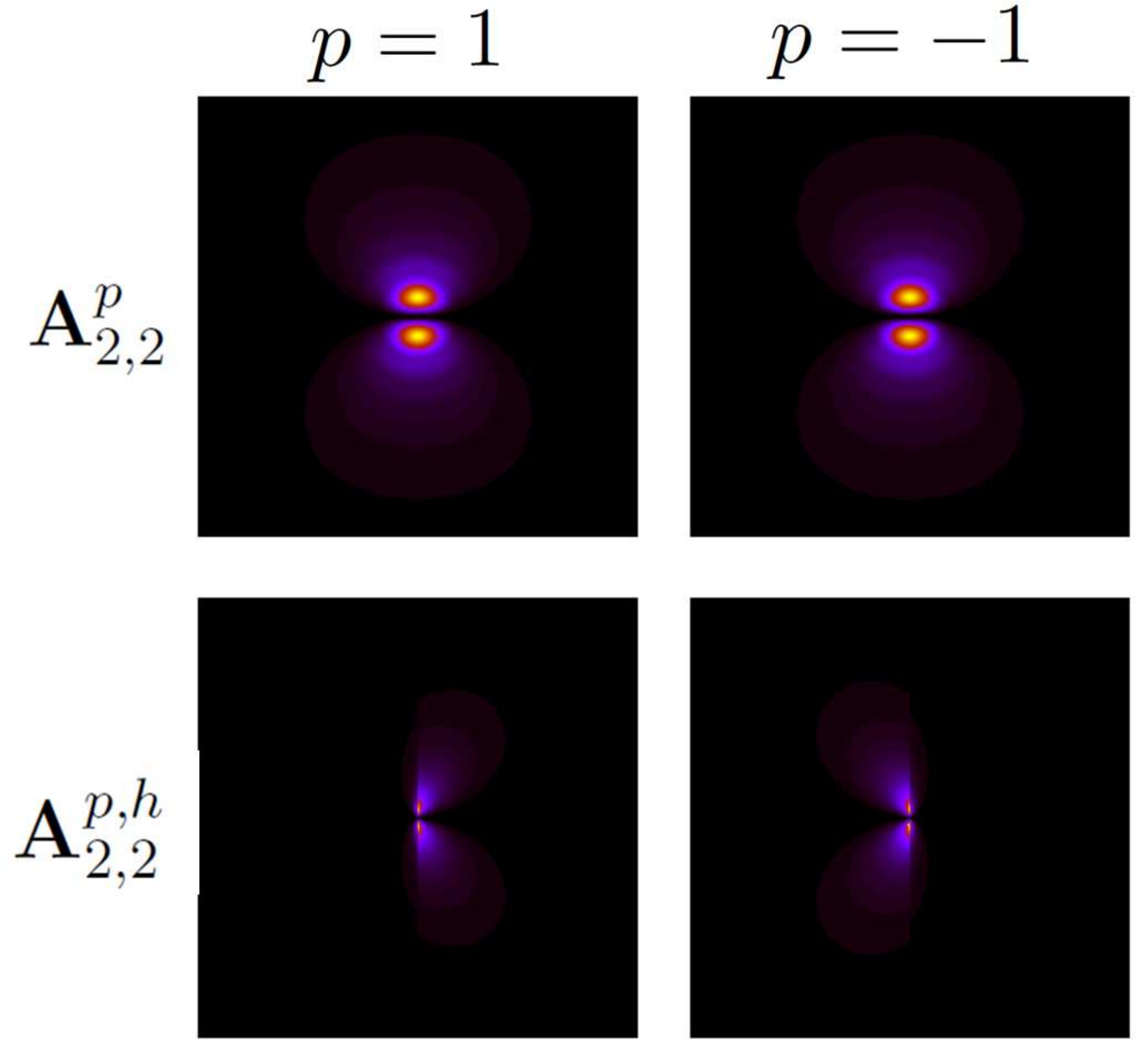} 
\caption{$\mathbf{A}_{2,2}^{p}$ and $\mathbf{A}_{2,2}^{p,h}$ for $p=\pm 1$. The plot area is a square of side $s=18\pi \lambda$. A contour of $0.5\pi \lambda$ around the origin has been removed to plot $\mathbf{A}_{2,2}^{p,h}$.\label{D5}}
\end{figure}
\begin{figure}[tbp]
\centering
\includegraphics[width=10.5cm]{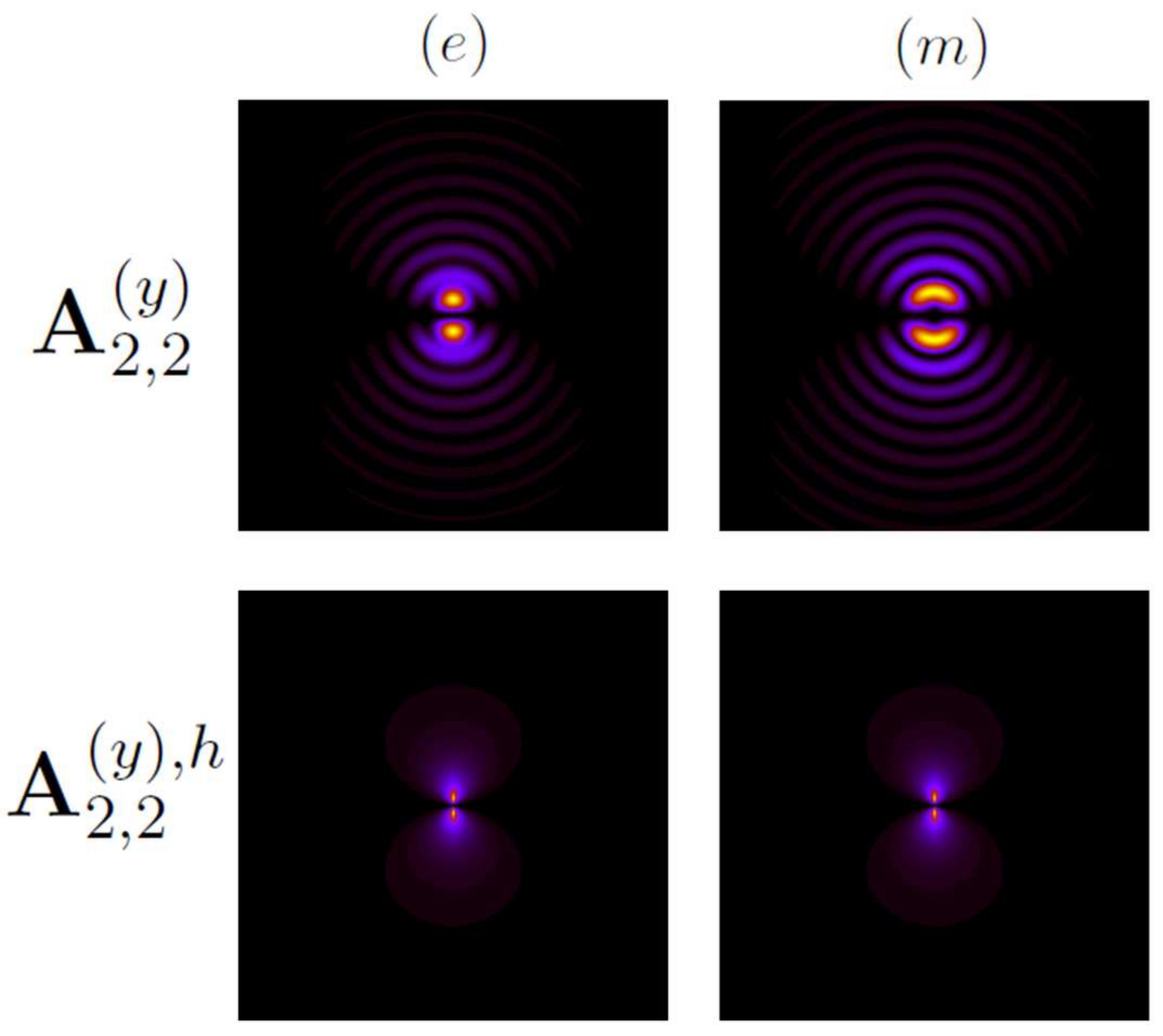} 
\caption{$\mathbf{A}_{2,2}^{(y)}$ and $\mathbf{A}_{2,2}^{(y),h}$ for $(y)=(e),(m)$.The plot area is a square of side $s=18\pi \lambda$. A contour of $0.5\pi \lambda$ around the origin has been removed to plot $\mathbf{A}_{2,2}^{(y),h}$. \label{D6}}
\end{figure}

\begin{figure}[tbp]
\centering
\includegraphics[width=10.5cm]{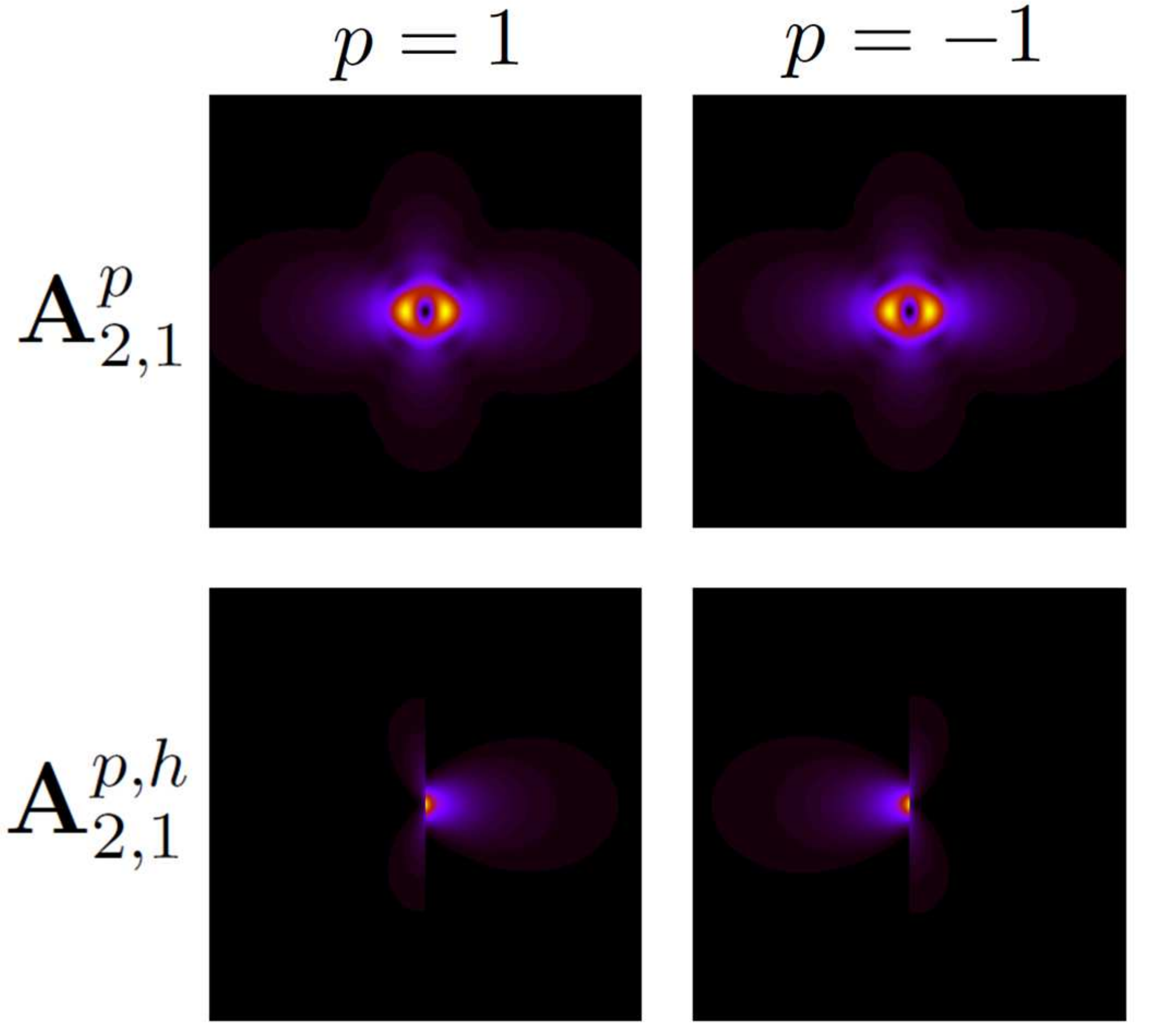} 
\caption{$\mathbf{A}_{2,1}^{p}$ and $\mathbf{A}_{2,1}^{p,h}$ for $p=\pm 1$. The plot area is a square of side $s=18\pi \lambda$. A contour of $0.8\pi \lambda$ around the origin has been removed to plot $\mathbf{A}_{2,1}^{p,h}$. \label{D7}}
\end{figure}
\begin{figure}[tbp]
\centering
\includegraphics[width=10.5cm]{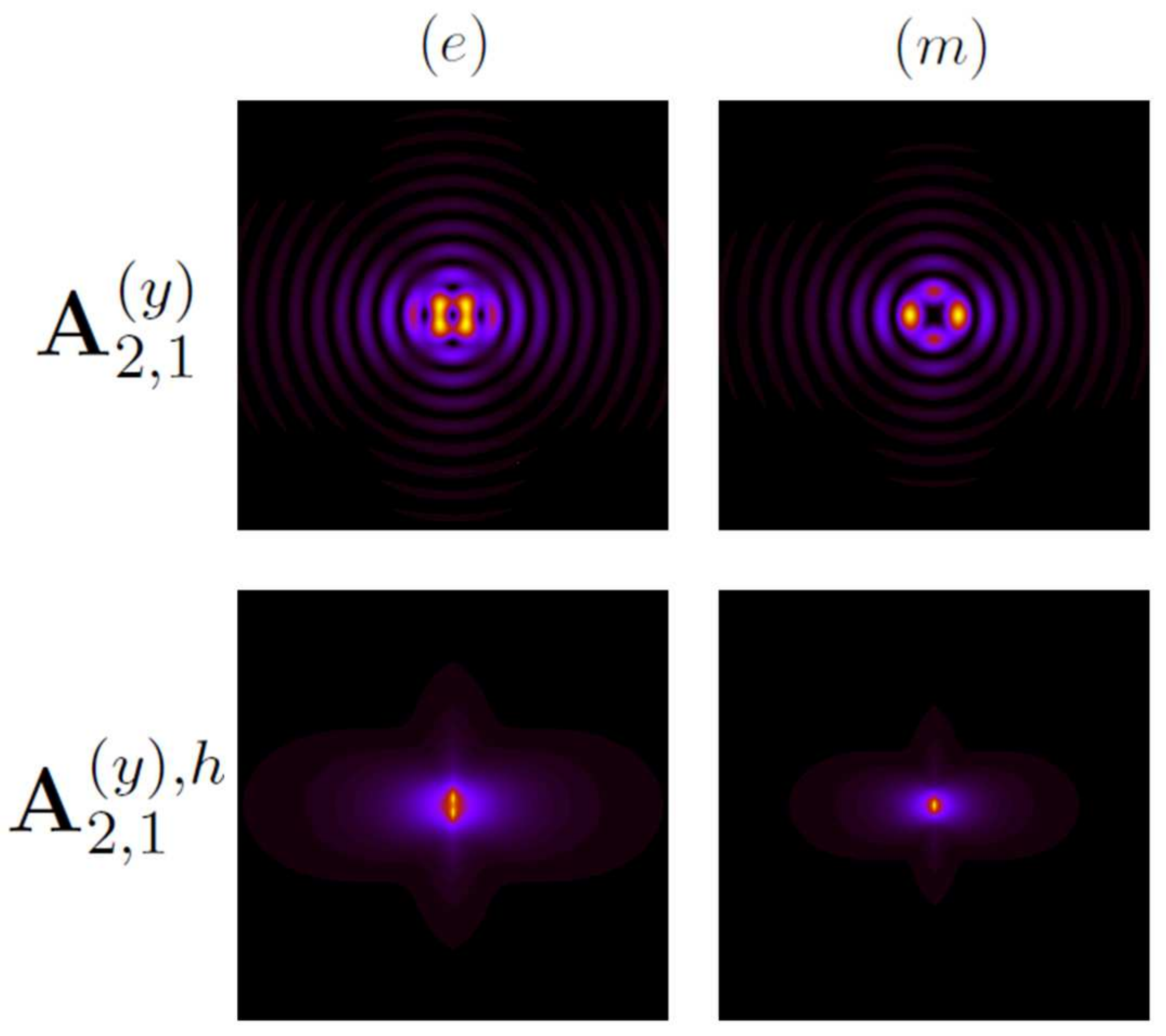} 
\caption{$\mathbf{A}_{2,1}^{(y)}$ and $\mathbf{A}_{2,1}^{(y),h}$ for $(y)=(e),(m)$. The plot area is a square of side $s=18\pi \lambda$. A contour of $0.8\pi \lambda$ around the origin has been removed to plot $\mathbf{A}_{2,1}^{(y),h}$. \label{D8}}
\end{figure}
\begin{figure}[tbp]
\centering
\includegraphics[width=10.5cm]{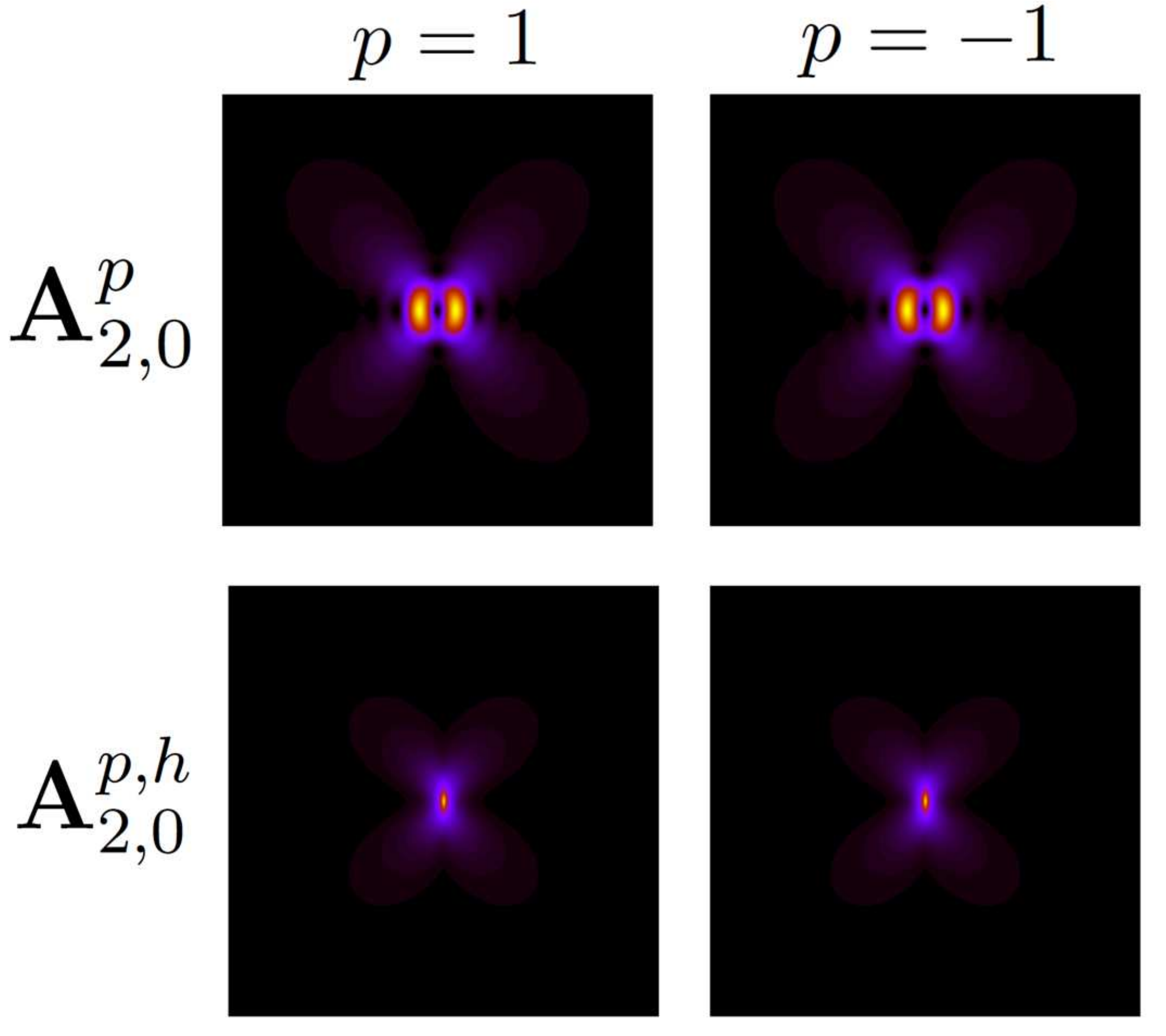} 
\caption{$\mathbf{A}_{2,0}^{p}$ and $\mathbf{A}_{2,0}^{p,h}$ for $p=\pm 1$. The plot area is a square of side $s=18\pi \lambda$. A contour of $0.8\pi \lambda$ around the origin has been removed to plot $\mathbf{A}_{2,0}^{p,h}$. \label{D9}}
\end{figure}
\begin{figure}[tbp]
\centering
\includegraphics[width=10.5cm]{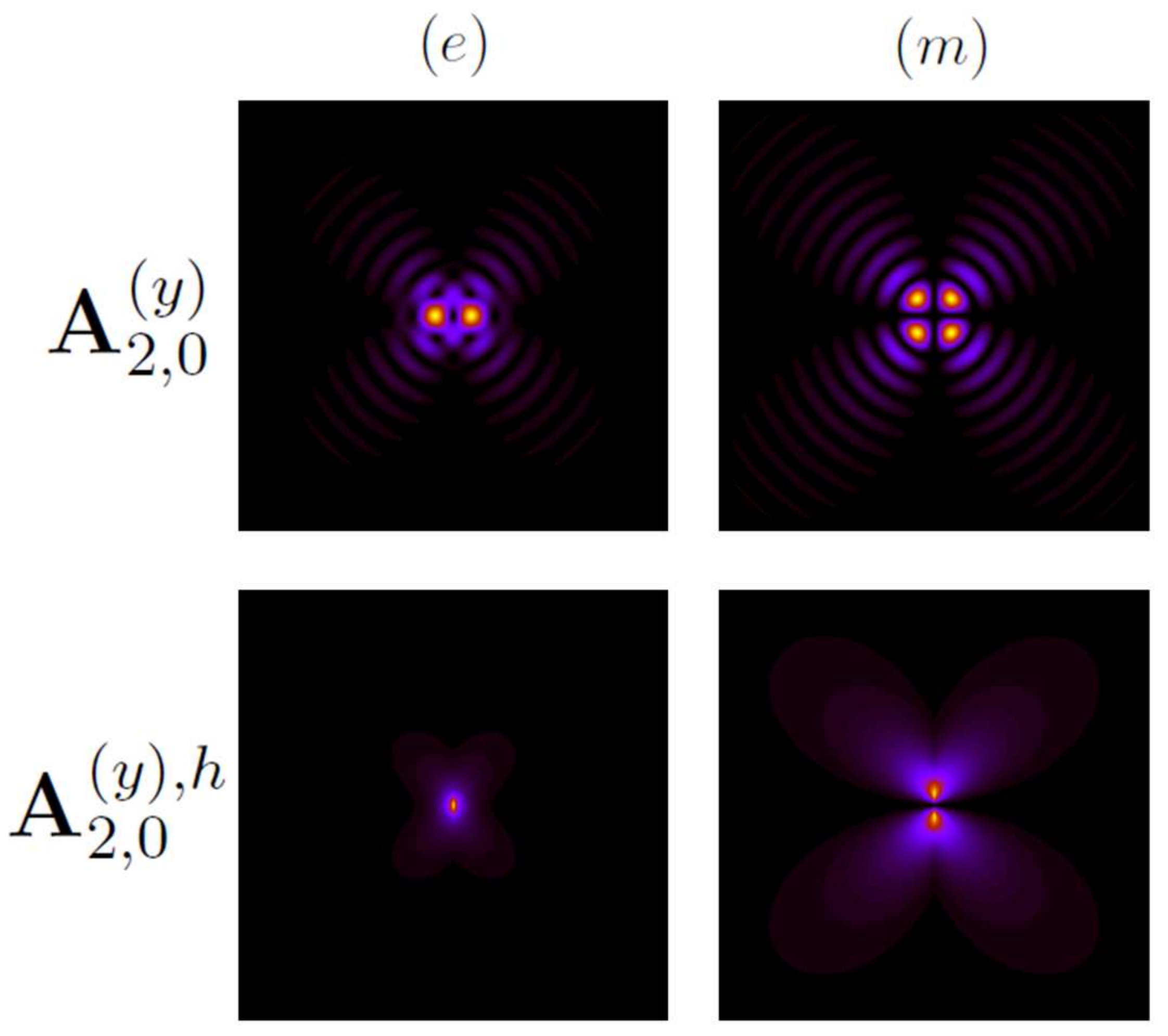} 
\caption{$\mathbf{A}_{2,0}^{(y)}$ and $\mathbf{A}_{2,0}^{(y),h}$ for $(y)=(e),(m)$. The plot area is a square of side $s=18\pi \lambda$. A contour of $0.8\pi \lambda$ around the origin has been removed to plot $\mathbf{A}_{2,0}^{(y),h}$. \label{D10}}
\end{figure}

\begin{figure}[tbp]
\centering
\includegraphics[width=10.3cm]{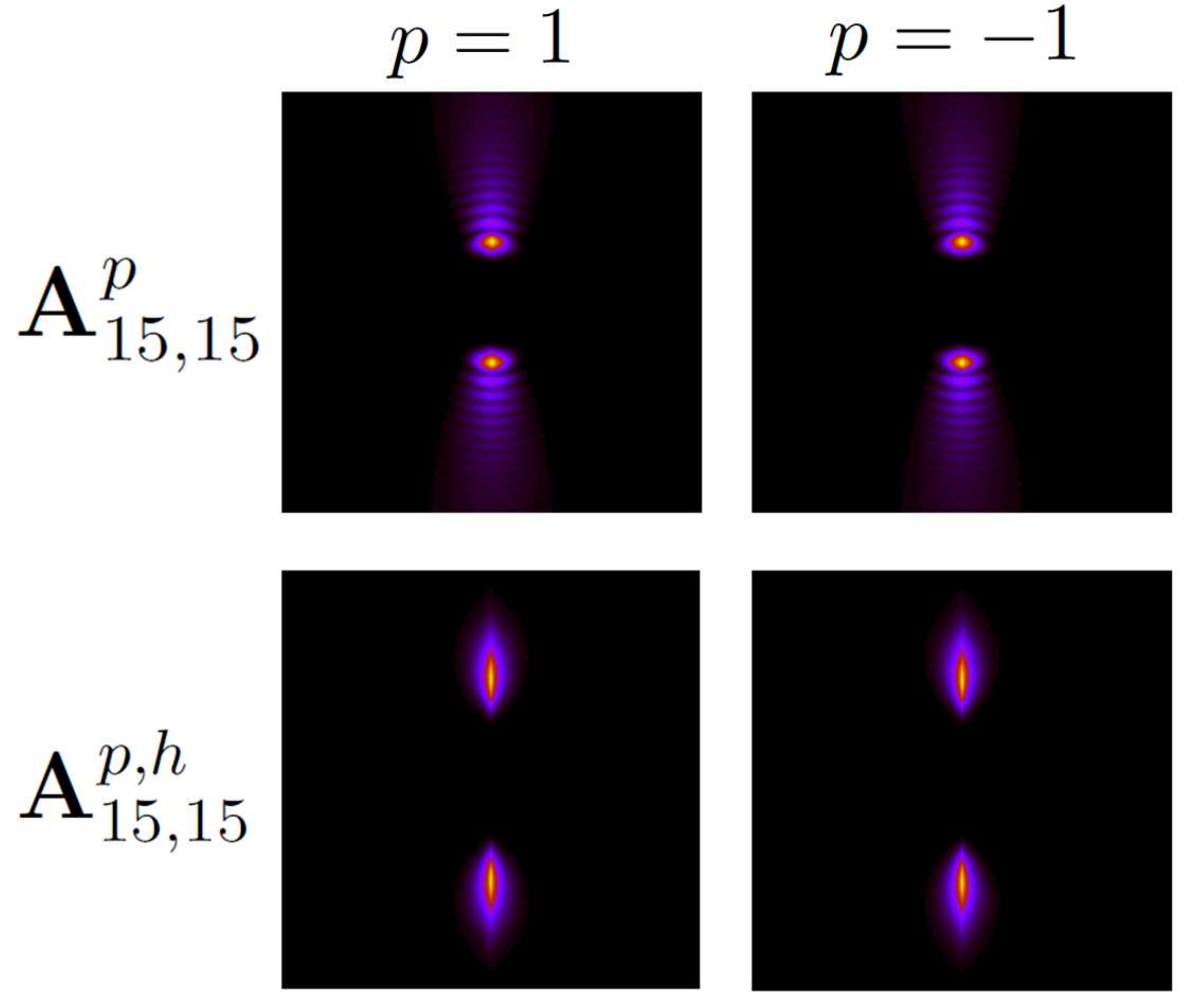} 
\caption{$\mathbf{A}_{15,15}^{p}$ and $\mathbf{A}_{15,15}^{p,h}$ for $p=\pm 1$. For $\mathbf{A}_{15,15}^{p}$, the plot area is a square of side $s=36\pi \lambda$. For $\mathbf{A}_{15,15}^{p,h}$, the plot area is a square of side $s=6\pi\lambda$, where a contour of $1.6\pi \lambda$ around the origin has been removed. \label{D11}}
\end{figure}
\begin{figure}[tbp]
\centering
\includegraphics[width=10.3cm]{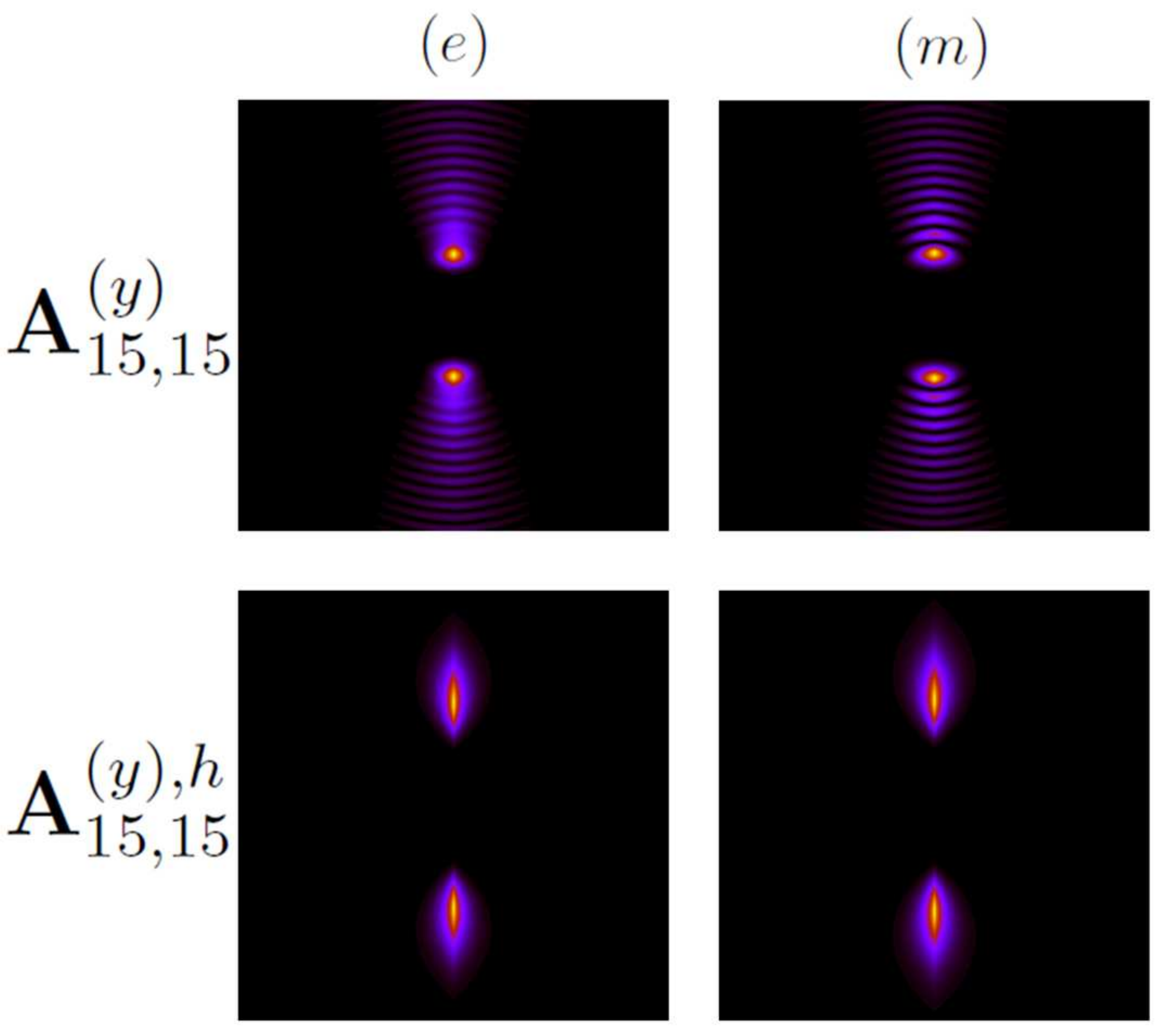} 
\caption{$\mathbf{A}_{15,15}^{(y)}$ and $\mathbf{A}_{15,15}^{(y),h}$ for $(y)=(e),(m)$. For $\mathbf{A}_{15,15}^{(y)}$, the plot area is a square of side $s=36\pi \lambda$. For $\mathbf{A}_{15,15}^{(y),h}$, the plot area is a square of side $s=6\pi\lambda$, where a contour of $0.8\pi \lambda$ around the origin has been removed. \label{D12}}
\end{figure}
\begin{figure}[tbp]
\centering
\includegraphics[width=10.3cm]{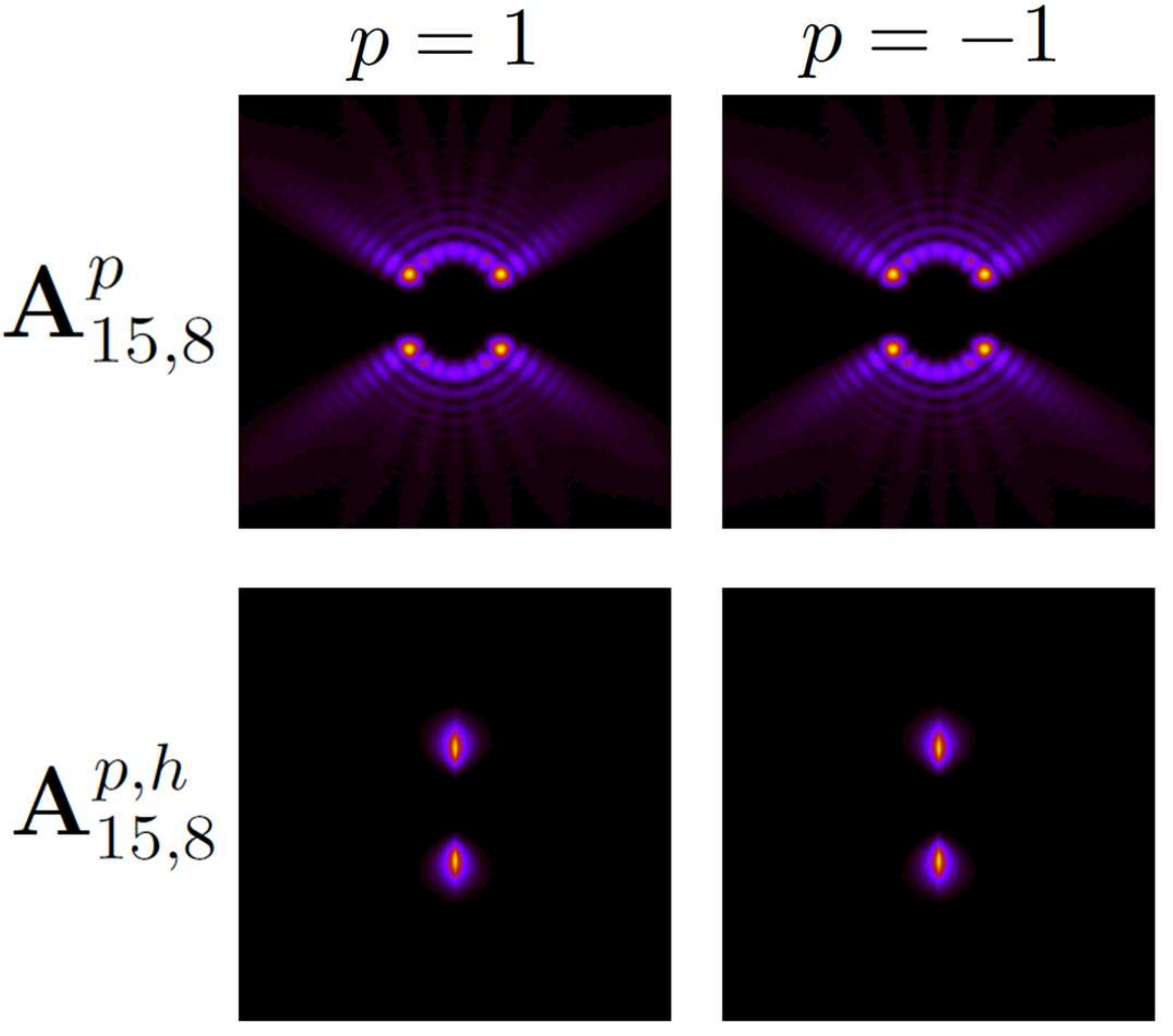} 
\caption{$\mathbf{A}_{15,8}^{p}$ and $\mathbf{A}_{15,8}^{p,h}$ for $p=\pm 1$. For $\mathbf{A}_{15,8}^{p}$, the plot area is a square of side $s=36\pi \lambda$. For $\mathbf{A}_{15,8}^{p,h}$, the plot area is a square of side $s=6\pi\lambda$, where a contour of $1.6\pi \lambda$ around the origin has been removed. \label{D13}}
\end{figure}
\begin{figure}[tbp]
\centering
\includegraphics[width=10.3cm]{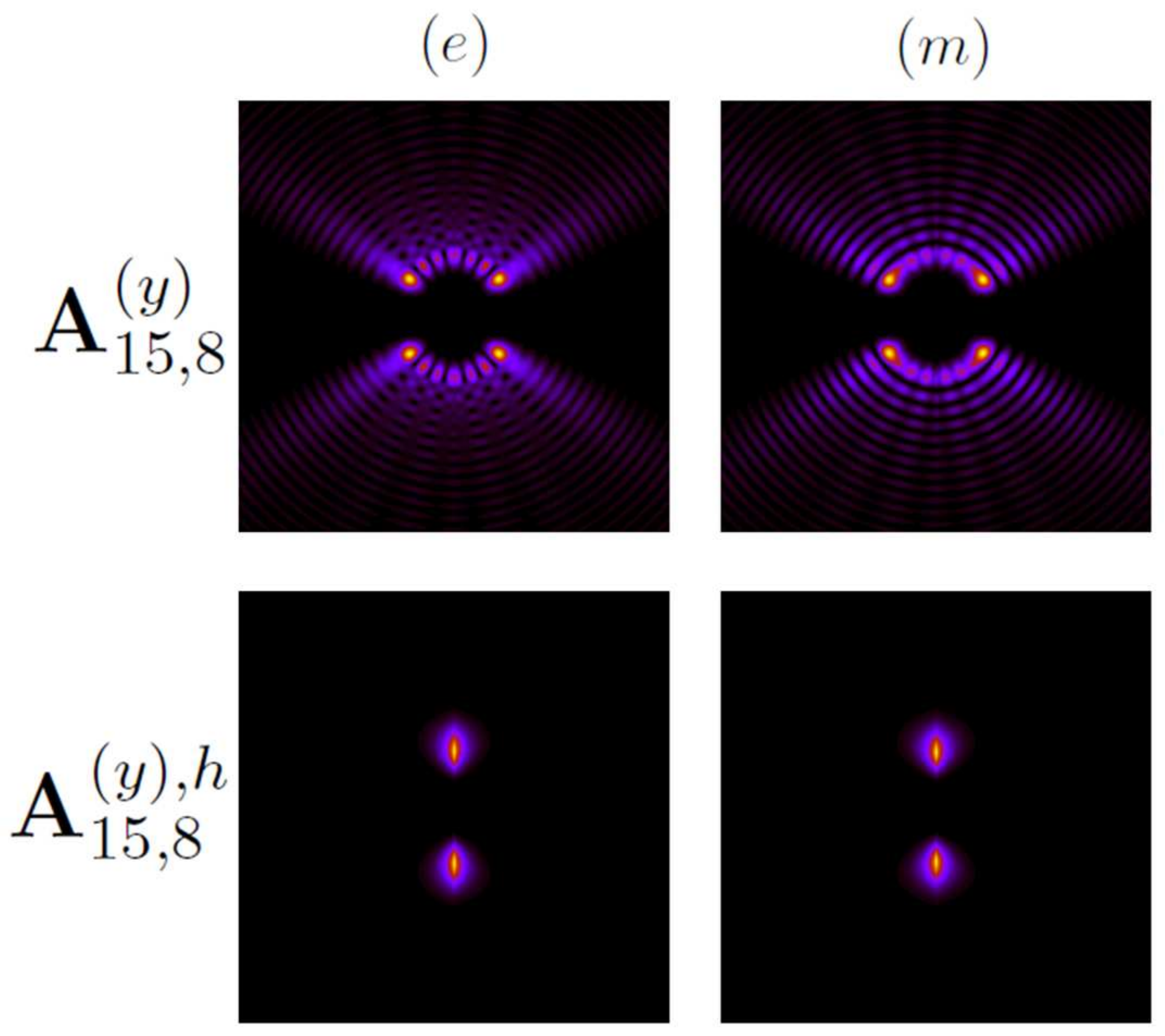} 
\caption{$\mathbf{A}_{15,8}^{(y)}$ and $\mathbf{A}_{15,8}^{(y),h}$ for $(y)=(e),(m)$. For $\mathbf{A}_{15,8}^{(y)}$, the plot area is a square of side $s=36\pi \lambda$. For $\mathbf{A}_{15,8}^{(y),h}$, the plot area is a square of side $s=6\pi\lambda$, where a contour of $0.8\pi \lambda$ around the origin has been removed. \label{D14}}
\end{figure}

\begin{figure}[tbp]
\centering
\includegraphics[width=10.3cm]{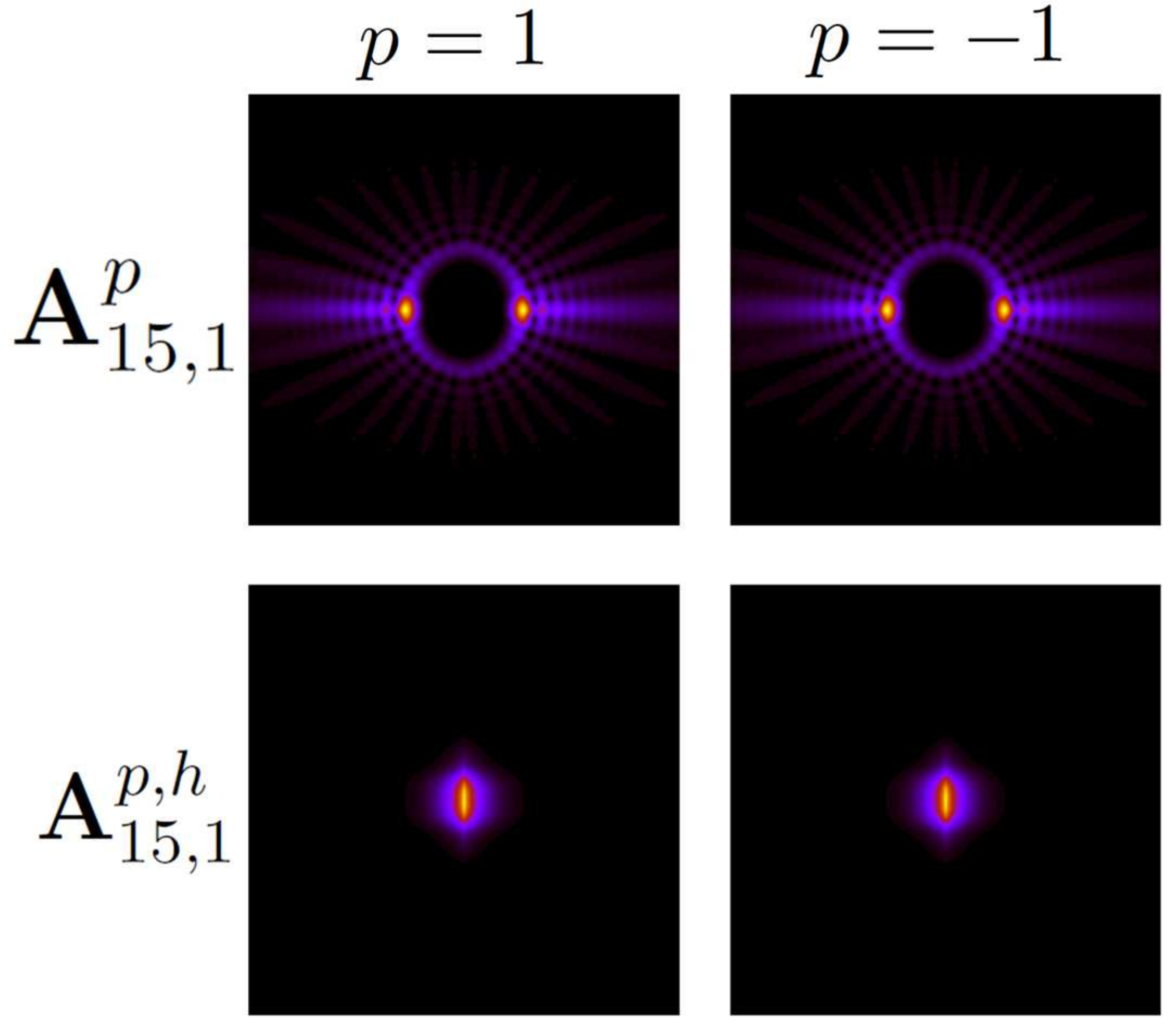} 
\caption{$\mathbf{A}_{15,1}^{p}$ and $\mathbf{A}_{15,1}^{p,h}$ for $p=\pm 1$. For $\mathbf{A}_{15,1}^{p}$, the plot area is a square of side $s=36\pi \lambda$. For $\mathbf{A}_{15,1}^{p,h}$, the plot area is a square of side $s=4.8\pi\lambda$, where a contour of $1.6\pi \lambda$ around the origin has been removed.\label{D15}}
\end{figure}
\begin{figure}[tbp]
\centering
\includegraphics[width=10.3cm]{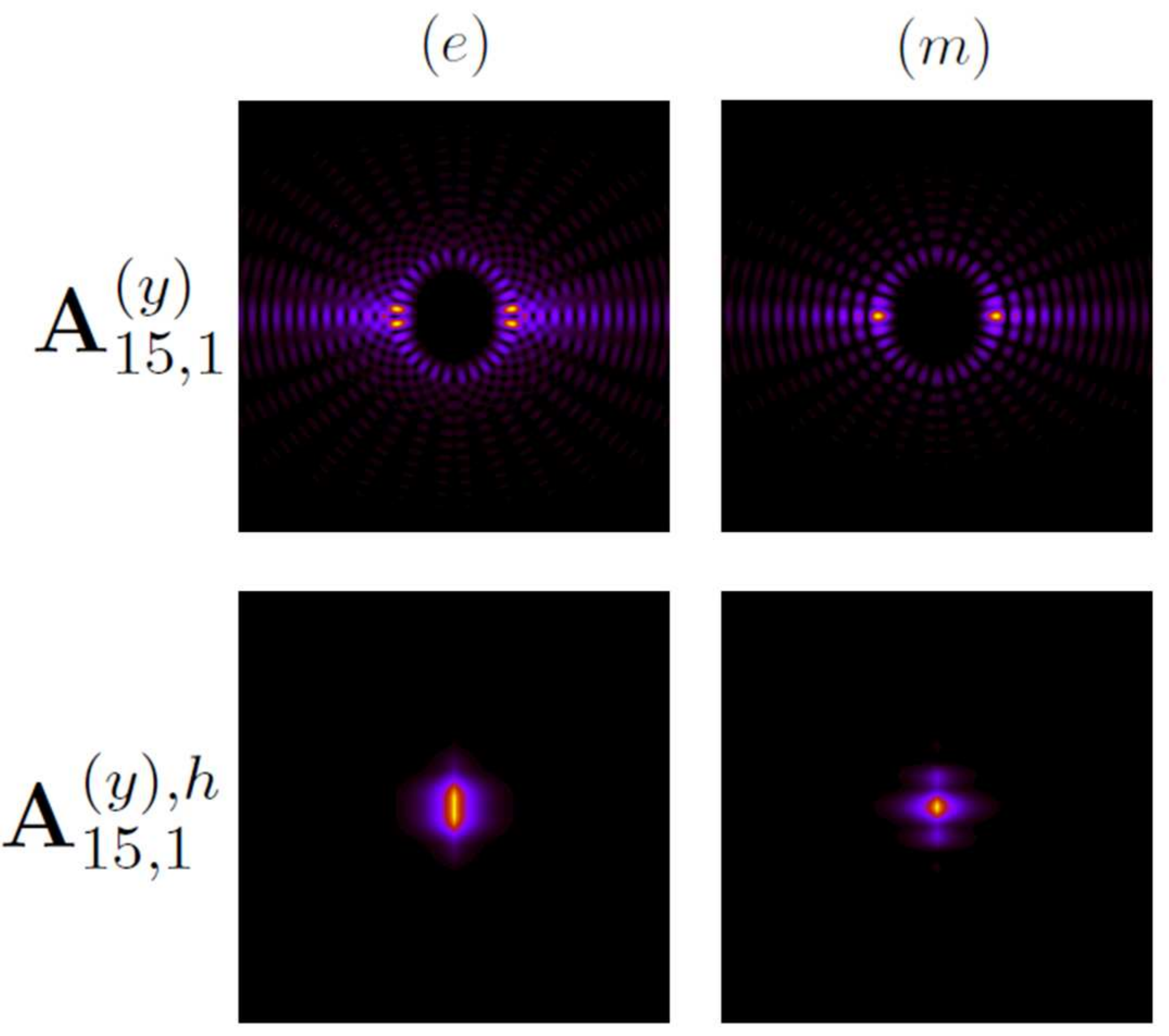} 
\caption{$\mathbf{A}_{15,1}^{(y)}$ and $\mathbf{A}_{15,1}^{(y),h}$ for $(y)=(e),(m)$. For $\mathbf{A}_{15,1}^{(y)}$, the plot area is a square of side $s=36\pi \lambda$. For $\mathbf{A}_{15,1}^{(y),h}$, the plot area is a square of side $s=4.8\pi\lambda$, where a contour of $0.8\pi \lambda$ around the origin has been removed. \label{D16}}
\end{figure}

\backmatter

\chapter{List of Symbols}
\underline{Fields}
\begin{list}{}{%
\setlength{\labelwidth}{30mm}
\setlength{\leftmargin}{32mm}}
\item[$ \psi(\mathbf{x},t)  $\dotfill] general EM field
\item[$\mathcal{D}(\mathbf{r},t)$\dotfill] electric displacement 
\item[$\mathcal{E}(\mathbf{r},t)$\dotfill] electric field 
\item[$\mathcal{B}(\mathbf{r},t)$\dotfill] magnetic induction
\item[$\mathcal{H}(\mathbf{r},t)$\dotfill] magnetic field 
\item[$\mathcal{P}(\mathbf{r},t)$\dotfill] polarization density
\item[$\mathcal{M}(\mathbf{r},t)$\dotfill] magnetization density
\item[$\mathbf{D}(\mathbf{r})$\dotfill] complex-amplitude electric displacement vector
\item[$\mathbf{E}(\mathbf{r})$\dotfill] complex-amplitude electric field vector
\item[$\mathbf{B}(\mathbf{r})$\dotfill] complex-amplitude magnetic induction vector
\item[$\mathbf{H}(\mathbf{r})$\dotfill] complex-amplitude magnetic field vector
\item[$\mathbf{E}^{\mathrm{foc}}$\dotfill] electric field at the focus of a lens
\item[$\Einf$\dotfill] angular spectrum of the electric field
\item[$\Ein$\dotfill] incoming electric field
\item[$(\Es, \Hs)$\dotfill] scattered EM field
\item[$(\Ei, \Hi)$\dotfill] incident EM field
\item[$(\Eint, \Hint)$\dotfill] interior EM field
\item[$(\Et, \Ht)$\dotfill] total exterior EM field
\item[$\Epq$\dotfill] electric field of an incident vortex beam with topological charge $l$ and helicity $p$
\item[$\Etpq$\dotfill] transmitted electric field when the incident beam is $\Epq$
\item[$\Es_p$\dotfill] direct helicity component of the scattered electric field
\item[$\Es_{-p}$\dotfill] crossed helicity component of the scattered electric field
\item[$\Atpqp$\dotfill] complex amplitude of the direct component of the transmitted electric field when the incident beam is $\Epq$
\item[$\Btpqpp$\dotfill] complex amplitude of the crossed component of the transmitted electric field when the incident beam is $\Epq$
\end{list}
\underline{Mathematical functions}
\begin{list}{}{%
\setlength{\labelwidth}{30mm}
\setlength{\leftmargin}{32mm}}
\item[$\delta(\mathbf{k}-\mathbf{k'}) $\dotfill] Dirac delta
\item[$\delta_{pp'}$\dotfill] Kronecker delta
\item[$J_n(\krho \rho)$\dotfill] Bessel function of the first kind and order $n$
\item[$\Djmp$\dotfill] Wigner rotation matrix
\item[$j_j(kr)$\dotfill] spherical Bessel function of the first kind and order $j$
\item[$P_j^{m_z}(\eta)$\dotfill] generalized Legendre polynomial, $\eta=\cos \theta$ 
\item[$h_j(kr)$\dotfill] spherical Hankel function of the first kind and order $j$
\item[$H(\thetak - \thetak^{M})$\dotfill] Heaviside step function
\item[$d_{m_z^*p}^j(\thetak)$\dotfill] reduced rotation matrix
\item[$\Gamma(z)$\dotfill] Gamma function
\item[$_2F_1(a,b,c:z)$\dotfill] hypergeometric function
\end{list}
\underline{Modes}
\begin{list}{}{%
\setlength{\labelwidth}{30mm}
\setlength{\leftmargin}{32mm}}
\item[$(\phat,\shat) e^{i \mathbf{k}\cdot \mathbf{r}}$ \dotfill] (p,s) plane-waves
\item[$(\epphat,\epmhat) e^{i \mathbf{k}\cdot \mathbf{r}}$\dotfill] plane wave with well-defined helicity $(+1,-1)$
\item[$\mathbf{A}_{jm_z}^{(m),h}, \mathbf{A}_{jm_z}^{(e),h}$\dotfill] Hankel multipolar fields (with well-defined parity)
\item[$\mathbf{A}_{jm_z}^{+,h}, \mathbf{A}_{jm_z}^{-,h}$\dotfill] Hankel multipolar fields with well-defined helicity
\item[$\Bmp$\dotfill] Bessel beam with helicity $p = \pm 1$
\item[$\Am, \Ae$\dotfill] multipolar fields (with well-defined parity)
\item[$\Ajmp, \Ajmm$\dotfill] multipolar fields with well-defined helicity
\item[$\Bx$\dotfill] Bessel beam with $(y)= \text{TE/TM}$ mirror symmetry
\item[$\mathbf{A}_{jj}^{(y)}$\dotfill] Whispering Gallery Mode
\end{list}
\underline{Operators}
\begin{list}{}{%
\setlength{\labelwidth}{30mm}
\setlength{\leftmargin}{32mm}}
\item[$H$\dotfill] hamiltonian
\item[$\TDt$\dotfill] temporal translation (or evolution) of magnitude $\Delta t$
\item[$\mathbf{P}$\dotfill] linear momentum vector operator
\item[$P_x,P_y,P_z$\dotfill] cartesian linear momentum operators
\item[$T_{\Delta \mathbf{r}}$\dotfill] linear translation of magnitude $\Delta \mathbf{r}$ 
\item[$\mathbf{J}$\dotfill] angular momentum vector operator
\item[$\mathbf{S}$\dotfill] spin angular momentum vector operator
\item[$\mathbf{L}$\dotfill] orbital angular momentum vector operator
\item[$J_x,J_y,J_z$\dotfill] cartesian angular momentum operators
\item[$R_{\nhat}(\varphi)$\dotfill] rotation operator of an angle $\varphi$ with respect to an axis given by $\nhat$
\item[$\mathbf{M}_{\nhat}(\varphi)$\dotfill] rotation matrix of an angle $\varphi$ with respect to an axis given by $\nhat$
\item[$\mathbf{R} (\phik, \thetak)$\dotfill] composition of three rotation operators. \\
$\mathbf{R} (\phik, \thetak)=R_z(\phik)R_y(\thetak)R_z(0)$
\item[$J^2$\dotfill] angular momentum squared operator
\item[$\Lambda$\dotfill] helicity operator
\item[$D_{\varphi}$\dotfill] generalized duality transformation of an angle $\varphi$
\item[$\Pi$\dotfill] parity operator
\item[$M_{\nhat}$\dotfill] Mirror symmetry operator with respect to a plane with normal vector $\nhat$
\item[$M_{\left\lbrace \nhat \right\rbrace }$\dotfill] Mirror symmetry operator with respect to a plane containing the vector $\nhat$
\end{list}
\underline{Coordinates and vectors}
\begin{list}{}{%
\setlength{\labelwidth}{30mm}
\setlength{\leftmargin}{32mm}}
\item[$t$\dotfill] time coordinate
\item[$\mathbf{r}=(x,y,z)$\dotfill] position vector in cartesian coordinates
\item[$\mathbf{r}=(\rho,\phi,z)$\dotfill] position vector in cylindrical coordinates
\item[$\mathbf{r}=(r,\phi,\theta)$\dotfill] position vector in spherical coordinates
\item[$\xhat,\yhat,\zhat$\dotfill] cartesian unitary vectors
\item[$\mathbf{\hat{\rho}} , \mathbf{\hat{\phi}} , \zhat $\dotfill] cylindrical unitary vectors
\item[$\rhat, \mathbf{\hat{\phi}}, \mathbf{\hat{\theta}} $\dotfill] spherical unitary vectors
\item[$\mathbf{k}=(k_x,k_y,k_z)$\dotfill] wave-vector in cartesian coordinates
\item[$\mathbf{k}=(\krho,\phik,k_z)$\dotfill] wave-vector in cylindrical coordinates
\item[$\mathbf{k}=(k,\phik,\thetak)$\dotfill] wave-vector in spherical coordinates
\item[$\nhat$\dotfill] unitary vector normal to a surface
\item[$\sphat,\smhat$\dotfill] left and right circular polarisation unitary vectors
\end{list}
\underline{Beam and material parameters}
\begin{list}{}{%
\setlength{\labelwidth}{30mm}
\setlength{\leftmargin}{32mm}}
\item[$\lambda$\dotfill] wavelength
\item[$\omega$\dotfill] frequency
\item[$c$\dotfill] speed of light
\item[$k$\dotfill] wave-number
\item[LG$_{l,q}$\dotfill] Laguerre-Gaussian beam 
\item[$q \geq 0$\dotfill] radial index
\item[$l $\dotfill] topological charge (azimuthal index)
\item[$w(z)$\dotfill] beam waist at a plane of $z$ constant
\item[$w_0$\dotfill] beam waist at $z=0$
\item[$p$\dotfill] helicity value
\item[$\epsilon_0, \mu_0$\dotfill] electric permittivity and magnetic permeability of vacuum   
\item[$\epsilon,\mu,n$\dotfill] electric electric permittivity, magnetic permeability, and refractive index of a linear, non-dispersive, homogeneous and isotropic medium
\item[$\sigma$\dotfill] charge density on a surface
\item[$\mathbf{j}_s$\dotfill] surface current density
\item[$\Sconv(\mathbf{g},\mathbf{m})$\dotfill] scattering matrix as a function of the geometrical and material properties
\item[$f$\dotfill] focal distance of a lens
\item[$t^s,t^p$\dotfill] Fresnel coefficients of the lens for $\shat$ and $\phat$ waves
\end{list}
\underline{GLMT}
\begin{list}{}{%
\setlength{\labelwidth}{30mm}
\setlength{\leftmargin}{32mm}}
\item[$\epsilon_1,\mu_1,n_1$\dotfill] electric electric permittivity, magnetic permeability, and refractive index of sphere
\item[$n_r=n_1/n$\dotfill] relative refractive index
\item[$R$\dotfill] radius of sphere
\item[$x=\dfrac{2\pi R}{\lambda}$\dotfill] size parameter
\item[$ a_j,b_j $\dotfill] scattering Mie coefficients of order $j$
\item[$c_j,d_j$\dotfill] interior Mie coefficients of order $j$
\item[$ \gjm, \gje $\dotfill] beam shape coefficients
\item[$\langle \St \rangle$\dotfill] time-averaged total exterior Poynting vector
\item[$ C_i $\dotfill] interior cross section
\item[$ C_s $\dotfill] scattering cross section
\item[$ C_{ext} $\dotfill] extinction cross section
\item[$ Q_i $\dotfill] interior efficiency factor
\item[$ Q_s $\dotfill] scattering efficiency factor
\item[$Q_{ext}$ \dotfill] extinction efficiency factor
\item[$\Cjmp$\dotfill] beam shape coefficients for beams with a well-defined helicity
\item[$\xi$\dotfill] root number for a Mie coefficient
\item[$Q$\dotfill] Q factor
\item[$\mathtt{w}$\dotfill] energy density
\item[$\mathtt{W}$\dotfill] energy of the EM per unit length
\item[$\mathtt{W}^\mathrm{s}$\dotfill] energy scattered
\item[$\mathtt{W}_p^\mathrm{s}$\dotfill] energy scattered in modes with the same helicity as the incident beam
\item[$\mathtt{W}_{-p}^\mathrm{s}$\dotfill] energy scattered in modes with opposite helicity to the incident field
\item[$T_{p}(R,n_r)$\dotfill] transfer function
\item[$\Tmp$\dotfill] transfer function for a cylindrically symmetric beam with $J_z=\ms$
\end{list}
\underline{Holography}
\begin{list}{}{%
\setlength{\labelwidth}{30mm}
\setlength{\leftmargin}{32mm}}
\item[$\Uu$\dotfill] complex wave-function
\item[$\U$\dotfill] complex amplitude
\item[$\I$\dotfill] intensity
\item[$t(x,y)$\dotfill] complex amplitude transmittance 
\item[$d(x,y)$\dotfill] thickness of an optical element
\item[$U_o$\dotfill] object wave
\item[$U_r$\dotfill] reference wave
\item[$\Delta x$\dotfill] width of a $2\pi$ phase ramp
\item[$\Delta x^{\lambda}$\dotfill] width of a $2\pi$ phase ramp for a given $\lambda$
\item[$\Delta \theta_x$\dotfill] diffraction angle given by a hologram
\end{list}
\underline{Experimental parameters}
\begin{list}{}{%
\setlength{\labelwidth}{30mm}
\setlength{\leftmargin}{32mm}}
\item[$\Phi$\dotfill] diameter
\item[$\Tt$\dotfill] target
\item[$\Top$\dotfill] target integro-differential operator
\item[$\text{CD}_l(\%)$\dotfill] CD given an incident vortex beam of order $l$
\item[$t_r$\dotfill] Extinction ratio of a linear polariser
\item[$\gamma$\dotfill] ratio of light scattered into crossed helicity component divided over the direct one
\item[$I_l^R$\dotfill] transmitted intensity for a RCP incident vortex beam of topological charge $l$
\item[$I_l^L$\dotfill] transmitted intensity for a LCP incident vortex beam of topological charge $l$
\item[$I^i$\dotfill] measured intensity of the incident EM field
\item[$I^s_{-p}$\dotfill] measured intensity of the crossed helicity component of the scattered EM field
\item[$I^s_{p}$\dotfill] measured intensity of the direct helicity component of the scattered EM field
\item[$\Ino$\dotfill] normalised (by the background) measured intensity 
\item[$\sigma_N(\%)$\dotfill] standard deviation with respect to the average 
\item[$I_{m,p}^{\text{norm}}(\lambda)$\dotfill] normalised (by the background) measured intensity at $\lambda$ given a monochromatic incident beam with $J_z=m$, $\Lambda=p$
\end{list}


\end{document}